\documentclass{osuthesis}

\providecommand{\UseBookClass}{
   \newcommand{\authordegrees}[1]{}
   \newcommand{\unit}[1]{}
   \newcommand{\advisername}[1]{}
   \newcommand{\coadvisername}[1]{}
   \newcommand{\member}[1]{}
   \newcommand{\skipcopyright}{\pagestyle{plain}}
   \newcommand{\makecopyright}{~\vfill\begin{center}%
                              Copyright \number\year
                              \end{center}\vfill}
   \newcommand{\dedication}[1]{\clearpage \vspace*{\fill}%
                               \begin{center} ##1 \end{center}%
                               \vspace*{\fill}}
   \newcommand{\dateitem}[2]{\noindent ##1 \dotfill \parbox[t]{2.75in}{##2}\par~\par}
   \pagestyle{empty}
   \newcommand{\startsinglespace}[1][\normalsize]{\renewcommand{\baselinestretch}{1}\small\large##1}
   \newcommand{\startdoublespace}[1][\normalsize]{\renewcommand{\baselinestretch}{2}\small\large##1}
}
\UseBookClass

\usepackage{graphicx}

\newcommand{\me}{\mathrm{e}}
\newcommand{\mi}{\mathrm{i}}

\begin{document}

% First, declare the parts of your title page 
\title{QUANTUM MECHANICAL THREE-BODY PROBLEM WITH SHORT-RANGE INTERACTIONS}
\author{Richard Frank Mohr, Jr.}
\authordegrees{B.S., M.S.}  % Degrees thus far, not including this one.
\unit{Department of Physics}
\advisername{Dr.\ Robert Perry}
\member{Dr.\ Richard Furnstahl}
\member{Dr.\ Eric Braaten}
\member{Dr.\ Thomas Humanic}

% Then create the title page
\maketitle

% Uncomment the following lines to produce an
% external abstract page.
%\begin{externalabstract}
%  \input{abstract}
%\end{externalabstract}

% Begin front matter
\frontmatter

% The following creates a page used to copyright your dissertation.
% This is optional, but if it is not used, then a blank page must
% be inserted.  To do this, comment out the first line, and uncomment
% the second one.
%\makecopyright
%\skipcopyright

% Abstract goes here.  The text is in the file "abstract.tex".
\chapter{Abstract}
%  The dissertation abstract can only be 350 words.

We have investigated S-wave bound states composed of three identical bosons interacting via regulated delta function potentials in non-relativistic quantum mechanics.  For low-energy systems, these short-range potentials serve as an approximation to the underlying physics, leading to an effective field theory.

A method for perturbatively expanding the three-body bound-state equation in inverse powers of the cutoff is developed.  This allows us to extract some analytical results concerning the behavior of the system.  Further results are obtained by solving the leading order equations numerically to 11 or 12 digits of accuracy.  The limit-cycle behavior of the required three-body contact interaction is computed, and the cutoff-independence of bound-state energies is shown.  By studying the relationship between the two- and three-body binding energies, we obtain a high accuracy numerical calculation  of Efimov's universal function.

Equations for the first order corrections, necessary for the study of cutoff dependence, are derived.  However, a numerical solution of these equations is not attempted.

% Dedication (this should be kept small).
\dedication{To my wife, my parents, and all my family and friends.}

% Bring in Acknowledgement and Vita from separate files named "ack.tex"
% and "vita.tex".
%
% The acknowledgment section
%
\chapter{Acknowledgments}

I wish to thank Robert Perry for all of the advice and guidance he has given to me, and Ken Wilson for laying the groundwork for what was to become my thesis topic.

I would like to thank Hans-Werner Hammer for helping me make sense of the physics involved in my research, and for providing references to current work related to my thesis.

Thanks also goes to  Eric Braaten and Dick Furnstahl not only for serving on my committee, but also for many useful discussions that helped to clear up some of my misunderstandings.

Finally, I wish to thank all of my family and friends.  I am grateful to my wife, Laura, for her constant support and belief in me; to my parents for instilling the work ethic, self-confidence, and love that I carry with me to this day;  to all the friends who have been there for me in good times and bad, and who helped to keep things in perspective when I could not.  To all of you, I am eternally grateful.

This work was supported in part by the National Science Foundation under grants PHY-9800964 and PHY-0098645.
\chapter{Vita}

\startsinglespace

\dateitem{September 13, 1974}{Born - Mariah Hill, Indiana}
\dateitem{1996}{B.S. Physics and Mathematics, Rose-Hulman Institute of Technology, Terre Haute, Indiana}
\dateitem{1996-1997}{University Fellow, The Ohio State University, Columbus, Ohio}
\dateitem{1996-1999}{Fowler Fellow, The Ohio State University, Columbus, Ohio}
\dateitem{1997-1998}{Teaching Assistant, The Ohio State University, Columbus, Ohio}
\dateitem{1999}{M.S. Physics, The Ohio State University, Columbus, Ohio}
\dateitem{2000-present}{Research Assistant, The Ohio State University, Columbus, Ohio}

\begin{center} PUBLICATIONS \end{center}

\noindent
Richard Mohr,
\newblock ``Nearness of Normals''.
\newblock {\em Pi Mu Epsilon Journal}, 10(4):257--264, 1996.
\par~~\par

\noindent
Richard Mohr,
\newblock ``CWATSETS: Weights, Cardinalities, and Generalizations''.
\newblock {\em Rose-Hulman Institute of Technology Technical Report Series}, MS TR 96-03, 1996.
\par~~\par

\noindent
Stephen S. Pinsky and Richard Mohr,
\newblock ``The Condensate for SU(2) Gauge Theory in 1+1 Dimensions Coupled to Massless Adjoint Fermions''.
\newblock {\em International Journal of Modern Physics A}, 12(6):1063--1073, 1997.
\par~~\par

\pagebreak
\begin{center} FIELDS OF STUDY \end{center}

\noindent Major Field:  Physics \par~\par

%% NOTE: If there is only one field of study, then the next
%% section might be better written simply as
%%
\hspace*{2em} Studies in three-body quantum mechanics: Professor Robert J.~Perry

% Make the Table of Contents, List of Figures and List of Tables
\tableofcontents
\listoffigures
\listoftables

% Start main text
\mainmatter

%
% The following is a list of chapters.  Each is brought in from a
% separate file using the \include{} command.  Notice that you do
% not need to include the ".tex" extension.
%

\chapter{Introduction}
\label{ch:intro}

Approximations abound in physics.  Classical mechanics is merely an approximation that works well at large distances and small speeds.  If distances become too short, we must revert to quantum mechanics.  If the speeds become too fast, we enter the realm of relativity.  If the distances are short and the speeds are fast, then the approximations supplied by quantum mechanics and relativity are no longer valid and we must combine the two to get quantum field theory.  Even quantum field theory, which is the basis for the Standard Model and nearly all of particle physics theory, is likely only an approximation. It may break down at even higher energies and need to be replaced by something else like string theory.

Just because an approximation is not correct {\em everywhere} does not mean it is useless and should be discarded.  In fact, the opposite is true.  Since it is correct {\em somewhere}, it should be embraced.  Approximations provide a way of isolating and controlling our ignorance. Identify only the necessary details and throw the garbage away!  The hard part, of course, is identifying what is necessary so we do not throw away the baby with the bath water.

The heart of this thesis is approximations.  We analyze the low-energy quantum mechanical three-body problem using delta function potentials.  For sufficiently low energies, the particles cannot discern the details of the interaction.  Therefore, any suitably-adjusted, short-range potential should provide an accurate approximation to the true underlying potential.  All knowledge of the true potential is characterized by a few parameters which can be fit to experimental data.  This allows us to study general behavior that is valid for any low energy system.

I shall first provide several examples of approximations in areas ranging from classical electrostatics to effective field theories.  This will hopefully illustrate the basic concepts used throughout the rest of our work.  Readers familiar with these ideas may wish to skip ahead to Section \ref{sec:mywork} where we discuss in more detail the problem considered in this dissertation.

%%%%%%%%%%%%%%%%%%%%%%%%%%%
%%  MULTIPOLE EXPANSION  %%
%%%%%%%%%%%%%%%%%%%%%%%%%%%
\section{Multipole Expansion}

The electrostatic potential created at some point $\vec{r}$ by a charge distribution is given by the formula
\begin{equation}
V(\vec{r}) = \frac{1}{4 \pi \epsilon_0} \int d^3\vec{r_{~}}' \, \frac{\rho(\vec{r}')}{\left| \vec{r} - {\vec{r_{~}}}' \right|} ,
\end{equation}
\noindent where $\rho$ is the charge density of the distribution.  Suppose that all of the charge is contained within some sphere of radius $l$ ($r' \le l$), and that the point at which we calculate the potential is very far away ($l \ll r$).  In this case, we can expand the denominator in the integral
\begin{equation}
\frac{1}{\left| \vec{r} - \vec{r_{~}}' \right|} = \frac{1}{r} \left[ 1 + \frac{r'}{r} \cos(\theta) + \left(\frac{r'}{r}\right)^2 \left(\frac{3}{2}\cos^2(\theta) - \frac{1}{2}\right) + \cdots \right] ,
\end{equation}
\noindent where $\theta$ is the angle between the vectors $\vec{r}$ and $\vec{r_{~}}'$.  This leads to approximations for the potential that result from truncating the multipole expansion \cite{Jackson:EM}:
\begin{equation}
V(\vec{r}) = \frac{1}{4 \pi \epsilon_0} \left[ \frac{1}{r} \int d^3\vec{r_{~}}' \, \rho(\vec{r_{~}}') + \frac{1}{r^2} \int d^3\vec{r_{~}}' \, r' \cos(\theta) \, \rho(\vec{r_{~}}') + \cdots \right] .
\end{equation}
\noindent The first term in the expansion is called the monopole term.  It represents the potential that would be created if all of the charge was concentrated at one point.  For large distances, it makes sense that the distribution would look like a point charge, and the first term reflects this.  However, if the total amount of charge is zero, then the monopole term is also zero.  Yet the charges must still create some potential.  This potential is approximated by the second piece called the dipole term.  It is the potential that would be created by a dipole at $\vec{r_{~}}' = 0$.  The full potential is built from all of these terms.  It behaves somewhat like a monopole, and somewhat like a dipole, and somewhat like a quadrupole, etc.

Notice that as we move farther away from the charges, all of the terms decrease in strength, but some decrease faster than others.  For a non-zero total charge, the dominant term is the monopole term at sufficiently large distance, so it is called the {\em leading order} term.  The dipole term is then called the {\em next-to-leading order} term, or equivalently the {\em first order correction} to the leading order term.  If the total charge is zero, then the dipole term is the leading order term with the quadrupole moment providing the first order correction.

Even if we stay at a fixed radius, the higher terms in the expansion still  contribute less and less.  Consider the dipole term.  The integration involves $r'$ which we know to be less than or equal to $l$.  We expect that the integral would be roughly equal to $l$ times some charge.  The whole term then looks like $l/r^2$.  This is smaller than the leading term, which behaves like $1/r$, by a factor of $l/r \ll 1$.  We say that this dipole term is of order $\mathcal{O}(l/r)$ compared to the leading term.

The small quantity $l/r$ acts as an expansion parameter for the potential.  Each additional term is smaller and smaller.  If we desire some accuracy $\epsilon$ in our calculation, we only need to include a finite number of terms from the expansion.  Let $N$ be the integer such that $(l/r)^N < \epsilon$.  Then any term in the expansion of $\mathcal{O}((l/r)^n)$ for $n > N$ may be dropped.  For example, if $N = 2$, we keep only the monopole and dipole terms.

The expansion hinges upon the fact that $l$ and $r$ are widely separated length scales.  If we considered distances where $r \sim l$, then each term of the expansion would be about as large as all the others.  They all contribute equally, and the expansion breaks down, reflecting the fact that we are now considering distances short enough to discern the details of the charge distribution.  Thus, there is a limit imposed on how small $r$ may be.  Exceed this limit, and the approximation is worthless.  Even without the $r > l$ restriction, we find that the individual terms in the multipole expansion diverge as $r \rightarrow 0$ even if the true potential never diverges.  This provides additional proof that the multipole approximation is no good at very short distances.

%%%%%%%%%%%%%%%%%%%%%%%%%
%%  QUANTUM MECHANICS  %%
%%%%%%%%%%%%%%%%%%%%%%%%%
\section{Quantum Mechanics}
\label{sec:introQM}

The Schr\"odinger equation in position space is
\begin{equation}
\left[ -\frac{\nabla^2}{2m} + V(\vec{r}) \right] \psi(\vec{r}) = E \, \psi(\vec{r}) .
\end{equation}
\noindent We will choose units so that $\hbar = 1$.  Let us consider the case of a spherically symmetric potential and look at low-energy S-wave scattering from this potential.  We will not cover scattering theory in great detail, but rather try to treat the subject quite simply.

If the potential is zero, then solutions to the equation for $E > 0$ are easily found.  There are two linearly independent solutions which we take to be incoming and outgoing spherical waves.  Any S-wave solution is written as a linear combination of these two solutions:
\begin{equation}
\psi(r) = A \frac{\me^{ikr}}{r} + B \frac{\me^{-ikr}}{r} .
\end{equation}
\noindent The energy for this wavefunction is $E = k^2/2m$.  For a non-zero short-range potential, this solution will still be valid at large distances where we can neglect the interaction.  So we can view scattering as a spherical wave approaching the potential from $r = \infty$, interacting with the potential, then leaving the potential and returning to infinity.  Since the probability associated with the incoming and outgoing waves must be conserved, the constants $A$ and $B$ can only differ by a phase, $A = -B \me^{2 \mi \delta_0(k)}$.  The function $\delta_0(k)$ is called the phase shift, and the $k$ dependence is shown to illustrate the fact that the phase can depend upon the energy of the wave.

Analytic properties of the phase shift $\delta_0(k)$ for very low momentum can be used to show that the quantity $k \, \cot(\delta_0(k))$ has a well-defined expansion about the point $k = 0$.  This expansion, known as the effective range expansion, is written as
\begin{equation}
k \, \cot(\delta_0(k)) = - \frac{1}{a} + \frac{1}{2} r_e k^2 + \cdots .
\end{equation}
\noindent For a more detailed derivation of this expansion, see Ref.~\cite{Taylor:scatter} or \cite{Newton:scatter}.

The parameters $a$, $r_e$, etc.~contain information about the details of the potential with which the spherical wave interacts.  The parameter $a$ is known as the scattering length, while $r_e$ is known as the effective range.  In general, an infinite number of terms appear in the exact expansion, but we can truncate the series for small $k$ to achieve a desired accuracy with a finite number of terms.

In a sense, we are using low-energy scattering to probe some of the properties of the potential.   While we can use this to determine some of the details, we cannot determine all of them (unless we measure the phase shift exactly for all $k$).  There will always be infinitely many potentials that have the same scattering length.  Even if we did distinguish some of them by different effective ranges, there would still be an infinite number of potentials that share both $a$ and $r_e$.  

Fortunately, this is an asset and not a hindrance.  What this means is that the low-energy behavior is insensitive to the detailed form of the potential, and any two potentials that give the same parameters are equally good approximations.  In fact, sometimes it is not even necessary to find a potential to use as an approximation.  The effective range expansion can be used to make predictions, and other derived quantities can be written in terms of the scattering length and other parameters.  By experimental measurements, we can determine $a$ and $r_e$ and then use them to predict other quantities.

Let us look at an example to make some of these ideas clearer.  We shall take the potential to be a square well
\begin{equation}
V(r) = -V_0 \, \theta(1/\Lambda - r).
\end{equation}
\noindent Inside the well, the wavefunction has the form
\begin{equation}
\psi(r) = A \sin(K r) ,
\end{equation}
\noindent where 
\begin{equation}
K = \sqrt{2 m (E + V_0)} .
\end{equation}
\noindent Outside the well, the solution takes the form 
\begin{equation}
\psi(r) = B \sin(k r + \delta_0) ,
\end{equation}
\noindent where 
\begin{equation}
k = \sqrt{2 m E} .
\end{equation}
\noindent Since the quantity $\partial \ln(\psi(r))/\partial r$ must be continuous, 
\begin{equation}
K \cot(K/\Lambda) = k \cot(k/\Lambda + \delta_0) \label{eqn:boundary}.
\end{equation}
\noindent For very low energies, we can expand both sides of Eq.~(\ref{eqn:boundary}) in powers of $k$ and then match terms.  Notice that the phase shift occurs on one side, and $V_0$ on the other.  This ensures that when we match expansions, the parameters for $\delta_0$ will be written in terms of the parameters for the potential.

Without going into details, we present the results of this expansion:
\begin{eqnarray}
\frac{1}{a} & = & - \frac{g \Lambda}{\tan(g) - g} \label{eqn:a},
\\
r_e & = & \frac{(1 - a \Lambda)^2}{a^2 \Lambda^3} \label{eqn:r2},
\end{eqnarray}
\noindent where $g = \sqrt{2 m V_0}/\Lambda$.  The one condition required to truncate the expansion is that $k \ll \Lambda$, which should come as no surprise.  A particle of energy $k^2/2m$ has a wavelength of about $1/k$.  If we demand insensitivity to the short distance behavior, then the wavelength must be much greater than the range of the interaction.  High energies probe small distances, but low energies do not.  The parameter $\Lambda$ acts as a momentum cutoff.  Incidently, this square well potential can be used to approximate any other short-range potential.

By choosing $g$, and hence $V_0$, to be a function of $\Lambda$, we can make the scattering length $a$ independent of the cutoff by always making sure that Eq.~(\ref{eqn:a}) is satisfied for the same constant $a$.  This defines a whole set of potentials that give identical scattering lengths.  The only difference between them is the value of $r_e$.  As $\Lambda$ changes, so will the effective range.  This is where the errors in a simple square well approximation come in.  If our potential had another free coupling to adjust, then we could typically make $r_e$ independent of the cutoff, and the error moves to the next parameter.  So we see that we can define many different potentials (one for each value of $\Lambda$) that can serve as equally accurate approximations.  We can even go so far as to let $\Lambda \rightarrow \infty$ and end up with a delta function potential well.

We should also take a moment to look at the relative size of our parameters.  The scattering length appears to be around $1/\Lambda$, and this would imply that $r_e$ is also about $1/\Lambda$.  [We say that $a$ and $r_e$ are both $\mathcal{O}(1/\Lambda)$.]  This is called a ``natural'' theory.  All of the parameters have their lengths set by the underlying length scale of the problem, in this case $1/\Lambda$.  However, it is possible that $g$ could be fine-tuned to make $a$ much larger than $1/\Lambda$.  Even though $a \Lambda \gg 1$, the size of $r_e$ would still be $\mathcal{O}(1/\Lambda)$.  The effective range is still set by the range of the potential.  However, this too could be made unnaturally large if there is a second parameter in our theory that we could fine-tune.

In addition to looking at low-energy scattering states, we could also look at low-energy bound states.  The condition for having a bound state of energy $-B$ is
\begin{equation}
\sqrt{g^2 \Lambda^2 - 2 m B} \cot\left(\sqrt{g^2 \Lambda^2 - 2 m B}/\Lambda\right) = - \sqrt{2 m B} \label{eqn:SWbound}.
\end{equation}
\noindent Treating $mB/\Lambda^2$ as a small quantity, we expand the left-hand side to leading order to obtain the relation
\begin{equation}
B = \frac{1}{2m} (g \Lambda \cot(g))^2 = \frac{1}{2m} \left(\frac{\Lambda}{1 - a\Lambda}\right)^2 \label{eqn:SWboundapprox}.
\end{equation}
\noindent  For very large $a$, the binding energy behaves like $1/ma^2$.  This shows the connection between large scattering lengths and shallow bound states.  If we wanted to expand the equation for $B$ to higher orders, we would find that for large $a$ it is expanded in powers of $1/a\Lambda \sim r_e/a$.  Once again, our expansion parameter is set by two widely separated length scales.

For a natural theory, the scattering length would be set by the scale $1/\Lambda$.  In this case, Eq.~(\ref{eqn:SWboundapprox}) implies that the binding energy is also set by the potential's length scale: $B \sim \Lambda^2/2m$.  This argument should not be taken too seriously because such a binding energy would violate the assumption that $mB/\Lambda^2 \ll 1$, which was required to expand Eq.~(\ref{eqn:SWbound}) in the first place.  The leading approximation for the binding energy also diverges when $a \Lambda = 1$ showing that it breaks down for a natural theory.  Nonetheless, the fact that $B \sim \Lambda^2/2m$ remains valid as can be seen by comparing Eqs.~(\ref{eqn:a}) and (\ref{eqn:SWbound}) for the case $a \sim 1/\Lambda$.

%%%%%%%%%%%%%%%%%%%%%%%%%%%%%%
%%  EFFECTIVE FIELD THEORY  %%
%%%%%%%%%%%%%%%%%%%%%%%%%%%%%%
\section{Effective Field Theory}

We now examine the interaction of two identical particles from the field theory point-of-view.  This requires some prior knowledge of field theory Lagrangians and how Feynman rules are derived from them.\footnote{See, for example, Ref.~\cite{Weinberg:QFT1}}  Once again, we will sacrifice rigor for simplicity and simply state many results so that readers unfamiliar with perturbative field theory can follow the discussion.  The important things at this point are not the mathematical steps, but the ideas behind them.

Suppose we have two identical particles interacting via some unknown short-range potential and we wish to approximate the behavior.  We will start the approximation with the Lagrangian
\begin{equation}
\mathcal{L} = \phi^*(\vec{x}) \left( i \frac{\partial}{\partial t} + \frac{\nabla^2}{2 m} \right) \phi(\vec{x}) + C_0 \left[\phi^*(\vec{x}) \phi(\vec{x})\right]^2 \label{eqn:L1} ,
\end{equation}
\noindent where $C_0$ is the coupling constant for the two-body interaction.  We require that all interactions in the Lagrangian be local operators.  That is, they are the product of fields at the same point.  These operators can be thought of as contact interactions since they act only at a single point.

If we expand the scattering amplitude perturbatively using this interaction, the first term will consist of a single vertex with a value of  $- \mi C_0$.  The next term will contain two vertices, two propagators, and a single loop integration.  Unfortunately, the loop integral behaves like $\int dk$ which diverges linearly.  If we try calculating the next term, we find that it diverges quadratically.  Each successive term has a more severe divergence than the last.  It appears that our hope of a perturbative expansion is lost.

Or is it?  The reason we encounter this divergence is because we have states with arbitrarily large differences in momenta coupled to one another.  However, our approximation is not meant to be used for arbitrarily high momentum.  It is only valid for low momenta that cannot resolve the internal details of the interaction.  For the moment, let us impose a cutoff $\Lambda$ on the momentum values.  This represents the point at which our approximation breaks down and knowledge of the true potential becomes necessary.  In the center-of-mass frame, the scattering amplitude as a function of the momentum $p$ for the first two perturbative terms turns out to be
\begin{equation}
\mathcal{A}(p) = -C_0 - \frac{C_0^2 m}{4 \pi^2} \left( \Lambda - \mi \pi p /2 \right) \label{eqn:Ap}.
\end{equation}
\noindent To calculate $\mathcal{A}$, we would need to know the values of $C_0$ and $\Lambda$.  Yet the reason we are using this approximation is because we do not know the underlying potential, and we want our results to be insensitive to such details as the exact value of $\Lambda$.

The solution is to let $C_0$ be a function of the cutoff.  At $p = 0$, 
\begin{equation}
\mathcal{A}(0) = -C_0 - \frac{C_0^2 m}{4 \pi^2} \Lambda ,
\end{equation}
\noindent which is related to the two-body scattering length $a$ by
\begin{equation}
a = - \frac{m \mathcal{A}(0)}{8 \pi} = \frac{m C_0}{8 \pi} \left( 1 + \frac{C_0 m}{4 \pi^2} \Lambda \right) \label{eqn:aC0expand}.
\end{equation}
\noindent If we invert this equation to get a perturbative expression for $C_0$ as a function of $a$, we obtain
\begin{equation}
C_0 = \frac{8 \pi a}{m} \left( 1 - \frac{2 a \Lambda}{\pi} \right) .
\end{equation}
\noindent This allows us to rewrite Eq.~(\ref{eqn:Ap}) as 
\begin{equation}
\mathcal{A}(p) = - \frac{8 \pi a}{m} \left( 1 - \mi a p\right) \label{eqn:Ap2}.
\end{equation}
\noindent Our originally divergent amplitude has now been cast in terms of the finite scattering length, which we can determine from experiment.

This procedure that we have just sketched is an example of {\em renormalization}.  We {\em regulate} any divergent integrals with some sort of cutoff, allow the coupling constants to be functions of this cutoff, and then choose the form of the couplings to remove this cutoff from physical quantities.  Because the coupling constants change as the cutoff changes, they are said to ``run with the cutoff'' and may be referred to as ``running couplings.''  In our example, all reference to $\Lambda$ is eliminated, at least up to the perturbative order in the couplings we have expanded in.  In general, this will not be the case.  Instead, there will remain some cutoff dependence involving inverse powers of $\Lambda$.  The cutoff can then be taken to infinity, leaving a finite result.

Notice that the second term in Eq.~(\ref{eqn:Ap2}) contains the product $a p$.  Our approximation is only valid for low momentum, such that $p^{-1}$ is much greater than the range of the underlying interaction.  If this is a natural theory, we expect that $a$ will be of the same order as the interaction range and hence $ap \ll 1$.  In other words, we can expand our amplitude as a power series in $ap$, and we have seen just that.  The $-\mi a p$ term is much smaller than the first term of $\mathcal{O}(1)$.

We have only included one interaction in the Lagrangian so far.  With just this, we cannot produce a theory with a fixed, non-zero effective range.  To control the effective range, we shall add another operator of the form $ C_2 \vec{\nabla}\left( \phi^*(\vec{x}) \phi(\vec{x})\right) \cdot \vec{\nabla}\left( \phi^*(\vec{x}) \phi(\vec{x})\right)$.  Since it contains derivatives of the fields, it is sometimes referred to as a ``derivative interaction.''  If we were to add its contributions to $\mathcal{A}(p)$, we would find divergences that need to be regulated.  This can be done once again by applying the cutoff $\Lambda$ and then properly choosing $C_2(\Lambda)$.  This allows us to control the effective range $r_e$, and after renormalization, the divergent quantities can be rewritten in powers of $r_e p$.  For a natural theory, this is the same order as $a p$ and can also be used as an expansion parameter.

This example has helped to introduce the general concepts of effective field theory (EFT) \cite{Weinberg:PhysicaA,Lepage:GeneralEFT,Kaplan:GeneralEFT}.  The main idea behind EFT is that the details of the underlying physics are embodied in the coupling constants of the theory.  By relating the couplings to physical quantities (like scattering length and effective range), other calculations can be written in terms of these physical parameters without any knowledge of the detailed underlying theory.  In addition, interactions containing more and more derivatives or fields contribute less and less to the overall result at low energies, just as higher order terms in the multipole expansion contribute less and less to the electrostatic potential far from the source.

To see this, we must look at the dimension of each operator.  We will assume that the range $R$ of the underlying interaction sets the scale of the terms in our Lagrangian density:
\begin{equation}
\mathcal{L} = \phi^*(\vec{x}) \left( i \frac{\partial}{\partial t} + \frac{\nabla^2}{2 m} \right) \phi(\vec{x}) + C_0 \left(\phi^*(\vec{x}) \phi(\vec{x})\right)^2 + C_2 \vec{\nabla}\left( \phi^*(\vec{x}) \phi(\vec{x})\right) \cdot \vec{\nabla}\left( \phi^*(\vec{x}) \phi(\vec{x})\right).
\end{equation}
\noindent If that is the case, then we would expect any length variables like $x$ to scale like $R$, which we shall write as $[x] = R$.  This implies that $[\partial/\partial x] = R^{-1}$ and hence $[\nabla] = R^{-1}$.  By these arguments, the kinetic energy operator would scale like $[\nabla^2/2m] = m^{-1} R^{-2}$.  A look at our Lagrangian density shows that the time derivative must also scale in the same way: $[\partial/\partial t] =  m^{-1} R^{-2}$.  This relation then implies that the time variable scales as $[t] = m R^2$. 

To find the scaling behavior of the fields and couplings, we observe that the action $S = \int dt \int d^3x \, \mathcal{L}$ is dimensionless.  Since $[dt \, d^3x] = [t x^3] = m R^5$, we must have $[\mathcal{L}] = m^{-1} R^{-5}$.  Therefore, every term that makes up a part of $\mathcal{L}$ must also scale in the same way:
\begin{eqnarray}
\left[\phi^*(\vec{x}) \left( i \frac{\partial}{\partial t} + \frac{\nabla^2}{2 m} \right) \phi(\vec{x})\right] & = & m^{-1} R^{-5} \nonumber,
\\
\left[C_0 \left(\phi^*(\vec{x}) \phi(\vec{x})\right)^2\right] & = & m^{-1} R^{-5} \nonumber,
\\
\left[C_2 \vec{\nabla}\left( \phi^*(\vec{x}) \phi(\vec{x})\right) \cdot \vec{\nabla}\left( \phi^*(\vec{x}) \phi(\vec{x})\right)\right] & = & m^{-1} R^{-5} \nonumber.
\end{eqnarray}
\noindent From this, we can easily draw the following relations:
\begin{eqnarray}
\left[ \phi \right] & = & R^{-3/2} \nonumber,
\\
\left[ C_0 \right] & = & m^{-1} R \nonumber,
\\
\left[ C_2 \right] & = & m^{-1} R^3 \nonumber.
\end{eqnarray}

Since the couplings are supposed to approximate the details of the short-range behavior, we expect that their scale should be set by the scale of the short-range interaction.  This is in agreement with the leading order approximation to the scattering length in Eq.~(\ref{eqn:aC0expand}) where $a \propto m C_0 \sim R$.  Notice the factor of $m^{-1}$ in both $C_0$ and $C_2$.  This same factor occurs in every coupling.  Every interaction operator will consist of fields ($\phi$), derivatives ($\nabla$), and some coupling.  Every field scales as $R^{-3/2}$, while every derivative scales like $R^{-1}$.  If the fields and derivatives combined scale like $R^{-N}$, then the coupling must scale like $m^{-1} R^{N-5}$.

Of course, any dimensionless quantity containing the couplings must have another scale to balance out the powers of $R$.  This scale comes from the momenta of the process being studied.  If the system has a typical momentum of $p$, then a dimensionless quantity would be a function of $p R$.  Hence, if $p R \ll 1$, we can expand in powers of $p R$.  An operator of dimension $R^N$ has a coupling proportional to $R^{N-5}$ and contributes at order $(pR)^{N-5}$.  This allows us to classify the operators and decide which Feynman diagrams are needed to achieve any given accuracy.

Returning to our example, a single $C_0$ vertex adds a term of order $R \sim a$ to $\mathcal{A}$.  Two $C_0$ vertices contribute $R^2 p \sim a^2 p$, while three add $R^3 p^2 \sim a^3 p^2$.  At this point, we notice that a single $C_2$ vertex is of order $R^3 p^2$, implying that we must include $C_2$ if we want to calculate $C_0$ effects beyond the first two orders.  Because the effects of the $C_2$ interaction do not contribute to leading order, we classify the operator $\vec{\nabla}\left( \phi^*(\vec{x}) \phi(\vec{x})\right) \cdot \vec{\nabla}\left( \phi^*(\vec{x}) \phi(\vec{x})\right)$ as {\em irrelevant}.  The ``irrelevant'' qualifier in the term irrelevant operator does not carry the usual meaning, implying that this operator is not needed at all.  As we have seen, it is vital if we desire a cutoff-independent effective range.  Rather, we use the term irrelevant operator to mean that the operator's effects are smaller than the leading order effects.  The renormalization group provides a more precise definition based on the scaling behavior of the operator about a fixed point, but this is more technical than required for our discussion.

We now have a well-defined expansion method.  We can calculate increasingly accurate results by adding additional operators, which means fitting more couplings to experimental data.  One might wonder why this is any different from curve fitting.  Why not create some arbitrary potential that includes a bunch of adjustable parameters, and then do some sort of least-squares fit to experimental data to determine the optimal parameter values?  The difference is that EFT provides a systematic way of improving the result, instead of merely guessing a potential that may or may not improve things.  For any given system, we can write down local operators that could appear as interactions, provided that they satisfy all symmetries of the system.  These operators can then be classified based on their dimension, and the point at which they each contribute to the expansion is known.

There are cases where the above expansion method is not as useful as might be expected.  The most obvious is when $a$ is very large.  Since we require $a p \ll 1$, a large scattering length severely limits the range of momentum in which our expansion might converge.  Another case would be the study of bound states.  The typical momentum in a bound state is of order $1/a$, so once again the expansion would not converge.  The solution in these cases is to treat $a p$ as $\mathcal{O}(1)$ and sum all terms.  If this is possible, then we can still treat powers of $r_e p$ perturbatively for the case where $r_e \ll a$, and we have an increased momentum range over which the expansion is valid.

Fortunately in our example, the contributions to $\mathcal{A}$ from terms containing only $C_0$ form a geometric series.  The result of the summation is 
\begin{equation}
\mathcal{A}(p) = - \left( \frac{8 \pi}{m} \right) \left[\left( \frac{8 \pi}{C_0 m} - \frac{2 \Lambda}{\pi} \right) + \mi p\right]^{-1} ,
\end{equation}
\noindent which leads to a scattering length of
\begin{equation}
a = \left( \frac{8 \pi}{C_0 m} - \frac{2 \Lambda}{\pi} \right)^{-1}.
\end{equation}
\noindent To keep the scattering length cutoff independent, we choose $C_0(\Lambda)$ to be
\begin{equation}
C_0(\Lambda) = \left(\frac{8 \pi a}{m}\right)\left(1 + \frac{2 a \Lambda}{\pi}\right)^{-1} .
\end{equation}
\noindent If we let $a \rightarrow \infty$ in the above equation, then we see that $C_0$ approaches $(4 \pi^2)/(m \Lambda)$.  It is by fine-tuning $C_0$ to be close to this value that we find a system with a very large scattering length.  This also illustrates that the running coupling $C_0$ possesses a ``fixed point.''  It is common to consider the behavior of the dimensionless versions of the couplings as $\Lambda \rightarrow \infty$. In this case, $\Lambda C_0 \rightarrow 4 \pi^2/m$ which is a fixed constant value.  This is a typical behavior for many couplings.  When we consider the three-body problem later, we will find that the three-body coupling exhibits another type of behavior: a limit cycle.  Limit-cycle behavior simply means that the coupling approaches a periodic function as $\Lambda \rightarrow \infty$.

%%%%%%%%%%%%%%%
%%  MY WORK  %%
%%%%%%%%%%%%%%%
\section{Three-Body Problem}
\label{sec:mywork}

The EFT approach has been applied with great success to many low-energy two-body systems \cite{Seki:EFT2body,Bedaque:EFT2body,VanKolck:EFT2body,Beane:EFT2body}.   A logical next step would be the application of EFT techniques to the three-body problem.  Recent investigations of such systems by Bedaque, Hammer, and van Kolck have turned up some surprising results \cite{hammer:orig}.  For three identical bosons, dimensional analysis suggests that only the leading order two-body interaction is needed to obtain the leading order three-body results.  While this is indeed sufficient to remove any integration divergences in the equations, cutoff sensitivity still remains in the form of a non-unique $\Lambda \rightarrow \infty$ limit.  The remedy is to introduce a three-body contact interaction of higher dimension, indicating that short-distance effects are playing an important role.  In addition, the coupling of the three-body interaction is found to exhibit a limit-cycle behavior, unlike the more common fixed-point behavior of running coupling constants.

I choose to investigate low-energy three-body systems in a slightly different manner from that used by Bedaque, Hammer, and van Kolck.  For sufficiently low energies, the potential $V(\vec{r})$ is replaced by a series of point-like potentials, namely products of $\delta^3(\vec{r})$ and its derivatives.  As in the case of EFT, the coupling constants multiplying these terms approximate the short-range details of $V(\vec{r})$ and must be fit to experimental quantities such as scattering length and effective range.

The focus of my calculation is on the bound-state equation, which provides a complimentary approach to the scattering methods employed in other works \cite{ksw:largea,hammer:orig,kolck:largea}.  My work closely follows unpublished calculations by Kenneth Wilson who developed the basic method and tools that I employ \cite{Wilson:notes}.  I use a combination of perturbative expansion and numerical methods that not only provides an independent verification of previous results but also offers several possible advantages.  There is no need to introduce a composite field, such as the ``dimeron'' used in Ref.~\cite{hammer:orig}, nor is it necessary to resum Feynman diagrams for the dimeron's propagator.  I introduce a technique for expanding the bound state equation in inverse powers of the cutoff, giving us the ability to examine cutoff-dependent behavior order by order.  Additionally, the numerical methods I use allow computation to high accuracy, typically 11 to 12 digits in the couplings and binding energies.

In this dissertation, I derive the expanded set of three-body integral equations and present the results of my numerical calculations.  I numerically ``prove'' that a three-body coupling term is necessary and sufficient to renormalize the three-body bound-state spectrum.  The three-body coupling exhibits limit-cycle behavior, and I present an argument for why this should be expected.  I also calculate the three-body bound-state spectrum and how it changes with the two-body bound-state energy.  This data is an essential part of computing Efimov's function for three-body bound states to high accuracy.  This function can be used to determine the binding energies of three-body bound states when there is a large scattering length \cite{efimov:2}.  Efimov's function is currently being used to study Feshbach resonances \cite{Braaten:2002er}.

%%%%%%%%%%%%%%%%%%%
%%  NEW RESULTS  %%
%%%%%%%%%%%%%%%%%%%
\section{New Results}

To the best of the author's knowledge, this dissertation contains several new results, some of which are extensions to work done by Wilson \cite{Wilson:notes}:
\begin{itemize}
\item Correction of minor errors in the leading order equations and the derivation of the equations for first order corrections.
\item Modification of Wilson's original leading order equations to maintain high accuracy when $B_2 \simeq B_3$ in numerical calculations.
\item Detailed explanation of the derivation of the three-body integral equations and the use of Wilson's methods in that derivation.
\item Proof of and corrections to Wilson's assertion that certain integrals which are encountered when computing corrections to the leading order results are identically zero.
\item Derivation of new integration limits to simplify numerical calculations.
\item Implementation of numerical methods to solve leading order equations.
\item Use of leading order equations to compute Efimov's universal function to high accuracy.
\item Use of leading order equations to analytically and numerically verify similar work by others.
\end{itemize}

%%%%%%%%%%%%%%%%
%%  OVERVIEW  %%
%%%%%%%%%%%%%%%%
\section{Organization of Dissertation}

After outlining the use of the $\delta^3(\vec{r})$ potential in the context of the two-body problem, I derive some formulas in Chapter 2 that will be of use later in the analysis of the three-body problem.  Chapter 3 gives the derivation of the three-body bound-state equation and outlines some assumptions made in my calculations.  I digress briefly in Chapter 4 to explain the method used for expanding the equations perturbatively and then apply the technique in Chapter 5.  This results in the final form of the equations used for numerical computation.  Chapter 6 outlines some details of the numerical technique and discusses some of the issues that arise from the need for high accuracy.  Following this is a discussion of analytic and numerical results in Chapter 7.  Chapter 8 is devoted to introducing Efimov's function and then numerically computing it using the leading order equations.  I conclude in Chapter 9 with a brief discussion of areas for future work. 
\chapter{The Two-Body Problem}
\label{ch:2body}

As an introduction, we first consider the two-body problem.  This provides a simple example that illustrates the approach we use with the three-body problem.  More importantly though, the results derived in this chapter will carry over unaffected to the three-body system and are in fact needed to attack that problem.  All equations refer to identical bosons of mass $1/2$ and S-wave bound states of such particles.  These restrictions simplify the equations while preserving the system characteristics we desire.

%%%%%%%%%%%%%%%%%%%%%%%
%%  BOUND STATE EQN  %%
%%%%%%%%%%%%%%%%%%%%%%%
\section{Bound State Equation}

The general two-body Schr\"odinger equation is
\begin{equation}
-\nabla_1^2 \, \psi(\vec{r_1}, \vec{r_2}) - \nabla_2^2 \, \psi(\vec{r_1}, \vec{r_2}) + \int d^3\vec{r_1}' \,  d^3\vec{r_2}' \, V_{12}(\vec{r_1}',\vec{r_2}'; \vec{r_1},\vec{r_2}) \, \psi(\vec{r_1}',\vec{r_2}') = E \, \psi(\vec{r_1},\vec{r_2}) ,
\end{equation}
\noindent where $V_{12}(\vec{r_1}',\vec{r_2}'; \vec{r_1},\vec{r_2}) = \langle\vec{r_1}, \vec{r_2}\,|\,V_{12}\,|\,\vec{r_1}', \vec{r_2}'\rangle$ represents the position-space matrix elements of the two-body interaction operator.  We choose the matrix elements of $V_{12}$ to be
\begin{equation}
\langle\vec{r_1}, \vec{r_2}\,|\,V_{12}\,|\,\vec{r_1}', \vec{r_2}'\rangle =  -g_2 \, \delta^3(\vec{r_1}-\vec{r_2}) \, \delta^3(\vec{r_1}'-\vec{r_2}') \, \delta^3 \left( \frac{\vec{r_1}+\vec{r_2}}{2} - \frac{\vec{r_1}'+\vec{r_2}'}{2} \right) \label{eqn:V12matrix} ,
\end{equation}
\noindent with $g_2$ denoting the two-body coupling constant.  Our choice ensures that
\begin{eqnarray}
\lefteqn{ \langle\vec{r_i} | V_{12} | \psi\rangle \; = \; \int d^3\vec{r_j}' \langle\vec{r_i}|V_{12}|\vec{r_j}'\rangle\langle\vec{r_j}'|\psi\rangle } \nonumber 
\\
 & = & -g_2 \int d^3\vec{r_j}' \; \delta^3(\vec{r_1}-\vec{r_2}) \delta^3(\vec{r_1}'-\vec{r_2}') \delta^3 \left( \frac{\vec{r_1}+\vec{r_2}}{2} - \frac{\vec{r_1}'+\vec{r_2}'}{2} \right) \psi(\vec{r_j}') \nonumber
\\
 & = & -g_2 \, \delta^3(\vec{r_1}-\vec{r_2}) \int d^3\vec{r_1}' \: d^3\vec{r_2}' \; \delta^3(\vec{r_1}'-\vec{r_2}') \delta^3 \left( \frac{\vec{r_1}+\vec{r_2}}{2} - \frac{\vec{r_1}'+\vec{r_2}'}{2} \right) \psi(\vec{r_1}',\vec{r_2}') \nonumber
\\
 & = & -g_2 \, \delta^3(\vec{r_1}-\vec{r_2}) \int d^3\vec{r_1}' \; \delta^3 \left( \frac{\vec{r_1}+\vec{r_2}}{2} - \vec{r_1}' \right) \psi(\vec{r_1}',\vec{r_1}') \nonumber
\\
 & = & -g_2 \, \delta^3(\vec{r_1}-\vec{r_2}) \psi \left( \frac{\vec{r_1}+\vec{r_2}}{2}, \frac{\vec{r_1}+\vec{r_2}}{2}\right) \nonumber
\\
 & = & -g_2 \, \delta^3(\vec{r_1}-\vec{r_2}) \psi(\vec{r_1},\vec{r_2}) ,
\end{eqnarray}
\noindent which is the form we desire.  In the center-of-mass frame, the bound-state equation for two identical bosons with binding energy $-B_2$ can be simplified to
\begin{equation}
-2 \, \nabla^2 \, \psi(\vec{r}) - g_2 \, \delta^3(\vec{r}) \, \psi(\vec{r}) = -B_2 \, \psi(\vec{r}) ,
\end{equation}
\noindent or in momentum space,
\begin{equation}
2 \, p^2 \, \phi(\vec{p}) - g_2 \, \int \frac{d^3\vec{q}}{(2 \pi)^3} \, \phi(\vec{q})  =  -B_2 \, \phi(\vec{p}) \label{eqn:pspace}.
\end{equation}
\noindent Here we have begun the convention of labeling two-body couplings and energies with a ``2'' subscript to distinguish them from the corresponding three-body parameters we will encounter later, which will carry a ``3'' subscript.

Equation (\ref{eqn:pspace}) has been previously studied and is known to suffer from divergences for any constant, finite value of $g_2$ \cite{jackiw:delta}.  To see this, define the constant 
\begin{equation}
\mathcal{I} \equiv \int \frac{d^3\vec{q}}{(2 \pi)^3} \phi(\vec{q})\label{eqn:Idef}.  
\end{equation}
\noindent Equation (\ref{eqn:pspace}) can then be written as
\begin{equation}
2 \, p^2 \, \phi(\vec{p}) - g_2 \, \mathcal{I} = -B_2 \, \phi(\vec{p}) \label{eqn:2bodyI}.
\end{equation}
\noindent This implies that
\begin{equation}
\phi(\vec{p}) = \frac{g_2 \, \mathcal{I}}{2 p^2 + B_2} \label{eqn:Iphi}.
\end{equation}
\noindent Using Eq.~(\ref{eqn:Iphi}) in Eq.~(\ref{eqn:Idef}) yields
\begin{equation}
\mathcal{I}  =  \int \frac{d^3\vec{q}}{(2 \pi)^3} \frac{g_2 \, \mathcal{I}}{2 q^2 + B_2} ,
\end{equation}
\noindent or equivalently
\begin{equation}
\frac{1}{g_2} = \int \frac{d^3\vec{q}}{(2 \pi)^3} \frac{1}{2 q^2 + B_2} = \int_0^{\infty} \frac{dq}{2 \pi^2} \frac{q^2}{2 q^2 + B_2}.
\end{equation}
\noindent Any finite value of $B_2$ results in a divergent integral, and hence a value of zero for $g_2$.  As we discussed in Chapter 1, the divergence is due to the coupling of states with arbitrarily large momentum differences.  In EFT terms, we are including high energy states that are beyond the region of validity for our approximation.  Therefore, it should come as no surprise that we encounter problems. 

To remedy the situation, we must find a way to regulate the problem.  Let us define
\begin{equation}
\tilde{U}_2(\vec{r}) = \int \frac{d^3\vec{p}}{(2 \pi)^3} \exp(i \vec{p} \cdot \vec{r}) \, U_2(\vec{p}),
\end{equation}
\noindent where
\begin{equation}
U_2(\vec{p}) = \exp(-p^2/\Lambda^2) \label{eqn:U2p}.
\end{equation}
\noindent For a finite value of $p$, allowing $\Lambda \rightarrow \infty$ results in $U_2(\vec{p}) = 1$ and hence $\tilde{U}_2(\vec{r}) = \delta^3(\vec{r})$.  In essence, this defines a regulated delta function that is ``smeared out'' in the sense that it now has a range of $\mathcal{O}(1/\Lambda)$.  This function can be used to regulate the two-body interactions by replacing the first two delta functions in Eq.~(\ref{eqn:V12matrix}) with $\tilde{U}_2$,
\begin{equation}
\langle\vec{r_1}, \vec{r_2}\,|\,V_{12}\,|\,\vec{r_1}', \vec{r_2}'\rangle =  -g_2 \, \tilde{U}_2(\vec{r_1}-\vec{r_2}) \, \tilde{U}_2(\vec{r_1}'-\vec{r_2}') \, \delta^3 \left( \frac{\vec{r_1}+\vec{r_2}}{2} - \frac{\vec{r_1}'+\vec{r_2}'}{2} \right) .
\end{equation}
\noindent The momentum space matrix elements become
\begin{eqnarray}
\langle\vec{p_i}\,|\,V_{12}\,|\,\vec{q_j}\rangle & = & \int d^3\vec{r_i} \: d^3\vec{r_j}' \langle\vec{p_i}|\vec{r_i}\rangle \langle\vec{r_i}\,|\,V_{12}\,|\,\vec{r_j}'\rangle \langle\vec{r_j}'|\vec{q_j}\rangle \nonumber
\\
 & = & -g_2 \int d^3\vec{r_i} \: d^3\vec{r_j}' \exp(-i \vec{p_i}\cdot\vec{r_i}) \exp(i \vec{q_j}\cdot\vec{r_j}') \, \tilde{U}_2(\vec{r_1}-\vec{r_2}) \nonumber
\\
 & & \times \tilde{U}_2(\vec{r_1}'-\vec{r_2}') \, \delta^3 \left( \frac{\vec{r_1}+\vec{r_2}}{2} - \frac{\vec{r_1}'+\vec{r_2}'}{2} \right) \nonumber
\\
 & = & -g_2 \int \frac{d^3\vec{k}}{(2 \pi)^3} \frac{d^3\vec{k}'}{(2 \pi)^3} U_2(\vec{k}) U_2(\vec{k}') \int d^3\vec{r_i} \: d^3\vec{r_j}' \exp(-i \vec{p_i}\cdot\vec{r_i}) \nonumber
\\
 & & \times \exp(i \vec{q_j}\cdot\vec{r_j}') \exp \left(i \vec{k}\cdot(\vec{r_1}-\vec{r_2}) \right) \exp \left( i \vec{k}'\cdot(\vec{r_1}'-\vec{r_2}') \right) \nonumber
\\
 & & \times \delta^3 \left( \frac{\vec{r_1}+\vec{r_2}}{2} - \frac{\vec{r_1}'+\vec{r_2}'}{2} \right)  \nonumber
\\
& = &  -g_2 \int \frac{d^3\vec{k}}{(2 \pi)^3} \frac{d^3\vec{k}'}{(2 \pi)^3} U_2(\vec{k}) U_2(\vec{k}') \int d^3\vec{r_i} \: d^3\vec{r_1}' \exp(-i \vec{p_i}\cdot\vec{r_i}) \nonumber
\\
 & & \times \exp \left( \vec{q_1}\cdot\vec{r_1}' + \vec{q_2}\cdot(\vec{r_1}+\vec{r_2}-\vec{r_1}') + \vec{k}\cdot(\vec{r_1}-\vec{r_2}) \right. \nonumber
\\
 & & +~ \left. \vec{k}'\cdot(2\vec{r_1}'-\vec{r_1}-\vec{r_2}) \right) \nonumber
\\
 & = & -g_2 \int \frac{d^3\vec{k}}{(2 \pi)^3} \frac{d^3\vec{k}'}{(2 \pi)^3} U_2(\vec{k}) U_2(\vec{k}') (2 \pi)^{9} \, \delta^3(-\vec{p_1}+\vec{q_2}+\vec{k}-\vec{k}') \nonumber
\\
 & & \times \delta^3(-\vec{p_2}+\vec{q_2}-\vec{k}-\vec{k}') \, \delta^3(\vec{q_1}-\vec{q_2}+2\vec{k}') \nonumber
\\
 & = & -g_2 \int \frac{d^3\vec{k}}{(2 \pi)^3} U_2(\vec{k}) U_2\left(\frac{\vec{q_2}-\vec{q_1}}{2}\right) (2 \pi)^{6} \, \delta^3\left(-\vec{p_1}+\mbox{$1 \over 2$}\vec{q_1} + \mbox{$1 \over 2$}\vec{q_2}+\vec{k}\right) \nonumber
\\
 & & \times \delta^3\left(-\vec{p_2}+\mbox{$1 \over 2$}\vec{q_1} + \mbox{$1 \over 2$}\vec{q_2}-\vec{k}\right) \nonumber
\\
 & = & -g_2 \, (2 \pi)^3 \, U_2\left(\mbox{$1 \over 2$}\vec{q_2} - \mbox{$1 \over 2$}\vec{q_1}\right) U_2\left(\mbox{$1 \over 2$}\vec{p_1} - \mbox{$1 \over 2$}\vec{p_2}\right) \nonumber
\\
 & & \times \delta^3(\vec{q_1}+\vec{q_2}-\vec{p_1}-\vec{p_2}) ,
\end{eqnarray}
\noindent which is used to calculate
\begin{eqnarray}
\langle\vec{p_i}|V_{12}|\phi\rangle & = & \int \frac{d^3\vec{q_j}}{(2 \pi)^6} \langle\vec{p_i}|V_{12}|\vec{q_j}\rangle\langle\vec{q_j}|\phi\rangle \nonumber
\\
 & = & -g_2 \: (2 \pi)^{-3} \int d^3\vec{q_j} \; U_2\left(\mbox{$1 \over 2$}\vec{q_2} - \mbox{$1 \over 2$}\vec{q_1}\right) U_2\left(\mbox{$1 \over 2$}\vec{p_1} - \mbox{$1 \over 2$}\vec{p_2}\right) \nonumber
\\
 & & \times \delta^3(\vec{q_1}+\vec{q_2}-\vec{p_1}-\vec{p_2}) \phi(\vec{q_j}) \nonumber
\\
 & = & -g_2 \: U_2\left(\mbox{$1 \over 2$}\vec{p_1} - \mbox{$1 \over 2$}\vec{p_2}\right) \int \frac{d^3\vec{q_1}}{(2 \pi)^{3}} U_2\left(\mbox{$1 \over 2$}\vec{p_1} + \mbox{$1 \over 2$}\vec{p_2} - \vec{q_1}\right) \nonumber
\\
 & & \times \phi(\vec{q_1},\vec{p_1}+\vec{p_2}-\vec{q_1}) .
\end{eqnarray}
\noindent In the center-of-mass frame $\vec{p_1} + \vec{p_2} = \vec{q_1} + \vec{q_2} = 0$, so we may write the eigenvalue equation as
\begin{equation}
\left( p_1^2 + (-p_1)^2 \right) \, \phi(\vec{p_1}, -\vec{p_1}) - g_2 \,U_2(\vec{p_1}) \, \int \frac{d^3\vec{q_1}}{(2 \pi)^3} \, U_2(\vec{q_1}) \, \phi(\vec{q_1},-\vec{q_1})  =  -B_2 \, \phi(\vec{p_1},-\vec{p_1}) .
\end{equation}
\noindent Making the substitutions $\vec{p_1} \rightarrow \vec{p}$, $\vec{q_1} \rightarrow \vec{q}$, and $\phi(\vec{q}, -\vec{q}) \rightarrow \phi(q)$ simplifies the equation further,
\begin{equation}
2 \, p^2 \, \phi(\vec{p}) - g_2 \,U_2(\vec{p}) \, \int \frac{d^3\vec{q}}{(2 \pi)^3} \, U_2(\vec{q}) \, \phi(\vec{q})  =  -B_2 \, \phi(\vec{p}) \label{eqn:regpspace}.
\end{equation}
\noindent The previous equation shows that the interaction term is of the form $\int dq \, V(p,q) \phi(q)$, where $V(p,q) \propto U_2(p) U_2(q)$.  Potentials that have the $p$ and $q$ dependence in separate factors are called ``separable potentials.''  The separable nature of this potential will be of use later, and this is why we have chosen this particular form for the regulated two-body interaction.\footnote{For an early example of the use of separable potentials in nuclear physics, see Refs.~\cite{Yamaguchi:1} and \cite{Yamaguchi:2}.}  It is important to note that how we regulate high-momentum states should not matter. This allows us to choose a regulator that is convenient for calculations.  Just as the details of the short-range interaction are unknown and unimportant, details of how we regulate the short-range interaction are not important for low-energy physics.

To find the new expression for $1/g_2$, we need only alter our previous arguments slightly.  Let
\begin{equation}
\mathcal{I}(\Lambda) = \int \frac{d^3\vec{q}}{(2 \pi)^3} U_2(\vec{q}) \, \phi(\vec{q}),
\end{equation}
\noindent so that Eq.~(\ref{eqn:regpspace}) becomes
\begin{equation}
2 \, p^2 \, \phi(\vec{p}) - g_2 \, U_2(\vec{p}) \, \mathcal{I}(\Lambda) = -B_2 \, \phi(\vec{p}),
\end{equation}
\noindent leading to
\begin{equation}
\phi(\vec{p}) = \frac{g_2 \, U_2(\vec{p}) \, \mathcal{I}(\Lambda)}{2 p^2 + B_2}.
\end{equation}
\noindent Once again, using the above form of $\phi(\vec{p})$ in the definition of $\mathcal{I}(\Lambda)$ leads to the formula for $g_2$:
\begin{eqnarray}
\frac{1}{g_2(\Lambda)} & = &  \int \frac{d^3\vec{q}}{(2 \pi)^3} \frac{U_2(\vec{q})^2}{2 q^2 + B_2} \label{eqn:g2inv}
\\
 & = & \int \frac{d^3\vec{q}}{(2 \pi)^3} \frac{\exp(-2 q^2/\Lambda^2)}{2 q^2 + B_2}.
\end{eqnarray}
The gaussian factor now renders the integration finite.  By choosing $g_2$ to satisfy this equation for any value of the cutoff, the bound-state energy $B_2$ is guaranteed to be independent of $\Lambda$.  For very large values of $\Lambda$, we may expand $g_2$ as a power series in $\sqrt{B_2}/\Lambda$:
\begin{equation}
g_2 = \frac{8 \sqrt{2} \pi^{3/2}}{\Lambda} \left( 1 + \sqrt{\pi} \frac{\sqrt{B_2}}{\Lambda} + \ldots \right) \label{eqn:g2expand}.
\end{equation}
\noindent The dimensionless two-body coupling $G_2 \equiv \Lambda \, g_2$ exhibits fixed-point behavior, approaching $8 \sqrt{2} \pi^{3/2}$ as $\Lambda \rightarrow \infty$.  The fact that this constant is not small is an indication that the problem is nonperturbative and an expansion in powers of $g_2$ is not feasible.

%%%%%%%%%%%%%%%%%%%%%%%%%%%%
%%  IRRELEVANT OPERATORS  %%
%%%%%%%%%%%%%%%%%%%%%%%%%%%%
\section{Irrelevant Operators}

As an EFT, the current theory is capable of removing all cutoff dependence from one physical observable which we have chosen to be the two-body bound-state energy.  However, any other observable quantity will contain cutoff dependence, and the accuracy can be no better that $\mathcal{O}(\sqrt{B_2}/\Lambda)$.\footnote{We will exhibit this shortly with the two-body effective range.}  Improving the accuracy and removing cutoff dependence from a second observable requires the introduction of other operators in the effective potential.

To demonstrate this, we attempt to remove the cutoff dependence from two parameters: the two-body scattering length $a$ and effective range $r_e$.  We can introduce a second operator by redefining $U_2(\vec{p})$ to be
\begin{equation}
U_2(\vec{p}) = \left(1 + h_2 \frac{p^2}{\Lambda^2}\right) \exp(-p^2/\Lambda^2) \label{eqn:U2h2}.
\end{equation}
\noindent We have chosen this form for convenience, but EFT states that the short-distance details are unimportant.  Another form could work equally well.  The steps leading to Eq.~(\ref{eqn:g2inv}) in the previous section remain unchanged, the only difference being a dependence on $h_2$ in the final result.

Roughly speaking, the introduction of this new term adds an irrelevant operator of the form $\nabla^2 \delta^3(\vec{r})$ to the effective potential with a coupling strength of $g_2 h_2$.  This operator is labelled as irrelevant because we will see that the value of $h_2$ has no effect on the leading order behavior of the two- or three-body system.

Scattering properties for two identical particles of mass $m$ \cite{Taylor:scatter,Newton:scatter} can be derived from the $T$ matrix, which is a solution to the Lippmann-Schwinger equation:
\begin{equation}
T(p',p;E) = V(p',p) + \int \frac{d^3\vec{q}}{(2 \pi)^3} \, \frac{V(p',q) T(q,p;E)}{E - q^2/m + \mi\epsilon} .
\end{equation}
\noindent The $+\mi\epsilon$ prescription implies boundary conditions for which there are no incoming spherical waves and is needed to avoid poles in the integrand.  If we use a principal value prescription instead, which corresponds to incoming and outgoing spherical waves, the solution would be the $K$ matrix:
\begin{equation}
K(p',p;E) = V(p',p) + \mathcal{P} \int \frac{d^3\vec{q}}{(2 \pi)^3} \, \frac{V(p',q) K(q,p;E)}{E - q^2/m} .
\end{equation}
\noindent The $K$ matrix contains the same scattering information as the $T$ matrix.  The reason for using the $K$ matrix is its simple relation to the effective range expansion
\begin{equation}
\frac{1}{K(p, p; p^2/m)} = - \frac{m}{4 \pi} p \cot(\delta_0(p)) = - \frac{m}{4 \pi} \left( -\frac{1}{a} + \frac{1}{2} r_e p^2 + \ldots \right) \label{eqn:Kmatrix} ,
\end{equation}
\noindent where $\delta_0(p)$ is the S-wave scattering phase shift.  Because we have a separable potential in momentum space, we can easily obtain an explicit expression for the $K$ matrix:
\begin{equation}
K(p', p; E) = - \frac{U_2(\vec{p_{~}}') U_2(\vec{p})}{D(E)},
\end{equation}
\noindent where
\begin{equation}
D(E) \equiv \frac{1}{g_2} - \mathcal{P} \int \frac{d^3\vec{q}}{(2 \pi)^3} \frac{U_2(\vec{q})^2}{2 q^2 - E} \label{eqn:defD(E)}.
\end{equation}
\noindent To relate the couplings $g_2$ and $h_2$ to the two-body parameters $a$ and $r_e$, we must expand $K(p,p;p^2/m)$ and hence $D(p^2/m)$ to the proper order in $p$.  Substituting $m = 1/2$ and writing the expansion of $D(2p^2)$ as $\mathcal{D}_0 + \mathcal{D}_1 p^2$ leads to the following expansion for $1/K$:
\begin{equation}
\frac{1}{K(p,p; 2p^2)} = -\mathcal{D}_0 - \left( \frac{2 \mathcal{D}_0 (1 - h_2)}{\Lambda^2} + \mathcal{D}_1 \right) p^2 \label{eqn:Kexpand},
\end{equation}
\noindent with
\begin{eqnarray}
\mathcal{D}_0 & = & \frac{1}{g_2} - \frac{\Lambda (3 h_2^2 + 8 h_2 + 16)}{128 \sqrt{2} \pi^{3/2}},
\\
\mathcal{D}_1 & = & - \frac{h_2^2 + 8 h_2 - 16}{32 \sqrt{2} \pi^{3/2} \Lambda}.
\end{eqnarray}
\noindent After equating Eqs.~(\ref{eqn:Kmatrix}) and (\ref{eqn:Kexpand}), some algebraic manipulation yields
\begin{equation}
\frac{1}{g_2} + \frac{1}{8 \pi a} - \frac{\Lambda (3 h_2^2 + 8 h_2 + 16)}{128 \sqrt{2} \pi^{3/2}}  =  0 \label{eqn:ar2g}.
\end{equation}
\begin{equation}
\frac{1}{4 \sqrt{2 \pi} \Lambda} h_2^2 + \left( \frac{2}{\sqrt{2 \pi} \Lambda} - \frac{2}{a \Lambda^2} \right) h_2 + \left( \frac{2}{a \Lambda^2} + \frac{1}{2} r_e - \frac{4}{\sqrt{2 \pi} \Lambda} \right)  =  0 \label{eqn:ar2h},
\end{equation}
\noindent These two equations allow us to choose the cutoff dependent forms of $g_2$ and $h_2$ in such a way that $a$ and $r_e$ are cutoff independent and match the experimental data.  In addition, the error in other $\Lambda$-dependent observable quantities will now be of $\mathcal{O}(B_2/\Lambda^2)$.  These equations also allow us to see what would happen if $h_2$ were equal to zero.  In that case, the effective range would be
\begin{equation}
r_e = \frac{8}{\sqrt{2 \pi} \Lambda} - \frac{4}{a \Lambda^2} = \frac{8}{\sqrt{2 \pi} \Lambda} \left[ 1 - \frac{\sqrt{2 \pi}}{2 a \Lambda} \right] ,
\end{equation}
\noindent showing that it now possesses cutoff dependence and vanishes as $\Lambda \rightarrow \infty$.  It also contains sub-leading corrections of $\mathcal{O}(1/a\Lambda) \sim \mathcal{O}(\sqrt{B_2}/\Lambda)$.

In order to examine the limiting behavior of the couplings, we must keep in mind one vital relation between $r_e$ and $\Lambda$.  For any finite value of $r_e$, at very large cutoffs Eq.~(\ref{eqn:ar2h}) looks like
\begin{equation}
h_2^2 + 8 h_2 + 2 \sqrt{2 \pi} \Lambda r_e -16 = 0 \label{eqn:h2largeCutoff}.
\end{equation}
\noindent There are now three cases to consider: $r_e = 0$, $r_e > 0$, and $r_e < 0$.

If $r_e = 0$, then Eq.~(\ref{eqn:h2largeCutoff}) reduces to 
\begin{equation}
h_2^2 + 8 h_2 - 16 = 0.
\end{equation}
\noindent The solution is  $h_2 = 4(-1 \pm \sqrt{2})$, indicating that $h_2$ is a fixed point.  The equation for $g_2$ has changed, but its qualitative behavior has not.  There remains a fixed point for the dimensionless coupling, but at a slightly different value:
\begin{equation}
G_2 = \frac{128 \sqrt{2} \pi^{3/2}}{16 + 8 h_2^* + 3 h_2^{*^2}}.
\end{equation}
\noindent Here, $h_2^*$ is the fixed point value of $h_2$.

If $r_e > 0$, the solution to Eq.~(\ref{eqn:h2largeCutoff}) for $h_2$ will only be real if $\Lambda \le 16/(\sqrt{2 \pi}r_e)$.  This implies that the cutoff has a limit on its value, and we cannot take the limit $\Lambda \rightarrow \infty$.  We must always have a finite value for the cutoff.

Finally, for the $r_e < 0$ case, there will always be a solution for $h_2$:
\begin{equation}
h_2 = -4 \pm \sqrt{32 - 2 \sqrt{2 \pi} r_e \Lambda}.
\end{equation}
\noindent  As $\Lambda \rightarrow \infty$, the solution for $h_2$ diverges like $\sqrt{\Lambda}$.  From Eq.~(\ref{eqn:ar2g}), we see that this behavior implies that $g_2 \rightarrow 0$.

If we had the desire to fix any more parameters in the effect range expansion, we would need more tunable couplings in our interaction.  This is easily done by extending the polynomial in Eq.~(\ref{eqn:U2h2}):
\begin{equation}
U_2(\vec{p}) = \left( 1 + h_2 \frac{p^2}{\Lambda^2} + h_2' \frac{p^4}{\Lambda^4} + \cdots \right) \exp(-p^2/\Lambda^2).
\end{equation}
\noindent Each additional term also reduces the errors in the other cutoff-dependent quantities, and there is a one-to-one correspondence between the terms in the expansion of $U_2$ and the terms in the effective range expansion in Eq.~(\ref{eqn:Kmatrix}).
\chapter{The Three-body Equation}
\label{ch:3body}

The previous chapter laid the foundation for a two-particle system.  We are now in a position to investigate the behavior of the system when a third identical boson is added.  The interaction between these particles will consist of all pair-wise two-body interactions, as well as a three-body contact interaction that takes the form of a product of two delta functions.

We begin with the position-space representation of the Hamiltonian.  The interactions are regulated by a cutoff $\Lambda$ to remove divergent behavior, with the three-body regulated interaction being modeled after the two-body version.  The bound-state equation is then converted to momentum space, where it is rewritten in the form of a one-dimensional integral equation.  At the end, we briefly discuss the necessity of including a three-body contact interaction.

%%%%%%%%%%%%%%%%%%%%
%% POSITION SPACE %%
%%%%%%%%%%%%%%%%%%%%
\section{Position Space Representation}

In position space, the three-body bound-state equation takes the form
\begin{eqnarray}
- B_3 \, \psi(\vec{r_1}, \vec{r_2}, \vec{r_3}) & = & \left[ -\nabla_1^2 - \nabla_2^2 - \nabla_3^2 \right. \nonumber
\\
 & & - \, g_2 \, \delta^3(\vec{r_1} - \vec{r_2}) - g_2 \, \delta^3(\vec{r_2} - \vec{r_3}) - g_2 \, \delta^3(\vec{r_3} - \vec{r_1}) \nonumber
\\
 & & \left. + \, g_3 \, \delta^3(\vec{r_1} - \vec{r_2}) \, \delta^3(\vec{r_2} - \vec{r_3}) \right] \, \psi(\vec{r_1}, \vec{r_2}, \vec{r_3}) \label{eqn:pos3bodyham}.
\end{eqnarray}
\noindent Here, $B_3$ is the three-body bound-state energy, and $g_3$ is the coupling strength of a three-body contact interaction which acts only when all three particles are at the same point.  The form of the three-body interaction is somewhat arbitrary.  We have chosen the product of two delta functions because it is the simplest one that is non-zero only when $\vec{r_1} = \vec{r_2} = \vec{r_3}$.  Other forms could be used, but the results will remain unchanged due to the principles of EFT.

In order to facilitate the transition to momentum space, we present a more detailed look at the position-space matrix elements of all interactions.   All two-body interactions are exactly the same as the ones used in Chapter \ref{ch:2body}, the only difference being the presence of a third position vector.  For example, the operator $V_{12}$ now has matrix elements
\begin{equation}
\langle\vec{r_1},\vec{r_2},\vec{r_3}\,|\,V_{12}\,|\,\vec{r_1}',\vec{r_2}',\vec{r_3}'\rangle = \langle\vec{r_1}, \vec{r_2}\,|\,V_{12}\,|\,\vec{r_1}', \vec{r_2}'\rangle \, \delta^3(\vec{r_3} - \vec{r_3}') ,
\end{equation}
\noindent where $\langle\vec{r_1}, \vec{r_2}\,|\,V_{12}\,|\,\vec{r_1}', \vec{r_2}'\rangle$ is defined in Eq.~(\ref{eqn:V12matrix}).  The addition of a third particle adds an extra delta function.  The expressions for $V_{23}$ and $V_{31}$ are obtained by a simple permutation of indices.  Because the two-body interaction remains unchanged, so does the renormalization of $g_2$.  Its behavior is fixed by the two-body sector and carried over directly to the three-body sector.

The three-body contact interaction, $V_{123}$, will have matrix elements of the form
\begin{equation}
\langle\vec{r_i}\,|\,V_{123}\,|\,\vec{r_j}'\rangle \; = \: g_3 \int d^3\vec{r_c} \prod_{i,j=1}^{3} \delta^3(\vec{r_i}-\vec{r_c}) \delta^3(\vec{r_j}'-\vec{r_c}) .
\end{equation}
\noindent This is chosen so that
\begin{eqnarray}
\lefteqn{\langle\vec{r_i}\,|\,V_{123}\,|\psi\rangle \;  = \: \int d^3\vec{r_j}' \langle\vec{r_i}\,|\,V_{123}\,|\,\vec{r_j}'\rangle \langle\vec{r_j}' | \psi \rangle } \nonumber
\\
& = & g_3 \int d^3\vec{r_j}'\: d^3\vec{r_c} \left[ \prod_{i,j=1}^{3} \delta^3(\vec{r_i}-\vec{r_c}) \delta^3(\vec{r_j}'-\vec{r_c}) \right] \psi(\vec{r_1}', \vec{r_2}', \vec{r_3}') \nonumber
\\
& = & g_3 \int d^3\vec{r_c} \; \delta^3(\vec{r_1}-\vec{r_c}) \delta^3(\vec{r_2}-\vec{r_c}) \delta^3(\vec{r_3}-\vec{r_c}) \psi(\vec{r_c}, \vec{r_c}, \vec{r_c}) \nonumber
\\
& = & g_3 \, \delta^3(\vec{r_1}-\vec{r_2}) \delta^3(\vec{r_2}-\vec{r_3}) \psi(\vec{r_1}, \vec{r_2}, \vec{r_3}).
\end{eqnarray}
\noindent We know from the previous chapter that this potential must be regulated to isolate and remove the spurious short-range features that cause divergences.  This is done in a manner similar to the two-body case.  We define a new set of functions  
\begin{eqnarray}
U_3(\vec{p}) & = & \left( 1 + h_3 \frac{p^2}{\Lambda^2} \right) \exp(-p^2/\Lambda^2) ,
\\
\tilde{U}_3(\vec{r}) & = & \int \frac{d^3\vec{p}}{(2 \pi)^3} \exp(i \vec{p} \cdot \vec{r}) \, U_3(\vec{p}) ,
\end{eqnarray}
\noindent to regulate the three-body interaction.  The regulated matrix elements of $V_{123}$ become
\begin{equation}
\langle\vec{r_i}\,|\,V_{123}\,|\,\vec{r_j}'\rangle \; = \: g_3 \int d^3\vec{r_c} \prod_{i,j=1}^{3} \tilde{U}_3(\vec{r_i}-\vec{r_c}) \tilde{U}_3(\vec{r_j}'-\vec{r_c}) .
\end{equation}
\noindent The parameter $h_3$ plays a role analogous to $h_2$, allowing the addition of a second three-body operator.  We can use $h_3$ to tune another three-body observable in much the same way that $h_2$ tunes the two-body effective range.  In both cases, the additional contribution is suppressed like an irrelevant operator.  Because of this, we will see that $h_3$ can be set to zero in the leading order calculation.  It is included here for completeness.

%%%%%%%%%%%%%%%%%%%%
%% MOMENTUM SPACE %%
%%%%%%%%%%%%%%%%%%%%
\section{Momentum Space Representation}

Knowing the complete regulated position-space representation of the Hamiltonian, we can derive the momentum-space Schr\"odinger equation that serves as the basis for all of our calculations.   The left-hand side of Eq.~(\ref{eqn:pos3bodyham}) and the kinetic terms on the right side are easily converted:
\begin{eqnarray}
- B_3 \, \psi(\vec{r_1}, \vec{r_2}, \vec{r_3}) & \rightarrow & -B_3 \, \phi(\vec{p_1},\vec{p_2},\vec{p_3}),
\\
\left( -\nabla_1^2 - \nabla_2^2 - \nabla_3^2 \right) \psi(\vec{r_1}, \vec{r_2}, \vec{r_3}) & \rightarrow & (p_1^2 + p_2^2 + p_3^2) \, \phi(\vec{p_1},\vec{p_2},\vec{p_3}).
\end{eqnarray}
\noindent For the two-body interaction $V_{12}$, the momentum-space matrix elements are
\begin{eqnarray}
\langle\vec{p_i}\,|\,V_{12}\,|\,\vec{q_j}\rangle & = & \int d^3\vec{r_i} \: d^3\vec{r_j}' \langle\vec{p_i}|\vec{r_i}\rangle \langle\vec{r_i}\,|\,V_{12}\,|\,\vec{r_j}'\rangle \langle\vec{r_j}'|\vec{q_j}\rangle \nonumber
\\
 & = & -g_2 \int d^3\vec{r_i} \: d^3\vec{r_j}' \exp(-i \vec{p_i}\cdot\vec{r_i}) \exp(i \vec{q_j}\cdot\vec{r_j}') \tilde{U}_2(\vec{r_1}-\vec{r_2}) \nonumber
\\
 & & \times \tilde{U}_2(\vec{r_1}'-\vec{r_2}') \delta^3 \left( \frac{\vec{r_1}+\vec{r_2}}{2} - \frac{\vec{r_1}'+\vec{r_2}'}{2} \right) \delta^3(\vec{r_3}-\vec{r_3}') \nonumber
\\
 & = & -g_2 \int \frac{d^3\vec{k}}{(2 \pi)^3} \frac{d^3\vec{k}'}{(2 \pi)^3} U_2(\vec{k}) U_2(\vec{k}') \int d^3\vec{r_i} \: d^3\vec{r_j}' \exp(-i \vec{p_i}\cdot\vec{r_i}) \nonumber
\\
 & & \times \exp(i \vec{q_j}\cdot\vec{r_j}') \exp \left(i \vec{k}\cdot(\vec{r_1}-\vec{r_2}) \right) \exp \left( i \vec{k}'\cdot(\vec{r_1}'-\vec{r_2}') \right) \nonumber
\\
 & & \times \delta^3 \left( \frac{\vec{r_1}+\vec{r_2}}{2} - \frac{\vec{r_1}'+\vec{r_2}'}{2} \right) \delta^3(\vec{r_3}-\vec{r_3}') \nonumber
\\
& = &  -g_2 \int \frac{d^3\vec{k}}{(2 \pi)^3} \frac{d^3\vec{k}'}{(2 \pi)^3} U_2(\vec{k}) U_2(\vec{k}') \int d^3\vec{r_i} \: d^3\vec{r_1}' \exp(-i \vec{p_i}\cdot\vec{r_i}) \nonumber
\\
 & & \times \exp \left( \vec{q_1}\cdot\vec{r_1}' + \vec{q_2}\cdot(\vec{r_1}+\vec{r_2}-\vec{r_1}') + \vec{q_3}\cdot\vec{r_3} + \vec{k}\cdot(\vec{r_1}-\vec{r_2}) \right. \nonumber
\\
 & & +~ \left. \vec{k}'\cdot(2\vec{r_1}'-\vec{r_1}-\vec{r_2}) \right) \nonumber
\\
 & = & -g_2 \int \frac{d^3\vec{k}}{(2 \pi)^3} \frac{d^3\vec{k}'}{(2 \pi)^3} U_2(\vec{k}) U_2(\vec{k}') (2 \pi)^{12} \, \delta^3(-\vec{p_1}+\vec{q_2}+\vec{k}-\vec{k}') \nonumber
\\
 & & \times \delta^3(-\vec{p_2}+\vec{q_2}-\vec{k}-\vec{k}') \delta^3(\vec{q_1}-\vec{q_2}+2\vec{k}') \delta^3(-\vec{p_3}+\vec{q_3}) \nonumber
\\
 & = & -g_2 \int \frac{d^3\vec{k}}{(2 \pi)^3} U_2(\vec{k}) U_2\left(\vec{q_2}/2 - \vec{q_1}/2\right) (2 \pi)^{9} \, \delta^3(-\vec{p_1}+\mbox{$1 \over 2$}\vec{q_1} + \mbox{$1 \over 2$}\vec{q_2}+\vec{k}) \nonumber
\\
 & & \times \delta^3(-\vec{p_2}+\mbox{$1 \over 2$}\vec{q_1} + \mbox{$1 \over 2$}\vec{q_2}-\vec{k}) \delta^3(-\vec{p_3}+\vec{q_3}) \nonumber
\\
 & = & -g_2 \, (2 \pi)^6 \, U_2\left(\mbox{$1 \over 2$}\vec{q_2} - \mbox{$1 \over 2$}\vec{q_1}\right) U_2\left(\mbox{$1 \over 2$}\vec{p_1} - \mbox{$1 \over 2$}\vec{p_2}\right) \delta^3(-\vec{p_3}+\vec{q_3}) \nonumber
\\
 & & \times \delta^3(\vec{q_1}+\vec{q_2}-\vec{p_1}-\vec{p_2}),
\end{eqnarray}
\noindent which is used to calculate
\begin{eqnarray}
\langle\vec{p_i}|V_{12}|\phi\rangle & = & \int \frac{d^3\vec{q_j}}{(2 \pi)^9} \langle\vec{p_i}|V_{12}|\vec{q_j}\rangle\langle\vec{q_j}|\phi\rangle \nonumber
\\
 & = & -g_2 \: (2 \pi)^{-3} \int d^3\vec{q_j} \; U_2\left(\mbox{$1 \over 2$}\vec{q_2} - \mbox{$1 \over 2$}\vec{q_1}\right) U_2\left(\mbox{$1 \over 2$}\vec{p_1} - \mbox{$1 \over 2$}\vec{p_2}\right) \nonumber
\\
 & & \times \delta^3(-\vec{p_3}+\vec{q_3}) \delta^3(\vec{q_1}+\vec{q_2}-\vec{p_1}-\vec{p_2}) \phi(\vec{q_j}) \nonumber
\\
 & = & -g_2 \: U_2\left(\mbox{$1 \over 2$}\vec{p_1} - \mbox{$1 \over 2$}\vec{p_2}\right) \int \frac{d^3\vec{q_1}}{(2 \pi)^{3}} U_2\left(\mbox{$1 \over 2$}\vec{p_1} + \mbox{$1 \over 2$}\vec{p_2} - \vec{q_1}\right) \nonumber
\\
 & & \times \phi(\vec{q_1},\vec{p_1}+\vec{p_2}-\vec{q_1},\vec{p_3}) .
\end{eqnarray}
\noindent  Because the bound state consists of identical bosons, in the center-of-mass frame we know that $\vec{p_1} + \vec{p_2} + \vec{p_3} = 0$ and $\phi(\vec{q_1},\vec{p_1}+\vec{p_2}-\vec{q_1},\vec{p_3}) = \phi(\vec{q_1},\vec{p_3},\vec{p_1}+\vec{p_2}-\vec{q_1})$.  This allows us to write
\begin{equation}
\langle\vec{p_i}|V_{12}|\phi\rangle \: = -g_2 \, U_2\left(\mbox{$1 \over 2$}\vec{p_1} - \mbox{$1 \over 2$}\vec{p_2}\right) \int \frac{d^3\vec{q_1}}{(2 \pi)^{3}} U_2\left(\vec{q_1} + \mbox{$1 \over 2$}\vec{p_3}\right) \phi(\vec{q_1},\vec{p_3},-\vec{q_1}-\vec{p_3}) \label{eqn:V12phi}.
\end{equation}
  For reasons that will become apparent shortly, we define the following ``pseudo-wavefunction'':
\begin{equation}
\Phi(\vec{p}) \equiv \int \frac{d^3\vec{q}}{(2 \pi)^3} U_2(\vec{q} + \mbox{$1 \over 2$}\vec{p}) \, \phi(\vec{q},\vec{p},-\vec{q}-\vec{p})\label{eqn:PHIdef}.
\end{equation}
\noindent Equation (\ref{eqn:V12phi}) is now written as
\begin{equation}
\langle\vec{p_i}|V_{12}|\phi\rangle \: = -g_2 \, U_2\left(\mbox{$1 \over 2$}\vec{p_1} - \mbox{$1 \over 2$}\vec{p_2}\right) \Phi(\vec{p_3}),
\end{equation}
with the other two-body interaction terms being obtained by a cyclic permutation of indices:
\begin{equation}
\langle\vec{p_i}|V_{23}|\phi\rangle \: = -g_2 \, U_2\left(\mbox{$1 \over 2$}\vec{p_2} - \mbox{$1 \over 2$}\vec{p_3}\right) \Phi(\vec{p_1}) = -g_2 \, U_2\left(\vec{p_2} + \mbox{$1 \over 2$}\vec{p_1}\right) \Phi(\vec{p_1})
\end{equation}
\begin{equation}
\langle\vec{p_i}|V_{31}|\phi\rangle \: = -g_2 \, U_2\left(\mbox{$1 \over 2$}\vec{p_3} - \mbox{$1 \over 2$}\vec{p_1}\right) \Phi(\vec{p_2}) = -g_2 \, U_2\left(\vec{p_1} + \mbox{$1 \over 2$}\vec{p_2}\right) \Phi(\vec{p_2})
\end{equation}
\noindent Turning to the three-body interaction, the matrix elements are
\begin{eqnarray}
\langle\vec{p_i}\,|\,V_{123}\,|\,\vec{q_j}\rangle & = & \int d^3\vec{r_i} \: d^3\vec{r_j}' <\vec{p_i}|\vec{r_i}> <\vec{r_i}\,|\,V_{123}\,|\,\vec{r_j}'> <\vec{r_j}'|\vec{q_j}> \nonumber
\\
 & = & g_3 \int d^3\vec{r_i} \: d^3\vec{r_j}' \exp(-i \vec{p_i}\cdot\vec{r_i}) \exp(i \vec{q_j}\cdot\vec{r_j}') \nonumber
\\
 & & \times \int d^3\vec{r_c} \; \prod_{i,j = 1}^{3} \tilde{U}_3(\vec{r_i}-\vec{r_c}) \, \tilde{U}_3(\vec{r_j}'-\vec{r_c}) \nonumber
\\
 & = & g_3 \int \frac{d^3\vec{k_i}}{(2 \pi)^9} \frac{d^3\vec{k_j}'}{(2 \pi)^9} \int d^3\vec{r_i} \: d^3\vec{r_j}' \: d^3\vec{r_c} \exp(-i \vec{p_i}\cdot\vec{r_i}) \exp(i \vec{q_j}\cdot\vec{r_j}') \nonumber
\\
 & & \times \exp\left(i \vec{k_i}\cdot(\vec{r_i}-\vec{r_c})\right) 
\exp\left(i \vec{k_j}'\cdot(\vec{r_j}'-\vec{r_c})\right) U_3(\vec{k_i}) U_3(\vec{k_j}')
\nonumber
\\
 & = & g_3 \int d^3\vec{k_i} \: d^3\vec{k_j}' U_3(\vec{k_i}) U_3(\vec{k_j}') (2 \pi)^{3} \: \delta^3(-\vec{p_1}+\vec{k_1}) \delta^3(-\vec{p_2}+\vec{k_2}) \nonumber
\\
 & & \times \delta^3(-\vec{p_3}+\vec{k_3}) \delta^3(\vec{q_1}+\vec{k_1}') \delta^3(\vec{q_2}+\vec{k_2}') \delta^3(\vec{q_3}+\vec{k_3}') \nonumber
\\
 & & \times \delta^3(\vec{k_1}+\vec{k_2}+\vec{k_3}+\vec{k_1}'+\vec{k_2}'+\vec{k_3}') \nonumber
\\
 & = & g_3 \, (2 \pi)^3 \: U_3(\vec{p_i}) U_3(\vec{q_j}) \delta^3(\vec{p_1}+\vec{p_2}+\vec{p_3}-\vec{q_1}-\vec{q_2}-\vec{q_3}).
\end{eqnarray}
\noindent This implies that
\begin{eqnarray}
\langle\vec{p_i}|V_{123}|\phi\rangle & = & \int \frac{d^3\vec{q_j}}{(2 \pi)^9} \langle\vec{p_i}|V_{123}|\vec{q_j}\rangle\langle\vec{q_j}|\phi\rangle \nonumber
\\
 & = & g_3 \, U_3(\vec{p_1}) U_3(\vec{p_2}) U_3(\vec{p_3}) \int \frac{d^3\vec{q_1}}{(2 \pi)^3} \frac{d^3\vec{q_2}}{(2 \pi)^3} U_3(\vec{q_1}) U_3(\vec{q_2}) \nonumber
\\
 & & \times U_3\left(\sum_{i=1}^{3}\vec{p_i}-\vec{q_1}-\vec{q_2}\right) \phi\left(\vec{q_1}, \vec{q_2}, \sum_{i=1}^{3}\vec{p_i}-\vec{q_1}-\vec{q_2}\right) \nonumber
\\
 & = & g_3 \, U_3(\vec{p_1}) U_3(\vec{p_2}) U_3(\vec{p_3}) \int \frac{d^3\vec{q_1}}{(2 \pi)^3} \frac{d^3\vec{q_2}}{(2 \pi)^3} U_3(\vec{q_1}) U_3(\vec{q_2}) \nonumber
\\
& & \times U_3(-\vec{q_1}-\vec{q_2}) \phi(\vec{q_1}, \vec{q_2}, -\vec{q_1}-\vec{q_2}) \label{eqn:V123phi},
\end{eqnarray}
\noindent where the last step uses the fact that $\sum_{i=1}^{3}\vec{p_i} = 0$.  If we define
\begin{equation}
\Phi_1 \equiv \int \frac{d^3\vec{q_1}}{(2 \pi)^3} \frac{d^3\vec{q_2}}{(2 \pi)^3} U_3(\vec{q_1}) U_3(\vec{q_2}) U_3(-\vec{q_1}-\vec{q_2}) \phi(\vec{q_1}, \vec{q_2}, -\vec{q_1}-\vec{q_2}),
\end{equation}
\noindent then Eq.~(\ref{eqn:V123phi}) may be written simply as
\begin{equation}
\langle\vec{p_i}|V_{123}|\phi\rangle \: = g_3 \, U_3(\vec{p_1}) U_3(\vec{p_2}) U_3(\vec{p_3}) \Phi_1.
\end{equation}
\noindent Overall, the momentum-space three-body bound-state equation takes the form
\begin{eqnarray}
-B_3 \, \phi(\vec{p_1},\vec{p_2},\vec{p_3}) & = & (p_1^2 + p_2^2 + p_3^2) \phi(\vec{p_1},\vec{p_2},\vec{p_3}) - g_2 \, U_2\left(\vec{p_2} + \mbox{$1 \over 2$}\vec{p_1}\right) \Phi(\vec{p_1}) \nonumber
\\
 & & -~g_2 \, U_2\left(\vec{p_1} + \mbox{$1 \over 2$}\vec{p_2}\right) \Phi(\vec{p_2}) - g_2 \, U_2\left(\mbox{$1 \over 2$}\vec{p_1} - \mbox{$1 \over 2$}\vec{p_2}\right) \Phi(\vec{p_3}) \nonumber
\\
 & & +~g_3 \, U_3(\vec{p_1}) U_3(\vec{p_2}) U_3(\vec{p_3}) \Phi_1 \label{eqn:pre3body}.
\end{eqnarray}

%%%%%%%%%%%%%%%%%%%%%%%%%%%
%%  PSEUDO-WAVEFUNCTION  %%
%%%%%%%%%%%%%%%%%%%%%%%%%%%
\section{Elimination of Wavefunction}

The bound-state equation can be simplified further by choosing to work entirely with the pseudo-wavefunction, $\Phi(\vec{p})$, rather than the wavefunction $\phi(\vec{p_1},\vec{p_2},\vec{p_3})$.  Because we limit our investigation to S-wave bound states, $\Phi(\vec{p})$ is only a function of the momentum's magnitude, not direction.  The result is a one-dimensional integral equation, as opposed to the multi-dimensional equation required for $\phi(\vec{p_1},\vec{p_2},\vec{p_3})$.

To eliminate the wavefunction, we solve (\ref{eqn:pre3body}) for $\phi$ and substitute the result into (\ref{eqn:PHIdef}):
\begin{eqnarray}
\Phi(\vec{p}\,) & = & \int \frac{d^3\vec{q}}{(2 \pi)^3} \: \frac{U_2(\vec{q}+\frac{1}{2}\vec{p}\,)}{p^2 + q^2 + (\vec{p}+\vec{q}\,)^2 + B_3} \left[ g_2 \, U_2\left(\vec{p} + \mbox{$1 \over 2$}\vec{q}\right) \Phi(\vec{q}\,) \right. \nonumber
\\
 & & +~g_2 \, U_2\left(\vec{q} + \mbox{$1 \over 2$}\vec{p}\right) \Phi(\vec{p}) + g_2 \, U_2\left(\mbox{$1 \over 2$}\vec{q} - \mbox{$1 \over 2$}\vec{p}\right) \Phi(-\vec{p}-\vec{q}\,) \nonumber
\\
 & & \left. -~g_3 \, U_3(\vec{q}\,) U_3(\vec{p}\,) U_3(-\vec{p}-\vec{q}\,) \Phi_1 \right],
\end{eqnarray}
\noindent which can be rewritten as
\begin{eqnarray}
\lefteqn{ \left[ 1 - g_2 \int \frac{d^3\vec{q}}{(2 \pi)^3} \frac{U_2(\vec{q}+\frac{1}{2}\vec{p}\,)^2}{p^2 + q^2 + (\vec{p}+\vec{q}\,)^2 + B_3} \right] \Phi(\vec{p}\,) = } \nonumber \hspace{2.5in}
\\
&& \hspace{-1.5in} g_2 \int \frac{d^3\vec{q}}{(2 \pi)^3} \: \frac{U_2(\vec{q}+\frac{1}{2}\vec{p}\,) U_2(\vec{p}+\frac{1}{2}\vec{q}\,)}{p^2 + q^2 + (\vec{p}+\vec{q}\,)^2 + B_3} \, \Phi(\vec{q}\,) \nonumber
\\
&& \hspace{-1.5in} {} + g_2 \int \frac{d^3\vec{q}}{(2 \pi)^3} \: \frac{U_2(\vec{q}+\frac{1}{2}\vec{p}\,) U_2(\frac{1}{2}\vec{q}-\frac{1}{2}\vec{p}\,)}{p^2 + q^2 + (\vec{p}+\vec{q}\,)^2 + B_3} \, \Phi(-\vec{p}-\vec{q}\,) \nonumber
\\
&& \hspace{-1.5in} {} - g_3  \int \frac{d^3\vec{q}}{(2 \pi)^3} \: \frac{U_2(\vec{q}+\frac{1}{2}\vec{p}\,) U_3(\vec{q}) U_3(\vec{p}\,) U_3(-\vec{p}-\vec{q}\,)}{p^2 + q^2 + (\vec{p}+\vec{q}\,)^2 + B_3} \, \Phi_1 \label{eqn:step2}.
\end{eqnarray}
\noindent By performing the change of variables $\vec{q} \rightarrow \vec{q} - \frac{1}{2}\vec{p}$, the left-hand side of the previous equation becomes
\begin{equation}
\left[ 1 - g_2 \int \frac{d^3\vec{q}}{(2 \pi)^3} \, \frac{U_2(\vec{q}\,)^2}{2q^2 + \frac{3}{2}p^2 + B_3} \right] \Phi(\vec{p}\,) = g_2 \, D\left(-\mbox{$3 \over 2$}p^2 - B_3\right) \Phi(\vec{p}\,) ,
\end{equation}
\noindent where $D(E)$ is defined in Eq.~(\ref{eqn:defD(E)}).

Next, consider the second integral on the right-hand side of Eq.~(\ref{eqn:step2}).  The change of variables $\vec{q} \rightarrow -\vec{q}-\vec{p}$ allows us to rewrite the integral as
\begin{equation}
g_2 \int \frac{d^3\vec{q}}{(2 \pi)^3} \, \frac{U_2(-\vec{q}-\frac{1}{2}\vec{p}\,) U_2(-\frac{1}{2}\vec{q}-\vec{p}\,)}{2p^2 + 2q^2 + 2\vec{p}\cdot\vec{q} + B_3} \, \Phi(\vec{q}\,) = \int \frac{d^3\vec{q}}{(2 \pi)^3} \, \frac{U_2(\vec{q}+\frac{1}{2}\vec{p}\,) U_2(\vec{p}+\frac{1}{2}\vec{q}\,)}{p^2 + q^2 + (\vec{p}+\vec{q}\,)^2 + B_3} \, \Phi(\vec{q}\,) ,
\end{equation}
\noindent which is exactly the same form as the first integral on the right-hand side of the same equation.  The two can therefore be combined to simplify the equation.  

Finally, we define
\begin{equation}
D_1(\vec{p}\,) \equiv \int \frac{d^3\vec{q}}{(2 \pi)^3} \, \frac{U_2(\vec{q}+\frac{1}{2}\vec{p}\,) U_3(\vec{q}\,) U_3(\vec{p}\,) U_3(-\vec{p}-\vec{q}\,)}{2p^2 + 2q^2 + 2\vec{p}\cdot\vec{q} + B_3}
\end{equation}
\noindent so that the three-body interaction in Eq.~(\ref{eqn:step2}) can be written as $-g_3 \, D_1(p) \, \Phi_1$.  Dropping any unnecessary vector notation, the bound-state equation takes its final form
\begin{eqnarray}
\Phi(p) & = & \frac{2}{D\left(-\frac{3}{2}p^2-B_3\right)} \int_0^{\infty} \frac{q^2 dq}{4 \pi^2} \int_{-1}^{1} dz \: \frac{U_2(\vec{q}+\frac{1}{2}\vec{p}\,) U_2(\vec{p}+\frac{1}{2}\vec{q}\,)}{2p^2 + 2q^2 + 2pqz + B_3} \Phi(q) \nonumber
\\
&& - \frac{g_3 \, D_1(p) \, \Phi_1}{g_2 \, D\left(-\frac{3}{2}p^2-B_3\right)} \label{eqn:final3body},
\end{eqnarray}
\noindent where $\vec{p}\cdot\vec{q} = pqz$. The definitions for quantities in the above equation are reproduced below:
\begin{eqnarray}
\Phi(p) & = & \int \frac{d^3\vec{q}}{(2 \pi)^3} \, U_2(\vec{q} + \mbox{$1 \over 2$}\vec{p}\,) \phi(\vec{q},\vec{p},-\vec{q}-\vec{p}\,) ,
\\
D(E) & = & \frac{1}{g_2} - \mathcal{P} \int \frac{d^3\vec{q}}{(2 \pi)^3} \, \frac{U_2(q)^2}{2q^2 - E} ,
\\
D_1(p) & = & \frac{1}{4 \pi^2} \int_0^{\infty} q^2 dq \int_{-1}^{1} dz \: \frac{U_2(\vec{q}+\frac{1}{2}\vec{p}\,) U_3(q) U_3(p) U_3(\vec{p}+\vec{q}\,)}{2p^2 + 2q^2 + 2pqz + B_3} ,
\\
\Phi_1 & = & \int \frac{d^3\vec{q}}{(2 \pi)^3} \frac{d^3\vec{p}}{(2 \pi)^3} \, U_3(q) U_3(p) U_3(\vec{q}+\vec{p}\,) \phi(\vec{q}, \vec{p}, -\vec{q}-\vec{p}\,) \label{eqn:defPhi1}.
\end{eqnarray}
\noindent At this point, the only thing preventing the exclusive use of $\Phi(p)$ in our calculations is the fact that the definition of $\Phi_1$ still involves $\phi$.  To remove this dependence, we insert the value of $\phi$ obtained from Eq.~(\ref{eqn:pre3body}) into (\ref{eqn:defPhi1}):
\begin{eqnarray}
\Phi_1 & = & \int \frac{d^3\vec{q}}{(2 \pi)^3} \frac{d^3\vec{p}}{(2 \pi)^3} \: \frac{U_3(q) U_3(p) U_3(\vec{q}+\vec{p}\,)}{2p^2 + 2q^2 + 2pqz + B_3} \bigg[ g_2 \, U_2(\vec{p} + \mbox{$1 \over 2$}\vec{q}\,) \Phi(q) \nonumber
\\
&& + g_2 \, U_2(\vec{q} + \mbox{$1 \over 2$}\vec{p}\,) \Phi(p) + g_2 \, U_2(\mbox{$1 \over 2$}\vec{q} - \mbox{$1 \over 2$}\vec{p}\,) \Phi(-\vec{p}-\vec{q}\,) \nonumber
\\
&& - g_3 \, U_3(q) U_3(p) U_3(\vec{q}+\vec{p}\,) \Phi_1 \bigg] \label{eqn:Phi1step2}.
\end{eqnarray}
\noindent Upon defining
\begin{equation}
D_2 \equiv \frac{1}{8 \pi^4} \int_0^{\infty} p^2 dp \int_0^{\infty} q^2 dq \int_{-1}^{1} dz \: \frac{U_3(q)^2 U_3(p)^2 U_3(\vec{q}+\vec{p}\,)^2}{2p^2 + 2q^2 + 2pqz + B_3},
\end{equation}
\noindent the $g_3$ term reduces to $-g_3 \, D_2 \, \Phi_1$.  The three $g_2$ terms in (\ref{eqn:Phi1step2}) can all be written in the same form, provided that we swap $\vec{p}$ and $\vec{q}$ in the second integral, and make the variable substitution $\vec{p} \rightarrow -\vec{p}-\vec{q}$ in the third.  This common form is
\begin{equation}
g_2 \int \frac{d^3\vec{q}}{(2 \pi)^3} \Phi(q) \int \frac{d^3\vec{p}}{(2 \pi)^3} \, \frac{U_3(q) U_3(p) U_3(\vec{q}+\vec{p}\,) U_2(\vec{p}+\frac{1}{2}\vec{q}\,)}{2p^2 + 2q^2 + 2pqz + B_3} = \frac{g_2}{2 \pi^2} \int_0^{\infty} dq \, q^2 \, D_1(q) \Phi(q).
\end{equation}
\noindent The equation for $\Phi_1$ can now be cast entirely in terms of $\Phi(p)$:
\begin{equation}
\Phi_1 = \frac{3 g_2}{2 \pi^2} \int_0^{\infty} dq \left[ q^2 D_1(q) \Phi(q) \right] - g_3 \, D_2 \Phi_1.
\end{equation}
\noindent Note that the quantities $\Phi(p)$, $\Phi_1$, $D_1(p)$, and $D_2$ contain dependence on $B_2$, $B_3$, and/or $\Lambda$ even though it is not explicitly indicated.

For the reader's convenience, the three-body bound state equation is given below along with any necessary supplementary equations.

\vspace{\baselineskip}
\fbox{
\begin{minipage}[t][3.3in][t]{5.5in}
\begin{eqnarray}
\Phi(p) & = & \frac{2}{D\left(-\frac{3}{2}p^2-B_3\right)} \int_0^{\infty} \frac{q^2 dq}{4 \pi^2} \int_{-1}^{1} dz \: \frac{U_2(\vec{q}+\frac{1}{2}\vec{p}\,) U_2(\vec{p}+\frac{1}{2}\vec{q}\,)}{2p^2 + 2q^2 + 2pqz + B_3} \Phi(q) \nonumber
\\
&& - \frac{g_3 \, D_1(p) \, \Phi_1}{g_2 \, D\left(-\frac{3}{2}p^2-B_3\right)} \nonumber
\\
D(E) & = & \frac{1}{g_2} - \mathcal{P} \int \frac{d^3\vec{q}}{(2 \pi)^3} \, \frac{U_2(q)^2}{2q^2 - E} \nonumber
\\
D_1(p) & = & \frac{1}{4 \pi^2} \int_0^{\infty} q^2 dq \int_{-1}^{1} dz \: \frac{U_2(\vec{q}+\frac{1}{2}\vec{p}\,) U_3(q) U_3(p) U_3(\vec{p}+\vec{q}\,)}{2p^2 + 2q^2 + 2pqz + B_3} \nonumber
\\
\Phi_1 & = & \frac{3 g_2}{2 \pi^2} \int_0^{\infty} dq \left[ q^2 D_1(q) \Phi(q) \right] - g_3 \, D_2 \Phi_1 \nonumber
\\
D_2 & = & \frac{1}{8 \pi^4} \int_0^{\infty} p^2 dp \int_0^{\infty} q^2 dq \int_{-1}^{1} dz \: \frac{U_3(q)^2 U_3(p)^2 U_3(\vec{q}+\vec{p}\,)^2}{2p^2 + 2q^2 + 2pqz + B_3} \nonumber
\end{eqnarray}
\end{minipage}
}

%%%%%%%%%%%%%%%%%%%%%%%
%%  NECESSITY OF g3  %%
%%%%%%%%%%%%%%%%%%%%%%%
\section{Necessity of Three-Body Interaction}

Up to this point, we have implicitly assumed that a three-body contact interaction is a necessary part of the bound-state equation without justifying why this may or may not be true.  In the absence of $g_3$, the bound-state equation involves only the quantities $B_2$, $B_3$, and $\Lambda$.\footnote{It also contains $g_2$, which is constrained by the two-body equation to be a function of $B_2$ and $\Lambda$.}  If the two-body equation has been renormalized by fixing $B_2$, then the only adjustable parameters are $B_3$ and $\Lambda$.  For any fixed value of $\Lambda$, the three-body bound-state equation becomes an eigenvalue equation for $B_3$.  This equation determines the entire discrete bound-state spectrum.  Fitting $B_3$ to experimental data without $g_3$ would require fine-tuning $\Lambda$ to an appropriate value.  The three-body bound-state energy would be inherently tied to the value of $\Lambda$.  For the additional case of $B_2 = 0$, the equation requires that $B_3/\Lambda^2$ be a constant, causing the three-body bound-state energies to diverge as $\Lambda^2$.\footnote{Taking the limit $\Lambda \rightarrow \infty$ results in a three-body spectrum that is unbounded from below.  This is exactly the behavior Thomas found for zero-range forces \cite{Thomas:shorta}.}

To remove cutoff dependence, we introduce $g_3$.  By allowing $g_3$ to be a function of $\Lambda$, any change in the cutoff can be compensated by a change in the three-body coupling, keeping the bound-state energy $B_3$ fixed.  Notice, however, that only one bound-state energy can be made cutoff independent in this manner.  Other binding energies in the spectrum will contain $\Lambda$ dependence.  Our ultimate goal is to study this dependence, and in the next chapter we develop a perturbative expansion technique for doing so.
\chapter{Perturbative Uniform Expansion Method}
\label{ch:pertmethod}

Our goal is to expand the three-body bound-state equation about an infinite cutoff and then study finite cutoff effects perturbatively.  This first requires a zeroth order solution to Eq.~(\ref{eqn:final3body}).  We assume that $B_2$ and $B_3$ are much less than $\Lambda^2$, making an expansion in powers of $\sqrt{B_2}/\Lambda$ and $\sqrt{B_3}/\Lambda$ possible.  However, the momentum covers the entire range of values from 0 to infinity.  If $p \sim \sqrt{B_3} \ll \Lambda$, we can obviously use $p/\Lambda$ as another expansion parameter.  But how should we handle the expansion when $\sqrt{B_3} \ll p \ll \Lambda$ or $\sqrt{B_3} \ll p \sim \Lambda$?  And how can we expand a function that will end up being integrated over all three regions?

The solution is to separately expand the function $\Phi(p)$ in each of the above regions.  Together, the three approximations can be combined into one function that is uniformly valid. This means that for any value of $p$ the approximation will be accurate to some chosen order in $\sqrt{B_2}/\Lambda$ or $\sqrt{B_3}/\Lambda$.  Although it may appear we are complicating matters by trading one integral equation for three coupled integral equations, we will see that to leading order the set of integral equations becomes uncoupled.  One of the equations can be solved analytically, while the other two must be solved numerically.  Because the expanded equations are simpler than the original, we can obtain highly accurate numerical results.  The high accuracy is necessary to ensure that any corrections to the leading order equations do not become obscured by numerical errors.  For certain cases of atomic scattering, this could be $\mathcal{O}(10^{-6})$, or even smaller if there is an external magnetic field tuned to a Feshbach resonance \cite{Verhaar:1,Verhaar:2}.

Towards this end, we develop a method for finding the {\em perturbative uniform expansion} of any function containing two widely separated energy scales, $\eta$ and $\Lambda$, such that $\eta \ll \Lambda$.\footnote{The procedure outlined here was provided by Kenneth Wilson. It is unknown to the author whether this represents original work.}  The following sections provide a general prescription for carrying out this expansion to any order in $\eta/\Lambda$.  It is applicable to functions involving one or two additional momentum variables, as well as integrals of such functions.

%%%%%%%%%%%%%%%%%%%%%%%%%%%%%%
%  THREE PARAMETER FUNCTION  %
%%%%%%%%%%%%%%%%%%%%%%%%%%%%%%
\section{Three Parameter Function}

Consider a function of three variables, $f(\eta,p,\Lambda)$, where $\eta \ll \Lambda$.  We wish to expand the function so that, for any given value of $p$, the approximation is equal to $f$ to within some given order of $\eta/\Lambda$.  In addition, the integral $\int dp \; f(\eta,p,\Lambda)$ should be accurate to the same order.

To achieve this, we expand $f(\eta,p,\Lambda)$ in three different regions.  The first region is $p \sim \eta \ll \Lambda$, where we expand in powers of $\eta/\Lambda$ and $p/\Lambda$.  We call this expansion $f_l(\eta,p,\Lambda)$ using the ``$l$'' subscript to designate the ``low-momentum'' region.  In the region $\eta \ll \Lambda \sim p$, the expansion parameters are $\eta/\Lambda$ and $\eta/p$.  We refer to this expansion as $f_h(\eta,p,\Lambda)$ using $h$ to indicate the ``high-momentum'' range.  The final region is $\eta \ll p \ll \Lambda$.  Here we expand in powers of $\eta/p$, $\eta/\Lambda$, and $p/\Lambda$.  The resulting function will be called $f_d(\eta,p,\Lambda)$, where $d$ stands for the ``double expansion'' region.  The approximation to the original function can then be written as
\begin{equation}
f_l(\eta,p,\Lambda) + f_h(\eta,p,\Lambda) - f_d(\eta,p,\Lambda) \simeq f(\eta,p,\Lambda) .
\end{equation}
We do not offer a formal proof  of this statement.  Instead, we illustrate the concepts with an example, and present arguments to convince the reader that the method is generally applicable.

Suppose we desire an approximation to the function
\begin{equation}
f(\eta,p,\Lambda) = \frac{1}{(p + \eta)(p + \Lambda)} 
\end{equation}
that is accurate to $\mathcal{O}(\eta^2/\Lambda^2)$.  To calculate $f_l$, all terms of order up to and including $\eta^2/\Lambda^2$, $\eta p/\Lambda$, and $p^2/\Lambda^2$ are kept:
\begin{equation}
f_l(\eta,p,\Lambda) = \frac{1}{\Lambda(p + \eta)} \left[ 1 - \frac{p}{\Lambda} + \frac{p^2}{\Lambda^2} \right] \label{eqn:example_fl}.
\end{equation}
\noindent For $f_h$, we must keep all terms smaller than or equal to orders $\eta^2/\Lambda^2$, $\eta^2/p^2$, and $\eta^2/p\Lambda$:
\begin{equation}
f_h(\eta,p,\Lambda) = \frac{1}{p(p + \Lambda)} \left[ 1 - \frac{\eta}{p} + \frac{\eta^2}{p^2} \right].
\end{equation}
\noindent To find the double expansion, $f_d$, we expand $f_l$ by treating $p$ as though it were of the same magnitude as $\Lambda$.  The result is
\begin{equation}
f_d(\eta,p,\Lambda) = \frac{1}{\Lambda} \left[ 1 - \frac{p}{\Lambda} + \frac{p^2}{\Lambda^2} \right] \frac{1}{p} \left[ 1 - \frac{\eta}{p} + \frac{\eta^2}{p^2} \right].
\end{equation}
\noindent Equally valid is expanding $f_h$ by treating $p$ as if it were of the same magnitude as $\eta$, leading to
\begin{equation}
f_d(\eta,p,\Lambda) = \frac{1}{p} \left[ 1 - \frac{\eta}{p} + \frac{\eta^2}{p^2} \right] \frac{1}{\Lambda} \left[ 1 - \frac{p}{\Lambda} + \frac{p^2}{\Lambda^2} \right].
\end{equation}
\noindent Notice that both approaches yield the same result, which should always be true.

To prove that the approximation is accurate to the order stated, we calculate the difference between the function and its approximation:
\begin{equation}
f - f_l - f_h + f_d = \frac{\eta^3}{\Lambda^3(p+\eta)(p+\Lambda)} = \frac{\eta^3}{\Lambda^3} f(\eta,p,\Lambda).
\end{equation}
\noindent The error is smaller than the function $f$ by a factor of $\eta^3/\Lambda^3$ for any value of $p$.

%%%%%%%%%%%%%%%%%%%%%%%%%%%%%%%%%%%%%%%%%%
%  INTEGRATING THREE PARAMETER FUNCTION  %
%%%%%%%%%%%%%%%%%%%%%%%%%%%%%%%%%%%%%%%%%%
\section{Integrating a Three Parameter Function \label{sec:int3var}}

The integral $\int_0^{\infty} dp \; f(\eta,p,\Lambda)$ will result in a function of $\eta$ and $\Lambda$. Integrating $\int_0^{\infty} dp \; (f_l + f_h - f_d)$ should result in the same function to within the accuracy of the approximation itself, which in our example is $\mathcal{O}(\eta^2/\Lambda^2)$.  The benefit of integrating each term individually is simplicity.  Obtaining a closed form solution for the integral of $f$ may prove difficult, if not impossible.  If one is obtained, then it must be expanded to the appropriate order.  On the other hand, $f_l$, $f_h$, and $f_d$ may have much simpler forms than the original function, making them easier to integrate.  After integration, no additional expansion is necessary since each term is already expanded to the proper order.

Unfortunately, a problem arises when integrating individual terms.  When integrating $f_l$ from Eq.~(\ref{eqn:example_fl}), we encounter the term
\begin{equation}
\int_0^{\infty} dp \; \frac{p^2}{\Lambda^3 (p + \eta)},
\end{equation}
\noindent which diverges as $\Lambda \rightarrow \infty$.  This quadratic divergence is canceled by an equal and opposite quadratic divergence from the $p/\Lambda^3$ term in $f_d$.  Even so, the resulting difference of these terms is $-(\eta p)/(\Lambda^3(p+\eta))$ which now produces a linear divergence.  This, in turn, is canceled by the $-\eta/\Lambda^3$ term from $f_d$, resulting in a difference of $\eta^2/(\Lambda^3(p+\eta))$ that is logarithmically divergent.  Once again, a term from $f_d$, namely $\eta^2/(p\Lambda^3)$, cancels this divergence leaving $-\eta^3/(p\Lambda^3(p+\eta))$.  This final term is finite when integrated to infinity.  Unfortunately, it is now afflicted with a divergence as $p \rightarrow 0$.  This divergence is again canceled (along with all other ultraviolet and infrared divergences) since the integral of the sum is assumed to converge.

In anticipation of the fact that all divergences must cancel, we define a standardized subtraction scheme that can be applied uniformly to every integration.  This will render all integrals finite by effectively adding and subtracting the same divergent term to different parts of the expansion.  The subtraction scheme we utilize is defined as follows:
\begin{itemize}
\item Integrals of positive powers of $p$ and constant terms independent of $p$ are dropped.
\item A $1/p$ integration that diverges as $p \rightarrow \infty$ will have an equal $1/p$ integration from $\Lambda$ to $\infty$ subtracted out.
\end{itemize}
\noindent These rules handle all $p \rightarrow \infty$ divergences.  For divergences resulting from $p \rightarrow 0$, the previous pair of rules is augmented with another similar pair:
\begin{itemize}
\item Integrals of negative powers of $p$ (other than $1/p$) are dropped.
\item A $1/p$ integration that diverges as $p \rightarrow 0$ will have an equal $1/p$ integration from $0$ to $\eta$ subtracted out.
\end{itemize}
\noindent If the integral $\int dp \: f(\eta,p,\Lambda)$ is finite, then any divergence occurring in one of the terms of $\int dp \: (f_l + f_h - f_d)$ must be canceled by an equal and opposite divergence in another term to guarantee a finite result.  By applying the subtraction scheme to the first divergence, we render it finite.  However, the same subtraction scheme would also need to be applied to its ``partner'' divergence resulting in an equal and opposite subtraction term.  The net result is that the sum of all terms before and after the subtractions are applied must be equal.  This same argument could be used to formulate alternate schemes that would work just as well, provided that they are applied uniformly to all integrations.  Of course, formulating and applying any subtraction scheme requires some analytic knowledge of the integrand.

As an illustration of these rules, an integral like
\begin{equation}
\int_0^{\infty} dp \; \frac{p^2}{\Lambda^3(p+\eta)} ,
\end{equation}
\noindent which diverges as $p \rightarrow \infty$ would now become
\begin{equation}
\int_0^{\infty} dp \: \left[ \frac{p^2}{\Lambda^3 (p + \eta)} - \frac{p}{\Lambda^3} + \frac{\eta}{\Lambda^3} \right] - \int_{\Lambda}^{\infty} dp \: \frac{\eta^2}{p \Lambda^3}.
\end{equation}
\noindent A $p \rightarrow 0$ divergent integral such as
\begin{equation}
\int_0^{\infty} dp \: \frac{\eta}{p^2 (p + \Lambda)},
\end{equation}
\noindent would be modified to
\begin{equation}
\int_0^{\infty} dp \: \left[ \frac{\eta}{p^2 (p + \Lambda)} - \frac{\eta}{p^2 \Lambda} \right] + \int_0^{\eta} dp \: \frac{\eta}{p \Lambda^2} .
\end{equation}

We can also apply the expansion method to the case of $g_2$.  From Eq.~(\ref{eqn:g2inv}) we have
\begin{equation}
\frac{1}{g_2} = \int \frac{d^3\vec{q}}{(2 \pi)^3} \: \frac{U_2(q)^2}{2q^2 + B_2} .
\end{equation}
\noindent Not only will this example help clarify the ideas presented, but it is also easily compared to the previously derived expansion for $g_2$.  We define
\begin{equation}
f(\eta,p,\Lambda) = \frac{p^2 \, U_2(p)^2}{2 \pi^2 \, (2p^2 + \eta^2)},
\end{equation}
\noindent so that $1/g_2 = \int_0^{\infty} dp \: f(\eta,p,\Lambda)$ with $\eta \equiv \sqrt{B_2}$.  We desire an answer accurate to $\mathcal{O}(\eta/\Lambda)$.  The values for $f_l$, $f_h$, and $f_d$ are:
\begin{eqnarray}
f_l(\eta,p,\Lambda) & = & \frac{p^2}{2 \pi^2 \, (2p^2 + \eta^2)} ,
\\
f_h(\eta,p,\Lambda) & = & \frac{1}{4 \pi^2} \, U_2(p)^2 ,
\\
f_d(\eta,p,\Lambda) & = & \frac{1}{4 \pi^2} .
\end{eqnarray}
\noindent Next, we integrate each of the above functions, applying the subtraction scheme as needed:
\begin{equation}
\int_0^{\infty} dp \: f_l(\eta,p,\Lambda) \rightarrow \int_0^{\infty} dp \: \frac{1}{2 \pi^2} \left[ \frac{p^2}{2p^2 + \eta^2} - \frac{1}{2} \right] = - \: \frac{\eta}{8 \sqrt{2} \, \pi} \label{eqn:int_fl},
\end{equation}
\begin{equation}
\int_0^{\infty} dp \: f_h(\eta,p,\Lambda) \rightarrow \int_0^{\infty} dp \: \frac{1}{4 \pi^2} \, U_2(p)^2 = \frac{\Lambda (3 h_2^2 + 8h_2 + 16)}{128 \sqrt{2} \, \pi^{3/2}} ,
\end{equation}
\begin{equation}
\int_0^{\infty} dp \: f_d(\eta,p,\Lambda) \rightarrow  \int_0^{\infty} dp \: \left[ \frac{1}{4 \pi^2} - \frac{1}{4 \pi^2} \right] = 0 \label{eqn:int_fd} .
\end{equation}
\noindent The final result is
\begin{equation}
\frac{1}{g_2} = \frac{\Lambda (3 h_2^2 + 8h_2 + 16)}{128 \sqrt{2} \, \pi^{3/2}} - \frac{\eta}{8 \sqrt{2} \, \pi} ,
\end{equation}
\noindent or upon inverting,
\begin{equation}
g_2 = \frac{128 \sqrt{2} \, \pi^{3/2}}{\Lambda (3 h_2^2 + 8h_2 + 16)} \left[ 1 + \frac{16 \sqrt{\pi} \eta}{\Lambda (3 h_2^2 + 8h_2 + 16)} \right] \label{eqn:fullg2expand}.
\end{equation}
\noindent This matches exactly with Eq.~(\ref{eqn:g2expand}) when $h_2 = 0$.  Notice that the subtraction scheme terms in Eqs.~(\ref{eqn:int_fl}) and (\ref{eqn:int_fd}) are equal and would cancel out if all terms were combined.  The simplicity of this approach is also evident considering that the alternative would mean integrating $f(\eta,p,\Lambda)$ to obtain (for $h_2 = 0$)
\begin{equation}
\frac{1}{8 \sqrt{2} \, \pi^{3/2}} \left[ \Lambda - \eta \sqrt{\pi} \exp\left(\frac{\eta^2}{\Lambda^2}\right) + \eta \sqrt{\pi} \exp\left(\frac{\eta^2}{\Lambda^2}\right) \mathrm{Erf}\left(\frac{\eta}{\Lambda}\right) \right] ,
\end{equation}
\noindent and then expanding this quantity in powers of $\eta/\Lambda$.  A non-zero value of $h_2$ would be more complicated still.

There is one important point to note.  When applying this expansion technique, it is possible that two terms of a given order can be multiplied to yield a term of the same order.  For example, when expanding to order $\eta/\Lambda$ in the region $\eta \ll p \ll \Lambda$, products of $\eta/p$ and $p/\Lambda$ may be obtained.  Each term individually must be kept because it is considered $\mathcal{O}(\eta/\Lambda)$.  However, the product $(\eta/p)(p/\Lambda)$ is also $\mathcal{O}(\eta/\Lambda)$ and cannot be dropped.  This is an important difference compared to most perturbative expansions where the product of two first order terms must be of second order.  Fortunately, this caveat only needs to be considered in the mid-momentum region, $\eta \ll p \ll \Lambda$.

%%%%%%%%%%%%%%%%%%%%%%%%%%%%%
%  FOUR PARAMETER FUNCTION  %
%%%%%%%%%%%%%%%%%%%%%%%%%%%%%
\section{Four Parameter Function \label{sec:4var}}

To expand our three-body equation, we need a more complex version of the above procedure that can be applied to functions of four variables.  We desire a uniform expansion of the function $u(\eta, p, q, \Lambda)$, where $\eta/\Lambda \ll 1$ and $p$ and $q$ are arbitrary.  The steps involved are very similar to the case of three variables: expand the function in several regions and combine the results.  Unlike the three variable case where we needed to expand in only three regions, we must now expand the function in eleven different regions.  The formula will be presented in a table format.  Each row in the table, which represents one region of the expansion, provides the following information:
\begin{enumerate}
\item A short-hand label, $u_i$, for the expansion of $u(\eta, p, q, \Lambda)$ in region \#$i$.
\item The sign to be used when adding the term to all others.
\item The range of values that describes the region.
\item The appropriate expansion parameters.
\end{enumerate}

\begin{table}
\begin{center}
\begin{tabular}{|c|c|c|c|}
\hline
Label & Sign & Region & Parameters
\\
\hline
\hline
$f_l$ & + & $(\eta, p) \ll \Lambda$ & $\eta/\Lambda, p/\Lambda$
\\
$f_d$ & $-$ & $\eta \ll p \ll \Lambda$ & $\eta/p, \eta/\Lambda, p/\Lambda$
\\
$f_h$ & + & $\eta \ll (p, \Lambda)$ & $\eta/p, \eta/\Lambda$
\\
\hline
\end{tabular}
\end{center}
\caption{\label{tab:3var}Expansion terms for a function of three variables.}
\end{table}

\noindent Table \ref{tab:3var} shows how this format would look for the previous case of three variables.  For a function of four variables, we have the terms listed in Table \ref{tab:4var}.  Once again, we offer no formal proof of our claim.

\begin{table}
\begin{center}
\begin{tabular}{|c|c|c|c|}
\hline
Label & Sign & Region & Parameters
\\
\hline
\hline
$u_1$ & + & $\eta \ll (p, q, \Lambda)$ & $\eta/p, \eta/q, \eta/\Lambda$
\\
$u_2$ & + & $(\eta, q) \ll (p, \Lambda)$ & $\eta/p, q/p, \eta/\Lambda, q/\Lambda$
\\
$u_3$ & $-$ & $\eta \ll q \ll (p, \Lambda)$ & $\eta/p, \eta/q, \eta/\Lambda, q/p, q/\Lambda$
\\
$u_4$ & + & $(\eta, p) \ll (q, \Lambda)$ & $\eta/q, \eta/\Lambda, p/q, p/\Lambda$
\\
$u_5$ & + & $(\eta, p, q) \ll \Lambda$ & $\eta/\Lambda, p/\Lambda, q/\Lambda$
\\
$u_6$ & $-$ & $(\eta, p) \ll q \ll \Lambda$ & $\eta/q, \eta/\Lambda, p/q, p/\Lambda, q/\Lambda$
\\
$u_7$ & $-$ & $\eta \ll p \ll (q, \Lambda)$ & $\eta/p, \eta/q, \eta/\Lambda, p/q, p/\Lambda$
\\
$u_8$ & $-$ & $\eta \ll (p, q) \ll \Lambda$ & $\eta/p, \eta/q, \eta/\Lambda, p/\Lambda, q/\Lambda$
\\
$u_9$ & $-$ & $(\eta, q) \ll p \ll \Lambda$ & $\eta/p, \eta/\Lambda, q/p, q/\Lambda, p/\Lambda$
\\
$u_{10}$ & + & $\eta \ll p \ll q \ll \Lambda$ & $\eta/p, \eta/q, \eta/\Lambda, p/q, p/\Lambda, q/\Lambda$
\\
$u_{11}$ & + & $\eta \ll q \ll p \ll \Lambda$ & $\eta/p, \eta/q, \eta/\Lambda, q/p, p/\Lambda, q/\Lambda$
\\
\hline
\end{tabular}
\end{center}
\caption{\label{tab:4var}Expansion terms for a function of four variables.}
\end{table}

One thing that becomes trickier when working with four variables is knowing which terms to keep at any given order.  Consider the region $\eta \ll p \ll q \ll \Lambda$, and assume we want the expansion accurate to order $\mathcal{O}(\eta^2/\Lambda^2)$.  Obviously, first order terms such as $\eta/q$, $p/\Lambda$, and $\eta/\Lambda$ are kept.  All products of first order terms that reduce to first order terms are also kept (e.g., $(\eta/p)(p/q) = \eta/q$ or $(\eta/p)(p/q)(q/\Lambda) = \eta/\Lambda$).

But what about $(\eta/p)(q/\Lambda)$?  It does not reduce to a first order term, so is it a second order term?  Is it the same order as $(\eta/q)(p/\Lambda)$?  To answer these questions, we present the following rule for determining the order of a product:
\begin{itemize}
\item In whichever region the expansion is being computed, replace $\ll$ with $\le$.  Minimize the term subject to these new constraints.  If this minimum value is greater than or equal to the desired order, the term must be kept. \footnote{This rule is only needed in two regions: $\eta \ll p \ll q \ll \Lambda$ and $\eta \ll q \ll p \ll \Lambda$.}
\end{itemize}
\noindent In this example, the region $\eta \ll p \ll q \ll \Lambda$ leads to the constraint $\eta \le p \le q \le \Lambda$.  A term such as $(\eta/p)(q/\Lambda)$ would be minimized by setting $p = q$.  The resulting minimum, $\eta/\Lambda$, is of the order we want.  This implies that any value of this product is always greater than or equal to the desired order and must be kept.  

On the other hand, the product $(\eta/q)(p/\Lambda)$ is minimized by setting $q = \Lambda$ and $p = \eta$.  This minimum is $\mathcal{O}(\eta^2/\Lambda^2)$.  However, the product is maximized by setting $p = q$, giving a product of $\eta/\Lambda$.  So we see that the order of this product is always bounded by $\mathcal{O}(\eta^2/\Lambda^2)$ and $\mathcal{O}(\eta/\Lambda)$.  Because its order at any point is less than the order we are expanding to, it may be safely dropped.  Keep in mind though that the product would be necessary for a calculation to $\mathcal{O}(\eta^2/\Lambda^2)$, which is why we classify the product according to its minimum value.

%%%%%%%%%%%%%%%%%%%%%%%%%%%%%%%%%%%%%%%%%
%  INTEGRATING FOUR PARAMETER FUNCTION  %
%%%%%%%%%%%%%%%%%%%%%%%%%%%%%%%%%%%%%%%%%
\section{Integrating a Four Parameter Function \label{sec:int4var}}

Just as we were able to integrate $f(\eta,p,\Lambda)$ to obtain a value accurate to the same order as $f(\eta,p,\Lambda)$, we can integrate $u(\eta,p,q,\Lambda)$ to obtain a function of three variables that is equally accurate.  Let
\begin{equation}
f(\eta,p,\Lambda) = \int_0^{\infty} dq \: u(\eta,p,q,\Lambda).
\end{equation}
\noindent We can now relate the functions $f_l$, $f_d$, and $f_h$ to the integration of the appropriate terms of $u$.  $f_l$ is determined in the region $(\eta,p) \ll \Lambda$.  Only the terms $u_4$, $u_5$, and $u_6$ satisfy this requirement.  Therefore,
\begin{equation}
f_l(\eta,p,\Lambda) = \int_0^{\infty} dq \: \left[ u_4(\eta,p,q,\Lambda) + u_5(\eta,p,q,\Lambda) - u_6(\eta,p,q,\Lambda) \right] \label{eqn:fl_int}.
\end{equation}
\noindent Likewise,
\begin{equation}
f_h(\eta,p,\Lambda) = \int_0^{\infty} dq \: \left[ u_1(\eta,p,q,\Lambda) + u_2(\eta,p,q,\Lambda) - u_3(\eta,p,q,\Lambda) \right] \label{eqn:fh_int} ,
\end{equation}
\begin{eqnarray}
f_d(\eta,p,\Lambda) & = & \int_0^{\infty} dq \: \left[ u_7(\eta,p,q,\Lambda) + u_8(\eta,p,q,\Lambda) + u_9(\eta,p,q,\Lambda) \right. \nonumber
\\
&& \left. ~-~u_{10}(\eta,p,q,\Lambda) - u_{11}(\eta,p,q,\Lambda) \right] \label{eqn:fd_int}.
\end{eqnarray}
\noindent As before, the possibility exists that the integration of any one term may be divergent, even though the entire quantity is finite.  Fortunately, the same subtraction scheme defined for the case of three variables can be applied here as well.  The result is that, for certain classes of functions, some terms in the expansion of $u(\eta,p,q,\Lambda)$ may be entirely removed.  In particular, the terms $u_3$, $u_6$, $u_{10}$, and $u_{11}$ are expanded in regions where $q$ is either much larger or much smaller than every other parameter.  If the function is such that $q$ only occurs in the form of pure powers in the expansion, the subtraction scheme demands that these terms be dropped.  This leaves us with the simplified formulas:
\begin{eqnarray}
f_l(\eta,p,\Lambda) & = & \int_0^{\infty} dq \: \left[ u_4(\eta,p,q,\Lambda) + u_5(\eta,p,q,\Lambda) \right] ,
\\
f_h(\eta,p,\Lambda) & = & \int_0^{\infty} dq \: \left[ u_1(\eta,p,q,\Lambda) + u_2(\eta,p,q,\Lambda) \right] ,
\\
f_d(\eta,p,\Lambda) & = & \int_0^{\infty} dq \: \left[ u_7(\eta,p,q,\Lambda) + u_8(\eta,p,q,\Lambda) + u_9(\eta,p,q,\Lambda) \right] .
\end{eqnarray}
\noindent It must be remembered that these equations are not universally applicable, and they must be restricted to functions whose expansions for $u_3$, $u_6$, $u_{10}$, and $u_{11}$ contain $q$ only as pure powers.  Later, we will see functions that do not allow this type of simplification.
\chapter{Three-Body Equation Expansion}
\label{ch:expand3body}

We are now prepared to apply the technique outlined in the previous chapter to the three-body bound-state equation.  The expansion will result in two sets of integral equations.  The first set provides the solution for the leading order behavior.  We refer to these equations as the LO (Leading Order) equations.  The second set provides corrections that are smaller than the LO results by a factor of $\sqrt{B_2}/\Lambda$ or $\sqrt{B_3}/\Lambda$.  These will be referred to as NLO (Next-to-Leading Order) equations.  Because there are many steps involved in the derivation of these equation sets, we provide a brief overview describing the contents of this chapter.

First, we begin by setting up some necessary information that will be needed for later steps.  This includes defining several new quantities and rewriting the bound state Eq.~(\ref{eqn:final3body}) in terms of these quantities.  Following this is an expansion of several functions that constitute parts of the bound-state equation, and a more detailed explanation of how they will be used.

Next, we use the previously derived results to obtain the equations describing the LO behavior.  The process is then continued to construct the set of NLO equations.  Any simplifications or exact solutions to these equations will be shown along the way.  Finally, we derive the relations needed to obtain values for the three-body coupling, $g_3$.  A summary of all vital equations is given at the end for use as a quick reference.

%%%%%%%%%%%%%%%%%%%
%%  PREPARATION  %%
%%%%%%%%%%%%%%%%%%%
\section{Preparation for Expansion}

To start, we define a few new quantities, including a function that will take the place of $\Phi(p)$.  This is done to facilitate the expansion later on.  Reasons for these new definitions will be given as the new quantities are defined.  We will also be repeatedly applying the expansion techniques in Chapter \ref{ch:pertmethod}, so a basic understanding of these methods is necessary to follow the derivation.  For readers not concerned with the details, a complete summary of all equations is provided in Sec.~\ref{sec:quickref}.

    %%%%%%%%%%%%%%%%%%%
    %%  DEFINITIONS  %%
    %%%%%%%%%%%%%%%%%%%
\subsection{Definitions}

Let us begin by defining
\begin{eqnarray}
\eta_2 & \equiv & \sqrt{B_2} ,
\\
\eta_3 & \equiv & \sqrt{B_3} .
\end{eqnarray}
\noindent Using $\eta_2$ and $\eta_3$ in our equations allows us to deal with quantities that have the same dimension as the momentum variable.  It also identifies their role in the expansion as the parameter $\eta$ from the previous chapter.  In fact, we will occasionally use $\eta$ to generically refer to $\eta_2$ and $\eta_3$ in places where either is a valid alternative.\footnote{For example, saying that something is $\mathcal{O}(\eta/\Lambda)$ can imply that it is $\mathcal{O}(\eta_2/\Lambda)$ or $\mathcal{O}(\eta_3/\Lambda)$.}

\noindent Dimensionless versions of the couplings $g_2$ and $g_3$ are defined by
\begin{eqnarray}
G_2 & \equiv & \Lambda \, g_2 ,
\\
G_3 & \equiv & \Lambda^4 \, g_3 .
\end{eqnarray}
\noindent These dimensionless couplings help to clarify the power counting arguments used later.  However, for the time being, we choose to replace $G_3$ with a new coupling term:\footnote{Although $\delta$ is not technically a coupling, we will often refer to it as such because it plays a role very similar to $G_3$.}
\begin{equation}
\delta \equiv \frac{G_3 \, \Phi_1}{\Lambda} .
\end{equation}
\noindent The reasons for doing so are subtle and not entirely obvious at first glance.  The quantity $\delta$ is finite and possesses a simple cosine behavior to leading order, whereas $G_3$ will be seen to exhibit divergent behavior at certain points.  If we are interested only in binding energies, we may solve the integral equation using $\delta$ and need not perform the additional integration necessary to compute $\Phi_1$ and hence $G_3$.  These statements will be justified later.

Another dimensionless quantity, and perhaps the most important, is the replacement of the pseudo-wavefunction $\Phi(p)$ with
\begin{equation}
f(\eta_2, \eta_3, p, \Lambda) \equiv (p^2 + \eta_3^2 - \eta_2^2) \Phi(p).
\end{equation}
\noindent This shares some of the benefits of all the previous definitions.  Like $\eta_2$ and $\eta_3$, its role in the expansion is easily seen to be the same as the function $f$ used in the previous chapter.  Like $G_2$ and $G_3$, the dimensionless nature of $f$ will simplify future power counting.  In addition, this function tends to be of $\mathcal{O}(1)$ throughout the entire range of $p$ and is less prone to large numerical fluctuations than $\Phi$ when $B_2 \simeq B_3$.\footnote{We cannot prove these statements  {\em a priori} and must verify them numerically.}

With these new definitions, the three-body bound-state equation (\ref{eqn:final3body}) now takes the form
\begin{eqnarray}
f(\eta_2, \eta_3, p, \Lambda) & = & \frac{p^2 + \eta_3^2 - \eta_2^2}{2 \pi^2 D\left(-\eta_3^2 - \frac{3}{2}p^2\right)} \int_0^{\infty} dq \: \left[ \frac{q^2}{q^2 + \eta_3^2 - \eta_2^2} \right. \nonumber
\\
&& \left. \times \int_{-1}^{1} dz \: \frac{U_2\left(\vec{q} + \frac{1}{2}\vec{p}\right) U_2\left(\vec{p} + \frac{1}{2}\vec{q}\right)}{\eta_3^2 + 2p^2 + 2q^2 + 2pqz}  f(\eta_2, \eta_3, q, \Lambda) \right] \nonumber
\\
&& - ~\delta \; \frac{(p^2 + \eta_3^2 - \eta_2^2) D_1(p)}{G_2 \Lambda^2 D\left( -\eta_3^2 - \frac{3}{2}p^2 \right)} \label{eqn:f3body}.
\end{eqnarray}

    %%%%%%%%%%%%%%%%%%%
    %%  CONVENTIONS  %%
    %%%%%%%%%%%%%%%%%%%
\subsection{Conventions}

The process of expanding Eq.~(\ref{eqn:f3body}) is done in a series of smaller steps.  Because the $\delta$ term is simply added to the integral term, each can be expanded individually and then added together at the end.  In addition, these terms are composed of other quantities like $D_1(p)$, $G_2$, etc. which have their own expansions in terms of $\eta/\Lambda$.  The goal is to split the equation for $f$ into separate equations for $f_l$, $f_d$, and $f_h$.  At the same time, we want to distinguish between the LO and NLO portions of these equations.  In order to categorize the contributions from every function in the three-body equation, we develop a labeling convention.

Just as the function $f$ can be divided into its low-, mid-, and high-momentum parts, so too can any other function of $\eta$, $p$, and $\Lambda$.  The same subscripts will be used to denote these terms.  For example, the approximation to $D(p)$ in the region $p \sim \eta \ll \Lambda$ will be written as $D_l(p)$.  This low-momentum version may also be expanded in powers of $\eta/\Lambda$ and $p/\Lambda$ when attempting to uniformly expand $D(p)$ to a given order in $\eta/\Lambda$.  To distinguish between orders, we will use a subscript $n$ to represent the $\mathcal{O}\left(\left(\eta/\Lambda\right)^n\right)$ portion.  Thus, $D_{l1}(p)$ denotes the part of $D_l(p)$ having the same magnitude as $\eta/\Lambda$ and $p/\Lambda$.  Likewise, the function $D_{1h0}(p)$ is the $\mathcal{O}(1)$ part of $D_1(p)$ in the region $\eta \ll p \sim \Lambda$.

The couplings are labeled in a similar manner.  Since they contain no $p$ dependence, the classification of $l$, $d$, or $h$ has no meaning.  Yet they may be expanded in powers of $\eta/\Lambda$, so the numerical subscript is useful.  For the coupling $\delta$, the order $n$ contribution will be identified as $\delta_n$, which should not be confused with the the phase shift $\delta_0(p)$.  For a coupling like $G_2$ which already possesses a subscript, we shall write it as $G_{2,n}$.  This comma delimiter takes on the role of '$l$', '$d$', or '$h$', acting as a placeholder between the two numbers.  This helps to avoid confusion in cases where the subscript may already consist of a double-digit number.

    %%%%%%%%%%%%%%%%%%%%%%%%%%%%%%
    %%  COMPONENTS' EXPANSIONS  %%
    %%%%%%%%%%%%%%%%%%%%%%%%%%%%%%
\subsection{Expansion of Components}

In order to achieve results accurate to $\mathcal{O}(\eta/\Lambda)$, the following quantities must also be expanded to the same order: $\delta$, $G_2$, $D\left(-\eta_3^2 - \frac{3}{2}p^2\right)$, and $D_1(p)$.  The value of $\delta$ is unknown prior to solving the equations, so we shall simply refer to the $\mathcal{O}(1)$ part as $\delta_0$ and the $\mathcal{O}(\eta/\Lambda)$ part as $\delta_1$.  Ultimately we must show that $\delta_0$ is indeed $\mathcal{O}(1)$, which is equivalent to showing that the 3-body contact interaction is required to renormalize the bound-state equation.

The terms in the expansion of $g_2$ have already been computed in Eq.~(\ref{eqn:fullg2expand}), making it easy to write the expansion for $G_2$:
\begin{equation}
G_{2,0} = \frac{128 \sqrt{2} \, \pi^{3/2}}{3 h_2^2 + 8h_2 + 16} ,
\end{equation}
\begin{equation}
G_{2,1} = \frac{2048 \sqrt{2} \, \pi^2}{(3 h_2^2 + 8h_2 + 16)^2} \left( \frac{\eta}{\Lambda} \right) .
\end{equation}
\noindent Here we have assumed that $h_2$ is of $\mathcal{O}(1)$, but the above equations will continue to hold provided that $h_2$ does not become large enough to negate the power counting arguments.  For $r_e < 0$, Eq.~(\ref{eqn:ar2h}) shows that $h_2$ will diverge like $\sqrt{\Lambda}$ as $\Lambda \rightarrow \infty$.  In this case, we would be required to limit the value of $\Lambda$ so that $h_2$ never becomes arbitrarily large.

Next we must expand
\begin{equation}
D\left(-\eta_3^2 - \mbox{$3 \over 2$} p^2\right) = \frac{\frac{3}{2}p^2 + \eta_3^2 - \eta_2^2}{2 \pi^2} \int_0^{\infty} dq \: \frac{q^2 \, U_2(q)^2}{\left( 2 q^2 + \eta_2^2 \right)\left( 2 q^2 + \frac{3}{2}p^2 + \eta_3^2 \right)} .
\end{equation}
\noindent This will be our first application of the methods in Secs.~\ref{sec:4var} and \ref{sec:int4var}.  We make $D$ fit the form of $\int dq \: u(\eta, p, q, \Lambda)$ by defining
\begin{equation}
u(\eta_2, \eta_3, p, q, \Lambda) =  \frac{q^2 \, \left( \frac{3}{2}p^2 + \eta_3^2 - \eta_2^2 \right) \, U_2(q)^2}{2 \pi^2 \left( 2 q^2 + \eta_2^2 \right)\left( 2 q^2 + \frac{3}{2}p^2 + \eta_3^2 \right)} . 
\end{equation}
\noindent The function $u$ must now be expanded in each of the eleven different regions up to and including first order corrections:
\begin{eqnarray}
u_1 & = & \frac{3 p^2}{8 \pi^2 \left( 2 q^2 + \frac{3}{2} p^2 \right)} \, U_2(q)^2 ,
\\
u_2 & = & \frac{q^2}{2 \pi^2 \left( 2 q^2 + \eta_2^2 \right)} ,
\\
u_3 & = & \frac{1}{4 \pi^2} ,
\\
u_4 & = &  \frac{1}{8 \pi^2 q^2} \left( \frac{3}{2}p^2 + \eta_3^2 - \eta_2^2 \right) U_2(q)^2 ,
\\
u_5 & = &  \frac{q^2 \, \left( \frac{3}{2}p^2 + \eta_3^2 - \eta_2^2 \right)}{2 \pi^2 \left( 2q^2 + \eta_2^2 \right)\left( 2q^2 + \frac{3}{2}p^2 + \eta_3^2 \right)} ,
\\
u_6 & = & \frac{\frac{3}{2}p^2 + \eta_3^2 - \eta_2^2}{8 \pi^2 q^2} ,
\\
u_7 & = & \frac{3 p^2}{16 \pi^2 q^2} \,  U_2(q)^2 ,
\\
u_8 & = & \frac{3 p^2}{8 \pi^2 \left( 2 q^2 + \frac{3}{2} p^2 \right)} ,
\\
u_9 & = & \frac{q^2}{2 \pi^2 \left( 2 q^2 + \eta_2^2 \right)} ,
\\
u_{10} & = & \frac{3 p^2}{16 \pi^2 q^2} ,
\\
u_{11} & = & \frac{1}{4 \pi^2} .
\end{eqnarray}
\noindent The approximations to $D$ in each of the momentum ranges are given by the following relations:
\begin{eqnarray}
D_l & = & \int_0^{\infty} dq \: \left[ u_4 + u_5 - u_6 \right] ,
\\
D_d & = & \int_0^{\infty} dq \: \left[ u_7 + u_8 + u_9 - u_{10} - u_{11} \right] ,
\\
D_h & = & \int_0^{\infty} dq \: \left[ u_1 + u_2 - u_3 \right] .
\end{eqnarray}
\noindent Now we must differentiate the LO behavior from the NLO behavior for each of the above integrands.  To do this, we integrate each term and look at the final result to determine the order at which it contributes.  This is much simpler than trying to determine {\em a priori} which integrands are LO or NLO and then performing the integrations, although we will see some cases where this is necessary.  The results are shown below.
\begin{eqnarray}
D_{l0}(\eta_2, \eta_3, p) & = & \frac{1}{2 \pi^2} \int_0^{\infty} dq \: \frac{q^2 \, \left( \frac{3}{2}p^2 + \eta_3^2 - \eta_2^2 \right)}{\left( 2q^2 + \eta_2^2 \right)\left( 2q^2 + \frac{3}{2}p^2 + \eta_3^2 \right)} \nonumber
\\
& = & \frac{\sqrt{\frac{3}{2}p^2 + \eta_3^2} - \eta_2}{8 \sqrt{2} \, \pi} \label{eqn:Dl0}
\\
D_{l1}(\eta_2, \eta_3, p, \Lambda) & = & \frac{ \frac{3}{2}p^2 + \eta_3^2 - \eta_2^2}{8 \pi^2} \int_0^{\infty} \frac{dq}{q^2} \: \left[ U_2(q)^2 - 1 \right] \nonumber
\\
& = & \left( \frac{h_2^2 + 8 h_2 - 16}{64 \sqrt{2} \pi^{3/2}} \right) \left( \frac{\frac{3}{2}p^2 + \eta_3^2 - \eta_2^2}{\Lambda} \right)
\\
D_{d0}(p) & = & \frac{3 p^2}{8 \pi^2} \int_0^{\infty} dq \: \frac{1}{ 2 q^2 + \frac{3}{2} p^2} \nonumber
\\
& = & \frac{\sqrt{3} \, p}{16 \pi}
\\
D_{d1}(\eta_2, p, \Lambda) & = & \int_0^{\infty} dq \: \left[  \frac{3 p^2}{16 \pi^2 q^2} \left( U_2(q)^2 - 1 \right)  + \left( \frac{q^2}{2 \pi^2 \left( 2 q^2 + \eta_2^2 \right)} - \frac{1}{4 \pi^2}  \right) \right] \nonumber
\\
& = & - \, \frac{\eta_2}{8 \sqrt{2} \, \pi} + \frac{3 \left( h_2^2 + 8 h_2 - 16 \right) p^2}{128 \sqrt{2} \, \pi^{3/2} \, \Lambda}
\\
D_{h0}(p, \Lambda) & = & \frac{3 p^2}{8 \pi^2} \int_0^{\infty} dq \:  \frac{U_2(q)^2}{ 2 q^2 + \frac{3}{2} p^2} \nonumber
\\
& = & \frac{p}{256 \pi^2 \Lambda^4} \left[ \sqrt{3} \, \pi \left( 4 \Lambda^2 - 3 h_2 p^2 \right)^2 \left( 1 - \mathrm{Erf}\left( \frac{\sqrt{3} \, p}{\sqrt{2} \, \Lambda} \right) \right) \exp\left( \frac{3 p^2}{2 \Lambda^2} \right) \right. \nonumber
\\
&& +~ 3 \sqrt{2 \pi} \, h_2 \Lambda p \left( \left(8 + h_2 \right) \Lambda^2 - 3 h_2 p^2 \right) \Bigg]
\\
D_{h1}(\eta_2) & = & \int_0^{\infty} dq \: \left[ \frac{q^2}{2 \pi^2 \left( 2 q^2 + \eta_2^2 \right)} - \frac{1}{4 \pi^2} \right] \nonumber
\\
& = & - \, \frac{\eta_2}{8 \sqrt{2} \, \pi} \label{eqn:Dh1}
\end{eqnarray}
\noindent These same steps are used to compute the expansion of $D_1(p)$.  In this case, the function $u$ takes the form
\begin{equation}
u(\eta_3, p, q, \Lambda) =  \frac{q^2}{4 \pi^2} \int_{-1}^{1} dz \: \frac{U_2(\vec{q}+\frac{1}{2}\vec{p}\,) U_3(q) U_3(p) U_3(\vec{p}+\vec{q}\,)}{2p^2 + 2q^2 + 2pqz + \eta_3^2} ,
\end{equation}
\noindent and yields the following expansions in each of the eleven regions:
\begin{eqnarray}
u_1 & = & \frac{q^2}{4 \pi^2} \int_{-1}^{1} dz \: \frac{U_2(\vec{q}+\frac{1}{2}\vec{p}\,) U_3(q) U_3(p) U_3(\vec{p}+\vec{q}\,)}{2p^2 + 2q^2 + 2pqz} ,
\\
u_2 & = & \frac{q^2}{4 \pi^2 p^2} \: U_3(p)^2 \, U_2(p/2) ,
\\
u_3 & = & \frac{q^2}{4 \pi^2 p^2} \: U_3(p)^2 \, U_2(p/2) ,
\\
u_4 & = & \frac{1}{4 \pi^2} \: U_3(q)^2 \, U_2(q) ,
\\
u_5 & = & \frac{q}{8 \pi^2 p} \ln\left( \frac{\eta_3^2 + 2p^2 + 2q^2 + 2pq}{\eta_3^2 + 2p^2 + 2q^2 - 2pq} \right) ,
\\
u_6 & = & \frac{1}{4 \pi^2} ,
\\
u_7 & = & \frac{1}{4 \pi^2} \: U_3(q)^2 \, U_2(q) ,
\\
u_8 & = & \frac{q}{8 \pi^2 p} \ln\left( \frac{p^2 + q^2 + pq}{p^2 + q^2 - pq} \right) ,
\\
u_9 & = & \frac{q^2}{4 \pi^2 p^2} ,
\\
u_{10} & = & \frac{1}{4 \pi^2} ,
\\
u_{11} & = & \frac{q^2}{4 \pi^2 p^2} .
\end{eqnarray}
\noindent Once again, the LO and NLO terms are identified after integration.  It is easily verified that the appropriate expansions are
\begin{eqnarray}
D_{1l0}(\Lambda) & = &  \frac{1}{4 \pi^2} \int_0^{\infty} dq \: U_3(q)^2 \,  U_2(q) \nonumber
\\
& = & \frac{\Lambda \left( 6 \left( h_3^2 + 4 h_3 + 12 \right) + h_2 \left( 5 h_3^2 + 12 h_3 + 12 \right) \right)}{576 \sqrt{3} \, \pi^{3/2}} ,
\\
D_{1l1}(\eta_3, p) & = & \frac{1}{8 \pi^2} \int_0^{\infty} dq \: \left[ \frac{q}{p} \ln\left( \frac{\eta_3^2 + 2p^2 + 2q^2 + 2pq}{\eta_3^2 + 2p^2 + 2q^2 - 2pq}\right) - 2 \right] \nonumber
\\
& = & - \, \frac{\pi}{2} \sqrt{2 \eta_3^2 + 3p^2} ,
\\
D_{1d0}(\Lambda) & = & \frac{1}{4 \pi^2} \int_0^{\infty} dq \: U_3(q)^2 \, U_2(q) \nonumber
\\
& = & \frac{\Lambda \left( 6 \left( h_3^2 + 4 h_3 + 12 \right) + h_2 \left( 5 h_3^2 + 12 h_3 + 12 \right) \right)}{576 \sqrt{3} \, \pi^{3/2}} ,
\\
D_{1d1}(p) & = & \frac{1}{4 \pi^2} \int_0^{\infty} dq \: \left[ \frac{q}{2 p} \ln\left( \frac{p^2 + q^2 + pq}{p^2 + q^2 - pq} \right) - 1 + \frac{q^2}{p^2} - \frac{q^2}{p^2} \right] \nonumber
\\
& = & - \, \frac{\sqrt{3} \, \pi}{2} p ,
\\
D_{1h0}(p, \Lambda) & = & \frac{1}{4 \pi^2} \int_0^{\infty} dq \: q^2 \, \int_{-1}^{1} dz \: \frac{U_2(\vec{q}+\frac{1}{2}\vec{p}) U_3(q) U_3(p) U_3(\vec{p}+\vec{q})}{2p^2 + 2q^2 + 2pqz} ,
\\
D_{1h1}(p) & = & 0 .
\end{eqnarray}
\noindent Notice that $D_{1h}$ is in some ways easier to analyze and in other ways harder.  Because $u_2 = u_3$, the expression $D_{1h} = \int dq \: \left[ u_1 + u_2 - u_3 \right]$ reduces to an integration of only one term.  Unfortunately, this integral has no simple analytic solution.  Dimensional analysis dictates that the final result must have dimensions of momentum, and since the only quantities involved in the integral are $p$ and $\Lambda$, the answer must be $\mathcal{O}(\Lambda)$ or $\mathcal{O}(p) = \mathcal{O}(\Lambda)$.  This is the same order as both $D_{1l0}$ and $D_{1d0}$, proving that this integral is a LO term.  This also means that there are no integrands that contribute to NLO, which is why $D_{1h1} = 0$.

    %%%%%%%%%%%%%%%%%%%%%%%%%%%%%%%%
    %%  MAIN INTEGRAND EXPANSION  %%
    %%%%%%%%%%%%%%%%%%%%%%%%%%%%%%%%
\subsection{Expansion of Main Integrand}

We are now in a position to expand the main integrand of Eq.~(\ref{eqn:f3body}).  If we ignore the three-body interaction for a moment, the bound-state equation can be written as
\begin{equation}
f(\eta_2, \eta_3, p, \Lambda) = \int_0^{\infty} dq \: Q(\eta_2, \eta_3, p, q, \Lambda) \, f(\eta_2, \eta_3, q, \Lambda) ,
\end{equation}
\noindent where
\begin{eqnarray}
Q(\eta_2, \eta_3, p, q, \Lambda) & = & \frac{q^2}{2 \pi^2 D\left(-\eta_3^2 - \frac{3}{2}p^2\right)} \left( \frac{p^2 + \eta_3^2 - \eta_2^2}{q^2 + \eta_3^2 - \eta_2^2} \right) \nonumber
\\
&& \hspace{1in} \times \int_{-1}^{1} dz \frac{U_2\left(\vec{q} + \frac{1}{2}\vec{p}\right) U_2\left(\vec{p} + \frac{1}{2}\vec{q}\right)}{\eta_3^2 + 2p^2 + 2q^2 + 2pqz} .
\end{eqnarray}
\noindent By equating $u(\eta_2, \eta_3, p, q, \Lambda) = Q(\eta_2, \eta_3, p, q, \Lambda) \, f(\eta_2, \eta_3, q, \Lambda)$, the bound-state equation becomes $f(\eta_2, \eta_3, p, \Lambda) = \int dq \: u(\eta_2, \eta_3, p, q, \Lambda)$.  This is just the form needed to expand the integral.

The function $Q$ can be expanded in the same 11 regions as $u$.  On the other hand, the function $f$ depends only upon three scales: $\eta$, $p$ (or $q$), and $\Lambda$.  It is expanded in only three regions to produce $f_l$, $f_d$, and $f_h$.  The relationship between the expansions of $u$, $Q$, and $f$ is easily determined by paying attention to the region of the expansion.  For instance, $u_1$ results from an expansion where $\eta \ll (p, q, \Lambda)$, as does $Q_1$.  In this same region, $f(q)$ would be expanded to $f_h(q)$.  Thus $u_1(\eta, p, q, \Lambda) = Q_1(\eta, p, q, \Lambda) f_h(\eta, q, \Lambda)$.  By this same reasoning, $u_6 = Q_6 \, f_d(q)$, $u_9 = Q_9 \, f_l(q)$, etc.

With this in mind, Eqs.~(\ref{eqn:fl_int}), (\ref{eqn:fh_int}), and (\ref{eqn:fd_int}) can be used to obtain the following relations:
\begin{eqnarray}
\hspace{-0.2in} f_l(p) & = & \int_0^{\infty} dq \: \left[ Q_4 \, f_h(q) + Q_5 \, f_l(q) - Q_6 \, f_d(q) \right] \label{eqn:fl_terms},
\\
\hspace{-0.2in} f_d(p) & = & \int_0^{\infty} dq \: \left[ Q_7 \, f_h(q) + Q_8 \, f_d(q) + Q_9 \, f_l(q) - Q_{10} \, f_d(q) - Q_{11} \, f_d(q) \right] ,
\\
\hspace{-0.2in} f_h(p) & = & \int_0^{\infty} dq \: \left[ Q_1 \, f_h(q) + Q_2 \, f_l(q) - Q_3 \, f_d(q) \right] \label{eqn:fh_terms}.
\end{eqnarray}
\noindent To avoid notational clutter, we have dropped the explicit dependence of the $f$ functions on $\eta_2$, $\eta_3$, and $\Lambda$.  We will continue to do so unless confusion would result or such dependence needs to be emphasized.

The task of expanding the integral equations now becomes the task of expanding $Q$.  Because $Q$ contains the function $D\left( -\eta_3^2 - \frac{3}{2}p^2 \right)$, each $Q_i$ will be written in terms of the expanded $D$ functions (\ref{eqn:Dl0}) through (\ref{eqn:Dh1}).  The eleven expansion terms are shown below. 
\begin{eqnarray}
Q_1 & = & \frac{p^2}{2 \pi^2 D_{h0}(p, \Lambda)} \left( 1 - \frac{D_{h1}(\eta_2)}{D_{h0}(p, \Lambda)} \right) \int_{-1}^{1} dz \, \frac{U_2\left(\vec{q} + \frac{1}{2}\vec{p}\right) U_2\left(\vec{p} + \frac{1}{2}\vec{q}\right)}{2p^2 + 2q^2 + 2pqz}
\\
Q_2 & = & \frac{q^2}{2 \pi^2 (q^2 + \eta_3^2 - \eta_2^2) D_{h0}(p, \Lambda)} \left( 1 - \frac{D_{h1}(\eta_2)}{D_{h0}(p, \Lambda)} \right) U_2(p) \, U_2\left( p/2 \right)
\\
Q_3 & = & \frac{1}{2 \pi^2 D_{h0}(p, \Lambda)} \left( 1 - \frac{D_{h1}(\eta_2)}{D_{h0}(p, \Lambda)} \right) U_2(p) \, U_2\left( p/2 \right)
\\
Q_4 & = & \frac{(p^2 + \eta_3^2 - \eta_2^2)}{2 \pi^2 q^2 D_{l0}(\eta_2,\eta_3,p)} \left( 1 - \frac{D_{l1}(\eta_2,\eta_3,p,\Lambda)}{D_{l0}(\eta_2,\eta_3,p)} \right)  U_2(q) \, U_2\left( q/2 \right)
\\
Q_5 & = & \frac{1}{4 \pi^2 D_{l0}(\eta_2,\eta_3,p)} \left( 1 - \frac{D_{l1}(\eta_2,\eta_3,p,\Lambda)}{D_{l0}(\eta_2,\eta_3,p)} \right) \frac{q \, (p^2 + \eta_3^2 - \eta_2^2)}{p \, (q^2 + \eta_3^2 - \eta_2^2)} \nonumber
\\
&& \hspace{1in} \times \ln\left(\frac{\eta_3^2 + 2p^2 + 2q^2 + 2pq}{\eta_3^2 + 2p^2 + 2q^2 - 2pq}\right)
\\
Q_6 & = & \frac{(p^2 + \eta_3^2 - \eta_2^2)}{2 \pi^2 q^2  D_{l0}(\eta_2,\eta_3,p)} \left( 1 - \frac{D_{l1}(\eta_2,\eta_3,p,\Lambda)}{D_{l0}(\eta_2,\eta_3,p)} \right)
\\
Q_7 & = & \frac{p^2}{2 \pi^2 q^2 D_{d0}(p)} \left( 1 - \frac{D_{d1}(\eta_2,p,\Lambda)}{D_{d0}(p)} \right) U_2(q) \, U_2\left( q/2 \right)
\\
Q_8 & = & \frac{p}{4 \pi^2 q D_{d0}(p)} \left( 1 - \frac{D_{d1}(\eta_2,p,\Lambda)}{D_{d0}(p)} \right) \ln\left(\frac{p^2 + q^2 + pq}{p^2 + q^2 - pq}\right)
\\
Q_9 & = & \frac{q^2}{2 \pi^2 (q^2 + \eta_3^2 - \eta_2^2) D_{d0}(p)} \left( 1 - \frac{D_{d1}(\eta_2,p,\Lambda)}{D_{d0}(p)} \right)
\\
Q_{10} & = & \frac{p^2}{2 \pi^2 q^2  D_{d0}(p)} \left( 1 - \frac{D_{d1}(\eta_2,p,\Lambda)}{D_{d0}(p)} \right)
\\
Q_{11} & = & \frac{1}{2 \pi^2 D_{d0}(p)} \left( 1 - \frac{D_{d1}(\eta_2,p,\Lambda)}{D_{d0}(p)} \right)
\end{eqnarray}
\noindent Of course, the three-body interaction has been neglected in Eqs.~(\ref{eqn:fl_terms}) to (\ref{eqn:fh_terms}), and it must likewise be expanded and included.

%%%%%%%%%%%%%%%%%
%%  L.O. EQNS  %%
%%%%%%%%%%%%%%%%%
\section{Leading Order Equations}

Most of the necessary work has been completed, allowing us to begin expanding the integral equations and separating the LO and NLO behavior.  The leading order is handled in this section, with next-to-leading order corrections being treated in Sec.~\ref{sec:NLO}.  

Discerning the order at which certain terms contribute now becomes more difficult.  Unlike previous expansions, we cannot integrate the terms to see the behavior of the final answer.   The reason is that some of the integrals involve $f$, which is the unknown function we are trying to solve for.  Therefore, we must resort to dimensional analysis and other methods that may require knowledge of the final solution.  We provide a detailed explanation of each technique the first time it is used, but the details are omitted for later uses in the interest of brevity.

    %%%%%%%%%%%%%%%%%%%%
    %%  LOW-MOMENTUM  %%
    %%%%%%%%%%%%%%%%%%%%
\subsection{Low-Momentum Region \label{sec:fl0}}

Consider the expansion terms comprising $f_l(p)$:
\begin{eqnarray}
f_l(p) & = & \frac{p^2 + \eta_3^2 - \eta_2^2}{2 \pi^2 D_{l0}(\eta_2,\eta_3,p)} \left( 1 - \frac{D_{l1}(\eta_2,\eta_3,p,\Lambda)}{D_{l0}(\eta_2,\eta_3,p)} \right)\int_0^{\infty} \frac{dq}{q^2} \: \Bigg[ U_2(q) U_2\left( q/2 \right) f_h(q)  \nonumber
\\
&& \left. +~\frac{q^3}{2 p (q^2 + \eta_3^2 - \eta_2^2)} \ln\left(\frac{\eta_3^2 + 2p^2 + 2q^2 + 2pq}{\eta_3^2 + 2p^2 + 2q^2 - 2pq}\right) f_l(q) - f_d(q) \right] \nonumber
\\
&& -~(\delta_0 + \delta_1) \, \frac{(p^2 + \eta_3^2 - \eta_2^2)(D_{1l0}(\Lambda) + D_{1l1}(\eta_3,p))}{\Lambda^2 \left( G_{2,0} + G_{2,1} \right) (D_{l0}(\eta_2,\eta_3,p) + D_{l1}(\eta_2,\eta_3,p,\Lambda))} .
\end{eqnarray}
\noindent For the leading order equation, we need to identify only the terms that are $\mathcal{O}(1)$.  Obviously, the ratio $D_{l1}/D_{l0}$ can be dropped from the factor in front of the integral because it only provides a $\eta/\Lambda$ correction compared to 1.  The remaining part of this factor behaves as $\mathcal{O}(\eta)$, and the integral itself must have the dimensions of inverse momentum to make the product dimensionless.

Next, we analyze each integrand to determine its leading order behavior.  The first integrand contains the functions $U_2$ and $f_h$.  For a leading order analysis, we only need to consider $f_{h0}(q)$.  The dimensionful parameters in this function are $q$ and $\Lambda$.\footnote{This statement will be verified later when we examine the equation for $f_{h0}$.}  The same holds true for $U_2$.  This means that after integrating over $q$, only $\Lambda$ will be left to provide the needed dimension to the result.  Consequently, the integral must behave as $\mathcal{O}(\Lambda^{-1})$.  With the factor in front of the integral, we are left with a term of $\mathcal{O}(\eta/\Lambda)$.  Thus, we can drop this term for now and use it later when we analyze the NLO behavior.   

Before continuing, notice that this integrand possesses a divergence as $q \rightarrow 0$.  In this limit, $U_2(q)$ approaches $1$, and $f_{h0}(q)$ approaches $f_{d0}(q)$.  We will see later that $f_{d0}$ is proportional to a cosine, making the integral diverge like $1/q$.  This is not a problem because the $f_{d0}$ term from the third integrand cancels this divergence.  Together they yield a finite result.  This allows us to classify the third integrand as a  $\mathcal{O}(\eta/\Lambda)$ contribution and drop it from immediate consideration.

The second integrand involves a term containing $\eta_2$, $\eta_3$, and $p$ along with the function $f_l(q)$.  Again, only the $\mathcal{O}(1)$ part of $f_l$, namely $f_{l0}$, needs to be considered.  As we will soon see, it has no $\Lambda$ dependence.  Therefore, the integral only involves quantities of order $\eta$ and must behave as $1/\eta$.  Combining this with the factor in front of the integral, we get the only $\mathcal{O}(1)$ contribution from the integration.

The last part to investigate is the three-body interaction term.  If we take only the leading order parts of each piece, we find that the entire term behaves like
\begin{equation}
\mathcal{O}(1) \, \frac{\mathcal{O}(\eta^2) \, \mathcal{O}(\Lambda)}{\Lambda^2 \, \mathcal{O}(1) \, \mathcal{O}(\eta)} \sim \mathcal{O}\left(\frac{\eta}{\Lambda}\right) .
\end{equation}
\noindent This can only make an addition to the NLO behavior.

Overall, we find that only the second integrand is used in the equation for $f_{l0}$.  The final equation is
\begin{eqnarray}
f_{l0}(\eta_2, \eta_3, p) & = & \frac{1}{4 \pi^2 D_{l0}(\eta_2,\eta_3,p)} \int_0^{\infty} dq \: \frac{q \, (p^2 + \eta_3^2 - \eta_2^2)}{p \, (q^2 + \eta_3^2 - \eta_2^2)} \nonumber
\\
&& \times \ln\left(\frac{\eta_3^2 + 2p^2 + 2q^2 + 2pq}{\eta_3^2 + 2p^2 + 2q^2 - 2pq}\right) f_{l0}(\eta_2,\eta_3,q) \label{eqn:int_fl0}.
\end{eqnarray}
\noindent From this equation, we see that our previous assumption that $f_{l0}$ contains no $\Lambda$ dependence is indeed correct.

    %%%%%%%%%%%%%%%%%%%%
    %%  MID-MOMENTUM  %%
    %%%%%%%%%%%%%%%%%%%%
\subsection{Mid-Momentum Region}

In the mid-momentum region, the integral equation is
\begin{eqnarray}
f_d(p) & = & \frac{1}{2 \pi^2 D_{d0}(p)}  \left( 1 - \frac{D_{d1}(\eta_2,p,\Lambda)}{D_{d0}(p)} \right)  \int_0^{\infty} dq \: \left[ \frac{p^2}{q^2} U_2(q) U_2\left(q/2\right) f_h(q) \right. \nonumber
\\
&& \left. +~\frac{p}{2q} \ln\left(\frac{p^2 + q^2 + pq}{p^2 + q^2 - pq}\right) f_d(q) + \frac{q^2}{(q^2 + \eta_3^2 - \eta_2^2)} f_l(q) - \frac{p^2}{q^2} f_d(q) - f_d(q) \right] \nonumber
\\
&& -~(\delta_0 + \delta_1) \, \frac{(p^2 + \eta_3^2 - \eta_2^2)(D_{1d0}(\Lambda) + D_{1d1}(p))}{\Lambda^2 \left( G_{2,0} + G_{2,1} \right) (D_{d0}(p) + D_{d1}(\eta_2,p,\Lambda))} \label{eqn:pre_int_fd0}.
\end{eqnarray}
\noindent The same reasoning used for $f_{l0}$ can be applied here as well.  The $D_{d1}/D_{d0}$ ratio is dropped from the integral prefactor, resulting in a $1/p$ term.  The first integrand has a $q \rightarrow 0$ divergence that is canceled by the fourth integrand.  Together they yield a finite integral that dimensional analysis shows to be $\mathcal{O}(p^2/\Lambda)$.  Combined with the prefactor, the end result is a $\mathcal{O}(p/\Lambda)$ term that does not contribute to leading order.  

The third integrand is divergent as $q \rightarrow \infty$, but the fifth integrand cancels this divergence.  The result is a finite term that behaves as $\eta/p$, also a NLO correction.  Only the second integrand contributes to leading order.  The function $f_d(q)$ is replaced by $f_{d0}(q)$, and the integral of this term is of order $p$.  With the prefactor, the term is $\mathcal{O}(1)$.

The leading behavior of the three-body interaction is
\begin{equation}
\mathcal{O}(1) \, \frac{p^2 \, \mathcal{O}(\Lambda)}{\Lambda^2 \, \mathcal{O}(1) \, \mathcal{O}(p)} \sim \mathcal{O}\left( \frac{p}{\Lambda} \right) ,
\end{equation}
\noindent which cannot contribute to leading order.  Thus, only one integrand adds to the leading order equation for $f_d$, giving us
\begin{equation}
f_{d0}(p) = \frac{p}{4 \pi^2 D_{d0}(p)} \int_0^{\infty} \frac{dq}{q} \, \ln\left(\frac{p^2 + q^2 + pq}{p^2 + q^2 - pq}\right) f_{d0}(q) \label{eqn:int_fd0}.
\end{equation}
\noindent This equation happens to have an analytic solution \cite{hammer:orig}, 
\begin{equation}
f_{d0}(p) = A \cos\left( s_0 \ln\left(\frac{p}{\Lambda}\right) + \theta \right) ,
\end{equation}
\noindent where $A$ is an arbitrary amplitude, $\theta$ is a phase determined by boundary conditions, and $s_0$ is the real, positive solution to the equation
\begin{equation}
\frac{8}{\sqrt{3}} \, \sinh\left(\frac{\pi s_0}{6}\right) = s_0 \, \cosh\left(\frac{\pi s_0}{2}\right) .
\end{equation}
\noindent The value of $s_0$ is approximately $1.006237825102$.

The $\Lambda$ dependence in this solution may be a bit misleading since Eq.~(\ref{eqn:int_fd0}) does not contain it.  The cutoff is included so that the argument of the logarithm is dimensionless.  Making this choice is a matter of preference, and it could just as easily have been $\eta_3$.

    %%%%%%%%%%%%%%%%%%%%%
    %%  HIGH-MOMENTUM  %%
    %%%%%%%%%%%%%%%%%%%%%
\subsection{High-Momentum Region}

The expansion terms for the high-momentum region take the form
\begin{eqnarray}
f_h(p) & = & \frac{\left( 1 - D_{h1}(\eta_2)/D_{h0}(p,\Lambda) \right)}{2 \pi^2 D_{h0}(p, \Lambda)} \int_0^{\infty} dq \: \left[ p^2 \int_{-1}^{1} dz \, \frac{U_2\left(\vec{q} + \frac{1}{2}\vec{p}\right) U_2\left(\vec{p} + \frac{1}{2}\vec{q}\right)}{2p^2 + 2q^2 + 2pqz} f_h(q) \right. \nonumber
\\
&& \left. +~U_2(p) U_2\left(p/2\right) \left( \frac{q^2}{(q^2 + \eta_3^2 - \eta_2^2)} f_l(q)  - f_d(q) \right) \right] \nonumber
\\
&& -~(\delta_0 + \delta_1) \, \frac{(p^2 + \eta_3^2 - \eta_2^2) D_{1h0}(p,\Lambda)}{\Lambda^2 \left( G_{2,0} + G_{2,1} \right) (D_{h0}(p,\Lambda) + D_{h1}(\eta_2))} \label{eqn:initfh}.
\end{eqnarray}
\noindent For leading order behavior, we drop the ratio $D_{h1}/D_{h0}$ from the integral's prefactor.  The remaining part of the prefactor behaves as $1/\Lambda$.  The only parameters in the first integrand with the same dimension as momentum are $p$ and $\Lambda$, both of which are of the same order.  Dimensional analysis requires that this integral be $\mathcal{O}(\Lambda)$, making its product with the prefactor $\mathcal{O}(1)$.  Thus, it contributes to the leading order equation.

The third integrand cancels the $q \rightarrow \infty$ divergence in the second, resulting in a finite term of order $\eta/\Lambda$ as required by dimensional analysis.  These terms are NLO contributions that we will need later.

Last of all, we examine the leading order behavior of the three-body interaction.  We find it to be
\begin{equation}
- \, \delta_0 \, \frac{p^2 D_{1h0}(p,\Lambda)}{\Lambda^2 G_{2,0} \,  D_{h0}(p,\Lambda)} \sim \mathcal{O}(1) \, \frac{p^2 \, \mathcal{O}(\Lambda)}{\Lambda^2 \, \mathcal{O}(1) \, \mathcal{O}(\Lambda)} \sim \mathcal{O}(1) ,
\end{equation}
\noindent making it necessary to include it in our equation for $f_{h0}$.  This is the only leading order equation that contains a three-body interaction term.  The exact form of this term is
\begin{equation}
- \delta_0 \, \frac{3 h_2^2 + 8 h_2 + 16}{128 \sqrt{2} \, \pi^{3/2}} \left( \frac{D_{1h0}(p,\Lambda)}{D_{h0}(p,\Lambda)}\right) \left( \frac{p^2}{\Lambda^2} \right) ,
\end{equation}
\noindent giving us the last piece we need to construct the complete integral equation for $f_{h0}$:
\begin{eqnarray}
f_{h0}(p, \Lambda) & = & \frac{p^2}{2 \pi^2 D_{h0}(p,\Lambda)} \int_0^{\infty} dq \, \int_{-1}^{1} dz \, \frac{U_2\left(\vec{q} + \frac{1}{2}\vec{p}\right) U_2\left(\vec{p} + \frac{1}{2}\vec{q}\right)}{2p^2 + 2q^2 + 2pqz} f_{h0}(q, \Lambda) \nonumber
\\
&& -~\delta_0 \, \frac{3 h_2^2 + 8 h_2 + 16}{128 \sqrt{2} \, \pi^{3/2}} \left( \frac{D_{1h0}(p,\Lambda)}{D_{h0}(p,\Lambda)} \right) \left( \frac{p^2}{\Lambda^2} \right) \label{eqn:int_fh0}.
\end{eqnarray}
\noindent Notice that this equation, and hence the solution $f_{h0}$, only contains dependence on $p$ and $\Lambda$.  This fact was used earlier in our derivation of the integral equation for $f_{l0}$, and now it has been justified.

%%%%%%%%%%%%%%%%%%%
%%  N.L.O. EQNS  %%
%%%%%%%%%%%%%%%%%%%
\section{Next-To-Leading Order Equations \label{sec:NLO}}

The leading order equations for all three pieces of $f$ have been derived.  Now we must go back and gather the NLO terms that were neglected in our first pass, as well as determine any sub-leading behavior to the leading order terms that may also contribute.  However, before doing so, we derive two relations that will be needed for our first order corrections.

    %%%%%%%%%%%%%%%%%%
    %%  IDENTITIES  %%
    %%%%%%%%%%%%%%%%%%
\subsection{Leading Order Identities}

As $p$ becomes much greater than $\eta$, the function $f_{l0}(p)$ must begin to equal $f_{d0}(p)$.  If we consider the leading order behavior of Eq.~(\ref{eqn:int_fl0}) when $p \gg \eta$, we obtain the relation
\begin{equation}
f_{d0}(p) = \frac{1}{4 \pi^2 D_{d0}(p)} \int_0^{\infty} dq \:  \frac{q p}{q^2 + \eta_3^2 - \eta_2^2}  \ln\left(\frac{p^2 + q^2 + pq}{p^2 + q^2 - pq}\right) f_{l0}(\eta_2,\eta_3,q)  .
\end{equation}
\noindent Since $f_{d0}(p)$ satisfies Eq.~(\ref{eqn:int_fd0}), the above relation can also be rewritten as
\begin{equation}
\frac{1}{4 \pi^2 D_{d0}(p)} \int_0^{\infty} dq \:  \ln\left(\frac{p^2 + q^2 + pq}{p^2 + q^2 - pq}\right) \left[ \frac{q p}{q^2 + \eta_3^2 - \eta_2^2} f_{l0}(\eta_2,\eta_3,q) - \frac{p}{q} f_{d0}(q) \right] = 0 .
\end{equation}
\noindent For values of $q \gg \eta$, the integrand will vanish.  Therefore, the previous relation must also hold when we restrict the $q$ integration to values of the same order as $\eta$.  Because $p$ is much greater than $\eta$, and hence $q$, the logarithm can be approximated as $2q/p$.  The relation now becomes
\begin{equation}
\frac{1}{4 \pi^2 D_{d0}(p)} \int_0^{\infty} dq \: \frac{2q}{p} \left[ \frac{q p}{q^2 + \eta_3^2 - \eta_2^2} f_{l0}(\eta_2,\eta_3,q) - \frac{p}{q} f_{d0}(q) \right] = 0 ,
\end{equation}
\noindent or more simply
\begin{equation}
\int_0^{\infty} dq \: \left[ \frac{q^2}{q^2 + \eta_3^2 - \eta_2^2} f_{l0}(\eta_2,\eta_3,q) -  f_{d0}(q) \right] = 0 \label{eqn:low_ident}.
\end{equation}

Just as $f_{l0}(p)$ equals $f_{d0}(p)$ for $p \gg \eta$, so must $f_{h0}(p)$ for $p \ll \Lambda$.  The leading order behavior of Eq.~(\ref{eqn:int_fh0}) for this range of $p$ is
\begin{eqnarray}
f_{d0}(p) & = & \frac{p^2}{2 \pi^2 D_{d0}(p)} \int_0^{\infty} dq \, \int_{-1}^{1} dz \, \frac{U_2(q) U_2(q/2)}{2p^2 + 2q^2 + 2pqz} f_{h0}(q, \Lambda) \nonumber
\\
&& -~\delta_0 \, \frac{3 h_2^2 + 8 h_2 + 16}{128 \sqrt{2} \, \pi^{3/2}} \left( \frac{D_{1d0}(\Lambda)}{D_{d0}(p)} \right)  \left( \frac{p^2}{\Lambda^2} \right).
\end{eqnarray}
\noindent Using Eq.~(\ref{eqn:int_fd0}) once again allows us to rewrite the previous equation in the form
\begin{eqnarray}
0 & = & \frac{p}{4 \pi^2 D_{d0}(p)} \int_0^{\infty} \frac{dq}{q} \,  \ln\left(\frac{p^2 + q^2 + pq}{p^2 + q^2 - pq}\right) \left[ U_2(q) U_2(q/2) f_{h0}(q, \Lambda) - f_{d0}(q) \right] \nonumber
\\
&& -~\delta_0 \, \frac{3 h_2^2 + 8 h_2 + 16}{128 \sqrt{2} \, \pi^{3/2}} \left( \frac{D_{1d0}(\Lambda)}{D_{d0}(p)} \right)  \left( \frac{p^2}{\Lambda^2} \right).
\end{eqnarray}
\noindent The integral vanishes for values of $q$ much less than $\Lambda$, so it may safely be restricted to values greater than or on the order of $\Lambda$.  This implies that $p \ll q$, permitting us to approximate the logarithm by $2p/q$.  Using this approximation gives us the integral
\begin{equation}
\frac{p^2}{2 \pi^2 D_{d0}(p)} \int_0^{\infty} \frac{dq}{q^2} \, \left[ U_2(q) U_2(q/2) f_{h0}(q, \Lambda) - f_{d0}(q) \right] .
\end{equation}
\noindent After dividing out the integral's $p$-dependent prefactor, we are left with the relation
\begin{eqnarray}
0 & = & \int_0^{\infty} \frac{dq}{q^2} \, \left[ U_2(q) U_2(q/2) f_{h0}(q, \Lambda) - f_{d0}(q) \right] \nonumber
\\
&& -{} \delta_0 \, \frac{\sqrt{\pi}}{64 \sqrt{2} \, \Lambda^2} \left( 3 h_2^2 + 8 h_2 + 16 \right) D_{1d0}(\Lambda) \label{eqn:high_ident}.
\end{eqnarray}
\noindent Equations (\ref{eqn:low_ident}) and (\ref{eqn:high_ident}) are the key relations, and in a moment we shall see why they are necessary.

    %%%%%%%%%%%%%%%%%%%%
    %%  LOW-MOMENTUM  %%
    %%%%%%%%%%%%%%%%%%%%
\subsection{Low-Momentum Region}

As shown before, the expansion for $f_l$ is
\begin{eqnarray}
f_l(p) & = & \frac{p^2 + \eta_3^2 - \eta_2^2}{2 \pi^2 D_{l0}(\eta_2,\eta_3,p)} \left( 1 - \frac{D_{l1}(\eta_2,\eta_3,p,\Lambda)}{D_{l0}(\eta_2,\eta_3,p)} \right) \int_0^{\infty} \frac{dq}{q^2} \: \Bigg[ U_2(q) U_2\left( q/2 \right) f_h(q) \nonumber
\\
&& \left. +~\frac{q^3}{2 p (q^2 + \eta_3^2 - \eta_2^2)} \ln\left(\frac{\eta_3^2 + 2p^2 + 2q^2 + 2pq}{\eta_3^2 + 2p^2 + 2q^2 - 2pq}\right) f_l(q) - f_d(q) \right] \nonumber
\\
&& -~(\delta_0 + \delta_1) \, \frac{(p^2 + \eta_3^2 - \eta_2^2)(D_{1l0}(\Lambda) + D_{1l1}(\eta_3,p))}{\Lambda^2 \left( G_{2,0} + G_{2,1} \right) (D_{l0}(\eta_2,\eta_3,p) + D_{l1}(\eta_2,\eta_3,p,\Lambda))} .
\end{eqnarray}
\noindent The leading order behavior has been extracted to get Eq.~(\ref{eqn:int_fl0}), and now we wish to examine the terms of order $\eta/\Lambda$ to obtain the NLO equation.

As noted is Sec.~\ref{sec:fl0}, the leading behavior of the first and third integrands is $\mathcal{O}(\eta/\Lambda)$.  Together, these terms provide an addition to the equation for first order corrections:
\begin{equation}
 \frac{p^2 + \eta_3^2 - \eta_2^2}{2 \pi^2 D_{l0}(\eta_2,\eta_3,p)} \int_0^{\infty} \frac{dq}{q^2} \: \left[ U_2(q) U_2\left( q/2 \right) f_{h0}(q) - f_d(q) \right]
\end{equation}
\noindent Using (\ref{eqn:high_ident}), this may be simplified to
\begin{eqnarray}
\frac{p^2 + \eta_3^2 - \eta_2^2}{2 \pi^2 D_{l0}(\eta_2,\eta_3,p)} \left[ \delta_0 \, \frac{\sqrt{\pi}}{64 \sqrt{2} \, \Lambda^2} \left( 3 h_2^2 + 8 h_2 + 16 \right) D_{1d0}(\Lambda) \right] \hspace{0.5in} \nonumber
\\
 = \delta_0 \, \frac{(p^2 + \eta_3^2 - \eta_2^2) D_{1d0}(\Lambda)}{\Lambda^2 G_{2,0} D_{l0}(\eta_2,\eta_3,p)} \label{eqn:fl1part1}.
\end{eqnarray}
\noindent Any other pieces of the first and third integrands are necessarily of higher order and may be ignored.

The leading order behavior of the second integrand was determined to be $\mathcal{O}(1)$, which means there may be sub-leading terms that are of order $\mathcal{O}(\eta/\Lambda)$.  In fact, there are two such terms.  One results from dropping the $D_{l1}/D_{l0}$ ratio from the integral's prefactor and using $f_{l1}$ in the integrand.  The other retains the prefactor ratio and uses $f_{l0}$ in the integrand.  Their combined contribution is
\begin{eqnarray}
\frac{p^2 + \eta_3^2 - \eta_2^2}{4 \pi^2 p D_{l0}(\eta_2,\eta_3,p)} \left[ \int_0^{\infty} dq \: \frac{q}{q^2 + \eta_3^2 - \eta_2^2} \ln\left(\frac{\eta_3^2 + 2p^2 + 2q^2 + 2pq}{\eta_3^2 + 2p^2 + 2q^2 - 2pq}\right) f_{l1}(q) \right. \nonumber
\\
- {}\left. \frac{D_{l1}(\eta_2,\eta_3,p,\Lambda)}{D_{l0}(\eta_2,\eta_3,p)}  \int_0^{\infty} dq \: \frac{q}{q^2 + \eta_3^2 - \eta_2^2} \ln\left(\frac{\eta_3^2 + 2p^2 + 2q^2 + 2pq}{\eta_3^2 + 2p^2 + 2q^2 - 2pq}\right) f_{l0}(q) \right]  ,
\end{eqnarray}
\noindent which may also be written as
\begin{eqnarray}
\frac{1}{4 \pi^2 D_{l0}(\eta_2,\eta_3,p)} \int_0^{\infty} dq \left[ \frac{q (p^2 + \eta_3^2 - \eta_2^2)}{p (q^2 + \eta_3^2 - \eta_2^2)} \ln\left(\frac{\eta_3^2 + 2p^2 + 2q^2 + 2pq}{\eta_3^2 + 2p^2 + 2q^2 - 2pq}\right) f_{l1}(q) \right] \nonumber
\\
-~\frac{D_{l1}(\eta_2,\eta_3,p,\Lambda)}{D_{l0}(\eta_2,\eta_3,p)} f_{l0}(p) \label{eqn:fl1part2},
\end{eqnarray}
\noindent by using Eq.~(\ref{eqn:int_fl0}).

The final source of possible corrections is the three-body interaction term.  Our previous analysis of $f_{l0}$ shows that the leading behavior of this term is $\mathcal{O}(\eta/\Lambda)$.  Therefore, it contributes a term to the NLO equation that looks like
\begin{equation}
- \delta_0 \, \frac{(p^2 + \eta_3^2 - \eta_2^2) D_{1l0}(\Lambda)}{\Lambda^2 G_{2,0} D_{l0}(\eta_2,\eta_3,p)} \label{eqn:fl1part3}.
\end{equation}
\noindent Combining the results of (\ref{eqn:fl1part1}), (\ref{eqn:fl1part2}), and (\ref{eqn:fl1part3}) yields
\begin{eqnarray}
\frac{1}{4 \pi^2 D_{l0}(\eta_2,\eta_3,p)} \int_0^{\infty} dq \left[ \frac{q (p^2 + \eta_3^2 - \eta_2^2)}{p (q^2 + \eta_3^2 - \eta_2^2)} \ln\left(\frac{\eta_3^2 + 2p^2 + 2q^2 + 2pq}{\eta_3^2 + 2p^2 + 2q^2 - 2pq}\right) f_{l1}(q) \right] \nonumber
\\
-~\frac{D_{l1}(\eta_2,\eta_3,p,\Lambda)}{D_{l0}(\eta_2,\eta_3,p)} f_{l0}(p) + \delta_0 \, \frac{(p^2 + \eta_3^2 - \eta_2^2) D_{1d0}(\Lambda)}{\Lambda^2 G_{2,0} D_{l0}(\eta_2,\eta_3,p)} - \delta_0 \, \frac{(p^2 + \eta_3^2 - \eta_2^2) D_{1l0}(\Lambda)}{\Lambda^2 G_{2,0} D_{l0}(\eta_2,\eta_3,p)} , \nonumber
\end{eqnarray}
\noindent which may be simplified by recognizing that $D_{1d0}(\Lambda) = D_{1l0}(\Lambda)$.  The final equation for the first order correction in the low-momentum region becomes
\begin{eqnarray}
f_{l1}(\eta_2, \eta_3, p, \Lambda) & = & \frac{1}{4 \pi^2 D_{l0}(\eta_2,\eta_3,p)} \int_0^{\infty} dq \left[ \frac{q (p^2 + \eta_3^2 - \eta_2^2)}{p (q^2 + \eta_3^2 - \eta_2^2)} \right. \nonumber
\\
&& \times \left. \ln\left(\frac{\eta_3^2 + 2p^2 + 2q^2 + 2pq}{\eta_3^2 + 2p^2 + 2q^2 - 2pq}\right) f_{l1}(\eta_2, \eta_3, q, \Lambda) \right] \nonumber
\\
&& - {} \frac{D_{l1}(\eta_2,\eta_3,p,\Lambda)}{D_{l0}(\eta_2,\eta_3,p)} f_{l0}(\eta_2,\eta_3,p) .
\end{eqnarray}

    %%%%%%%%%%%%%%%%%%%%
    %%  MID-MOMENTUM  %%
    %%%%%%%%%%%%%%%%%%%%
\subsection{Mid-Momentum Region}

The first order correction terms for $f_{d1}$ are obtained in a manner very similar to $f_{l1}$.  The leading order analysis of Eq.~(\ref{eqn:pre_int_fd0}) showed that the first, third, fourth, and fifth integrands have an $\mathcal{O}(\eta/\Lambda)$ behavior for their leading terms.   This provides a contribution of
\begin{equation}
\frac{1}{2 \pi^2 D_{d0}(p)} \int_0^{\infty} dq \, \left[ \frac{p^2}{q^2} U_2(q) U_2\left(q/2\right) f_{h0}(q) - \frac{p^2}{q^2} f_{d0}(q) + \frac{q^2}{q^2 + \eta_3^2 - \eta_2^2} f_{l0}(q)  -  f_{d0}(q) \right] .
\end{equation}
\noindent The identities (\ref{eqn:low_ident}) and (\ref{eqn:high_ident}) can be used to reduce these terms to
\begin{equation}
\delta_0 \, \frac{p^2 \, D_{1d0}(\Lambda)}{\Lambda^2 \, G_{2,0} D_{d0}(p)} .
\end{equation}
\noindent The three-body interaction also provides a term of the same order,
\begin{equation}
- \delta_0 \, \frac{p^2 \, D_{1d0}(\Lambda)}{\Lambda^2 \, G_{2,0} D_{d0}(p)} ,
\end{equation}
\noindent which exactly cancels the previous integral terms.  The only term left to provide corrections is the second integrand involving $f_d(q)$.  There are two obvious corrections: one that results from dropping the $D_{d1}/D_{d0}$ ratio in the prefactor and using $f_{d1}(q)$ in the integral, and the other that results from keeping the ratio in the prefactor and using $f_{d0}(q)$ in the integral.  However, there is a third contribution that involves keeping the prefactor ratio and using $f_{d1}(q)$ in the integral.  For the case of $f_{l1}$, no such term is encountered because the product would be $\mathcal{O}(\eta^2/\Lambda^2)$.  Yet, in this case, it is entirely possible that two first order corrections can have a product that is also a first order correction.\footnote{The simplest example is $(\eta/p)(p/\Lambda) = (\eta/\Lambda)$.}

The final integral equation is now written as
\begin{equation}
f_{d1}(p) = \frac{p}{4 \pi^2 D_{d0}(p)} \left( 1 - \frac{D_{d1}(\eta_2,p,\Lambda)}{D_{d0}(p)} \right) \int_0^{\infty} \frac{dq}{q} \, \ln\left(\frac{p^2 + q^2 + pq}{p^2 + q^2 - pq}\right) \left( f_{d0}(q) + f_{d1}(q) \right) ,
\end{equation}
\noindent and we must remember to keep only the terms that are the same order as $f_{d1}$.  

Like the leading order function, $f_{d0}$, this equation has an analytic solution:
\begin{eqnarray}
f_{d1}(\eta_2,p,\Lambda) & = & A_{l1} \frac{\eta_2}{p} \cos\left(s_0 \ln\left(\frac{p}{\Lambda}\right) + \theta_{l1} \right) \nonumber
\\
&& +~A_{h1} \frac{p}{\Lambda} \cos\left(s_0 \ln\left(\frac{p}{\Lambda}\right) + \theta_{h1} \right) \nonumber
\\
&& +~A_{d10} \frac{\eta_2}{\Lambda} \cos\left(s_0 \ln\left(\frac{p}{\Lambda}\right) + \theta_{d10} \right) \nonumber
\\
&& +~A_{d11} \frac{\eta_2}{\Lambda} \ln\left(\frac{p}{\Lambda}\right) \cos\left(s_0 \ln\left(\frac{p}{\Lambda}\right) + \theta_{d11} \right) .
\end{eqnarray}
\noindent The details of the derivation leading to this form are given in Appendix \ref{app:fd1form}.

    %%%%%%%%%%%%%%%%%%%%%
    %%  HIGH-MOMENTUM  %%
    %%%%%%%%%%%%%%%%%%%%%
\subsection{High-Momentum Region}

During the analysis of Eq.~(\ref{eqn:initfh}) to determine $f_{h0}$, we discovered that the second and third integrands have a leading behavior that can only contribute to the first order corrections.  Upon closer inspection, we see that the sum of these integrands fits the form of Eq.~(\ref{eqn:low_ident}) and must therefore equal zero.

The first integrand provides two additions to the NLO result,
\begin{equation}
\frac{p^2}{2 \pi^2 D_{h0}(p,\Lambda)} \int_0^{\infty} dq \, \int_{-1}^{1} dz \, \frac{U_2\left(\vec{q} + \frac{1}{2}\vec{p}\right) U_2\left(\vec{p} + \frac{1}{2}\vec{q}\right)}{2p^2 + 2q^2 + 2pqz} f_{h1}(q) \label{eqn:fh1part1},
\end{equation}
\noindent and
\begin{equation}
 - \, \frac{p^2 D_{h1}(\eta_2)}{2 \pi^2 D_{h0}(p,\Lambda)^2} \int_0^{\infty} dq \, \int_{-1}^{1} dz \, \frac{U_2\left(\vec{q} + \frac{1}{2}\vec{p}\right) U_2\left(\vec{p} + \frac{1}{2}\vec{q}\right)}{2p^2 + 2q^2 + 2pqz} f_{h0}(q) \label{eqn:fh0addfh1}.
\end{equation}
\noindent Using the integral equation for $f_{h0}(p)$, equation (\ref{eqn:fh0addfh1}) can be simplified:
\begin{equation}
- \frac{D_{h1}(\eta_2)}{D_{h0}(p,\Lambda)} \left[ f_{h0}(p) + \delta_0 \, \frac{p^2 \, D_{1h0}(p,\Lambda)}{\Lambda^2 \, G_{2,0} \, D_{h0}(p,\Lambda)} \right] \label{eqn:fh1part2} .
\end{equation}

Finally, we must consider the three-body interaction term.  The leading behavior of this term is $\mathcal{O}(1)$, but there are three sub-leading terms that must be added to our first order corrections.  These terms are
\begin{equation}
- \delta_1 \, \frac{p^2 D_{1h0}(p,\Lambda)}{\Lambda^2 \, G_{2,0} \, D_{h0}(p,\Lambda)} \label{eqn:fh1part3.1},
\end{equation}
\begin{equation}
 \delta_0 \, \frac{p^2 D_{1h0}(p,\Lambda) \, G_{2,1}}{\Lambda^2 \, G_{2,0}^2 \, D_{h0}(p,\Lambda)} \label{eqn:fh1part3.2},
\end{equation}
\noindent and
\begin{equation}
\delta_0 \, \frac{p^2 D_{1h0}(p,\Lambda) D_{h1}(\eta_2)}{\Lambda^2 \, G_{2,0} \, D_{h0}(p,\Lambda)^2} \label{eqn:fh1part3.3}.
\end{equation}
\noindent Notice that Eq.~(\ref{eqn:fh1part3.3}) cancels the second part of Eq.~(\ref{eqn:fh1part2}).  The final integral equation is
\begin{eqnarray}
&& \hspace{-0.25in} f_{h1}(\eta_2,\eta_3,p,\Lambda) = \frac{p^2}{2 \pi^2 D_{h0}(p,\Lambda)} \int_0^{\infty} dq \, \int_{-1}^{1} dz \, \frac{U_2\left(\vec{q} + \frac{1}{2}\vec{p}\right) U_2\left(\vec{p} + \frac{1}{2}\vec{q}\right)}{2p^2 + 2q^2 + 2pqz} f_{h1}(\eta_2,\eta_3,q,\Lambda) \nonumber
\\
&& \hspace{0.5in} - {} \frac{D_{h1}(\eta_2)}{D_{h0}(p,\Lambda)} f_{h0}(p,\Lambda) - \delta_0 \,  \frac{p^2 D_{1h0}(p,\Lambda)}{\Lambda^2 \, G_{2,0} \, D_{h0}(p,\Lambda)} \left[ \frac{\delta_1}{\delta_0} - \frac{G_{2,1}}{G_{2,0}} \right] .
\end{eqnarray}

%%%%%%%%%%%%%%%%%%%%
%%  PSEUDO => G3  %%
%%%%%%%%%%%%%%%%%%%%
\section{Obtaining Values for $G_3$}

With the integral equations derived so far, it is possible to compute solutions for all of the $f$ functions as well as the value of $\delta$.  The next step is to determine the necessary equations so that the value of $\delta$ can be used to compute the value of $G_3$.  The basic equations needed are:
\begin{equation}
\delta = G_3 \, \Phi_1 / \Lambda ,
\end{equation}
\begin{equation}
\Phi_1 = \frac{3 G_2}{2 \pi^2 \Lambda} \int_0^{\infty} dp \left[ p^2 \, D_1(p) \, \Phi(p) \right] - \frac{G_3 \, D_2 \, \Phi_1}{\Lambda^4}.
\end{equation}
\noindent By defining the dimensionless quantity
\begin{equation}
\mathcal{I} \equiv \frac{3}{2 \pi^2 \Lambda^2} \int_0^{\infty} dp \left[ p^2 \, D_1(p) \, \Phi(p) \right] ,
\end{equation}
\noindent the equation for $\Phi_1$ may be written as
\begin{equation}
\Phi_1 = G_2 \, \Lambda \, \mathcal{I} - G_3 \, D_2 \, \Phi_1/\Lambda^4 ,
\end{equation}
\noindent or equivalently
\begin{equation}
\frac{\delta \, \Lambda}{G_3} = G_2 \, \Lambda \, \mathcal{I} - \frac{D_2 \, \delta}{\Lambda^3} \label{eqn:preDeltaG3}.
\end{equation}
\noindent The goal is to expand both sides of this equation in powers of $\eta/\Lambda$ and then equate terms of the same order on both sides.  The result will be a relationship between $\delta_0, \delta_1$ and $G_{3,0}, G_{3,1}$.  From there, the value of $g_3$ is easily reconstructed.

    %%%%%%%%%%%%%%%%
    %%  MATCHING  %%
    %%%%%%%%%%%%%%%%
\subsection{Matching Conditions}

We know the expansion for every quantity in Eq.~(\ref{eqn:preDeltaG3}) except for two: $\mathcal{I}$ and $D_2$.  We will expand $\mathcal{I}$ in a moment, but for now we simply write the terms of its expansion as $\mathcal{I}_0$ and $\mathcal{I}_1$.  The remaining quantity is defined by
\begin{equation}
D_2 = \frac{1}{8 \pi^4} \int_0^{\infty} p^2 dp \int_0^{\infty} q^2 dq \int_{-1}^{1} dz \: \frac{U_3(q)^2 U_3(p)^2 U_3(\vec{q}+\vec{p}\,)^2}{2p^2 + 2q^2 + 2pqz + \eta_3^2} .
\end{equation}
\noindent The order $(\eta_3/\Lambda)^0$ term is obtained simply by setting $\eta_3$ equal to zero.  This results in
\begin{equation}
D_{2,0} = \frac{1}{8 \pi^4} \int_0^{\infty} p^2 dp \int_0^{\infty} q^2 dq \int_{-1}^{1} dz \: \frac{U_3(q)^2 U_3(p)^2 U_3(\vec{q}+\vec{p}\,)^2}{2p^2 + 2q^2 + 2pqz} ,
\end{equation}
\noindent which by dimensional analysis must be proportional to $\Lambda^4$.   Hence, $D_{2,0}/\Lambda^4$ is a constant.  The next term in the expansion of $D_2$ behaves like $(\eta_3/\Lambda)^2$, so we find that $D_{2,1} = 0$.

If we multiply both sides of Eq.~(\ref{eqn:preDeltaG3}) by $G_3/\Lambda$, we obtain
\begin{equation}
\delta = G_3 \left( G_2 \, \mathcal{I} - \frac{D_2}{\Lambda^4} \, \delta \right) .
\end{equation}
\noindent By expanding both sides,
\begin{equation}
\delta_0 + \delta_1 = \left( G_{3,0} + G_{3,1} \right) \left[ \left( G_{2,0} + G_{2,1} \right) \left( \mathcal{I}_0 + \mathcal{I}_1 \right) - \frac{D_{2,0}}{\Lambda^4} \left( \delta_0 + \delta_1 \right) \right] ,
\end{equation}
\noindent and matching corresponding orders,
\begin{eqnarray}
\delta_0 & = & G_{3,0} \left( G_{2,0} \, \mathcal{I}_0  - \frac{D_{2,0}}{\Lambda^4} \, \delta_0 \right) ,
\\
\delta_1 & = & G_{3,0} \left( G_{2,0} \, \mathcal{I}_1 + G_{2,1} \, \mathcal{I}_0  - \frac{D_{2,0}}{\Lambda^4} \, \delta_1 \right) + G_{3,1} \left( G_{2,0} \, \mathcal{I}_0  - \frac{D_{2,0}}{\Lambda^4} \, \delta_0 \right) ,
\end{eqnarray}
\noindent we obtain the relations
\begin{eqnarray}
G_{3,0} & = & \frac{\delta_0}{G_{2,0} \, \mathcal{I}_0 - D_{2,0} \, \delta_0/\Lambda^4 } ,
\\
G_{3,1} & = & \frac{\delta_1}{\delta_0} G_{3,0} -  \frac{1}{\delta_0} \left( G_{2,0} \, \mathcal{I} + G_{2,1} \, \mathcal{I}_0  - \frac{D_{2,0}}{\Lambda^4} \, \delta_1 \right) G_{3,0}^2 .
\end{eqnarray}

    %%%%%%%%%%%%%%%%%%%
    %%  I EXPANSION  %%
    %%%%%%%%%%%%%%%%%%%
\subsection{Expansion of $\mathcal{I}$}

To expand $\mathcal{I}$, we utilize the expansion technique of Sec.~\ref{sec:int3var}.  The definition
\begin{equation}
u(\eta_2,\eta_3,p,\Lambda) = \frac{3 p^2}{2 \pi^2 \Lambda^2 (p^2 + \eta_3^2 - \eta_2^2)} D_1(p) f(p)
\end{equation}
\noindent lets us write $\mathcal{I} = \int dp \left[ u_l + u_h - u_d \right]$.  The function $u$ must be expanded to $\mathcal{O}(\eta/\Lambda)$ in each of the three regions.  These expansions are shown below:
\begin{eqnarray}
u_l & = & \frac{3 p^2}{2 \pi^2 \Lambda^2 (p^2 + \eta_3^2 - \eta_2^2)} D_{1l0}(\Lambda) f_{l0}(p) \left[ 1 + \frac{D_{1l1}(\eta_3,p)}{D_{1l0}(\Lambda)} + \frac{f_{l1}(p)}{f_{l0}(p)} \right] ,
\\
u_d & = & \frac{3}{2 \pi^2 \Lambda^2} D_{1d0}(\Lambda) f_{d0}(p) \left[ 1 + \frac{D_{1d1}(p)}{D_{1d0}(\Lambda)} + \frac{f_{d1}(p)}{f_{d0}(p)} + \frac{D_{1d1}(p)}{D_{1d0}(\Lambda)}\frac{f_{d1}(p)}{f_{d0}(p)} \right] ,
\\
u_h & = & \frac{3}{2 \pi^2 \Lambda^2} D_{1h0}(p,\Lambda) f_{h0}(p) \left[ 1 + \frac{f_{h1}(p)}{f_{h0}(p)} \right] .
\end{eqnarray}
\noindent Note that the last term in $u_d$ contains the product of two first order corrections.  Only other first order corrections resulting from this product should be kept.

By using dimensional analysis, we can determine the pieces contributing to $\mathcal{I}_0$ and those contributing to $\mathcal{I}_1$.  When integrated, the leading term in $u_l$ is $\mathcal{O}(\eta/\Lambda)$ and thus becomes part of the equation for $\mathcal{I}_1$.  The other terms in $u_l$ may be dropped because they are $\mathcal{O}(\eta^2/\Lambda^2)$ or higher.   These statements also apply to $u_d$.  Therefore, only the leading terms of $u_l$ and $u_d$ need be considered.  Together they provide a contribution of
\begin{equation}
\int_0^{\infty} dp \, \left[ \frac{3 p^2}{2 \pi^2 \Lambda^2 (p^2 + \eta_3^2 - \eta_2^2)} D_{1l0}(\Lambda) f_{l0}(p) - \frac{3}{2 \pi^2 \Lambda^2} D_{1d0}(\Lambda) f_{d0}(p) \right] .
\end{equation}
\noindent Because $D_{1d0}(\Lambda) = D_{1l0}(\Lambda)$, we can simplify the equation further to obtain
\begin{equation}
\frac{3}{2 \pi^2 \Lambda^2} D_{1d0}(\Lambda) \int_0^{\infty} dp \, \left[ \frac{p^2}{p^2 + \eta_3^2 - \eta_2^2} f_{l0}(p) -  f_{d0}(p) \right] ,
\end{equation}
\noindent which we know from Eq.~(\ref{eqn:low_ident}) is zero.

The only contributions to $\mathcal{I}$ must then come from $u_h$.  The first term is $\mathcal{O}(1)$, making the second $\mathcal{O}(\eta/\Lambda)$.  These two terms provide us with the final equations:
\begin{eqnarray}
\mathcal{I}_0 & = & \frac{3}{2 \pi^2 \Lambda^2} \int_0^{\infty} dp \, D_{1h0}(p,\Lambda) \, f_{h0}(p,\Lambda) ,
\\
\mathcal{I}_1 & = & \frac{3}{2 \pi^2 \Lambda^2} \int_0^{\infty} dp \, D_{1h0}(p,\Lambda) \, f_{h1}(\eta_2,\eta_3,p,\Lambda) .
\end{eqnarray}
\noindent With these equations, we now have everything needed to convert values of $\delta$ to values of $G_3$.

%%%%%%%%%%%%%%%%%%%%%%%
%%  QUICK REFERENCE  %%
%%%%%%%%%%%%%%%%%%%%%%%
\section {Quick Reference \label{sec:quickref}}

All previously derived equations are provided here for convenience.  Related functions are grouped together.
\\
\noindent {\bfseries Pseudo-Wavefunctions:}
\begin{eqnarray}
 f_{l0}(\eta_2, \eta_3, p) & = & \frac{1}{4 \pi^2 D_{l0}(\eta_2,\eta_3,p)} \int_0^{\infty} dq \: \left[ \frac{q \, (p^2 + \eta_3^2 - \eta_2^2)}{p \, (q^2 + \eta_3^2 - \eta_2^2)} \right.  \nonumber
\\
 && \times \left. \ln\left(\frac{\eta_3^2 + 2p^2 + 2q^2 + 2pq}{\eta_3^2 + 2p^2 + 2q^2 - 2pq}\right) f_{l0}(\eta_2,\eta_3,q) \right] \label{eqn:QRfl0}
\\
&&~\nonumber
\\
f_{l1}(\eta_2, \eta_3, p, \Lambda) & = & \frac{1}{4 \pi^2 D_{l0}(\eta_2,\eta_3,p)} \int_0^{\infty} dq \left[ \frac{q (p^2 + \eta_3^2 - \eta_2^2)}{p (q^2 + \eta_3^2 - \eta_2^2)} \right. \hspace{1.5in} \nonumber
\\
&& \hspace{0.5in} \times \left. \ln\left(\frac{\eta_3^2 + 2p^2 + 2q^2 + 2pq}{\eta_3^2 + 2p^2 + 2q^2 - 2pq}\right) f_{l1}(\eta_2, \eta_3, q, \Lambda) \right]  \nonumber
\\
&& \hspace{1.5in} - {} \frac{D_{l1}(\eta_2,\eta_3,p,\Lambda)}{D_{l0}(\eta_2,\eta_3,p)} f_{l0}(\eta_2,\eta_3,p) 
\\
&&~\nonumber
\\
f_{d0}(p) & = &\frac{p}{4 \pi^2 D_{d0}(p)} \int_0^{\infty} \frac{dq}{q} \, \ln\left(\frac{p^2 + q^2 + pq}{p^2 + q^2 - pq}\right) f_{d0}(q)
\\
&&~\nonumber
\\
f_{d1}(p) & = &  \frac{p}{4 \pi^2 D_{d0}(p)} \left( 1 - \frac{D_{d1}(\eta_2,p,\Lambda)}{D_{d0}(p)} \right) \int_0^{\infty} \frac{dq}{q} \left[ \ln\left(\frac{p^2 + q^2 + pq}{p^2 + q^2 - pq}\right) \right. \nonumber
\\
&& \hspace{1.0in} \times \left( f_{d0}(q) + f_{d1}(q) \right) \Bigg]
\\
&&~\nonumber
\\
f_{h0}(p, \Lambda) & = & \frac{p^2}{2 \pi^2 D_{h0}(p,\Lambda)} \int_0^{\infty} dq \, \int_{-1}^{1} dz \, \frac{U_2\left(\vec{q} + \frac{1}{2}\vec{p}\right) U_2\left(\vec{p} + \frac{1}{2}\vec{q}\right)}{2p^2 + 2q^2 + 2pqz} f_{h0}(q, \Lambda) \nonumber
\\
 && \hspace{0.5in} -~\delta_0 \, \frac{3 h_2^2 + 8 h_2 + 16}{128 \sqrt{2} \, \pi^{3/2}} \left( \frac{D_{1h0}(p,\Lambda)}{D_{h0}(p,\Lambda)} \right) \left( \frac{p^2}{\Lambda^2} \right)
\\
&&~\nonumber
\\
f_{h1}(\eta_2,\eta_3,p,\Lambda) & = & \frac{p^2}{2 \pi^2 D_{h0}(p,\Lambda)} \int_0^{\infty} dq \left[ \int_{-1}^{1} dz  \frac{U_2\left(\vec{q} + \frac{1}{2}\vec{p}\right) U_2\left(\vec{p} + \frac{1}{2}\vec{q}\right)}{2p^2 + 2q^2 + 2pqz} \right. \nonumber
\\
&& \times f_{h1}(\eta_2,\eta_3,q,\Lambda) \Bigg] - \frac{D_{h1}(\eta_2)}{D_{h0}(p,\Lambda)} f_{h0}(p,\Lambda) \nonumber
\\
&&  - \delta_0 \, \frac{p^2 D_{1h0}(p,\Lambda)}{\Lambda^2 \, G_{2,0} \, D_{h0}(p,\Lambda)} \left[ \frac{\delta_1}{\delta_0} - \frac{G_{2,1}}{G_{2,0}} \right]
\end{eqnarray}

\noindent {\bfseries Analytic Solutions:}
\begin{eqnarray}
f_{d0}(p) & = & A \cos\left( s_0 \ln\left(\frac{p}{\Lambda}\right) + \theta \right)
\\
&&~\nonumber
\\
f_{d1}(\eta_2,p,\Lambda) & = & A_{l1} \frac{\eta_2}{p} \cos\left(s_0 \ln\left(\frac{p}{\Lambda}\right) + \theta_{l1} \right) \nonumber
\\
&& +~A_{h1} \frac{p}{\Lambda} \cos\left(s_0 \ln\left(\frac{p}{\Lambda}\right) + \theta_{h1} \right) \nonumber
\\
&& +~A_{d10} \frac{\eta_2}{\Lambda} \cos\left(s_0 \ln\left(\frac{p}{\Lambda}\right) + \theta_{d10} \right) \nonumber
\\
&& +~A_{d11} \frac{\eta_2}{\Lambda} \ln\left(\frac{p}{\Lambda}\right) \cos\left(s_0 \ln\left(\frac{p}{\Lambda}\right) + \theta_{d11} \right)
\end{eqnarray}

\noindent {\bfseries $\mathbf{D}$ Functions:}
\begin{eqnarray}
D_{l0}(\eta_2, \eta_3, p) & = & \frac{\sqrt{\frac{3}{2}p^2 + \eta_3^2} - \eta_2}{8 \sqrt{2} \, \pi}
\\
&& \nonumber
\\
D_{l1}(\eta_2, \eta_3, p, \Lambda) & = & \frac{h_2^2 + 8 h_2 - 16}{64 \sqrt{2} \pi^{3/2}} \left( \frac{\frac{3}{2}p^2 + \eta_3^2 - \eta_2^2}{\Lambda} \right)
\\
&& \nonumber
\\
D_{d0}(p) & = & \frac{\sqrt{3} \, p}{16 \pi}
\\
&& \nonumber
\\
D_{d1}(\eta_2, p, \Lambda) & = & - \, \frac{\eta_2}{8 \sqrt{2} \, \pi} + \frac{3 \left( h_2^2 + 8 h_2 - 16 \right) p^2}{128 \sqrt{2} \, \pi^{3/2} \, \Lambda}
\\
&& \nonumber
\\
D_{h0}(p, \Lambda) & = & \frac{p}{256 \pi^2 \Lambda^4} \left[ \sqrt{3} \, \pi \left( 4 \Lambda^2 - 3 h_2 p^2 \right)^2 \left( 1 - \mathrm{Erf}\left( \frac{\sqrt{3} \, p}{\sqrt{2} \, \Lambda} \right) \right) \exp\left( \frac{3 p^2}{2 \Lambda^2} \right) \right. \nonumber
\\
&& +~ 3 \sqrt{2 \pi} \, h_2 \Lambda p \left( \left(8 + h_2 \right) \Lambda^2 - 3 h_2 p^2 \right) \Bigg]
\\
&& \nonumber
\\
D_{h1}(\eta_2) & = & - \, \frac{\eta_2}{8 \sqrt{2} \, \pi}
\end{eqnarray}

\noindent {\bfseries $\mathbf{D_1}$ Functions:}
\begin{eqnarray}
D_{1l0}(\Lambda) & = & \frac{\Lambda \left( 6 \left( h_3^2 + 4 h_3 + 12 \right) + h_2 \left( 5 h_3^2 + 12 h_3 + 12 \right) \right)}{576 \sqrt{3} \, \pi^{3/2}}
\\
&& \nonumber
\\
D_{1l1}(\eta_3, p) & = & - \, \frac{\pi}{2} \sqrt{2 \eta_3^2 + 3p^2}
\\
&& \nonumber
\\
D_{1d0}(\Lambda) & = & \frac{\Lambda \left( 6 \left( h_3^2 + 4 h_3 + 12 \right) + h_2 \left( 5 h_3^2 + 12 h_3 + 12 \right) \right)}{576 \sqrt{3} \, \pi^{3/2}}
\\
&& \nonumber
\\
D_{1d1}(p) & = & - \, \frac{\sqrt{3} \, \pi}{2} p
\\
&& \nonumber
\\
D_{1h0}(p, \Lambda) & = & \frac{1}{4 \pi^2} \int_0^{\infty} dq \: q^2 \, \int_{-1}^{1} dz \: \frac{U_2(\vec{q}+\frac{1}{2}\vec{p}\,) U_3(q) U_3(p) U_3(\vec{p}+\vec{q}\,)}{2p^2 + 2q^2 + 2pqz}
\\
&& \nonumber
\\
D_{1h1}(p) & = & 0
\end{eqnarray}

\noindent {\bfseries Two-Body Coupling:}
\begin{eqnarray}
G_2 & \equiv & \Lambda g_2
\\
G_{2,0} & = & \frac{128 \sqrt{2} \, \pi^{3/2}}{3 h_2^2 + 8h_2 + 16}
\\
&& \nonumber
\\
G_{2,1} & = & \frac{2048 \sqrt{2} \, \pi^2}{(3 h_2^2 + 8h_2 + 16)^2} \left( \frac{\eta}{\Lambda} \right)
\end{eqnarray}

\noindent {\bfseries Three-Body Coupling:}
\begin{eqnarray}
G_3 & \equiv & \Lambda^4 g_3
\\
G_{3,0} & = & \frac{\delta_0}{G_{2,0} \, \mathcal{I}_0 - D_{2,0}\,\delta_0/\Lambda^4 } \label{eqn:QRG30}
\\
&& \nonumber
\\
G_{3,1} & = & \frac{\delta_1}{\delta_0} G_{3,0} -  \frac{1}{\delta_0} \left( G_{2,0} \, \mathcal{I} + G_{2,1} \, \mathcal{I}_0  - \frac{D_{2,0}}{\Lambda^4} \, \delta_1 \right) G_{3,0}^2
\end{eqnarray}

\noindent {\bfseries $\mathbf{\mathcal{I}}$ Functions:}
\begin{eqnarray}
\mathcal{I}_0 & = & \frac{3}{2 \pi^2 \Lambda^2} \int_0^{\infty} dp \, D_{1h0}(p,\Lambda) \, f_{h0}(p,\Lambda)
\\
&& \nonumber
\\
\mathcal{I}_1 & = & \frac{3}{2 \pi^2 \Lambda^2} \int_0^{\infty} dp \, D_{1h0}(p,\Lambda) \, f_{h1}(\eta_2,\eta_3,p,\Lambda)
\end{eqnarray}

\noindent {\bfseries $\mathbf{D_2}$ Functions:}
\begin{eqnarray}
D_{2,0} & = & \frac{1}{8 \pi^4} \int_0^{\infty} p^2 dp \int_0^{\infty} q^2 dq \int_{-1}^{1} dz \: \frac{U_3(q)^2 U_3(p)^2 U_3(\vec{q}+\vec{p})^2}{2p^2 + 2q^2 + 2pqz}
\\
&& \nonumber
\\
D_{2,1} & = & 0
\end{eqnarray}
\chapter{Solution Methodology}
\label{ch:solnmethod}

Once the integral equations have been derived, an approach for solving them must be decided upon.  The approach used here is a combination of analytical and numerical methods.  By analyzing the equations, we discover certain properties that help make the numerical calculations more efficient.  This results in higher accuracy and lower computational costs.

This chapter begins with a brief discussion of the steps needed to solve the integral equations.  We point out quantities that one might compute and what information is necessary to carry out that computation.  We also attempt to illustrate the relationships between the parameters in the integral equations.

After this discussion, details of the numerical solution technique are given.  The method involves discretizing the integral equations with an exponential spacing of points to significantly improve convergence.  Ways of reducing the number of necessary points by decreasing the range of integration are also discussed.   The final discretized equations are exhibited.

Our analysis will be limited to the set of leading order integral equations.  These are the equations relevant to the $\Lambda \rightarrow \infty$ limit.  The equations for the first order corrections must be solved if one is interested in the cutoff-dependent behavior.  Although we do not attempt to numerically solve for these corrections, the equations have been derived and all that remains is implementing a program to solve them.  Many, if not all, of the issues discussed in this chapter are relevant to their solution.

%%%%%%%%%%%%%%%%%%%%%%%%
%%  GENERAL APPROACH  %%
%%%%%%%%%%%%%%%%%%%%%%%%
\section{General Approach}

One of the leading order equations has already been solved, namely the one for $f_{d0}(p)$.  The solution has been determined to be of the form $A \, \cos(s_0 \ln(p/\Lambda) + \theta)$.  Because any solution to the integral equations can be rescaled to give another equally valid solution, we choose to always scale our solutions so that $A = 1$.

The solutions for $f_{l0}(p)$ and $f_{h0}(p)$ are inherently linked by the solution for $f_{d0}(p)$.  For $p \gg (\eta_2,\eta_3)$, the function $f_{l0}$ must approach $f_{d0}$, and for $p \ll \Lambda$, $f_{h0}$ must approach $f_{d0}$. In essence, these requirements act as boundary conditions for our solutions. 

Consider the case where the values of $B_2$ and $B_3$ are known for a given cutoff $\Lambda$.  With just the energy values, we have enough information to solve the equation for $f_{l0}(p)$.  Once a solution is found, we can examine its behavior at large momentum.  It must have a cosine behavior, and we can use this fact to fit the function $\cos(s_0 \ln(p/\Lambda) + \theta)$ to the solution in this region.  This allows the determination of the phase, which we will call $\theta_{l0}$.  

For the function $f_{h0}(p)$, we see from the integral equation that the solution depends upon two parameters, $\Lambda$ and $\delta$.  Since $\Lambda$ is known, we can choose a value for $\delta$ and then solve the equation.  The result is a solution for $f_{h0}$ that, for very small values of $p$, has a cosine behavior like $f_{d0}$.  Call the phase of this solution $\theta_{h0}$.

Our boundary condition can then be stated more precisely as
\begin{equation}
\theta_{l0} = \theta_{h0} \pm n \pi \label{eqn:thetaBC},
\end{equation}
\noindent for some positive integer $n$.\footnote{The reason why this is $\pi$ and not $2 \pi$ has to do with the equation for $f_{l0}$.  For any low-momentum solution with a phase of $\theta$, we can multiply the function by -1 to obtain a solution with phase $\theta + \pi$.  The phases are only unique up to a difference of $\pi$.}  This shows that the value chosen for $\delta$ cannot be arbitrary as it must be a value that allows (\ref{eqn:thetaBC}) to be satisfied.  We see that choosing the bound-state energies determines a unique value of $\theta$, which in turn determines a unique value for $\delta$.  The value of $G_3$ can then be computed from $\delta$.

Now suppose that we would like to compute the other three-body bound-state energies in the spectrum.  The three-body coupling has been fixed by our initial data, which is equivalent to saying that the phase has been fixed.  The problem then becomes one of finding other values for $B_3$ that will yield a solution for $f_{l0}$ with the same asymptotic phase $\theta_{l0}$.  Thus we find that we can calculate a bound-state spectrum by solving the equation for $f_{l0}$ and matching phases, never once needing to calculate $f_{h0}$.

All other calculations are simply a matter of relating $G_3$ and $B_3$ via the intermediate phase $\theta$.

%%%%%%%%%%%%%%%%%%%%%%%%%
%%  NUMERICAL METHODS  %%
%%%%%%%%%%%%%%%%%%%%%%%%%
\section{Numerical Methods}

Because a closed-form solution for either $f_{l0}$ or $f_{h0}$ is unknown, we must resort to numerical calculations of these functions as well as any energies or couplings.  A common method for solving an integral equation involves changing the integration into a sum over discrete points.  The integral equation then becomes a matrix equation easily solved by standard methods.

While this may appear straightforward, there are several practical issues to consider.  For example, new limits on the integral equation must be determined.  It is impossible to numerically integrate to infinity, so a suitable upper bound must be chosen.  In our case, we will use a logarithmic integration range so a new lower bound to replace $0$ must also be determined.  How the discrete points are chosen must be carefully considered.  Should we use the basic Simpson's Rule for numerical integration, or perhaps a more complicated quadrature method?

Each choice is a compromise between accuracy and size.  For example, we could choose an upper limit that is extremely large (ensuring that we are ``close'' to infinity) and discrete points that are closely spaced (minimizing errors in the sum).  The trade-off is that using more points requires using a larger matrix.  Using $N$ discrete points will result in an $N \times N$ matrix with $N^2$ elements.  If all numbers are double precision decimals, even 8000 points would be enough to overwhelm a computer with 512 MB of memory.  This does not even take into account the time needed to process such a matrix.  Obviously, the goal is to obtain the desired accuracy with a minimal number of points.  We will discuss a few methods that drastically reduce the number of points we need.

In the following sections, we will assume that our goal is about 12 digits of accuracy.  This high accuracy may not be necessary for most leading order calculations, but it is essential when including the first order corrections.  Besides directly obtaining the equations for the $\Lambda \rightarrow \infty$ limit, one principle reason for expanding the three-body equation in powers of $\eta/\Lambda$ is to analyze the cutoff dependence.  If we are attempting to study this behavior, we must be certain that our numerical errors are not larger than the corrections being studied,  otherwise there is no way to distinguish the small corrections from the numerical ``noise.''  By laying the groundwork for high accuracy in the leading order calculation, higher order calculations should be easier to implement.

    %%%%%%%%%%%%%%%%%%%%%%%%%
    %%  TRANSITION TO fd0  %%
    %%%%%%%%%%%%%%%%%%%%%%%%%
\subsection{Transition to Mid-Momentum Function}

One way to limit the size of the matrix is to limit the range over which the function must be integrated.  We know that $f_{l0}$ and $f_{h0}$ approach $f_{d0}$ for $p \gg \eta$ and $p \ll \Lambda$ respectively.  This allows us to replace either function with $\cos\left(s_0 \ln(p/\Lambda) + \theta\right)$ in the appropriate range.  The point at which we can make the switch is determined by the accuracy we desire.  These limits are derived for the case of 12 digits of accuracy.

For $p \gg \eta$, the equation for $f_{l0}$ can be written as
\begin{eqnarray}
&& \hspace{-0.25in} f_{l0}(\eta_2, \eta_3, p) = \frac{4 (1 + \eta_3^2/p^2 - \eta_2^2/p^2)}{\sqrt{3} \pi \left( \sqrt{1 + (2 \eta_3^2)/(3 p^2)} - (\sqrt{2} \eta_2)/(\sqrt{3} p) \right)} \int_0^{\infty} dq \: \frac{q}{q^2 + \eta_3^2 - \eta_2^2} \nonumber
\\
&& \hspace{-0.25in} \times \left[ \ln\left(\frac{p^2 + q^2 + pq}{p^2 + q^2 - pq}\right) + \frac{\eta_3^2}{2 p^2 + 2 q^2 + 2 p q} - \frac{\eta_3^2}{2 p^2 + 2 q^2 - 2 p q} \right] f_{l0}(\eta_2,\eta_3,q) .
\end{eqnarray}
\noindent Here we have treated $\eta/p$ as a small quantity and perturbatively expanded all factors.  As long as both $\eta_2/p$ and $\eta_3/p$ are less than $10^{-12}$, this equation will match the one for $f_{d0}$ to 12 digits.  Of course, $\eta_3$ must be greater than $\eta_2$ since we are considering only stable bound states.  This means that $p \simeq 10^{12} \, \eta_3$ sets the limit above which $f_{l0}$ can be replaced by $f_{d0}$.  In practice, we must have enough data points above this limit to ensure that our cosine fit is accurate to 12 digits also.  Therefore, we will use an actual limit of $p = 10^{15} \, \eta_3$.

Similarly, we can expand the equation for $f_{h0}$ in the region $p \ll \Lambda$.  Notice that in this region $D_{1h0}(p,\Lambda)$ approaches $D_{1d0}(\Lambda)$ which is proportional to $\Lambda$, and the function $D_{h0}(p,\Lambda)$ becomes equal to $D_{d0}(p) = \sqrt{3} \, p/16 \pi$.  In the $f_{h0}$ integral equation, the momentum-dependent part of the three-body interaction becomes
\begin{equation}
  \left( \frac{p^2}{\Lambda^2} \right) \left( \frac{D_{1h0}(p,\Lambda)}{D_{h0}(p,\Lambda)} \right) \stackrel{p \ll \Lambda}{\longrightarrow} \left( \frac{p^2}{\Lambda^2} \right) \left( \frac{D_{1d0}(\Lambda)}{D_{d0}(p)} \right) \propto \frac{p}{\Lambda} .
\end{equation}
\noindent Since the leading order mid-momentum equation has no three-body interaction term, the above term will equal zero to 12 digits if we choose $p \sim 10^{-12} \, \Lambda$.

The integral part for $f_{h0}$ looks like
\begin{equation}
\frac{p^2}{2 \pi^2 D_{h0}(p,\Lambda)} \int_0^{\infty} dq \, \int_{-1}^{1} dz \, \frac{U_2\left(\vec{q} + \frac{1}{2}\vec{p}\right) U_2\left(\vec{p} + \frac{1}{2}\vec{q}\right)}{2p^2 + 2q^2 + 2pqz} f_{h0}(q, \Lambda) .
\end{equation}
\noindent We have already stated that $D_{h0}(p,\Lambda)$ approaches $D_{d0}(p)$, but more importantly, it approaches like
\begin{equation}
D_{h0}(p,\Lambda) \stackrel{p \ll \Lambda}{\longrightarrow} D_{d0}(p) \left[ 1 + \left( \frac{\sqrt{3} \, (h_2^2 + 8 h_2 - 16)}{8 \sqrt{2 \pi}} \right) \frac{p}{\Lambda} \right] .
\end{equation}
\noindent It will therefore equal $D_{d0}$ to 12 digits if $p \sim 10^{-12} \, \Lambda$.  This limit also applies to the integrand itself, so we may replace $f_{h0}$ with $f_{d0}$ for values of $p$ less than $\mathcal{O}(10^{-12} \, \Lambda)$.  In practice however, we use a limit of $p = 10^{-17} \, \Lambda$ to ensure that we have enough points below this region to fit the cosine behavior.

    %%%%%%%%%%%%%%%%%%
    %%  NEW LIMITS  %%
    %%%%%%%%%%%%%%%%%%
\subsection{New Integration Limits}

For the case of $f_{h0}$, we have limited the range of integration to be $10^{-17} \, \Lambda$ to $\infty$.  (Below this range, we use $f_{d0}$.)  Naturally, we cannot integrate to infinity and instead must find a new limit to replace it.  Let us call this limit $\lambda$.  We choose $\lambda$ such that
\begin{equation}
\frac{p^2}{2 \pi^2 D_{h0}(p,\Lambda)} \int_{\lambda}^{\infty} dq \, \int_{-1}^{1} dz \, \frac{U_2\left(\vec{q} + \frac{1}{2}\vec{p}\right) U_2\left(\vec{p} + \frac{1}{2}\vec{q}\right)}{2p^2 + 2q^2 + 2pqz} f_{h0}(q, \Lambda) < \mathcal{O}(10^{-12}).
\end{equation}

The exponentials in $U_2$ suggest that the integrand should die off quickly, allowing us to make an initial guess for $\lambda$ using
\begin{equation}
\me^{- \lambda^2/\Lambda^2} = 10^{-12} .
\end{equation}
\noindent This gives an initial value of $\lambda \simeq 5.25 \, \Lambda$.  However, the double integral makes the analysis harder since we cannot determine the exact behavior.  We must resort to numerical computation, and some sample calculations reveal that a limit of $\lambda = 10 \, \Lambda$ is sufficient for our purposes.

From $10^{-17} \, \Lambda$ down to 0, $f_{h0}$ is replaced by $f_{d0}$.  We would like to replace the $0$ limit with a larger value that still maintains our desired accuracy.  Even though we know the analytic solution for $f_{d0}$, narrowing the range of integration will reduce our computational effort.  Call this new lower limit $\epsilon$, which is chosen so that
\begin{equation}
\frac{p^2}{2 \pi^2 D_{h0}(p,\Lambda)} \int_{0}^{\epsilon} dq \, \int_{-1}^{1} dz \, \frac{U_2\left(\vec{q} + \frac{1}{2}\vec{p}\right) U_2\left(\vec{p} + \frac{1}{2}\vec{q}\right)}{2p^2 + 2q^2 + 2pqz} f_{d0}(q, \Lambda) < \mathcal{O}(10^{-12}) .
\end{equation}
\noindent We assume that $\epsilon \ll 10^{-17} \, \Lambda < p$.  To within 12 digits of accuracy, the $z$ integration can be written as
\begin{eqnarray}
\int_{-1}^{1} dz \, \frac{U_2\left(\vec{q} + \frac{1}{2}\vec{p}\right) U_2\left(\vec{p} + \frac{1}{2}\vec{q}\right)}{2p^2 + 2q^2 + 2pqz} & = & \int_{-1}^{1} dz \, \frac{U_2\left(\vec{p}/2\right) U_2\left(\vec{p}\right)}{2p^2 + 2pqz} \nonumber
\\
& = & \frac{1}{2pq} \, U_2\left(\vec{p}/2\right) U_2\left(\vec{p}\right) \, \ln\left(\frac{p^2 + pq}{p^2 - pq}\right) .
\end{eqnarray}
\noindent Since $q \ll p$, the logarithm can be approximated as $2q/p$.  Our constraint for $\epsilon$ now becomes
\begin{equation}
\frac{1}{2 \pi^2 D_{h0}(p,\Lambda)} \,  U_2\left(\vec{p}/2\right) U_2\left(\vec{p}\right) \, \int_0^{\epsilon} dq \, f_{d0}(q) < \mathcal{O}(10^{-12}).
\end{equation}
\noindent The values of $p$ are of the same order as $\Lambda$, so we expect the value of $U_2$ to be $\mathcal{O}(1)$.  The function $f_{d0}$ is also $\mathcal{O}(1)$, so it is simply replaced by $1$ in this approximation.  This leaves an integral with a value of $\epsilon$.  When combined with $D_{h0}(p,\Lambda) \sim \mathcal{O}(p)$, we find that $\epsilon/p < \mathcal{O}(10^{-12})$.  The smallest value for $p$ is $10^{-17} \, \Lambda$, implying that $\epsilon \simeq 10^{-29} \, \Lambda$.  In practice, this value is sufficient for 12 digits of accuracy.

Having replaced the limits for $f_{h0}$, we move on to $f_{l0}$.  Again, we must find a finite upper limit to substitute for infinity.  For $p > 10^{15} \, \eta_3$, $f_{l0}$ is replaced by $f_{d0}$.  This new limit, $\lambda$, is determined by the condition
\begin{eqnarray}
&& \hspace{-1.5in} \frac{(p^2 + \eta_3^2 - \eta_2^2)}{4 \pi^2 p \, D_{l0}(\eta_2,\eta_3,p)} \int_{\lambda}^{\infty} dq \: \left[ \frac{1}{q} \ln\left(\frac{p^2 + q^2 + pq}{p^2 + q^2 - pq}\right) f_{d0}(q) \right] \nonumber
\\
&& \simeq  \frac{(p^2 + \eta_3^2 - \eta_2^2)}{4 \pi^2 p \, D_{l0}(\eta_2,\eta_3,p)} \int_{\lambda}^{\infty} \frac{dq}{q} \: \left(\frac{2p}{q}\right) f_{d0}(q) \nonumber
\\
&& = \frac{(p^2 + \eta_3^2 - \eta_2^2)}{2 \pi^2 \, D_{l0}(\eta_2,\eta_3,p)} \int_{\lambda}^{\infty} \frac{dq}{q^2} \: f_{d0}(q) \nonumber
\\
&& \sim \frac{(p^2 + \eta_3^2 - \eta_2^2)}{2 \pi^2 \, D_{l0}(\eta_2,\eta_3,p)} \int_{\lambda}^{\infty} \frac{dq}{q^2} \nonumber
\\
&& = \frac{(p^2 + \eta_3^2 - \eta_2^2)}{2 \pi^2 \, D_{l0}(\eta_2,\eta_3,p)} \, \left(\frac{1}{\lambda}\right)  \le \mathcal{O}(10^{-12}) \label{eqn:fl0UL},
\end{eqnarray}
\noindent where we have used the fact that $q \gg p \sim \eta_3$.  For values of $p$ much larger than $\eta_3$, the term in Eq.~(\ref{eqn:fl0UL}) is proportional to $p/\lambda$.  The largest value $p$ can obtain is $10^{15} \, \eta_3$, implying $\lambda = 10^{27} \, \eta_3$.  Numerical calculations verify that this limit is sufficient to assure the desired accuracy.

Finally, we must replace the lower limit for $f_{l0}$ with a non-zero value $\epsilon$ that we assume to be much smaller than $\eta_3$.  Our requirement is that
\begin{equation}
 \frac{(p^2 + \eta_3^2 - \eta_2^2)}{4 \pi^2 p (\eta_3^2 - \eta_2^2) D_{l0}(\eta_2,\eta_3,p)} \int_0^{\epsilon} dq \, q \ln\left(\frac{\eta_3^2 + 2p^2 + 2pq}{\eta_3^2 + 2p^2 - 2pq}\right) f_{l0}(\eta_2,\eta_3,q) < \mathcal{O}(10^{-12}) .
\end{equation}
\noindent Expanding the logarithm yields
\begin{equation}
 \frac{(p^2 + \eta_3^2 - \eta_2^2)}{\pi^2 (\eta_3^2 - \eta_2^2) (\eta_3^2 + 2 p^2) D_{l0}(\eta_2,\eta_3,p)} \int_0^{\epsilon} dq \, q^2  \, f_{l0}(\eta_2,\eta_3,q) < \mathcal{O}(10^{-12}) \label{eqn:fl0LL}.
\end{equation}
\noindent For small values of $q$, we will find that $f_{l0}$ is approximately constant and of $\mathcal{O}(1)$.  Therefore, the integral is roughly equal to $\epsilon^3$.  If $p \ll \eta_3$ or $p \sim \eta_3$, then Eq.~(\ref{eqn:fl0LL}) is $\mathcal{O}(\epsilon^3/\eta^3)$.  If $p \gg \eta_3$, then it is $\mathcal{O}(\epsilon^3/p\eta_3^3) < \mathcal{O}(\epsilon^3/\eta^3)$.  This seems to imply that a value of $\epsilon \simeq 10^{-4} \, \eta_3$ is adequate.

Unfortunately, using this value of $\epsilon$ will result in poor accuracy when $\eta_2 \simeq \eta_3$.  This is a result of the $(\eta_3^2 - \eta_2^2)$ term in the denominator of Eq.~(\ref{eqn:fl0LL}).  Originally, this was our approximation to the term $(q^2 + \eta_3^2 - \eta_2^2)$ in the integral equation.  When the energies are nearly equal, our approximation needs to be $q^2$.  The condition on $\epsilon$ should now become
\begin{equation}
 \frac{(p^2 + \eta_3^2 - \eta_2^2)}{\pi^2 (\eta_3^2 + 2 p^2) D_{l0}(\eta_2,\eta_3,p)} \int_0^{\epsilon} dq   \, f_{l0}(\eta_2,\eta_3,q) < \mathcal{O}(10^{-12}) \label{eqn:newfl0LL}.
\end{equation}
\noindent The integral is roughly equal to $\epsilon$, and the entire term is $\mathcal{O}(\epsilon/\eta)$. This means we must use the lower limit $\epsilon = 10^{-12} \, \eta_3$.

These derivations are general enough to determine the appropriate limits for other cases of desired accuracy.  Keep in mind however that the limits must always be tested numerically to ensure that they are indeed sufficient.

    %%%%%%%%%%%%%%%%%%%%%%%%
    %%  DISCRETE SPACING  %%
    %%%%%%%%%%%%%%%%%%%%%%%%
\subsection{Discrete Point Spacing}

Now that we have stricter limits in place, we must choose the discrete values of $q$ within these limits at which to evaluate our functions.  We employ discrete points that are equally spaced on a logarithmic scale.  There are two main reasons for making this choice.

First, we have already seen that $f_{d0} = \cos\left(s_0 \ln(p/\Lambda) + \theta \right)$ is periodic on a logarithmic scale.  In fact, other functions have similar behavior, including $G_3$.  It makes sense that a logarithmic spacing is better suited to capturing the behavior of the system.

Second, by choosing points in this manner, an integration of $q$ from $0$ to $\infty$ becomes an integration of $\ln(q)$ from $-\infty$ to $\infty$.  By equally spacing points on this log scale, we can achieve convergence that improves exponentially\footnote{For a basic mathematical proof of this claim, see Appendix \ref{app:kenproof}.  Appendix \ref{app:erroranalysis} has a short error analysis that numerically shows the exponential convergence.} with the spacing, as opposed to the more typical power law convergence.  This drastically reduces the number of points needed to achieve our desired accuracy.  For instance, to cover the range of $f_{l0}$ from $10^{-12} \, \eta_3$ to $10^{15} \, \eta_3$, and maintain 12 digits of accuracy, we can space our points by $p_{n+1} = p_n \me^{0.2}$.  This requires only about 300 points.

%%%%%%%%%%%%%%%%%%%%%%%%%%%%%
%%  DISCRETIZED EQUATIONS  %%
%%%%%%%%%%%%%%%%%%%%%%%%%%%%%
\section{Discretized Equations}

The discretized integral equations are shown below.  Here, $\Delta$ represents the momentum spacing, and the momentum values are related by the equation $p_{n+1} = p_{n} \, \me^{\Delta}$.  The identity matrix is represented by $I_{nm}$, and the ceiling function represented by $\lceil x \rceil$ returns an integer value $n$ such that $(n-1) < x \le n$.

{\bf Low-Momentum:} The value of $n$ for $p_n$ ranges from 0 to $N_{mid}$, while the value of $m$ for $q_m$ ranges from 0 to $N_{max}$. These are defined as: $p_0 = 10^{-12} \, \eta_3$,  $N_{mid} = \lceil \ln(10^{27})/\Delta \rceil$, and $N_{max} = \lceil \ln(10^{39})/\Delta \rceil$ .

\begin{equation}
 \sum_{m = 0}^{N_{mid}} \left( M_{nm} - I_{nm} \right) \, f_{l0}(p_m) = b_n 
\end{equation}
\begin{equation}
M_{nm} = \frac{\Delta}{4 \pi^2 D_{l0}(\eta_2,\eta_3,p_n)} \frac{q_m^2 (p_n^2 + \eta_3^2 - \eta_2^2)}{p_n (q_m^2 + \eta_3^2 - \eta_2^2)} \ln\left(\frac{\eta_3^2 + 2 p_n^2 + 2 q_m^2 + 2 p_n q_m}{\eta_3^2 + 2 p_n^2 + 2 q_m^2 + 2 p_n q_m}\right) 
\end{equation}
\begin{eqnarray}
&& \hspace{-0.5in} b_n = - \frac{\Delta (p_n^2 + \eta_3^2 - \eta_2^2)}{4 \pi^2 p_n D_{l0}(\eta_2,\eta_3,p_n)} \sum_{m = N_{mid}+1}^{N_{max}} \left[ \ln\left(\frac{\eta_3^2 + 2 p_n^2 + 2 q_m^2 + 2 p_n q_m}{\eta_3^2 + 2 p_n^2 + 2 q_m^2 + 2 p_n q_m}\right) \right. \nonumber
\\
&& \hspace{2in} \left. \times \cos\left( s_0 \ln\left(\frac{q_m}{\Lambda}\right) + \theta \right) \right]
\end{eqnarray}

{\bf High-Momentum:} The value of $n$ for $p_n$ ranges from 0 to $N_{max}$, while the value of $m$ for $q_m$ ranges from $-N_{min}$ to $N_{max}$.  These are defined as: $p_0 = 10^{-17} \, \Lambda$,  $N_{max} = \lceil \ln(10^{18})/\Delta \rceil$, and $N_{min} = \lceil \ln(10^{12})/\Delta \rceil$ .

\begin{equation}
 \sum_{m = 0}^{N_{max}} \left( M_{nm} - I_{nm} \right) \, f_{h0}(p_m) = a_n + \delta_0 \, b_n
\end{equation}
\begin{equation}
M_{nm} = \frac{q_m p_n^2 \Delta}{2 \pi^2 D_{h0}(p_n,\Lambda)} K(p_n, q_m)
\end{equation}
\begin{equation}
a_n = - \frac{p_n^2 \Delta}{2 \pi^2 D_{h0}(p_n,\Lambda)} \sum_{m = -N_{min}}^{-1} q_m \, K(p_n,q_m) \cos\left( s_0 \ln\left(\frac{q_m}{\Lambda}\right) + \theta \right)
\end{equation}
\begin{equation}
b_n = \frac{p_n^2 D_{1h0}(p_n,\Lambda)}{\Lambda^2 G_{2,0} D_{h0}(p_n,\Lambda)}
\end{equation}
\begin{eqnarray}
K(p_n, q_m) & = & \frac{\Delta}{2 \Lambda^2} \sum_{k = -200}^{200} \left[ (1 - z_k^2) \left( 1 + h_2 \left( \frac{q_m^2}{\Lambda^2} + \frac{q_m p_n z_k}{\Lambda^2} + \frac{p_n^2}{4 \Lambda^2} \right)\right) \right. \nonumber
\\
&& \times \left( 1 + h_2 \left( \frac{p_n^2}{\Lambda^2} + \frac{q_m p_n z_k}{\Lambda^2} + \frac{q_m^2}{4 \Lambda^2} \right)\right) \nonumber
\\
&& \left. \times \frac{\exp\left( -(5 p_n^2 + 8 p_n q_m z_k + 5 q_m^2)/(4 \Lambda^2) \right)}{2\left( p_n^2/\Lambda^2 + q_m^2/\Lambda^2 + p_n q_m z_k/\Lambda^2 \right)} \right]
\end{eqnarray}
\begin{equation}
z_k = \frac{\me^{0.2 \, k} - 1}{\me^{0.2 \, k} + 1}
\end{equation}
\chapter{Analytic and Numerical Results}
\label{ch:results}

A computer program to solve the discretized integral equations can be implemented with the help of some basic numerical algorithms.  Once completed, it can be used to generate highly accurate results and analyze the three-body system.  Nonetheless, attempts to extract analytic results should not be overlooked.  Even some of the most general properties of the integral equations allow us to draw conclusions about the behavior of the system.

We begin this chapter with analytic results obtained from studying the leading order integral equations.  These results include statements about the cutoff dependence of bound-state energies and the phase for $f_{d0}$.  A proof for the cyclic behavior of $\delta_0$ is given, which is then used to infer similar behavior for $G_3$.

Following the analytic results is a section containing numerical solutions to the integral equations.  Here we examine behavior that cannot be determined analytically.  Solutions for the functions $f_{l0}$ and $f_{h0}$ are shown, and the cutoff dependence of $G_3$ is calculated.  Some relations that are proven analytically are also verified numerically.

%%%%%%%%%%%%%%%%%%%%%%%%
%%  ANALYTIC RESULTS  %%
%%%%%%%%%%%%%%%%%%%%%%%%

\section{Analytic Results}\label{sec:analytic}

In Chapter \ref{ch:expand3body}, we saw that Eq.~(\ref{eqn:QRfl0}) contains no $\Lambda$ dependence, making the function $f_{l0}(p)$ cutoff-independent.  However, for values of $p \gg \eta_3$, the solution must behave like $f_{d0}(p) = \cos\left(s_0 \, \ln\left(p/\Lambda\right) + \theta \right)$, which explicitly has the cutoff in it.  The only way these two equations can be reconciled is if $\theta$ contains cutoff dependence of the form $s_0 \, \ln(\Lambda)$.  Any remaining part of $\theta$ must be a function of $\eta_2$ and $\eta_3$.  Therefore, we shall write
\begin{equation}
\theta = s_0 \, \ln\left(\Lambda/\eta_3\right) + \tilde{\theta}\left(\eta_2/\eta_3\right) \label{eqn:theta_tilde},
\end{equation}
\noindent where $\tilde{\theta}$ is a dimensionless function of the ratio $\eta_2/\eta_3$.  This relation holds for any values of $\eta_2$ and $\eta_3$, including all $\eta_3$ values corresponding to multiple bound states with the same $\eta_2$.  In the next chapter, we shall see that $\tilde{\theta}$ is closely related to Efimov's universal function \cite{efimov:2}.    Using $\eta_3$ in the ratio with $\Lambda$ is a matter of choice. Any quantity composed of $\eta_2$ and $\eta_3$ with the same dimension as $\Lambda$ would work just as well; however, we need to allow $\eta_2 = 0$, so $\eta_2$ alone is a poor choice.

While Eq.~(\ref{eqn:theta_tilde}) is quite simple, it has many interesting consequences.  As we mentioned earlier, two different three-body bound states in the same spectrum must have the same phase $\theta$.  Suppose that we choose some fixed values for $\eta_2$ and $\eta_3$ and make them cutoff-independent by choosing the appropriate $\Lambda$ dependence for $G_2$ and $G_3$.  This might be desirable if we are trying to match those energies to experimental data.  The phase for the solution in this case is
\begin{equation}
\theta_{\Lambda} =  s_0 \, \ln\left(\Lambda/\eta_3\right) + \tilde{\theta}\left(\eta_2/\eta_3\right) ,
\end{equation}
\noindent for some given cutoff $\Lambda$.  If the cutoff is changed, $G_2$, $G_3$, and $\theta_{\Lambda}$ will all change with it, but $\eta_2$, $\eta_3$, and $\tilde{\theta}(\eta_2/\eta_3)$ will not.

Imagine now that we find a second three-body bound-state solution with the same phase.  Let us call its energy $\bar{\eta}_3$.  The phase for this solution is
\begin{equation}
\bar{\theta}_{\Lambda} =  s_0 \, \ln\left(\Lambda/\bar{\eta}_3\right) + \tilde{\theta}\left(\eta_2/\bar{\eta}_3\right) ,
\end{equation}
\noindent which must be equal to $\theta_{\Lambda}$ by assumption.  This results in the relation
\begin{equation}
s_0 \, \ln\left(\bar{\eta}_3/\eta_3\right) = \tilde{\theta}\left(\eta_2/\bar{\eta}_3\right) - \tilde{\theta}\left(\eta_2/\eta_3\right) .
\end{equation}
\noindent If the cutoff is now changed to a new value $\Lambda'$, the original data gives a phase of
\begin{equation}
\theta_{\Lambda'} =  s_0 \, \ln\left(\Lambda'/\eta_3\right) + \tilde{\theta}\left(\eta_2/\eta_3\right) .
\end{equation}
\noindent The question is whether $\bar{\eta}_3$ is still a valid solution.  Its new phase is
\begin{eqnarray}
\bar{\theta}_{\Lambda'} & = & s_0 \, \ln\left(\Lambda'/\bar{\eta}_3\right) + \tilde{\theta}\left(\eta_2/\bar{\eta}_3\right) \nonumber
\\
& = & s_0 \, \ln\left(\Lambda'/\bar{\eta}_3\right) + \left[ s_0 \, \ln\left(\bar{\eta}_3/\eta_3\right) + \tilde{\theta}\left(\eta_2/\eta_3\right) \right] \nonumber
\\
& = & s_0 \, \ln\left(\Lambda'/\eta_3\right) + \tilde{\theta}\left(\eta_2/\eta_3\right) \nonumber
\\
& = & \theta_{\Lambda'} .
\end{eqnarray}
\noindent Since the phases still match, $\bar{\eta}_3$ is still a bound-state solution.  This remains true for any cutoff, implying that $\bar{\eta}_3$ is also cutoff-independent like $\eta_3$.  Of course, the same statement applies to any other three-body bound state {\em making the entire spectrum completely independent of $\Lambda$}.  Such behavior should come as no surprise since the leading order equations represent the $\Lambda \rightarrow \infty$ limit.  Keep in mind that this is true for any other physical quantity, but does not apply to the couplings.  Obviously, it is the cutoff dependence of the couplings that enables the bound states to be cutoff independent.  We have shown that a single three-body contact interaction allows us to renormalize the entire three-body bound-state spectrum.

Just as $f_{l0}$ has no dependence on $\Lambda$, neither does it have any dependence on $h_2$.  As a consequence, $h_2$ has no effect at leading order on the binding energies or other physical quantities.  It does appear in the equation for $f_{l1}$ however, showing that it is needed when considering first order corrections.  Because we are working only to leading order, we shall use $h_2 = 0$ in our calculations from now on unless otherwise specified.

We now turn to the equation for $f_{h0}(p)$:
\begin{eqnarray}
f_{h0}(p, \Lambda) = \frac{p^2}{2 \pi^2 D_{h0}(p,\Lambda)} \int_0^{\infty} dq \, \int_{-1}^{1} dz \, \frac{U_2\left(\vec{q} + \frac{1}{2}\vec{p}\right) U_2\left(\vec{p} + \frac{1}{2}\vec{q}\right)}{2p^2 + 2q^2 + 2pqz} f_{h0}(q, \Lambda) \nonumber
\\
 \hspace{0.5in} -~\delta_0 \, \frac{3 h_2^2 + 8 h_2 + 16}{128 \sqrt{2} \, \pi^{3/2}} \left( \frac{D_{1h0}(p,\Lambda)}{D_{h0}(p,\Lambda)} \right) \left( \frac{p^2}{\Lambda^2} \right) \label{eqn:thisfh0}.
\end{eqnarray}
\noindent The dependence on $p$ and $\Lambda$ has been explicitly shown.  Since $f_{h0}$ is dimensionless, only the ratio $p/\Lambda$ can occur in the function.  Furthermore, there is a dependence upon $h_2$ and $\theta$.  The $h_2$ dependence comes from its appearance in the function $U_2(p)$, either directly in the integral or indirectly in $D_{h0}$ and $D_{1h0}$.  The $\theta$ dependence is a result of $f_{h0}$ approaching $f_{d0}$ for $p \ll \Lambda$.

Assuming that $\Lambda$ is held fixed, choosing a value for $\delta_0$ will uniquely determine the phase.  Thus, we can view the phase as a function of the coupling, $\theta(\delta_0)$.  Conversely, choosing a phase determines the coupling, so it is equally valid to treat the coupling as a function of the phase, $\delta_0(\theta)$.  The coupling $\delta_0$ can also have a dependence upon $h_2$ but not upon $\Lambda$.  The reason is that $\delta_0$ is dimensionless and there is no other quantity available to form a dimensionless ratio with $\Lambda$.

With this in mind, consider a solution to Eq.~(\ref{eqn:thisfh0}) with a coupling of $\delta_0(\theta)$ and a function $f_{h0}(p)$ that behaves asymptotically as $\cos\left(s_0 \, \ln\left(p/\Lambda\right) + \theta \right)$ for small $p$.  For a phase of $\theta + 2 \pi$, the cosine behavior will remain unchanged.  We may therefore conclude that 
\begin{equation}
\delta_0(\theta + 2 \pi) = \delta_0(\theta) ,
\end{equation}
\noindent showing that the coupling exhibits periodic behavior.  If we apply the operator $-\partial^2/\partial\theta^2$ to both sides of (\ref{eqn:thisfh0}), we find once again that the asymptotic cosine behavior is the same.  This implies
\begin{equation}
-\frac{\partial^2 \delta_0}{\partial\theta^2} = \delta_0(\theta) .
\end{equation}
\noindent The conclusion to be drawn is that the coupling may be written as a cosine function with some amplitude $\mathcal{A}$ and phase $\phi$.  These two parameters will contain any $h_2$ dependence that $\theta$ may possess, so we shall write
\begin{equation}
\delta_0(\theta) = \mathcal{A}(h_2) \, \cos\left( \theta + \phi(h_2) \right) \label{eqn:PseudoCosine}.
\end{equation}

The equation relating $\delta_0$ and $G_3$,
\begin{equation}
G_{3,0}  =  \frac{\delta_0}{G_{2,0} \, \mathcal{I}_0 - D_{2,0}\,\delta_0/\Lambda^4 } ,
\end{equation}
\noindent shows that whenever $\delta_0$ is zero, so is $G_3$.\footnote{The only problem that might arise is if $\mathcal{I}_0 = 0$ in such a way that the ratio is non-zero, but we have found no parameters for which this is true.}  Since two adjacent zeros of $\delta_0$ occur when the cutoffs satisfy
\begin{equation}
\theta_{\Lambda'} - \theta_{\Lambda} = s_0 \, \ln\left(\frac{\Lambda'}{\Lambda}\right) = \pi ,
\end{equation}
\noindent the adjacent zeros of $G_3(\Lambda)$ should be spaced by a cutoff factor of $\Lambda'/\Lambda = \exp(\pi/s_0) \simeq 22.69438259536$.  This suggests that $G_3(\Lambda)$ may also possess cyclic behavior, but this must ultimately be verified numerically.

%%%%%%%%%%%%%%%%%%%%%%%
%%  NUMERIC RESULTS  %%
%%%%%%%%%%%%%%%%%%%%%%%

\section{Numerical Results}

We begin our numerical investigation by considering solutions for the functions $f_{l0}$ and $f_{h0}$ which lead to an approximation for the complete function $f(p)$.  Next, the cutoff independence of the three-body spectrum is verified, followed by an analysis of the coupling constants.  These numerical results are confirmed by Wilson's calculations \cite{Wilson:notes}.

    %%%%%%%%%%%%%%%%%%%%%
    %%  WAVEFUNCTIONS  %%
    %%%%%%%%%%%%%%%%%%%%%
\subsection{Low- and High-Momentum Functions}

\begin{figure}
\begin{center}
\includegraphics[height=4.0in]{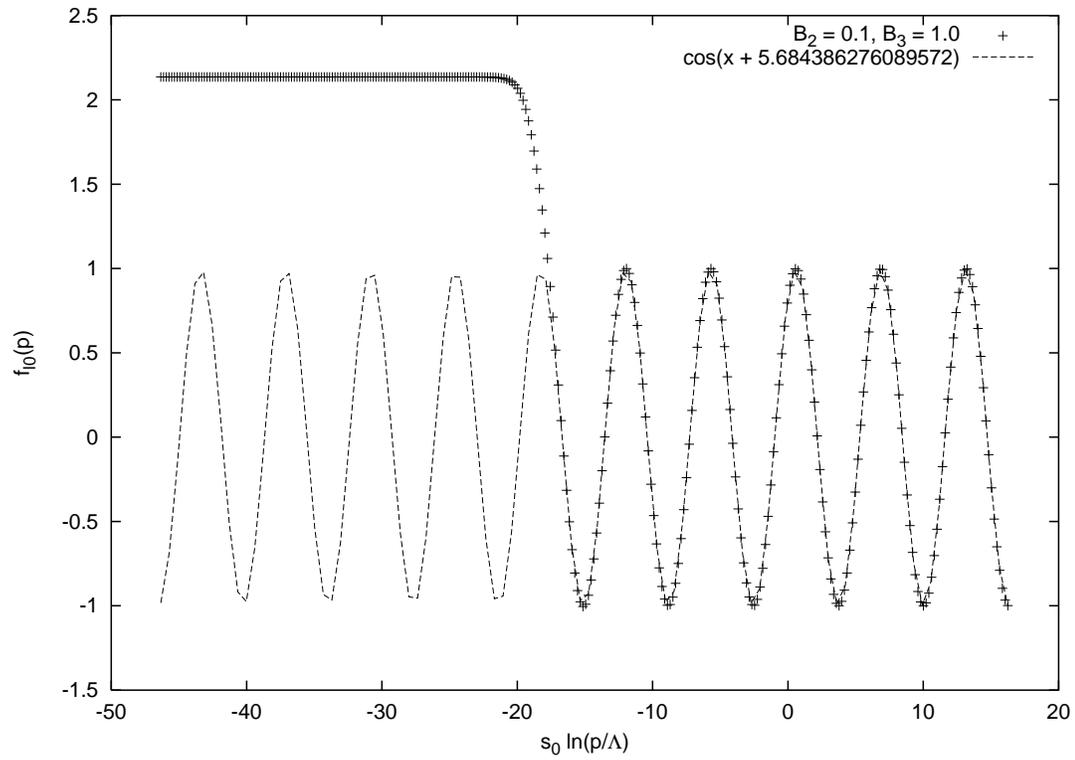}
\end{center}
\caption[Numerical solution of $f_{l0}(p)$ for the case of $B_2 = 0.1$, $B_3 = 1.0$, and $\Lambda = 10^8$.]{\label{fig:FL0-2}Numerical solution of $f_{l0}(p)$ for the case of $B_2 = 0.1$, $B_3 = 1.0$, and $\Lambda = 10^8$.  The dashed line is the best-fit cosine curve that matches the high-momentum behavior.}
\end{figure}

\begin{figure}
\begin{center}
\includegraphics[height=4.0in]{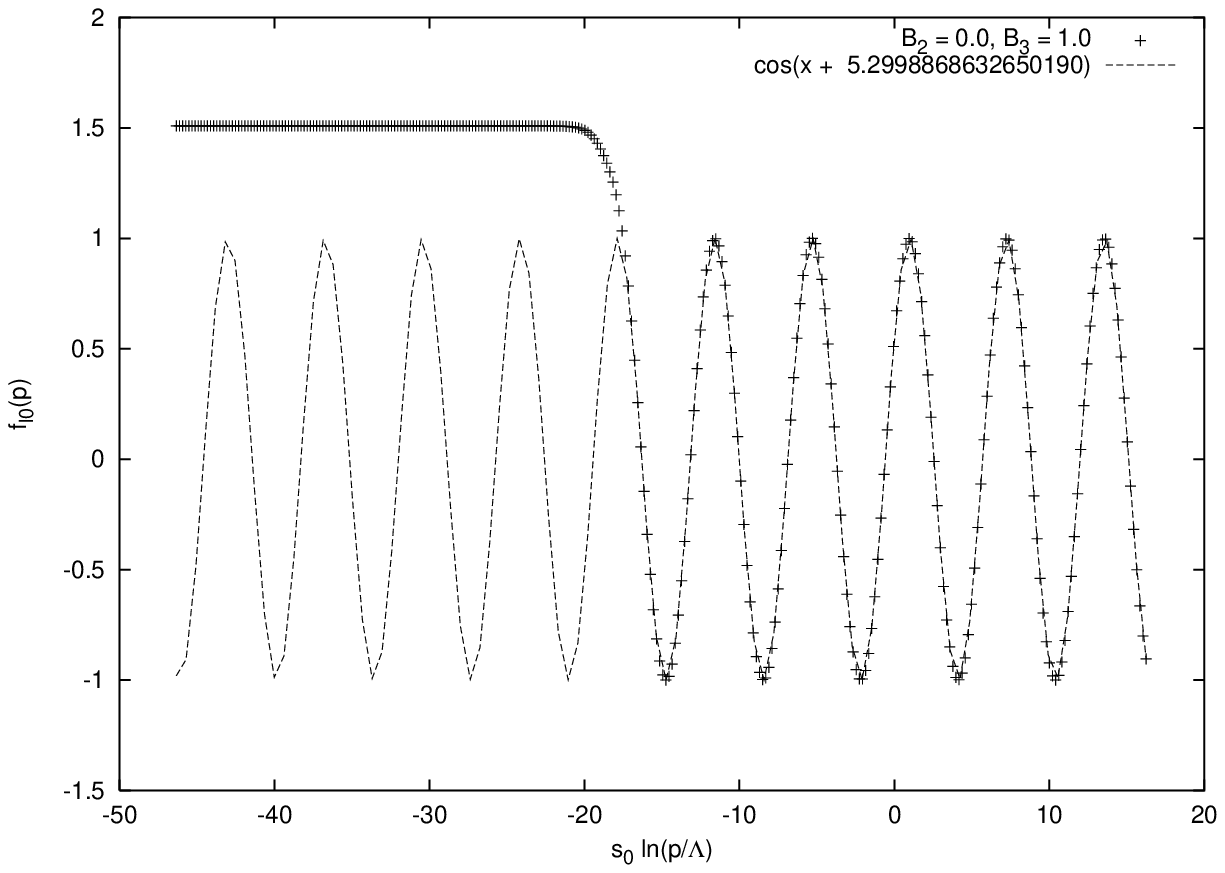}
\end{center}
\caption[Numerical solution of $f_{l0}(p)$ for the case of $B_2 = 0.0$, $B_3 = 1.0$, and $\Lambda = 10^8$.]{\label{fig:FL0-1}Numerical solution of $f_{l0}(p)$ for the case of $B_2 = 0.0$, $B_3 = 1.0$, and $\Lambda = 10^8$.  The dashed line is the best-fit cosine curve that matches the high-momentum behavior.}
\end{figure}

\begin{figure}
\begin{center}
\includegraphics[height=4.0in]{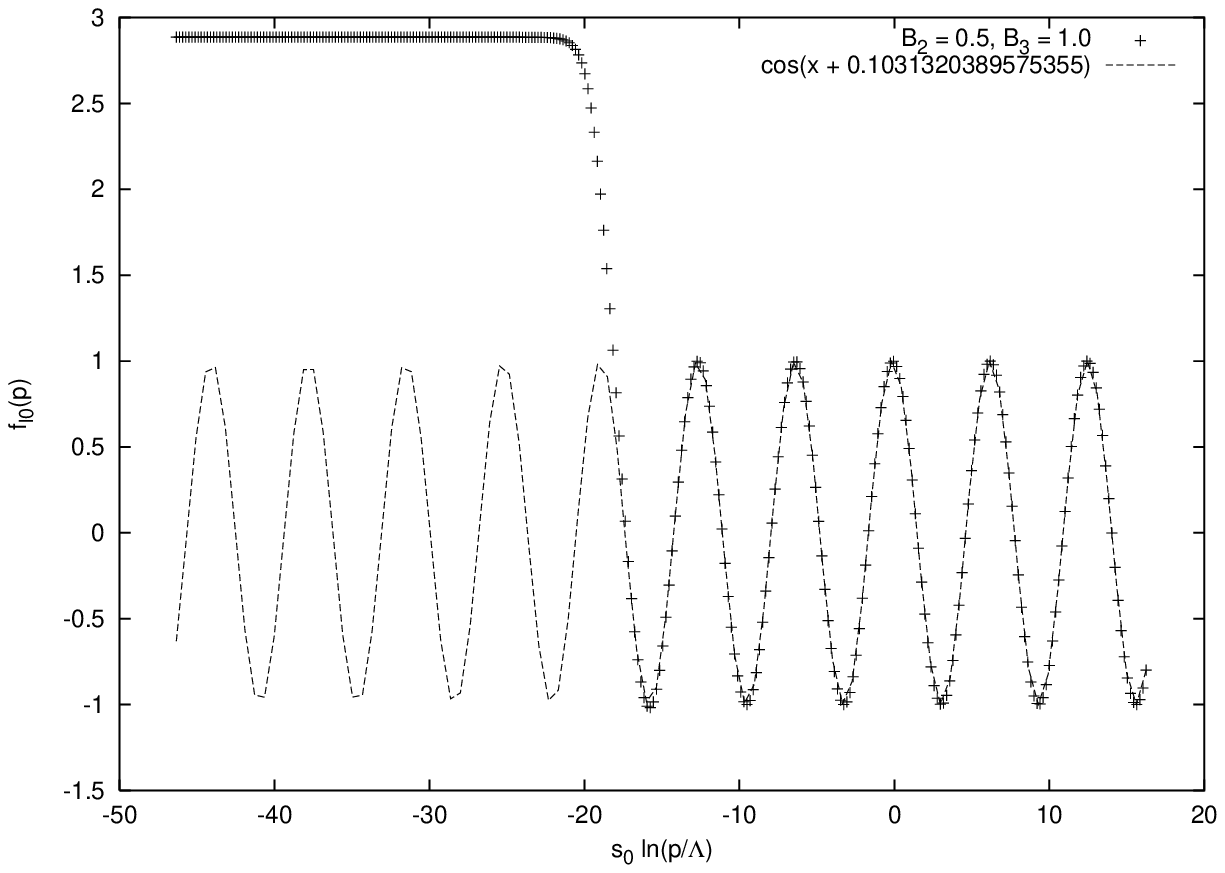}
\end{center}
\caption[Numerical solution of $f_{l0}(p)$ for the case of $B_2 = 0.5$, $B_3 = 1.0$, and $\Lambda = 10^8$.]{\label{fig:FL0-3}Numerical solution of $f_{l0}(p)$ for the case of $B_2 = 0.5$, $B_3 = 1.0$, and $\Lambda = 10^8$.  The dashed line is the best-fit cosine curve that matches the high-momentum behavior.}
\end{figure}

\begin{figure}
\begin{center}
\includegraphics[height=4.0in]{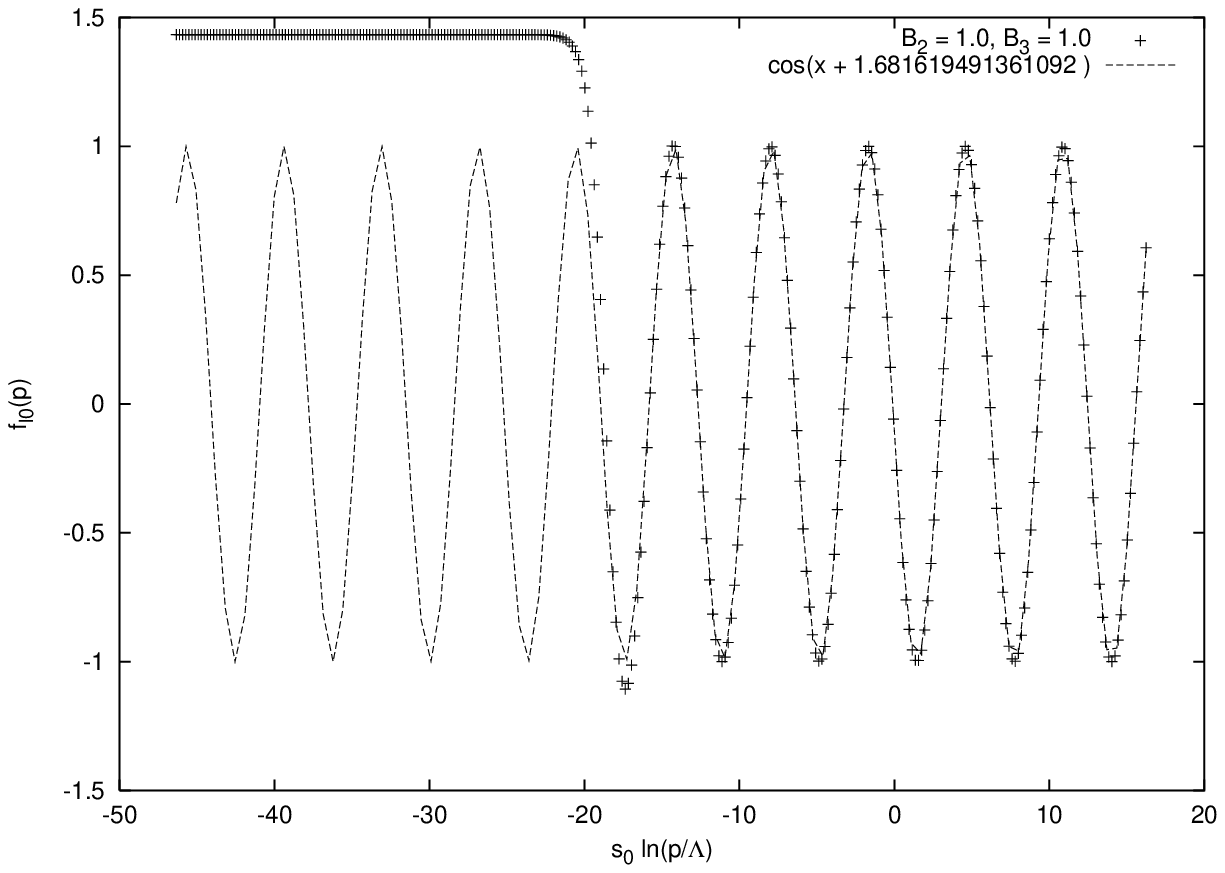}
\end{center}
\caption[Numerical solution of $f_{l0}(p)$ for the case of $B_2 = 1.0$, $B_3 = 1.0$, and $\Lambda = 10^8$.]{\label{fig:FL0-4}Numerical solution of $f_{l0}(p)$ for the case of $B_2 = 1.0$, $B_3 = 1.0$, and $\Lambda = 10^8$.  The dashed line is the best-fit cosine curve that matches the high-momentum behavior.}
\end{figure}

\begin{figure}
\begin{center}
\includegraphics[height=4.0in]{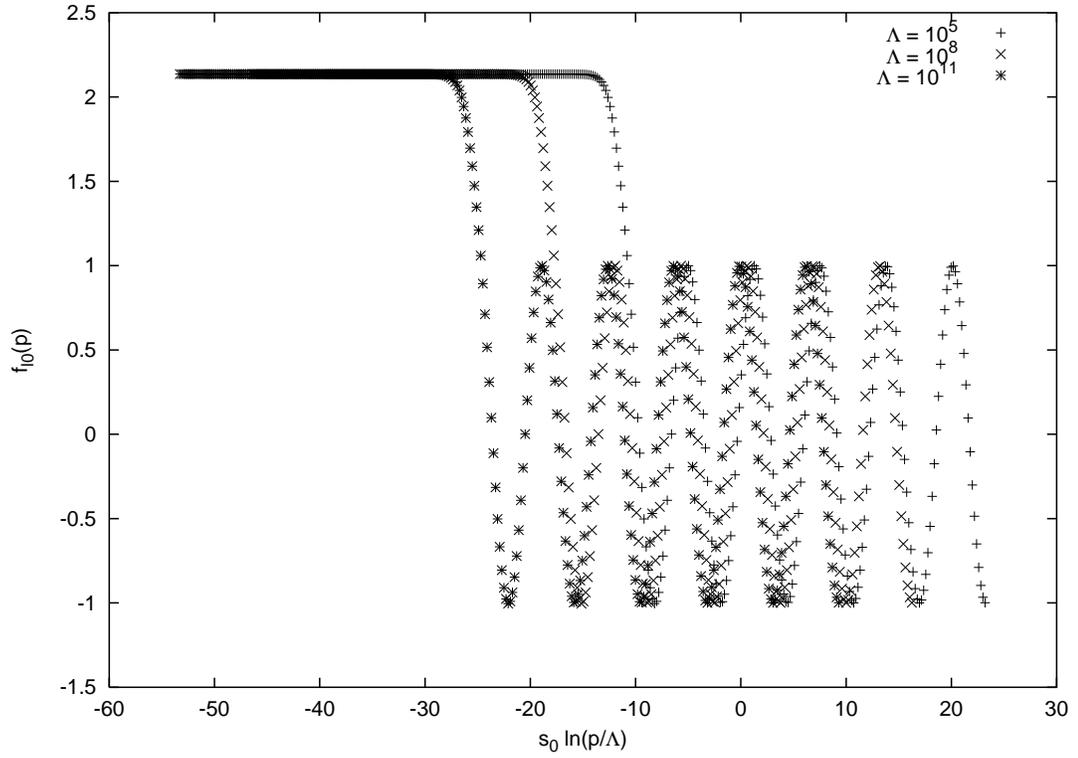}
\end{center}
\caption[Numerical solution of $f_{l0}(p)$ for the cutoffs $\Lambda = 10^5, 10^8,$ and $10^{11}$ with energies $B_2 = 0.1$ and $B_3 = 1.0$.]{\label{fig:FL0-256}Numerical solution of $f_{l0}(p)$ for the cutoffs $\Lambda = 10^5, 10^8,$ and $10^{11}$ with energies $B_2 = 0.1$ and $B_3 = 1.0$.  The phases corresponding to these cutoffs are 5.016726935003937, 5.684386276089572, and 0.06886030999563049 respectively.}
\end{figure}

Figure \ref{fig:FL0-2} shows the numerical solution for $f_{l0}(p)$ in the case of $B_2 = 0.1$, $B_3 = 1.0$, and $\Lambda = 10^8$.  Notice that it is constant for small values of momentum and then takes on the cosine behavior as $p$ becomes large.  This behavior is typical of all solutions for $f_{l0}$.  For reference, we have included numerical solutions for the additional cases of $B_2 = 0.0, 0.5,$ and $1.0$ in Figs.~\ref{fig:FL0-1}, \ref{fig:FL0-3}, and \ref{fig:FL0-4} respectively.  Since the behavior of the solution to the low-momentum equation is determined by the ratio $B_2/B_3$, these figures provide a representative sample of all solutions.  We can also study the effects of changing the cutoff while keeping the energies fixed.  Equation (\ref{eqn:theta_tilde}) shows that changing the value of $\Lambda$ will only result in a shift in the value of $\theta$, which is exactly the behavior seen in Fig.~\ref{fig:FL0-256}.  Note that we follow the convention of restricting all phases to be within the range 0 to $2 \pi$, which explains why one of the phase values is close to zero.

Several bound states may exist for the same values of $B_2$ and $\Lambda$.  In the case of $B_2 = 1.0$ and $B_3 = 1.0$ for $\Lambda = 10^8$, there exist states of energy $B_3 = 6.7502901502599$ and $B_3 = 1406.130393204$.  The solutions of $f_{l0}$ for these energies are shown in Fig.~\ref{fig:FL0-Spectrum2}.  All of the functions are very similar.  The only real difference is in the length of the initial ``plateau.''  As the bound-state energies become larger, the flat region becomes longer, joining the cosine part near later peaks.  An additional illustration of this behavior is shown in Fig.~\ref{fig:FL0-Spectrum1} using different bound-state energies.

\begin{figure}
\begin{center}
\includegraphics[height=4.0in]{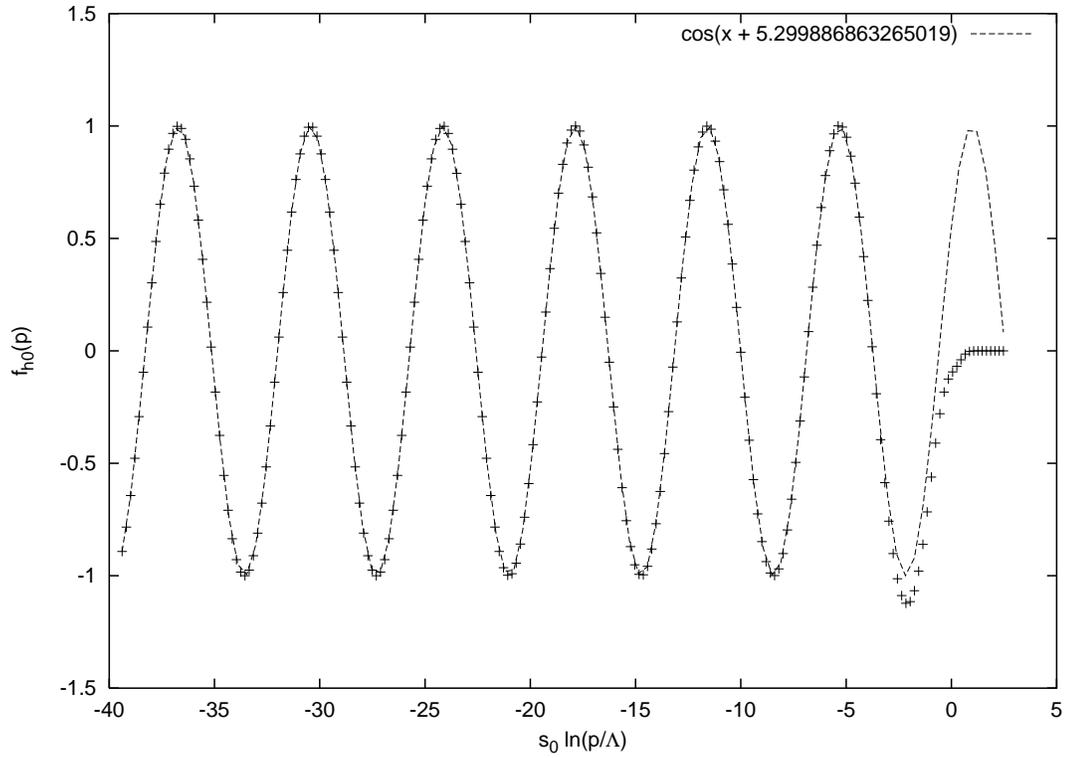}
\end{center}
\caption[Numerical solution of $f_{h0}(p)$ for the case $\theta = 5.299886863265019$ and $\Lambda = 10^8$.]{\label{fig:FH0-1}Numerical solution of $f_{h0}(p)$ for the case $\theta = 5.299886863265019$ and $\Lambda = 10^8$.  The dashed line is the best-fit cosine curve that matches the low-momentum behavior.}
\end{figure}

\begin{figure}
\begin{center}
\includegraphics[height=4.0in]{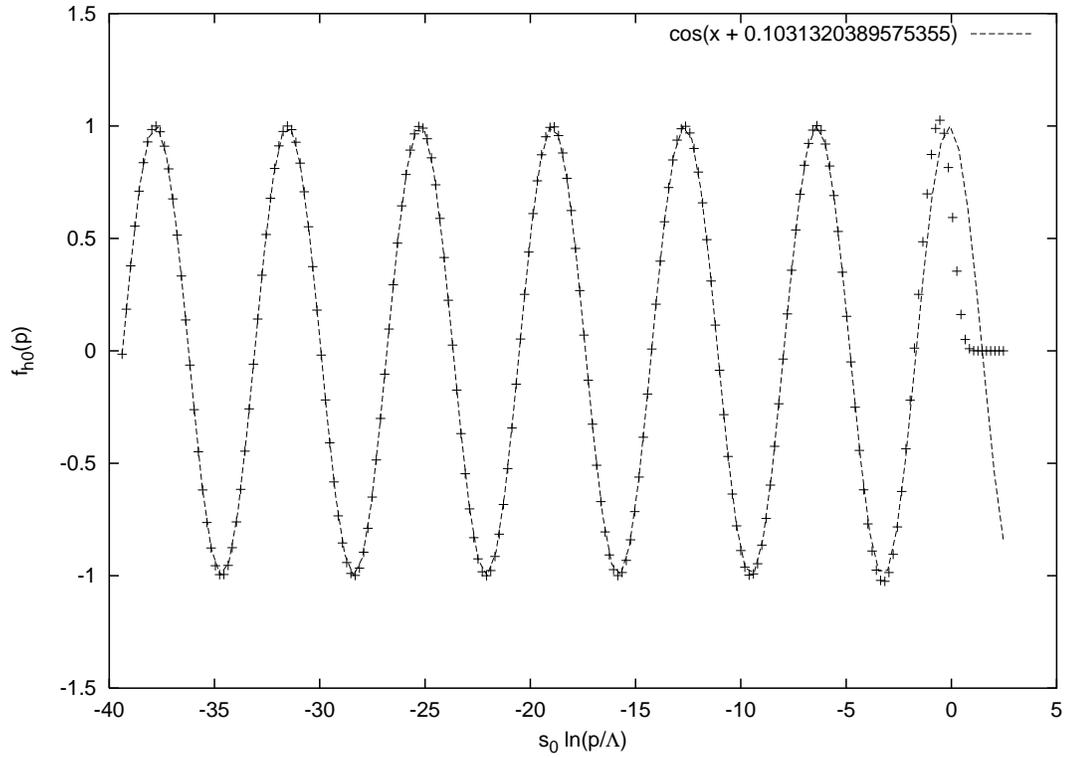}
\end{center}
\caption[Numerical solution of $f_{h0}(p)$ for the case $\theta = 0.1031320389575355$ and $\Lambda = 10^8$.]{\label{fig:FH0-3}Numerical solution of $f_{h0}(p)$ for the case $\theta = 0.1031320389575355$ and $\Lambda = 10^8$.  The dashed line is the best-fit cosine curve that matches the low-momentum behavior.}
\end{figure}

\begin{figure}
\begin{center}
\includegraphics[height=4.0in]{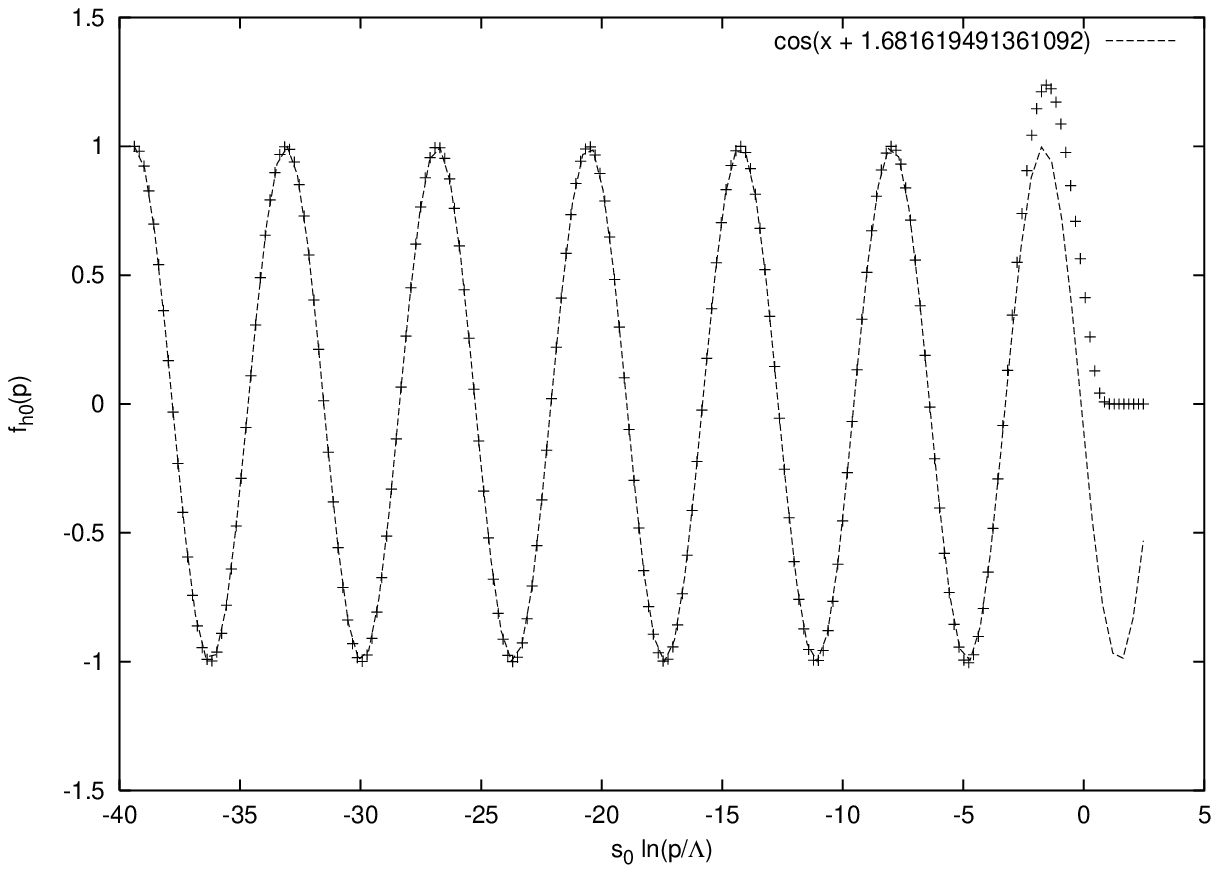}
\end{center}
\caption[Numerical solution of $f_{h0}(p)$ for the case $\theta = 1.681619491361092$ and $\Lambda = 10^8$.]{\label{fig:FH0-4}Numerical solution of $f_{h0}(p)$ for the case $\theta = 1.681619491361092$ and $\Lambda = 10^8$.  The dashed line is the best-fit cosine curve that matches the low-momentum behavior.}
\end{figure}

\begin{figure}
\begin{center}
\includegraphics[height=4.0in]{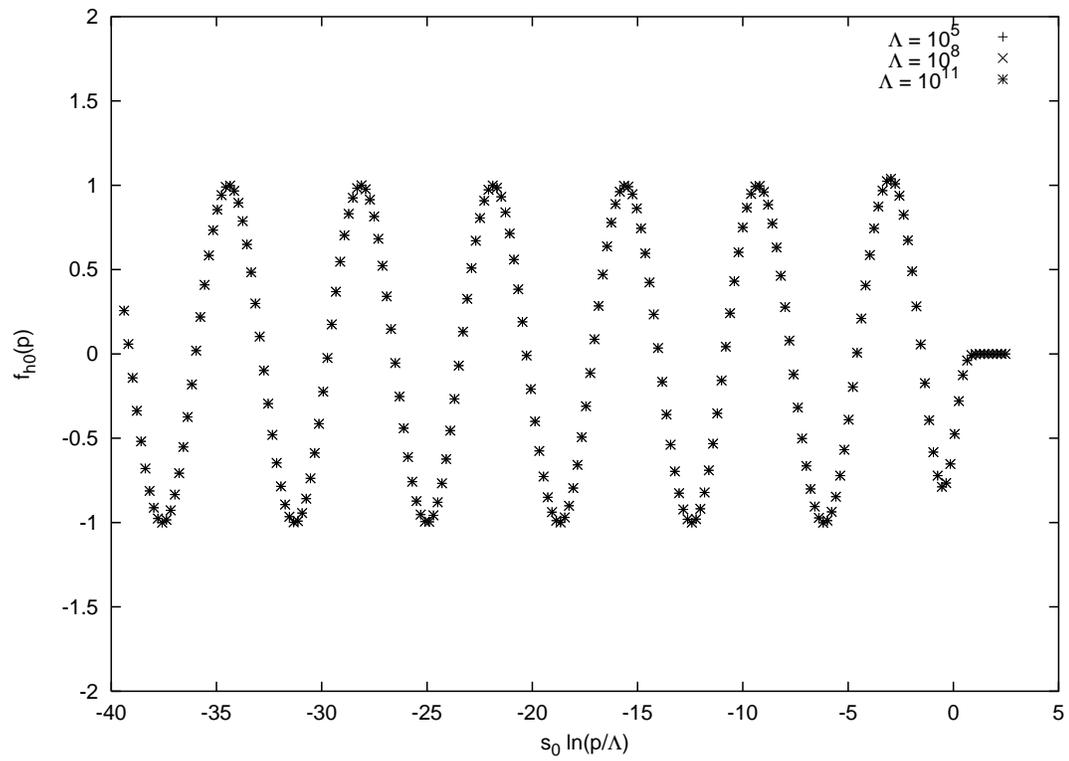}
\end{center}
\caption{\label{fig:FH0-567}Numerical solution of $f_{h0}(p)$ with $\theta = 3.0$ for cutoffs $\Lambda = 10^5, 10^8,$ and $10^{11}$.}
\end{figure}

Using the parameters $\theta = 5.684386276089572$ and $\Lambda = 10^8$, we find the numerical solution for the function $f_{h0}(p)$ shown in Fig.~\ref{fig:FH0-2}.  This is the high-momentum function corresponding to the low-momentum function in Fig.~\ref{fig:FL0-2}.  Figures \ref{fig:FH0-1} to \ref{fig:FH0-4} display the high-momentum counterparts of Figs.~\ref{fig:FL0-1} to \ref{fig:FL0-4}.  For all $f_{h0}$ solutions, we see the cosine behavior for $p \ll \Lambda$ and a suppression of large momentum values when $p > \Lambda$.  This suppression is an effect of the gaussian behavior of $U_2$.  The behavior of $f_{h0}$ is determined by the phase.  Different values of the cutoff for the same phase should result in the same function.  An example of this is shown in Fig.~\ref{fig:FH0-567} where $\theta = 3.0$ for $\Lambda = 10^5, 10^8,$ and $10^{11}$.  Notice that all three solutions coincide.

\begin{figure}
\begin{center}
\includegraphics[height=4.0in]{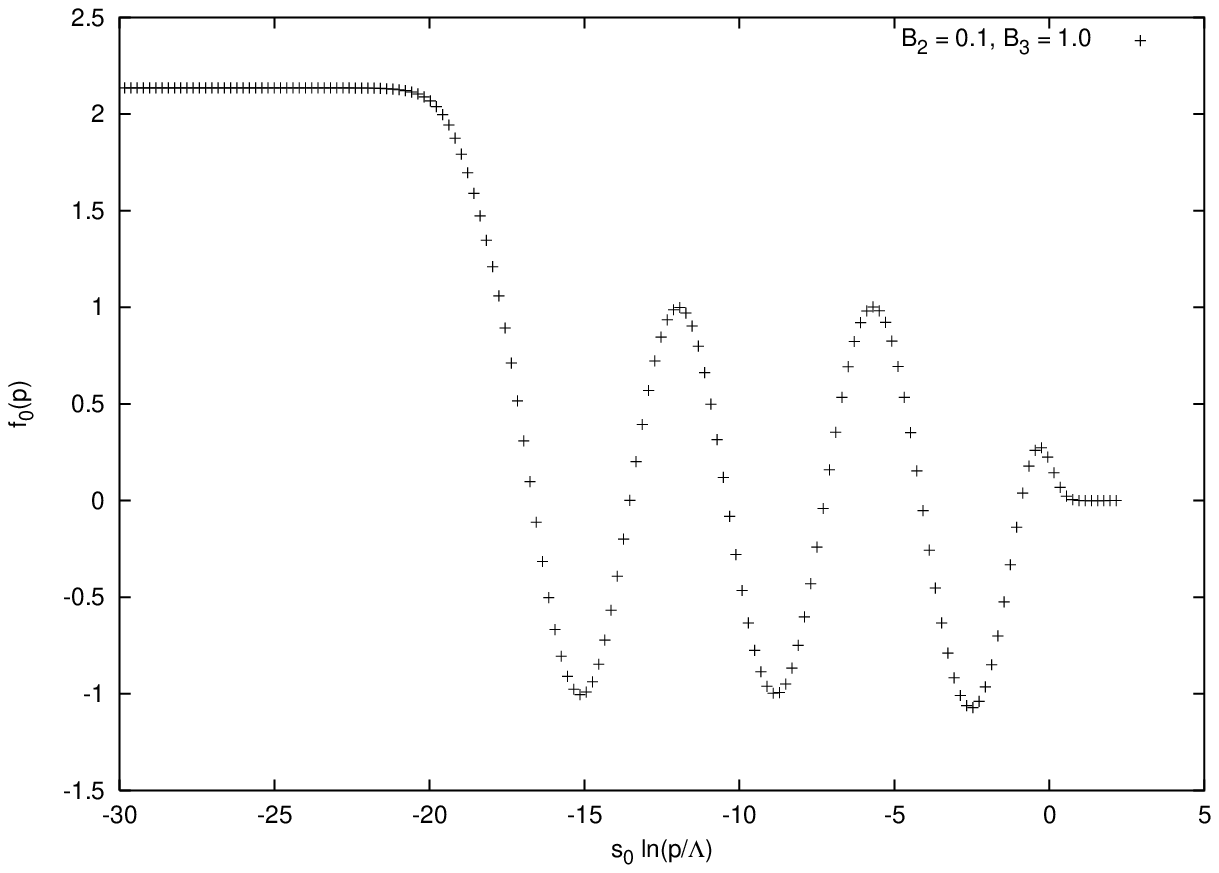}
\end{center}
\caption{\label{fig:F0-2}Numerical solution of $f_0(p)$ for the case of $B_2 = 0.1$, $B_3 = 1.0$, and $\Lambda = 10^8$.}
\end{figure}

\begin{figure}
\begin{center}
\includegraphics[height=4.0in]{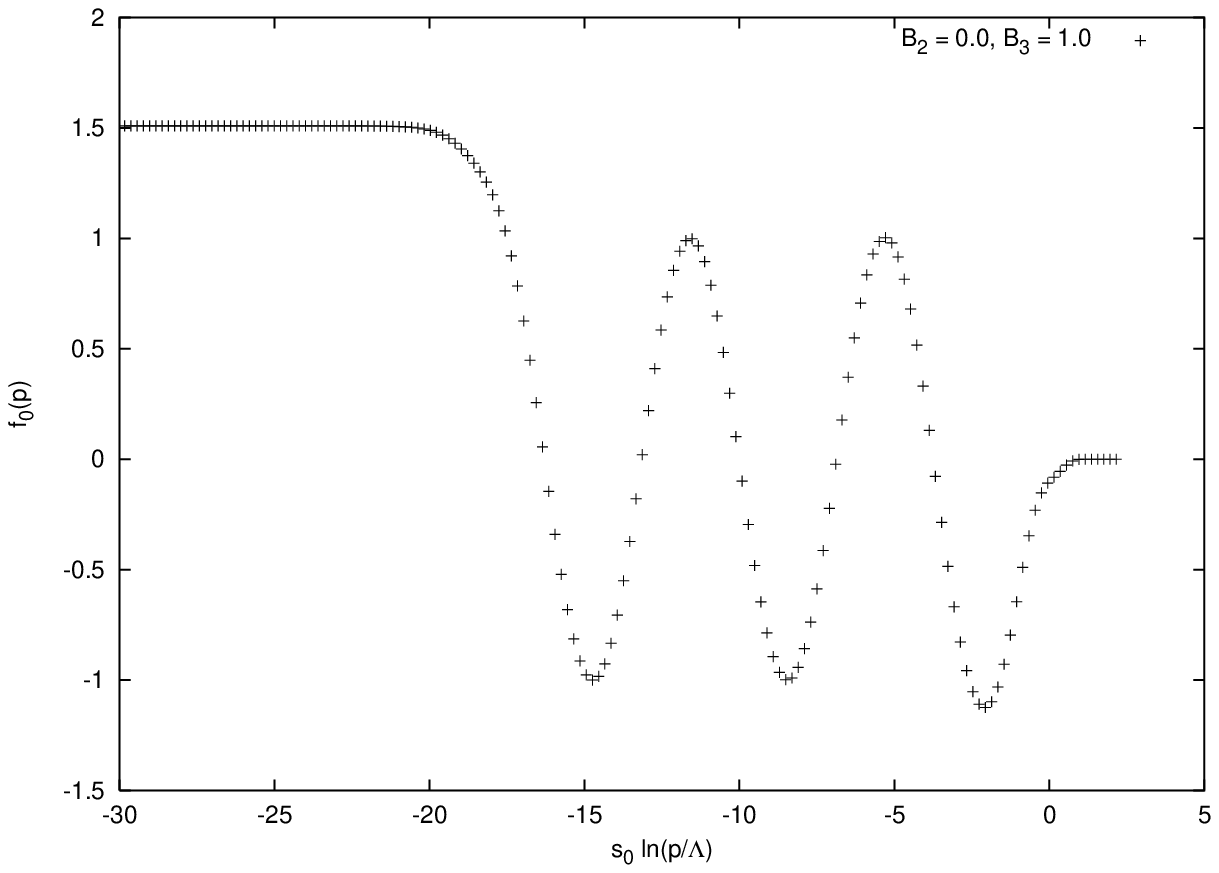}
\end{center}
\caption{\label{fig:F0-1}Numerical solution of $f_0(p)$ for the case of $B_2 = 0.0$, $B_3 = 1.0$, and $\Lambda = 10^8$.}
\end{figure}

\begin{figure}
\begin{center}
\includegraphics[height=4.0in]{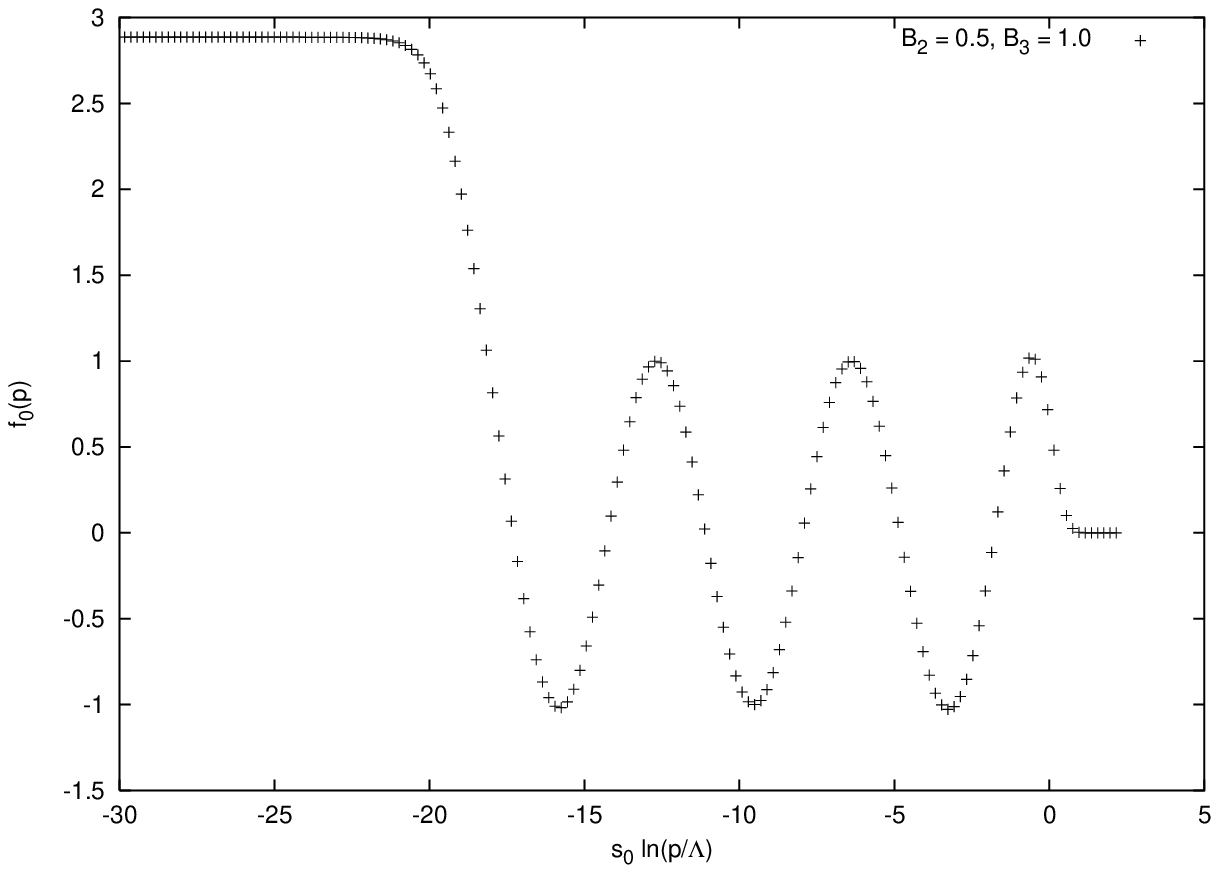}
\end{center}
\caption{\label{fig:F0-3}Numerical solution of $f_0(p)$ for the case of $B_2 = 0.5$, $B_3 = 1.0$, and $\Lambda = 10^8$.}
\end{figure}

\begin{figure}
\begin{center}
\includegraphics[height=4.0in]{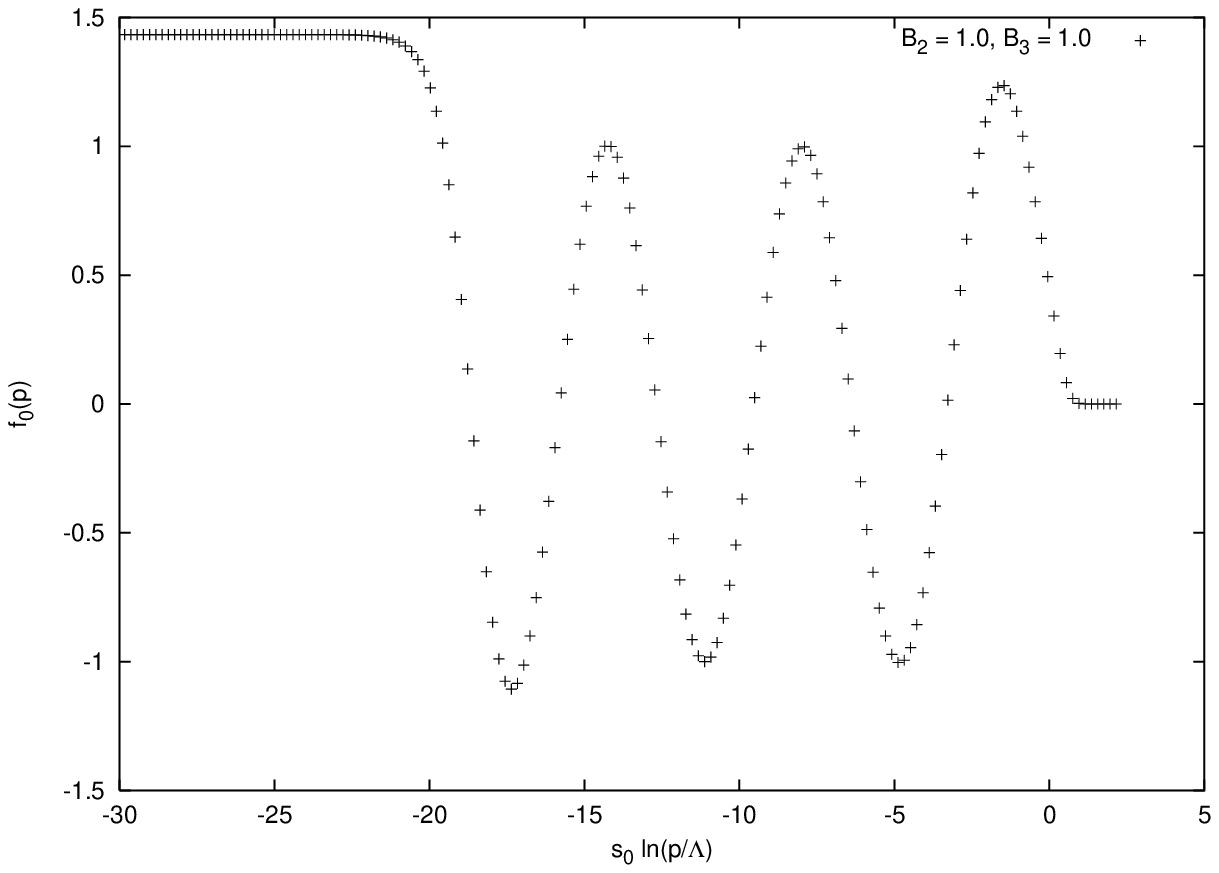}
\end{center}
\caption{\label{fig:F0-4}Numerical solution of $f_0(p)$ for the case of $B_2 = 1.0$, $B_3 = 1.0$, and $\Lambda = 10^8$.}
\end{figure}

\begin{figure}
\begin{center}
\includegraphics[height=4.0in]{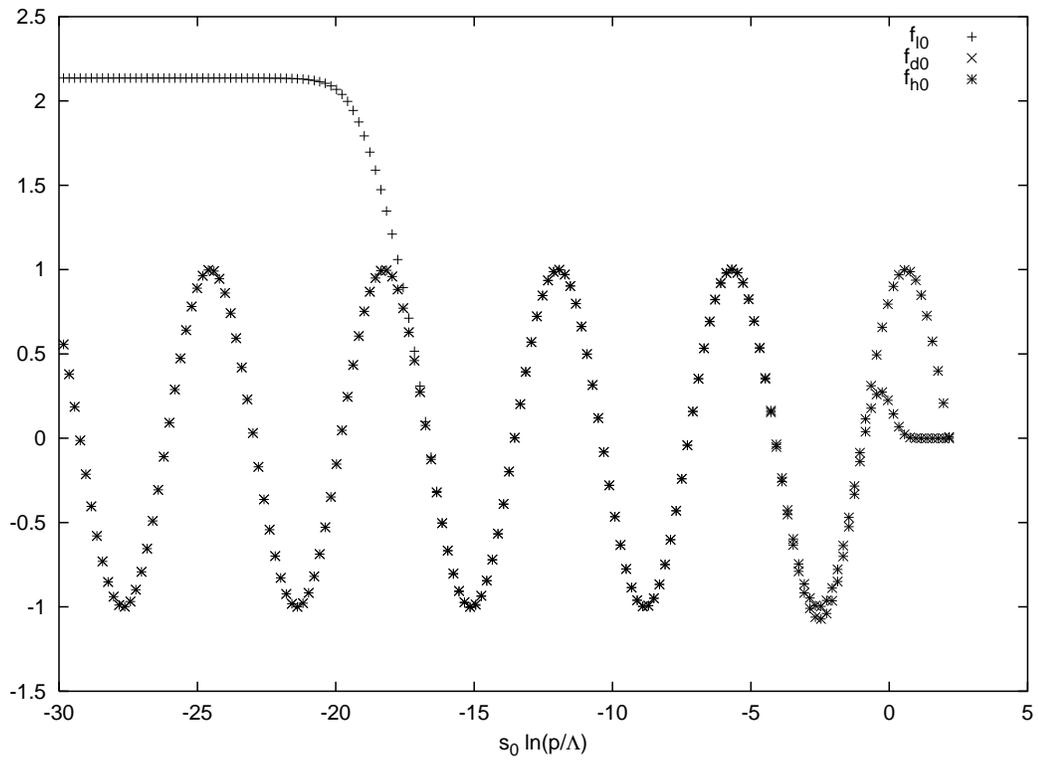}
\end{center}
\caption{\label{fig:F0-2LDH}Numerical solutions of $f_{l0}$, $f_{d0}$, and $f_{h0}$ for the case of $B_2 = 0.1$, $B_3 = 1.0$, and $\Lambda = 10^8$.}
\end{figure}

Combining $f_{l0}$, $f_{d0}$ and $f_{h0}$, we can see the overall leading order behavior for the function $f(p)$.  Plots of $f_{l0}+ f_{h0} - f_{d0}$ are shown in Figs.~\ref{fig:F0-2} to \ref{fig:F0-4} for the four low- and high-momentum solutions previously considered.  These solutions span several values of $B_2/B_3$ and show the typical behavior of any leading order solution.  Figure \ref{fig:F0-2LDH} shows the three individual components for $B_2 = 0.1$ on the same plot.  From this, it easy to see how cancellations occur in each region making it clear why $f_l + f_h - f_d$ is a good approximation to $f$ everywhere.

    %%%%%%%%%%%%%%%%
    %%  ENERGIES  %%
    %%%%%%%%%%%%%%%%
\subsection{Three-Body Binding Energies}

\begin{figure}
\begin{center}
\includegraphics[height=4.0in]{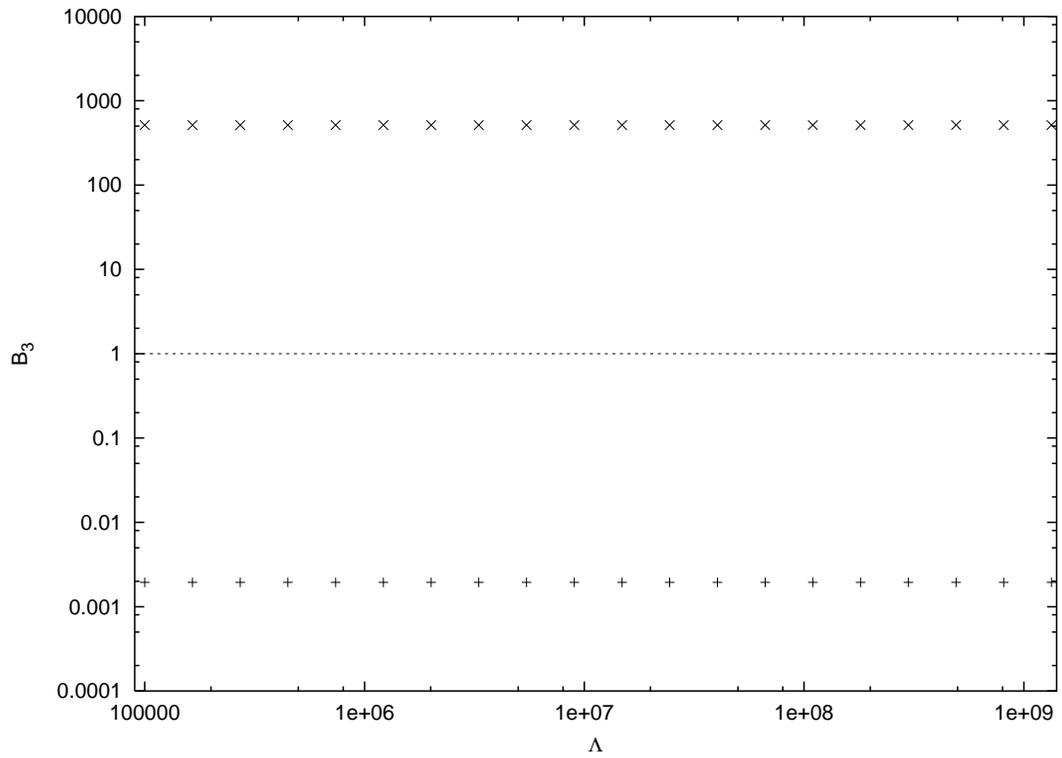}
\end{center}
\caption{\label{fig:B3vsLAM}Binding energies of the next shallowest and next deepest 3-body bound states as a function of the cutoff $\Lambda$ for $B_2 = 0$ and  $B_3 = 1.0$.}
\end{figure}

In Sec.~\ref{sec:analytic} we proved that the three-body bound-state spectrum is cutoff independent to leading order.  This is verified numerically in Fig.~\ref{fig:B3vsLAM}.  Here we have chosen $B_2 = 0$ and let $G_3$ change with $\Lambda$ so that the state $B_3 = 1.0$ is held constant.  Two other states, one shallower and one deeper, are calculated as the cutoff changes.  Since very small fluctuations are impossible to see in the plot, we have included Table \ref{tab:B3vsLAMBDA-1}, which shows the calculated energies for several values of $\Lambda$.  Using Efimov's result that the ratio of adjacent binding energies is $\exp(2 \pi/s_0)$ when $B_2 = 0$ \cite{efimov:2,efimov:4}, the relative error for each calculation can be determined and is also given in the table.  This illustrates that each energy is cutoff-independent to about 12 digits and also matches the true value to the same accuracy.  As an additional example, Table \ref{tab:B3vsLAMBDA-2} shows the case of $B_2 = B_3 = 1.0$ and considers the next two deeper states as $\Lambda$ changes.  The binding energies are approximately 6.75029 and 1406.13.  These energies have been previously calculated by Braaten, Hammer. and Kusunoki \cite{Braaten:uefes} using a method that gives at most two digits of numerical precision.  Their results are 6.8 and $1.4\times 10^3$, which match the results given here to within relative errors of $\mathcal{O}(10^{-3})$.  Note that our values for these states are independent of $\Lambda$ to about 12 digits.

\begin{table}
\begin{center}
\begin{tabular}{|c|c|c|c|c|}
\hline
$\Lambda$ & $B_3$ (Shallow) & Error & $B_3$ (Deep) & Error \\
\hline
\hline
100000.00000000 & 0.0019416156131338 & 6.7e-13 & 515.03500138461 & 5.2e-13 \\
738905.60989306 & 0.0019416156131358 & 3.2e-13 & 515.03500138403 & 1.6e-12 \\
5459815.0033144 & 0.0019416156131435 & 4.3e-12 & 515.03500138287 & 3.9e-12 \\
109663315.84284 & 0.0019416156131358 & 3.2e-13 & 515.03500138520 & 6.0e-13 \\
3631550267.4246 & 0.0019416156131435 & 4.3e-12 & 515.03500138520 & 6.0e-13 \\
\hline
\end{tabular}
\end{center}
\caption[Binding energies of the next shallowest and next deepest 3-body bound states for $B_2 = 0.0$, $B_3 = 1.0$ and various cutoffs.]{\label{tab:B3vsLAMBDA-1}Binding energies of the next shallowest and next deepest 3-body bound states for $B_2 = 0.0$, $B_3 = 1.0$ and various cutoffs.  Several data points from Fig.~\ref{fig:B3vsLAM} are shown along with the relative error of each result.}
\end{table}

\begin{figure}
\begin{center}
\includegraphics[height=4.0in]{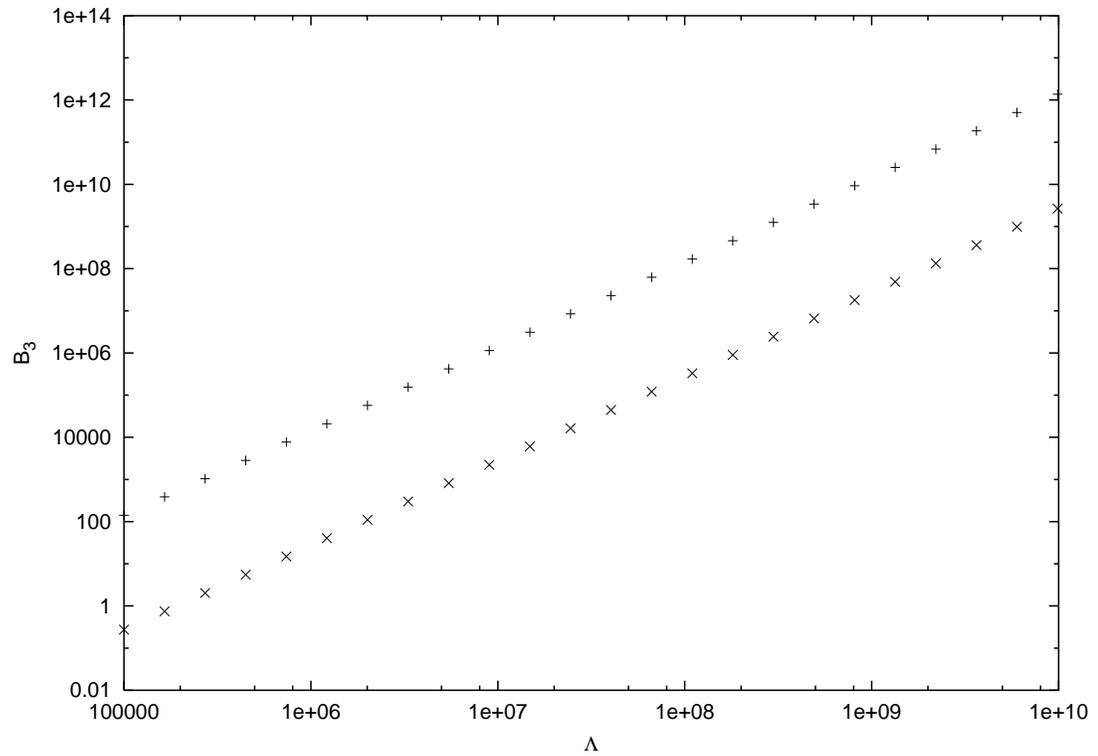}
\end{center}
\caption[Binding energies of two three-body bound states for the case of $B_2 = 0$ and $G_3 = 0$.]{\label{fig:B3diverge}Binding energies of two three-body bound states for the case of $B_2 = 0$ and $G_3 = 0$.  Notice that without a three-body interaction, all the binding energies diverge as the cutoff is increased.}
\end{figure}

We have shown that a three-body interaction is capable of fixing a three-body binding energy.  However, this does not prove that such an interaction is necessary.  To investigate this claim, let us assume that $G_3 = 0$ for any value of $\Lambda$, so that the three-body interaction is removed from our equations.  Using $B_2 = 0$, we calculate two values of $B_3$ for a cutoff of $\Lambda = 10^5$.  The energy of each state is tracked as the value of $\Lambda$ changes.  The results are shown in Fig.~\ref{fig:B3diverge}.

\begin{table}
\begin{center}
\begin{tabular}{|c|c|c|}
\hline
$\Lambda$ & $B_3$ \#1 & $B_3$ \#2 \\
\hline
\hline
100000.00000000 & 6.750290150257678 & 1406.13039320296 \\
738905.60989306 & 6.750290150257678 & 1406.13039320593 \\
2008553.6923187 & 6.750290150255419 & 1406.13039320345 \\
14841315.910257 & 6.750290150268966 & 1406.13039320345 \\
298095798.70417 & 6.750290150257678 & 1406.13039320593 \\
5987414171.5197 & 6.750290150259935 & 1406.13039320345 \\
44241339200.892 & 6.750290150257678 & 1406.13039320296 \\
\hline
\end{tabular}
\end{center}
\caption{\label{tab:B3vsLAMBDA-2}Binding energies of the two next deeper 3-body bound states for $B_2 = B_3 = 1.0$ and various cutoffs.}
\end{table}

Both binding energies diverge like $\Lambda^2$.  This is easily seen using equation (\ref{eqn:theta_tilde}).  Since $G_3$ = 0, the phase must be fixed.  This means
\begin{equation}
\theta = s_0 \, \ln\left(\frac{\Lambda}{\sqrt{B_3}}\right) + \tilde{\theta}(0) = \mathrm{constant}, 
\end{equation}
\noindent from which we can easily derive the relation $B_3 \propto \Lambda^2$.  This divergent behavior is an example of the unbounded three-body spectrum predicted by L.~H.~Thomas for short-ranged interactions \cite{Thomas:shorta}.  While this alone is not a proof that a three-body contact interaction is required, it should be possible to show that any other local interaction does not enter the leading order equation.  This would imply $G_3$ is the only coupling capable of making any binding energy independent of the cutoff.

    %%%%%%%%%%%%%%%%%
    %%  COUPLINGS  %%
    %%%%%%%%%%%%%%%%%
\subsection{Couplings}

\begin{figure}
\begin{center}
\includegraphics[height=4.0in]{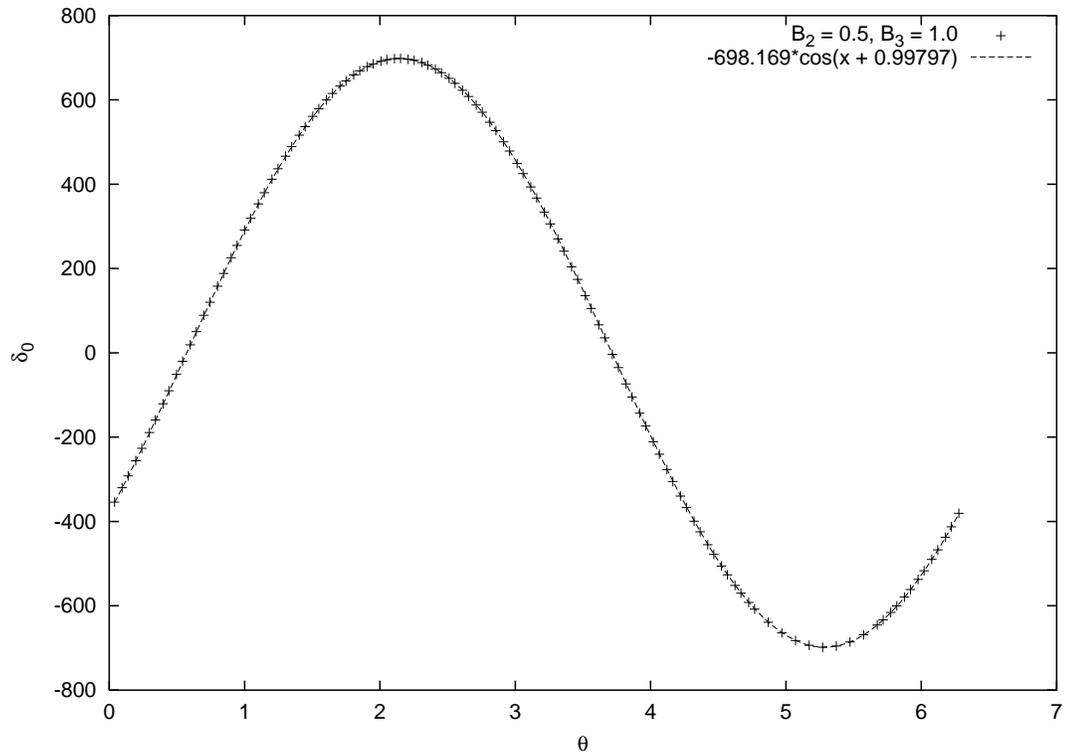}
\end{center}
\caption[The coupling $\delta_0$ as a function of the phase $\theta$ for the case $B_2 = 0.5$ and $B_3 = 1.0$.]{\label{fig:PSEUDOvsTHETA-2}The coupling $\delta_0$ as a function of the phase $\theta$ for the case $B_2 = 0.5$ and $B_3 = 1.0$.  The dashed line is the best-fit cosine curve.  The data matches to within an error of about $10^{-10}$.}
\end{figure}

\begin{figure}
\begin{center}
\includegraphics[height=4.0in]{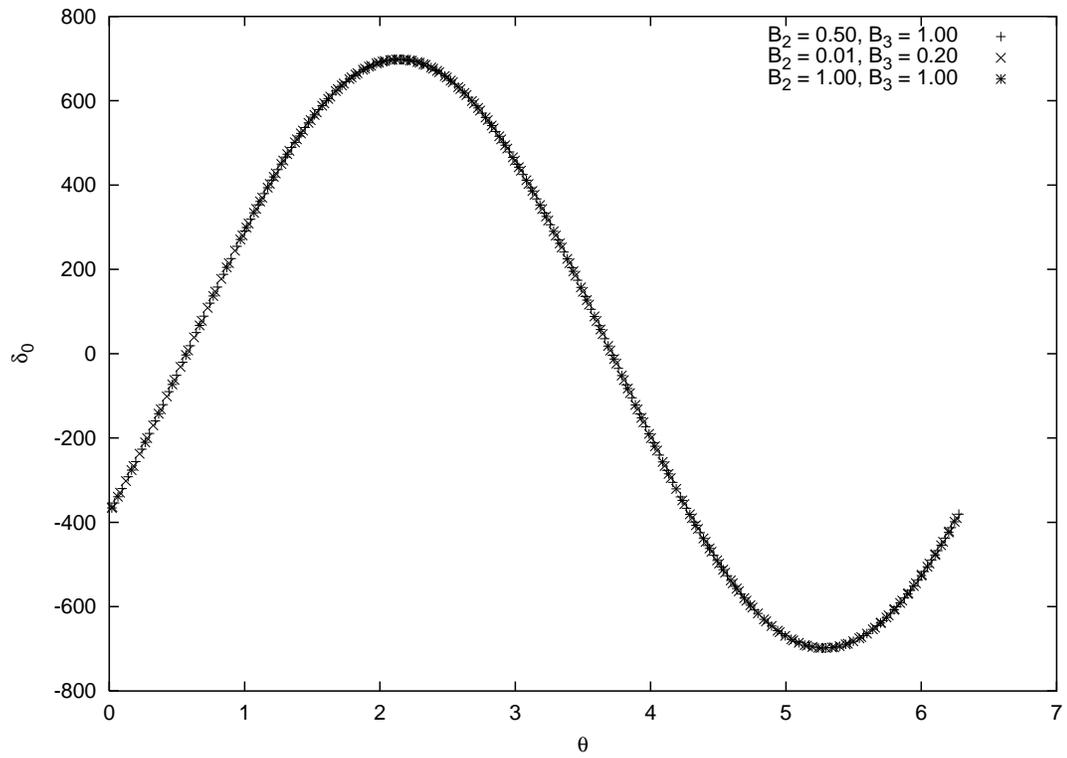}
\end{center}
\caption[The coupling $\delta_0$ as a function of the phase $\theta$ for three different sets of $B_2$ and $B_3$.]{\label{fig:PSEUDOvsTHETA-123}The coupling $\delta_0$ as a function of the phase $\theta$ for three different sets of $B_2$ and $B_3$.  Notice that all curves are identical to within numerical error.}
\end{figure}

Equation (\ref{eqn:PseudoCosine}) shows that the coupling $\delta_0$ should have a cosine dependence on the phase $\theta$, which is defined by $f_{d0} = \cos(s_0\ln(p/\Lambda) + \theta)$.  Numerical data for $\delta_0$ as a function of the phase is shown in Fig.~\ref{fig:PSEUDOvsTHETA-2}, along with a best-fit cosine curve to verify the behavior.  This behavior is independent of $B_2$ and $B_3$.  Figure \ref{fig:PSEUDOvsTHETA-123} proves this by plotting the same relation for different energies.  Keep in mind that $h_2$ can affect the amplitude and/or phase shift of the cosine behavior.

\begin{figure}
\begin{center}
\includegraphics[height=4.0in]{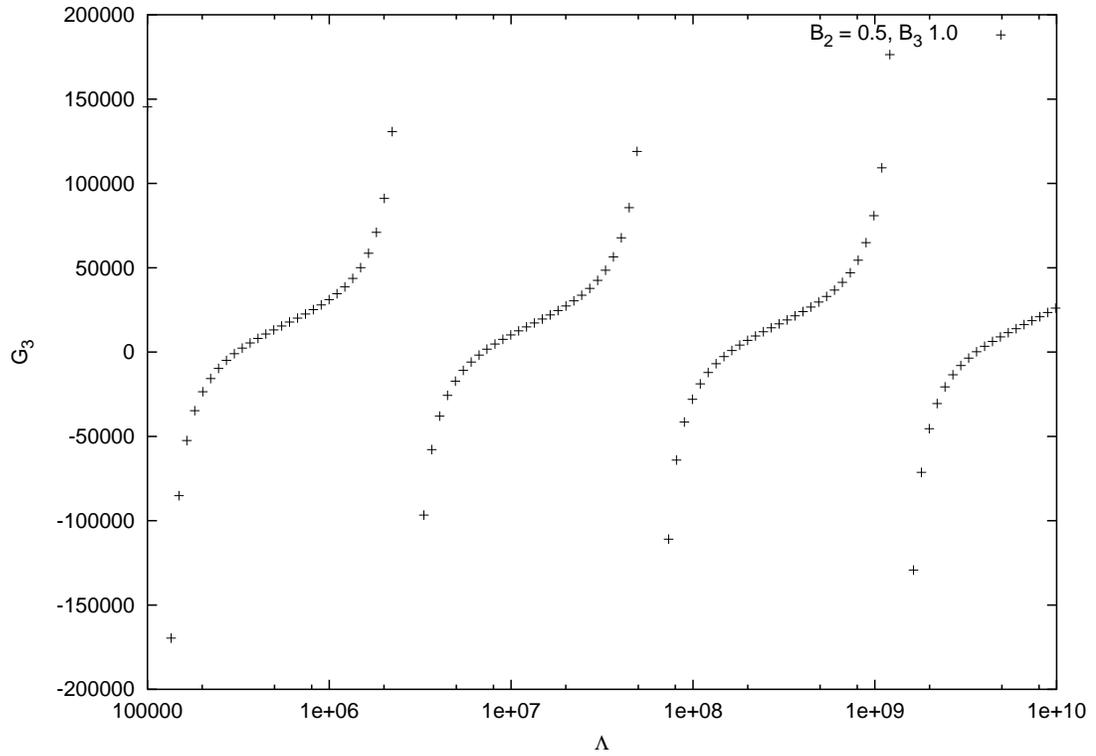}
\end{center}
\caption[The dimensionless three-body coupling $G_3$ as a function of the cutoff $\Lambda$ for $B_2 = 0.5$ and $B_3 = 1.0$.]{\label{fig:G3vsLAMBDA-2}The dimensionless three-body coupling $G_3$ as a function of the cutoff $\Lambda$ for $B_2 = 0.5$ and $B_3 = 1.0$.  The limit-cycle behavior of $G_3$ is evident from the fact that it is periodic in $\ln(\Lambda)$.}
\end{figure}

\begin{figure}
\begin{center}
\includegraphics[height=4.0in]{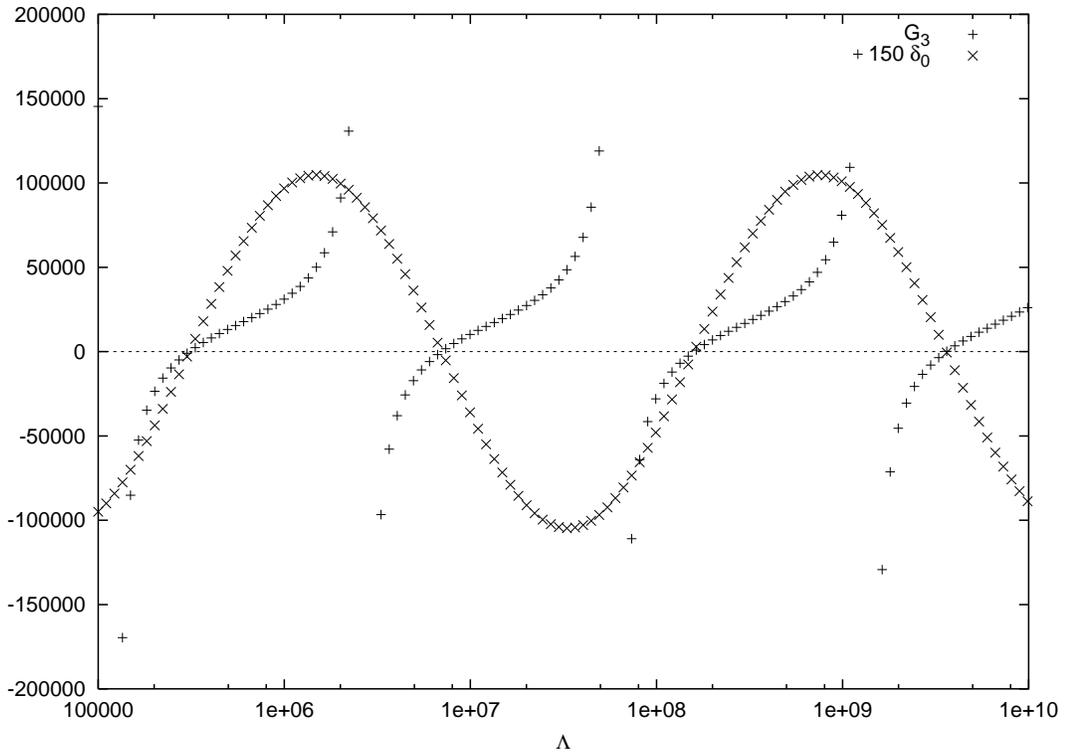}
\end{center}
\caption[Dependence of $G_3$ and $\delta_0$ on the cutoff $\Lambda$ for the case of $B_2 = 0.5$ and $B_3 = 1.0$.]{\label{fig:G3andPSEUDOvsLAMBDA-2}Dependence of $G_3$ and $\delta_0$ on the cutoff $\Lambda$ for the case of $B_2 = 0.5$ and $B_3 = 1.0$. The values for $\delta_0$ have been multiplied by a constant factor of 150 to make the cosine behavior visible.  Both couplings are shown to emphasize that they equal zero at the same points.}
\end{figure}

\begin{figure}
\begin{center}
\includegraphics[height=4.0in]{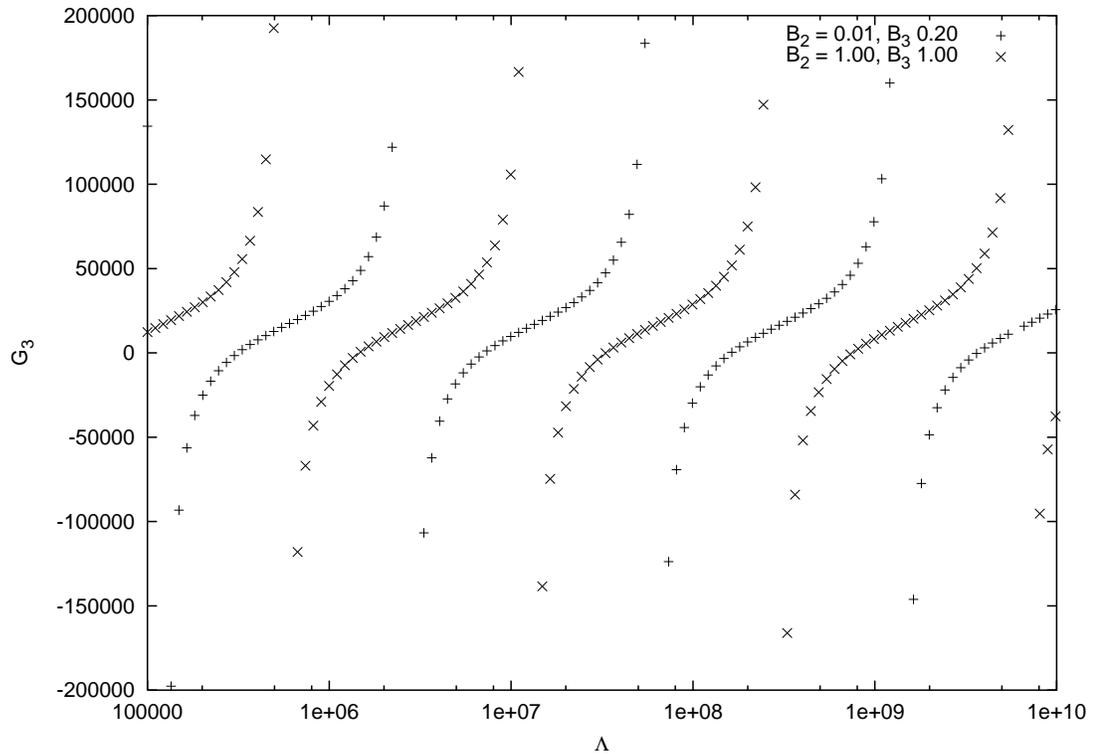}
\end{center}
\caption[$G_3$ as a function of the cutoff $\Lambda$ for two different sets of $B_2$ and $B_3$.]{\label{fig:G3vsLAMBDA-13}$G_3$ as a function of the cutoff $\Lambda$ for two different sets of $B_2$ and $B_3$.  Both curves have the same limit-cycle behavior, but the difference in energies causes one to be shifted relative to the other.}
\end{figure}

We suggested in Sec.~\ref{sec:analytic} that this periodic behavior should carry over to $G_3$.  Figure \ref{fig:G3vsLAMBDA-2} displays a plot of $G_3$ as a function of $\Lambda$.  This data exhibits the limit-cycle behavior of the three-body coupling.  As the cutoff increases, $G_3$ becomes larger and larger, eventually diverging to infinity.  It then jumps to negative infinity and continues increasing.  The data for $\delta_0$ and $G_3$ are combined in Fig.~\ref{fig:G3andPSEUDOvsLAMBDA-2} to show that $\delta_0 = 0$ implies $G_3 = 0$, a claim also made in Sec.~\ref{sec:analytic}.  Figure \ref{fig:G3vsLAMBDA-13} uses two different sets of energies to show that the limit-cycle behavior is not dependent upon any specific bound-state values, but the positioning of the cycle is dependent upon the energies.

If the scattering length $a$ is large compared to the effective range, the zero angular momentum sector of the three-body problem with short-ranged interactions reduces to a one-dimensional problem with a potential of $1/r^2$ in the region $r_e \ll r \ll a$ \cite{Nielsen:review}.  The $1/r^2$ potential is studied in \cite{Beane:2001} where a short-range interaction is used to regulate the divergences from the long-range $1/r^2$ interaction.  While one might expect to find cutoff-dependent behavior similar to what we see, they instead find a monotonically increasing coupling with no divergences.  They suggest that the limit-cycle behavior is not a universal aspect of the renormalization group flow but rather an artifact of the regularization method used.

However, this cyclic behavior has been previously observed \cite{hammer:orig} using a sharp cutoff that simply discards all momenta above $\Lambda$.  Our method does not discard such momenta.  It merely suppresses them by a gaussian factor.  Yet the limit cycle still remains.  This evidence seems to indicate that the limit-cycle behavior is universal, and the issue deserves further investigation.

\begin{figure}
\begin{center}
\includegraphics[height=4.0in]{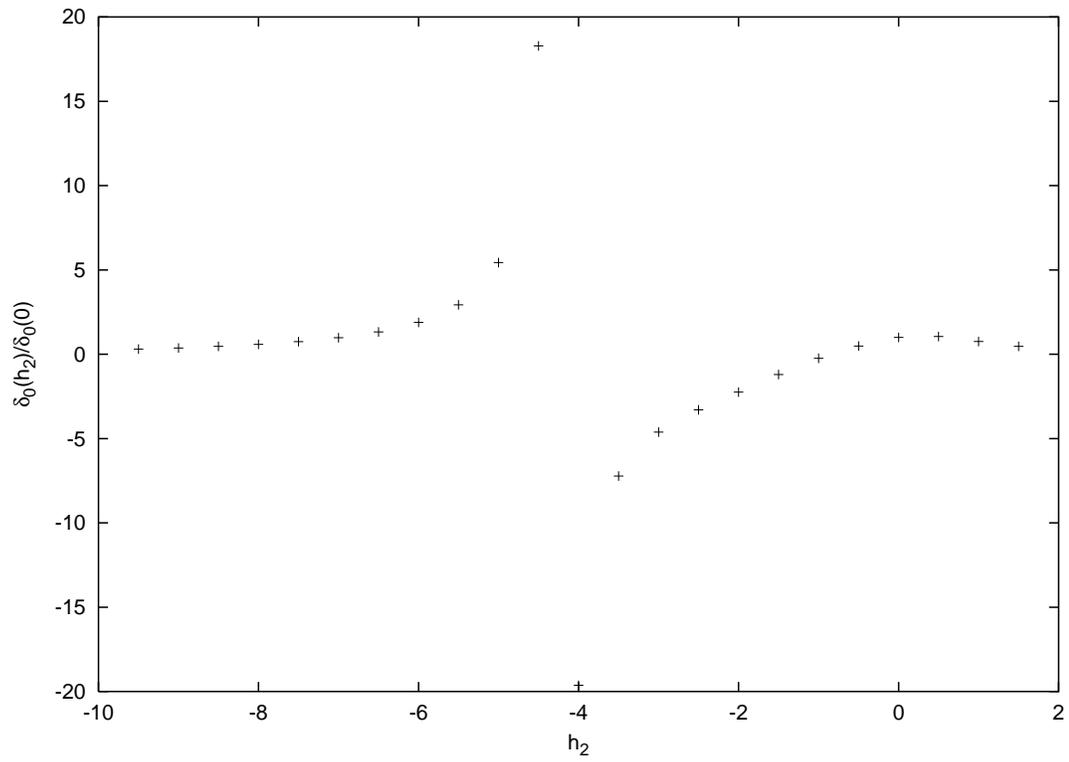}
\end{center}
\caption{\label{fig:PSEUDOvsH2}The ratio $\delta_0(h_2)/\delta_0(0)$ as a function of $h_2$ for $\theta = 1.0$ and $\Lambda = 10^5$.}
\end{figure}

\begin{figure}
\begin{center}
\includegraphics[height=4.0in]{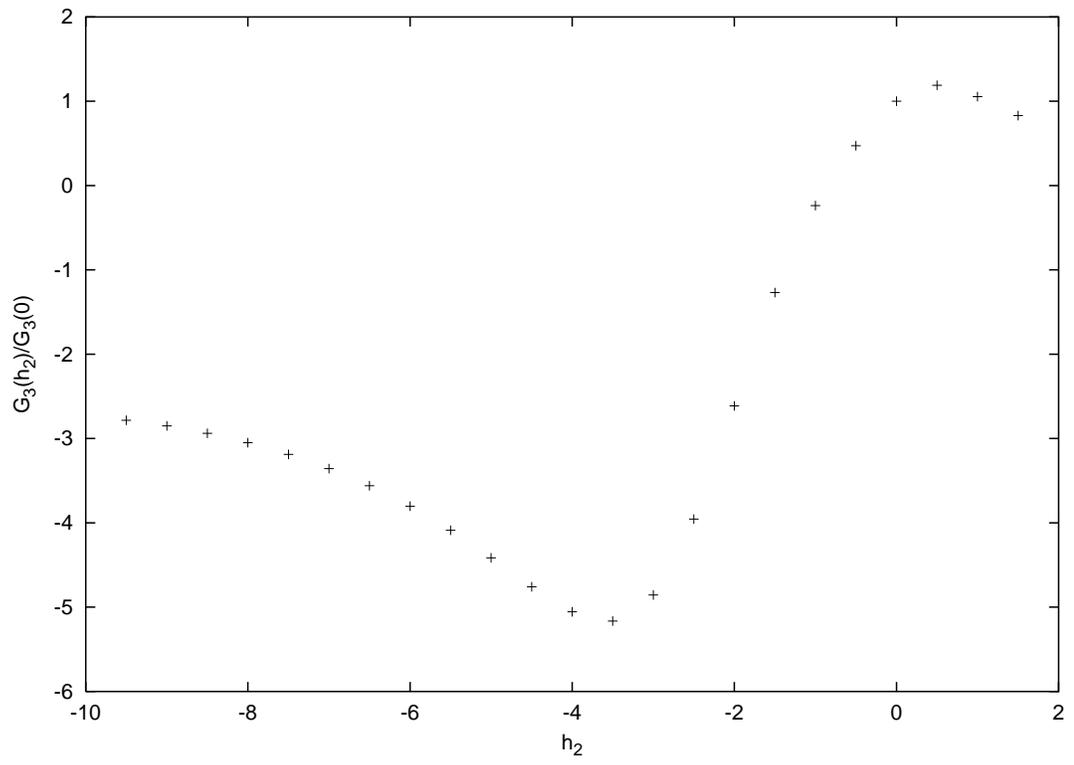}
\end{center}
\caption{\label{fig:G3vsH2}The ratio $G_3(h_2)/G_3(0)$ as a function of $h_2$ for $\theta = 1.0$ and $\Lambda = 10^5$.}
\end{figure}

 Another coupling constant we can study is $h_2$.  It is shown above that $h_2$ has no effect on the binding energies to leading order, but it may affect other quantities like $G_3$ and $\delta_0$.  In fact, Eq.~(\ref{eqn:PseudoCosine}) explicitly exhibits such dependence.  In Figs.~\ref{fig:PSEUDOvsH2} and \ref{fig:G3vsH2}, we plot the effect of changing $h_2$ upon $\delta_0$ and $G_3$ respectively.  The range of $h_2$ values is determined from Eq.~(\ref{eqn:ar2h}), reproduced here for convenience:
\begin{equation}
\frac{1}{4 \sqrt{2 \pi} \Lambda} h_2^2 + \left( \frac{2}{\sqrt{2 \pi} \Lambda} - \frac{2}{a \Lambda^2} \right) h_2 + \left( \frac{2}{a \Lambda^2} + \frac{1}{2} r_e - \frac{4}{\sqrt{2 \pi} \Lambda} \right)  =  0
\end{equation}
\noindent For very large cutoffs, the effective range as computed from $h_2$ takes the form
\begin{equation}
r_e = - \frac{1}{2 \sqrt{2 \pi} \Lambda} \left( h_2^2 + 8 h_2 - 16 \right) .
\end{equation}
\noindent For illustrative purposes, we restrict ourselves to the case $r_e > 0$ so that the values of $h_2$ are limited to the range $-4(1 + \sqrt{2}) \le h_2 \le -4(1 - \sqrt{2})$.  This example is chosen simply to provide a window of $h_2$ values to concentrate on.

\begin{figure}
\begin{center}
\includegraphics[height=4.0in]{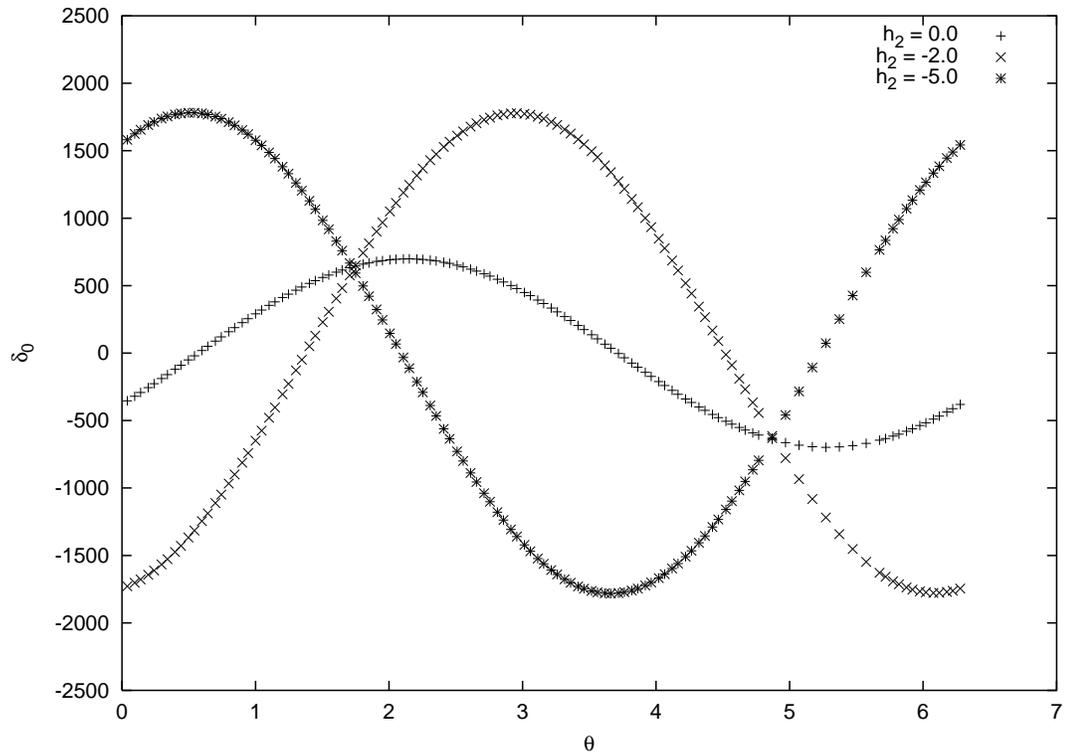}
\end{center}
\caption[The coupling $\delta_0$ as a function of $\theta$ for several values of $h_2$.]{\label{fig:PSEUDOvsTHETA-245}The coupling $\delta_0$ as a function of $\theta$ for several values of $h_2$.  The curves are all cosines with different amplitudes and phases.}
\end{figure}

The cosine behavior of $\delta_0(\theta)$ should remain unchanged if $h_2$ is changed, but the amplitude and phase of the curve may differ.  This is exactly what we find in Fig.~\ref{fig:PSEUDOvsTHETA-245} for the cases $h_2 = 0.0, -2.0,$ and $-5.0$.  It appears as though all curves intersect at the same two points.  However, we do not know if this can be explicitly proven from the leading order equations or what its significance may be.

\begin{figure}
\begin{center}
\includegraphics[height=4.0in]{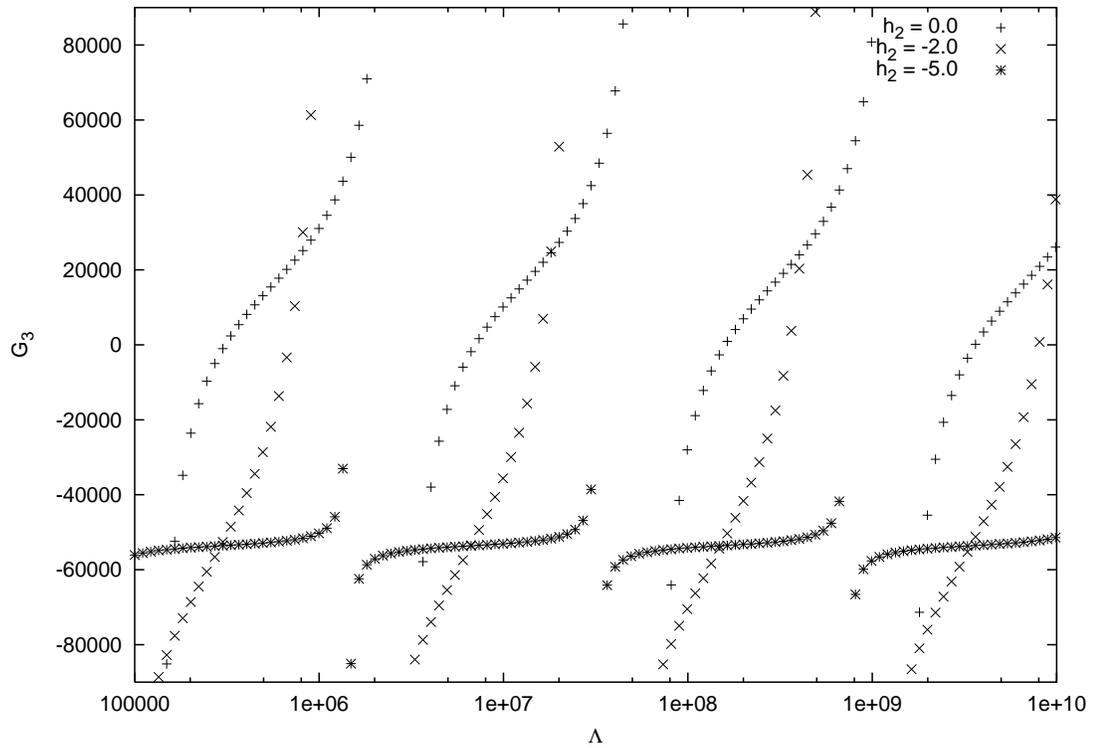}
\end{center}
\caption[$G_3$ as a function of $\Lambda$ for several values of $h_2$.]{\label{fig:G3vsLAMBDA-245}$G_3$ as a function of $\Lambda$ for several values of $h_2$. A change in $h_2$ results in a shift and/or flattening of the limit-cycle curve, but the general periodic form is the same.}
\end{figure}

The general limit-cycle behavior of $G_3$ also stays the same for $h_2 = 0.0, -2.0,$ and $-5.0$.  These curves are shown together in Fig.~\ref{fig:G3vsLAMBDA-245}.  Changing $h_2$ results in a shift and/or flattening of the limit cycles, but the periodic behavior is still apparent.

%%%%%%%%%%%%%%%
%%  FIGURES  %%
%%%%%%%%%%%%%%%

\begin{figure}
\begin{center}
\includegraphics[height=4.0in]{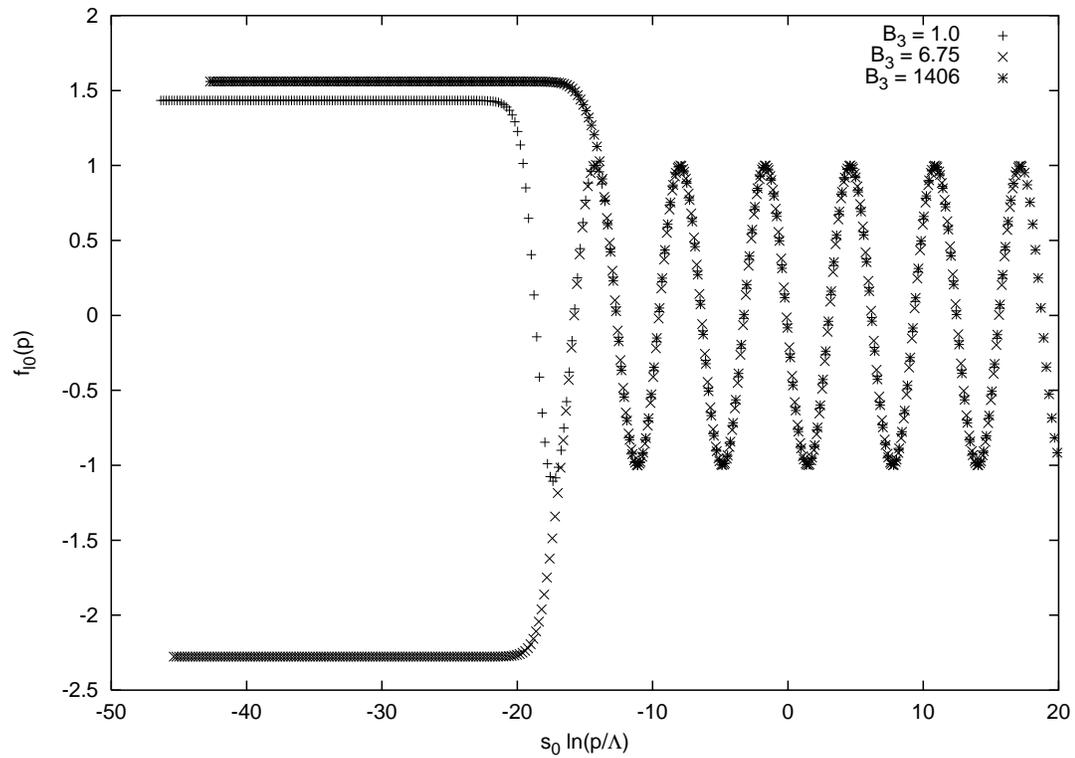}
\end{center}
\caption[Numerical solutions of $f_{l0}$ for several bound states using $B_2 = 1.0$ and $\Lambda = 10^8$.]{\label{fig:FL0-Spectrum2}Numerical solutions of $f_{l0}$ for several bound states using $B_2 = 1.0$ and $\Lambda = 10^8$.  The three-body bound-state energies are 1.0, 6.7502901502599, and 1406.1303932044.}
\end{figure}

\begin{figure}
\begin{center}
\includegraphics[height=4.0in]{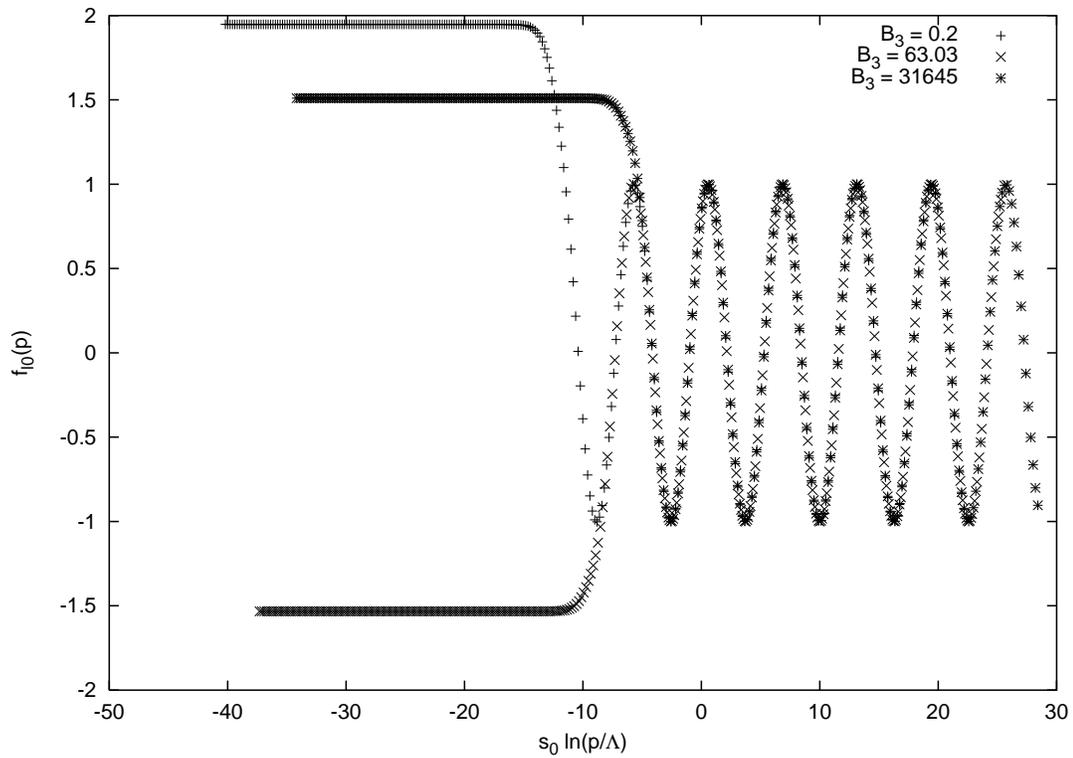}
\end{center}
\caption[Numerical solutions of $f_{l0}$ for several bound states using $B_2 = 0.01$ and $\Lambda = 10^5$.]{\label{fig:FL0-Spectrum1}Numerical solutions of $f_{l0}$ for several bound states using $B_2 = 0.01$ and $\Lambda = 10^5$.  The three-body bound-state energies are 0.2, 63.033762419242, and 31645.59444559555.}
\end{figure}

\clearpage

\begin{figure}
\begin{center}
\includegraphics[height=4.0in]{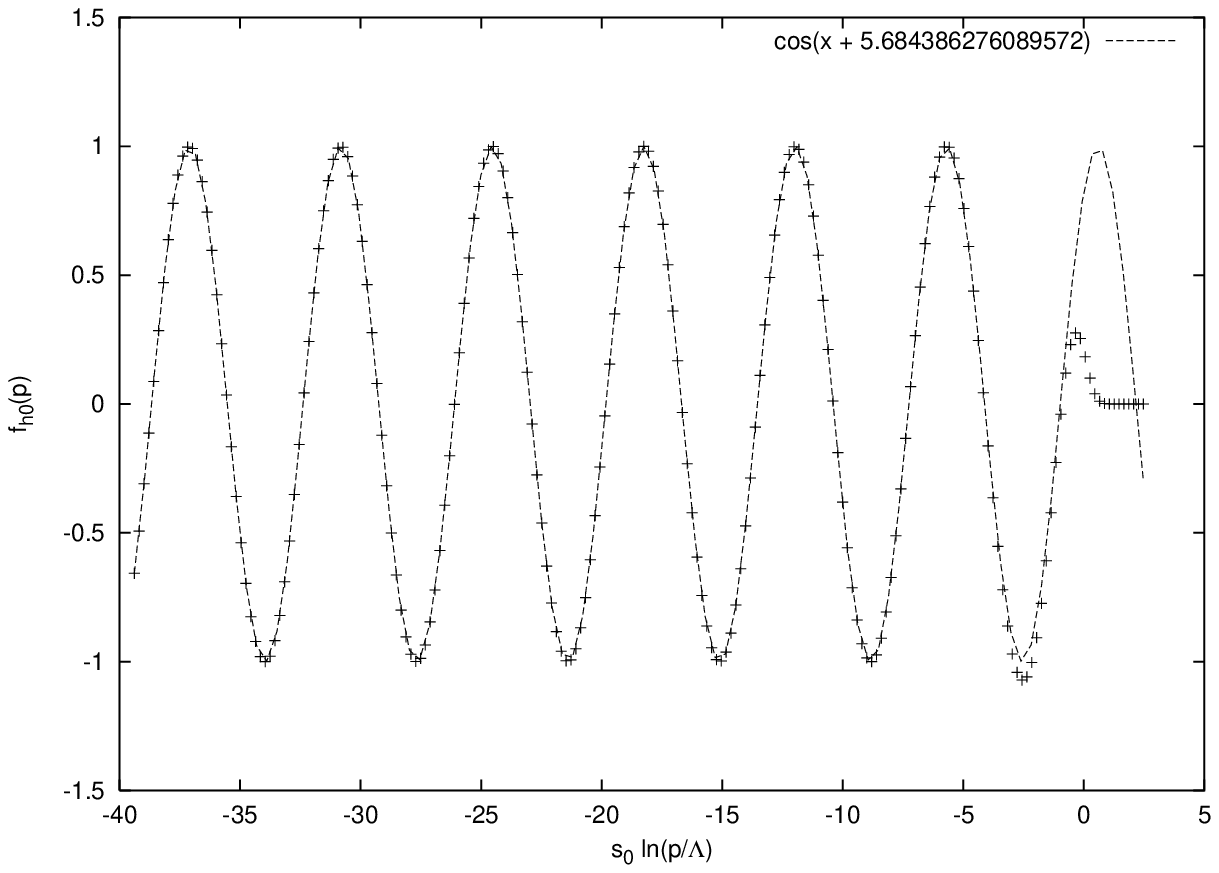}
\end{center}
\caption[Numerical solution of $f_{h0}(p)$ for the case $\theta = 5.684386276089572$ and $\Lambda = 10^8$.]{\label{fig:FH0-2}Numerical solution of $f_{h0}(p)$ for the case $\theta = 5.684386276089572$ and $\Lambda = 10^8$.  The dashed line is the best-fit cosine curve that matches the low-momentum behavior.}
\end{figure}

\clearpage

\clearpage

\chapter{Efimov's Function}
\label{ch:efimov}

In the 1970's, Vitaly Efimov investigated the behavior of three-body systems whose particles interacted via pair-wise ``resonant'' forces.   A resonant force is one that supports a shallow two-body bound state or a virtual state close to the two-body threshold, resulting in a scattering length whose absolute value is much larger than the range of the interaction.  This disparity in length scales allowed him to discover properties that were independent of the exact form of the two-body interaction.  One of the most useful results is the relation of the entire three-body spectrum to a single universal function.

This chapter begins with a brief introduction to Efimov's work.  We outline the arguments used to derive Efimov's universal function and exhibit the discrete scaling symmetry it possesses.  Some uses of his work in the fields of nuclear and atomic physics are discussed.  Since our three-body problem satisfies the constraints used by Efimov, we use the leading order integral equations to numerically compute Efimov's function to high accuracy.  We also analytically derive the relationship between Efimov's function and the function $\tilde{\theta}$ defined in Eq.~(\ref{eqn:theta_tilde}).

%%%%%%%%%%%%%%%%%%
%%  BACKGROUND  %%
%%%%%%%%%%%%%%%%%%
\section{Background}

Efimov's work hinges upon the use of a resonant two-body force.  Assuming a positive scattering length, this simply means that the potential associated with this force has a shallow two-body bound state.  Since $B_2 = \hbar/ma^2$ to leading order, this is the same as saying the potential has a large scattering length.  Such a situation may arise either naturally (e.g., the deuteron) or artificially (e.g., alkali atoms by fine-tuning an external magnetic field).  In any case, it will be assumed that the scattering length is much larger than the range of the interaction, which we call $r_0$.  Efimov also assumed that there are no deeply-bound states in the two-body spectrum.

We begin with the adiabatic hyperspherical representation of the three-body equation \cite{Nielsen:review,Federov:1993}.  The hyperspherical radius $R$ is defined by
\begin{equation}
R^2 = \frac{1}{3} \left( r_{12}^2 + r_{13}^2 + r_{23}^2 \right) ,
\end{equation}
\noindent where $r_{ij} = |\vec{r_i} - \vec{r_j}|$.  In the region $r_0 \ll R \ll a$, the three-body Schr\"odinger equation for a zero angular momentum state takes the form
\begin{equation}
- \frac{\hbar^2}{2 m} \left[ \frac{\partial^2}{\partial R^2} + \frac{s_0^2 + 1/4}{R^2} \right] f(R) = E \, f(R) \label{eqn:hamil}.
\end{equation}
\noindent Here, $f(R)$ is the radial wavefunction and $s_0 \approx 1.00624$ is the same constant seen in previous chapters.  The goal is to construct solutions for the bound state energies $E = -B_3$.

Since we are interested only in low-energy states ($|E| \sim 1/ma^2$), the requirement $R \ll a$ implies that the $1/R^2$ potential is much larger than $E$.  Therefore, we can ignore the right-hand side of Eq. (\ref{eqn:hamil}).  The resulting equation has a general solution of the form
\begin{equation}
f(R) = \sqrt{H R} \left[ A\,\me^{\mi s_0 \ln(H R)} + B\,\me^{-\mi s_0 \ln(H R)} \right] ,
\end{equation}
\noindent where $H$ has dimensions of momentum, and $A$ and $B$ are dimensionless.  For bound state solutions, these quantities must be functions of the bound-state energy $B_3$ and the two-body scattering length $a$.  Let us define
\begin{eqnarray}
H & = & \sqrt{\frac{m B_3}{\hbar^2} + \frac{1}{a^2}} \label{eqn:Hdef},
\\
\xi & = & - \arctan(a \sqrt{m B_3}/\hbar) \label{eqn:xidef}.
\end{eqnarray}
\noindent This definition of $\xi$ is valid only for positive scattering lengths.  For $a < 0$, we must use the definition $\xi = - \arctan(a \sqrt{m B_3}/\hbar) - \pi$.  In either case though, the dimensionless quantities $A$ and $B$ can depend upon $\xi$.

The radial wavefunction in the region $r_0 \ll R \ll a$ is a sum of incoming and outgoing hyperspherical waves.  At short distances ($R \sim r_0$) and long distances ($R \sim a$) the wavefunction is more complicated, and it is the boundary conditions at these points that determine the actual bound-state energies.

We first consider the constraints from unitarity in the short-distance region $R \sim r_0$.  Because there are no deeply-bound two-body states, all the probability associated with the incoming hyperspherical wave must be reflected into the outgoing wave.  This implies that $A$ and $B$ can only differ by a phase: $B = -A \, \me^{\mi \theta}$.  The wavefunction in the region $r_0 \ll R \ll a$ is now
\begin{equation}
f(R) \propto \sqrt{H R} \, \sin\left( s_0 \, \ln(H R) - \theta/2 \right),
\end{equation}
\noindent so we may express the phase as
\begin{equation}
- \frac{\theta}{2} = -s_0 \, \ln(HR) + \mathrm{arccot}\left[ \frac{1}{s_0}\left(R \frac{f'(R)}{f(R)} - \frac{1}{2}\right) \right] .
\end{equation}
\noindent  The phase is determined by specifying the derivative $R_0 f'(R_0)/f(R_0)$ at any point $R_0$ satisfying $r_0 \ll R_0 \ll a$.  We will simplify the form by writing it as
\begin{equation}
\theta = 2 s_0 \ln\left(H / \Lambda_*\right) .
\end{equation}
\noindent The quantity $\Lambda_*$ is a complicated function of $R_0$ and $R_0 f'(R_0)/f(R_0)$, the details of which are unimportant.  What is important is that the derivative at $R_0$ is ultimately determined by the short-distance behavior.  Essentially, $\Lambda_*$ parameterizes the effects of all short-distance interactions without knowledge of the detailed forms.

We now consider the constraints from unitarity in the long-distance region $R \sim a$.  Because we are considering three-body bound states, all of the probability associated with the outgoing hyperspherical wave must be reflected into the incoming wave.  This implies that $A$ and $B$ differ only by a phase: $B = -A \, \me^{\mi \Delta}$.  The phase $\Delta$ is determined by the long-range parameters $B_3$ and $a$.  We will write this phase as $\Delta(\xi)$ to explicitly show its dependence upon $B_3$ and $a$ through the variable $\xi$.

The two phases $\theta$ and $\Delta$ must match to within an additive multiple of $2 \pi$, so
\begin{equation}
2 s_0 \ln\left(\frac{H}{\Lambda_*}\right) = \Delta(\xi) + 2 \pi n ,
\end{equation}
\noindent where $n$ is an integer.  Using Eq.~(\ref{eqn:Hdef}), we rewrite the relation in the form
\begin{equation}
B_3 + \frac{\hbar^2}{m a^2} = \frac{\hbar^2 \Lambda_*^2}{m} \me^{2 \pi n/s_0} \exp[ \Delta(\xi) / s_0 ] \label{eqn:efunc}.
\end{equation}
\noindent The values for $n$ correspond to different 3-body bound states.  Once the function $\Delta(\xi)$ is known, we are able to calculate the entire spectrum by solving Eq.~(\ref{eqn:efunc}) for various $n$.  $\Delta(\xi)$ is what we have referred to as Efimov's universal function.

\begin{figure}
\begin{center}
\setlength{\unitlength}{1.0in}
\begin{picture}(4,4)
\thicklines
\put(0.5,4.0){\vector(0,-1){4.0}}
\put(0.0,3.5){\vector(1,0){4.0}}
\put(2.15,1.0){\circle*{0.1}}
\put(0.5,3.5){\vector(2,-3){1.6}}
\qbezier(1.0,3.5)(1.0,3.22)(0.73,3.2)
\put(3.5,3.3){$a^{-1}$}
\put(0.7,0.2){$-\sqrt{m B_3/\hbar^2}$}
\put(1.8,1.8){$H = \sqrt{m B_3/\hbar^2 + a^{-2}}$}
\put(1.1,3.1){$\xi$}
\put(0.0,3.7){$(0,0)$}
\end{picture}
\end{center}
\caption[Relation between bound-state energy and the variables $H$ and $\xi$.]{\label{fig:HvsXi}Relation between bound-state energy and the variables $H$ and $\xi$.}
\end{figure}
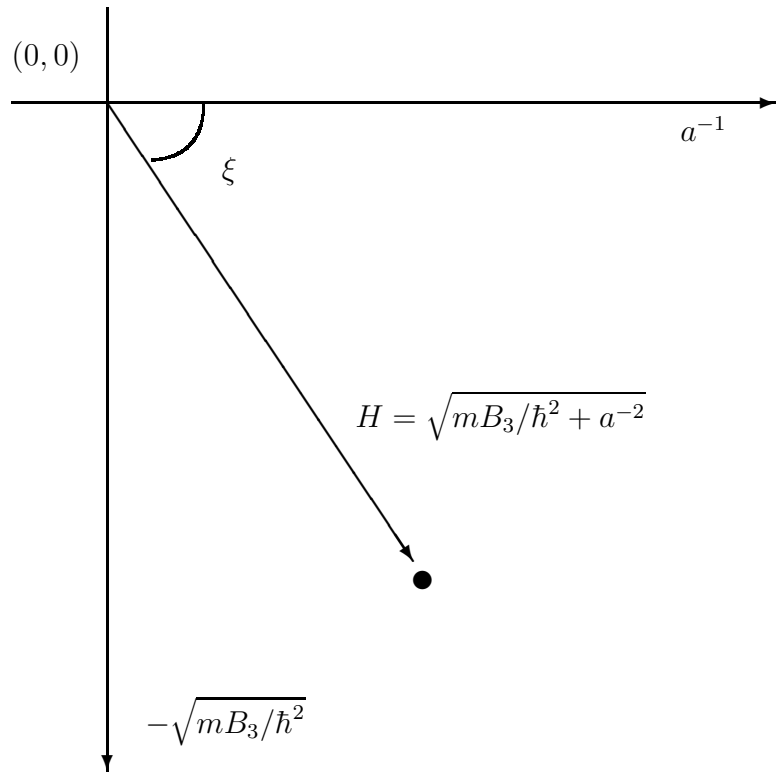

The variables $H$ and $\xi$ in our equations can be treated like a ``radius'' and an ``angle'' respectively.  Suppose we plot a 3-body bound-state energy on a graph, choosing the x-axis to be $1/a$ and the y-axis to be $- \sqrt{m B_3/\hbar^2}$ (See Fig.~\ref{fig:HvsXi}).  Then $H$ is the distance of a line from that bound-state point to the origin, and $\xi$ is the angle that line makes with the x-axis.  Figure \ref{fig:EfimovStates} shows two Efimov states as a function of the 2-body binding energy which is proportional to $1/ma^2$ to leading order.  In this figure, we have chosen to label the axes differently so that the behavior of both states can be better seen.

\begin{figure}
\begin{center}
\includegraphics[height=4.0in]{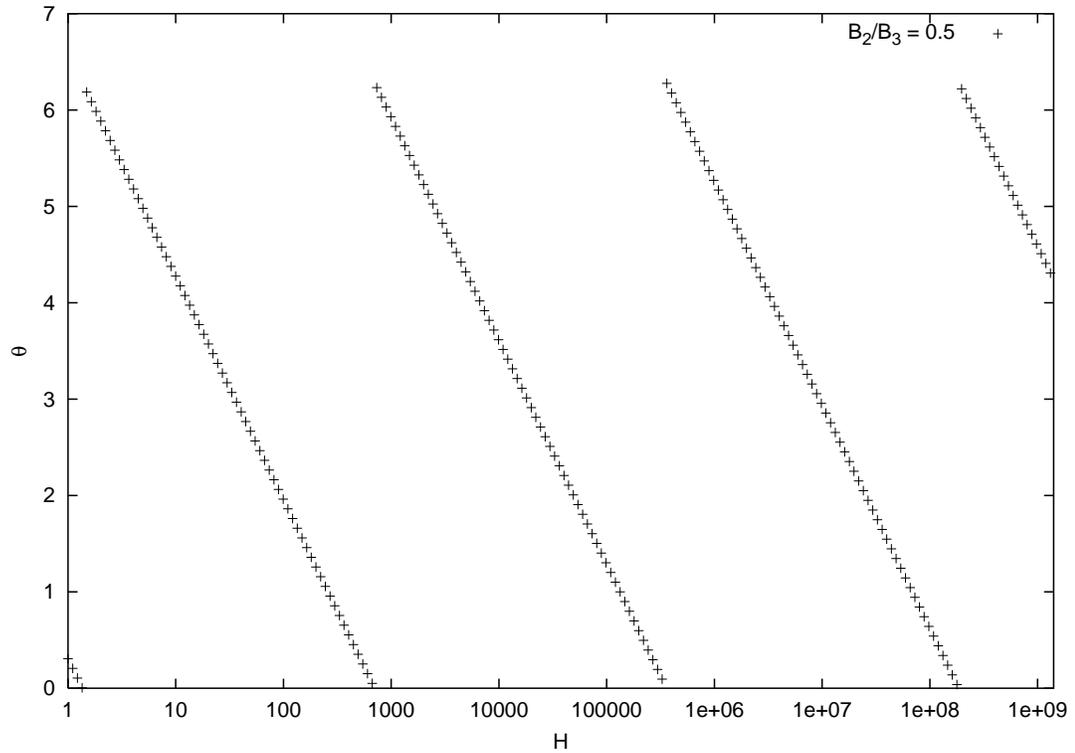}
\end{center}
\caption[Phase $\theta$ for three-body bound states as a function of $H$ with $B_2/B_3$ = 0.5.]{\label{fig:H-1}Phase $\theta$ for three-body bound states as a function of $H$ with $B_2/B_3$ = 0.5.  The periodic behavior as $H$ increases is a direct result of the discrete scaling symmetry in Efimov's equation.  For a constant ratio $B_2/B_3$, the phase is linear in $\ln(H)$.}
\end{figure}

Using these parameters, we can demonstrate that the solutions to Efimov's equation have a discrete scaling symmetry.  If there exists a bound state with binding energy $B_3$ for the parameters $a$ and $\Lambda_*$, then there will also be a bound state with binding energy $\lambda^2 B_3$ for the parameters $\lambda^{-1} a$ and $\Lambda_*$ provided that $\lambda$ is of the form $\exp[n' \pi / s_0]$ for some integer $n'$.  This is equivalent to saying that for any given value of $\xi$, the values of $H$ for any successive bound states differ by a multiplicative factor of $\exp(\pi/s_0) \simeq 22.7$.  Since we are considering only the case $a > 0$ with three-body binding energies that satisfy $B_3 > B_2$, the variable $\xi$ is restricted to the range $-\pi/2 \le \xi \le -\pi/4$.  This periodic behavior can be easily seen by examining the value of the phase in $f_{d0}$ along a path of constant $\xi$.  Figure \ref{fig:H-1} shows the value of $\theta$ for bound states with $B_2/B_3 = 0.5$ as $H$ increases.  Successive Efimov states must have the same phase which is periodic in $\ln(H)$.

It should be noted that Eq.~(\ref{eqn:efunc}) is an approximation based upon the limit $r_0/a = 0$, and as such is exact only for zero-range theories.  Any calculated energies can be expected to have errors of $\mathcal{O}(r_0/a)$.  Some first order corrections due to a non-zero effective range have been calculated by Efimov \cite{efimov:3,efimov:7}.

%%%%%%%%%%%%%%%%%%%%
%%  APPLICATIONS  %%
%%%%%%%%%%%%%%%%%%%%
\subsection{Applications}

Efimov's general framework may be applied whenever there is a resonant two-body interaction.  Several instances where it is applicable:

\begin{itemize}
\item The two-nucleon system has a shallow bound state, the deuteron, in the spin-triplet channel and a large negative scattering length in the spin-singlet channel.  This led Efimov to suggest that the few-nucleon system could be described using zero-range potentials with the effective range treated as a perturbation.  The three-nucleon system includes two bound states: the triton (a $pnn$ bound state) and ${}^3\mathrm{He}$ (a $ppn$ bound state).  The equations are more complicated since the nucleons possess spin and isospin, and the $ppn$ state is further complicated by the Coulomb interaction between the two protons.   A leading order analysis with the Coulomb interaction neglected has been carried out by Efimov \cite{efimov:2,efimov:5} and revisited in the EFT framework by Bedaque, Hammer, and van Kolck \cite{hammer:orig}.  An analysis at NLO in the effective range was carried out by Efimov \cite{efimov:3} and repeated by Hammer and Mehen using EFT \cite{Mehen:r0}.

\item ${}^4\mathrm{He}$ atoms have a large two-body scattering length and a shallow two-body bound state.  The ground state of the ${}^4\mathrm{He}$ trimer has been observed, but its binding energy has not been measured.  The Schr\"odinger equation for the three-body bound states has been solved accurately for potential models of the interaction between ${}^4\mathrm{He}$ atoms.  In addition to the ground state trimer, there is an excited state this is shallower by a factor of about 50 to 70.  It has been found that Efimov's function can be used to predict the binding energy of one of the two 3-body bound states using the other as input \cite{hammer:orig,Braaten:HeAtoms,Frederico:HeAtoms}.

\item For alkali atoms, the atom-atom scattering length can be made large by tuning and external magnetic field to a Feshbach resonance \cite{Verhaar:1,Verhaar:2}.  One complication in this case is that the alkali atoms have many deeply-bound two-body states, so an Efimov state can decay into an atom and a deep two-body bound state.  As a consequence, the Efimov states are resonances with a binding energy $B_3$ and a width $\Gamma_3$.  The complex energies $-(B_3 + \mi \Gamma_3)$ still satisfy Efimov's equation, but it requires the analytic continuation of the function $\Delta(\xi)$ to complex values of the angle $\xi$ \cite{Braaten:uefes}.

\end{itemize}

%%%%%%%%%%%%%%%%%%%
%%  COMPUTATION  %%
%%%%%%%%%%%%%%%%%%%
\section{Computation of Efimov's Function}

Although Efimov explicitly considered only resonant two-body interactions, his conclusions result from applying a boundary condition on the three-body wavefunction at short distances.  This encompasses any short-range interactions, including the three-body contact interaction used in our work.  The $g_3$ potential in our equations only acts when all three particles are very close together.  Using our previous notation, this would correspond to the region $R \sim r_0$, which is inside the radius where the boundary condition matching occurs.  Its influence is combined with that of the two-body interaction, and the total effect is seen only through the parameter $\Lambda_*$.

    %%%%%%%%%%%%%%
    %%  B2 = 0  %%
    %%%%%%%%%%%%%%
\subsection{$B_2 = 0$}
As a verification of this argument, let us consider the case $B_2 = 0$.  According to Eq.~(\ref{eqn:efunc}), the bound-state energies are given by
\begin{equation}
B_3 = \frac{\hbar^2 \Lambda_*^2}{m} \me^{2 \pi n/s_0} \exp[ \Delta(-\pi/2) / s_0 ] ,
\end{equation}
\noindent showing that the ratio of energies for adjacent states is $\exp(2 \pi/s_0)$.  To 13 digits, this ratio is 515.0350013848.  In Table \ref{tab:EfimovSpacing}, we have computed several of the energies using $B_3 = 1.0$ as a reference point.  The table also shows the ratios between adjacent states, all of which equal the predicted value to 11 digits.

\begin{table}
\begin{center}
\begin{tabular}{|c|l|l|}
\hline
Level & Energy & Ratio \\
\hline
\hline
$+3$  & 1.366187266197138e+08 & 515.0350013845557 \\
$+2$  & 2.652610526516549e+05 & 515.0350013850982 \\
$+1$  & 5.150350013849171e+02 & 515.0350013849171 \\
$\hspace{8pt}0$  & 1.000000000000000     & 515.0350013849775 \\
$-1$ & 1.941615613134847e-03 & 515.0350013851568 \\
$-2$ & 3.769871189167697e-06 & 515.0350013846162 \\
$-3$ & 7.319640760400369e-09 & - \\
\hline
\end{tabular}
\end{center}
\caption[Efimov state energies for $B_2 = 0$.]{\label{tab:EfimovSpacing}Efimov state energies for $B_2 = 0$.  The value 1.0 is fixed as a starting point.  Notice that the ratio of energies for adjacent states equals $\exp(2\pi/s_0)$ to 11 digits.}
\end{table}

We can prove that the ratio of adjacent binding energies is $\exp(2 \pi/s_0)$ using Eq.~(\ref{eqn:theta_tilde}).  Two adjacent bound states, $B_3$ and $B_3'$, must have phases that differ by $\pi$.  Therefore,
\begin{eqnarray}
&& s_0 \ln\left( \Lambda/\sqrt{B_3}\right) = s_0 \ln\left( \Lambda/\sqrt{B_3'}\right) + \pi \nonumber
\\
& \Longrightarrow & s_0 \ln\left( \sqrt{B_3'}/\sqrt{B_3}\right) = \pi \nonumber
\\
& \Longrightarrow & \frac{B_3'}{B_3} = \me^{2 \pi/s_0} .
\end{eqnarray}

    %%%%%%%%%%%%%%%%%%%%%%%%%%
    %%  Phase to Delta(xi)  %%
    %%%%%%%%%%%%%%%%%%%%%%%%%%
\subsection{From $\theta$ to $\Delta$}

Before we can compute Efimov's function, we must find a way to relate the parameter $\Lambda_*$ to the parameters in our equations.  We will do this by considering the behavior of the phase $\theta$ from the perspective of both the high- and low-momentum equations.

First consider the high-momentum perspective.  We have seen in Sec.~\ref{sec:analytic} that it is possible to view $\theta$ as a function of $\delta_0$, but it can just as easily be viewed as a function of $G_3$.  For a given value of $G_3$, we write the phase generated by this coupling as $\theta_h(G_3)$.

From the low-momentum perspective, the phase is determined only by the parameters $B_2$, $B_3$, and $\Lambda$.  We also know explicitly the cutoff dependence of this phase:
\begin{equation}
\theta_l(B_2, B_3, \Lambda) = s_0 \ln\left(\Lambda/\sqrt{B_3}\right) + \tilde{\theta}(B_2/B_3).
\end{equation}
\noindent Because these phases must match to within a multiple of $\pi$,
\begin{equation}
\theta_h(G_3) = \theta_l(B_2, B_3, \Lambda) + n \pi = s_0 \ln\left(\Lambda/\sqrt{B_3}\right) + \tilde{\theta}(B_2/B_3) + n \pi \label{eqn:ThetaHL}.
\end{equation}
\noindent The purpose of $G_3$ is to make some value of $B_3$ cutoff-independent.  As the cutoff is changed, $G_3$ is adjusted to ensure that $B_3$ remains the same.  Removing $\Lambda$ from Eq.~(\ref{eqn:ThetaHL}) can only be done if $\theta_h(G_3)$ contains $\Lambda$ dependence of the form $s_0 \ln(\Lambda)$ implicitly through $G_3$.  We use this fact to define a new parameter $\Lambda_*$ via the equation
\begin{equation}
\theta_h(G_3) = s_0 \ln\left(\Lambda/\Lambda_*\right) .
\end{equation}
\noindent Eq.~(\ref{eqn:ThetaHL}) can then be written as
\begin{equation}
s_0 \ln\left(\Lambda/\Lambda_*\right) = s_0 \ln\left(\Lambda/\sqrt{B_3}\right) + \tilde{\theta}(B_2/B_3) + n \pi ,
\end{equation}
\noindent which implies
\begin{equation}
B_3 = \Lambda_*^2 \me^{2 \pi n/s_0} \me^{2 \tilde{\theta}/s_0} \label{eqn:B3Delta}.
\end{equation}
\noindent The ratio $B_2/B_3$ can be written in terms of $\xi$:
\begin{equation}
B_2/B_3 = 1/\tan^2(-\xi) \label{eqn:ratioxi}.
\end{equation}
\noindent Since $\tilde{\theta}$ is a function of $B_2/B_3$, it is also a function of $\xi$, and from now on we will write $\tilde{\theta}(\xi)$ to emphasize this.  Multiplying both sides of Eq.~(\ref{eqn:B3Delta}) by $(1 + B_2/B_3)$ yields
\begin{equation}
B_3 + B_2 = \Lambda_*^2 \me^{2 \pi n/s_0} \exp\left[ \tilde{\theta}(\xi) - 2\ln(\sin(-\xi))\right] \label{eqn:myefunc}.
\end{equation}
\noindent We see that we can match Efimov's relation if we choose
\begin{eqnarray}
\Lambda_* & = & \Lambda \me^{-\theta/s_0} \label{eqn:LamStarDef},
\\
\Delta(\xi) & = & \tilde{\theta}(\xi) - 2\ln(\sin(-\xi)) \label{eqn:DeltaDef}.
\end{eqnarray}
\noindent We should note that this choice for $\Lambda_*$ is proportional to the quantity $\Lambda_*$ defined in Ref.~\cite{hammer:orig} and will result in a function $\Delta(\xi)$ that differs from the one found in \cite{Braaten:uefes} by an inconsequential additive constant.

    %%%%%%%%%%%%%%%%%%%%%%%%
    %%  NUMERICAL VALUES  %%
    %%%%%%%%%%%%%%%%%%%%%%%%
\subsection{Numerical Values}

\begin{figure}
\begin{center}
\includegraphics[height=4.0in]{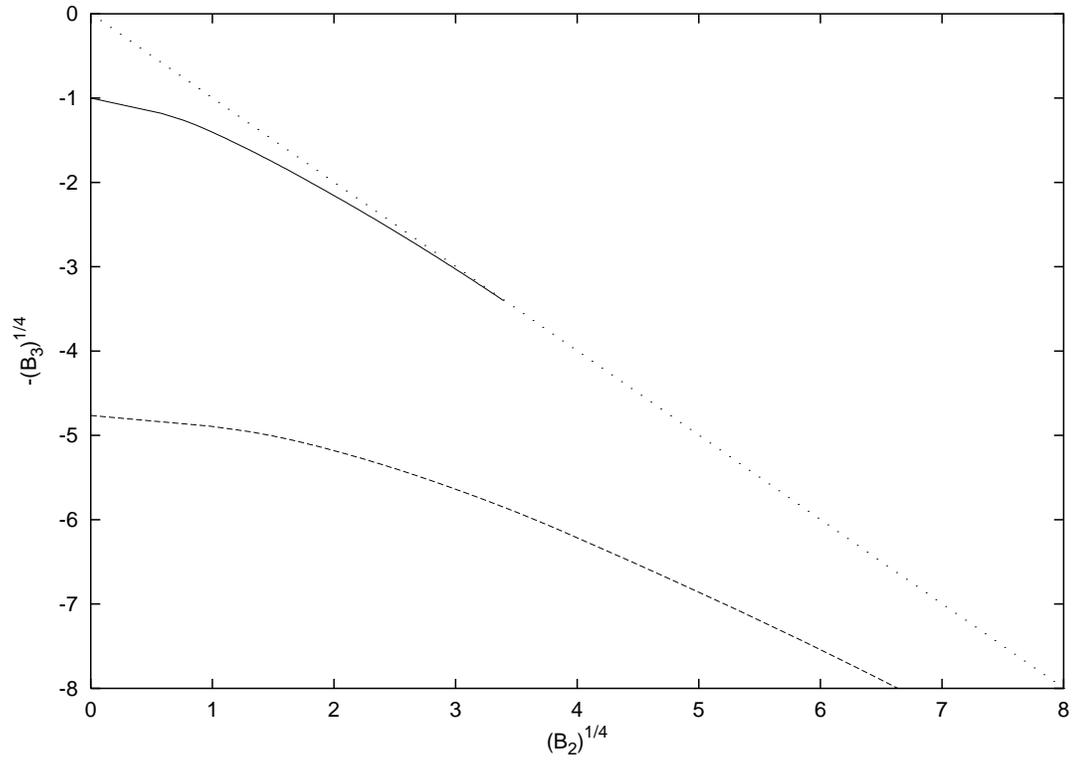}
\end{center}
\caption[Binding energies of Efimov states as functions of the two-body binding energy.]{\label{fig:EfimovStates}Binding energies of Efimov states as functions of the two-body binding energy.  The dotted line is the scattering threshold.}
\end{figure}

\begin{figure}
\begin{center}
\includegraphics[height=4.0in]{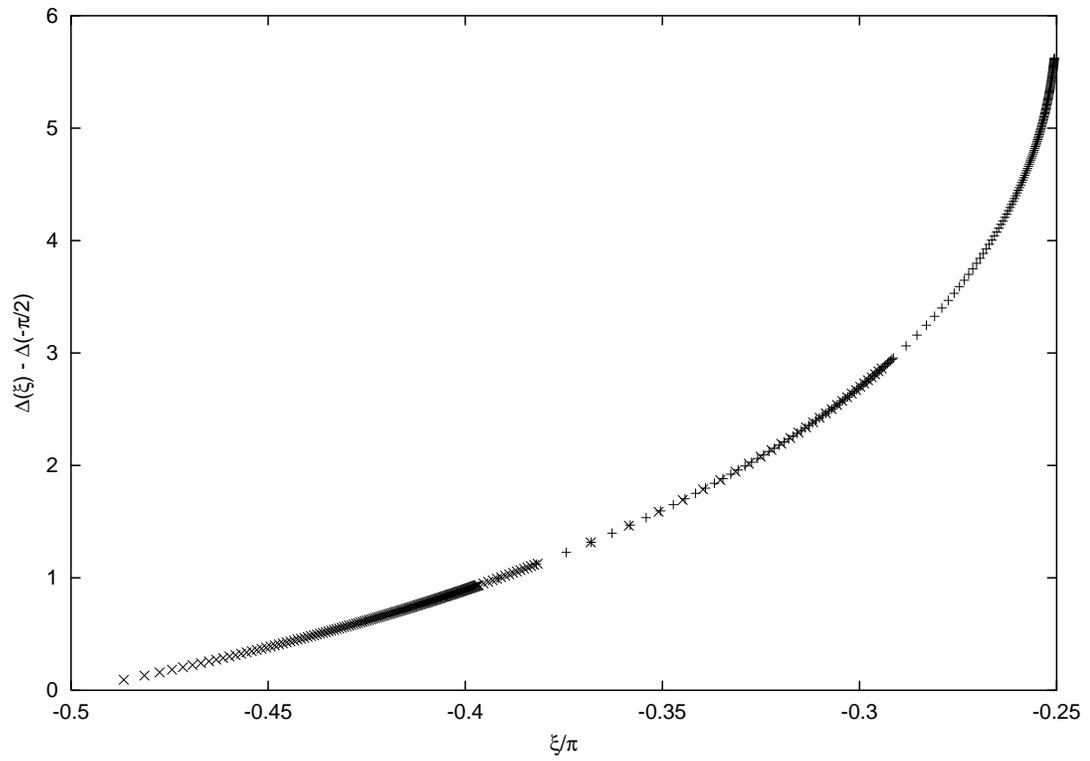}
\end{center}
\caption[The Efimov function $\Delta(\xi)$ as a function of $\xi$ computed using three-body bound states with constant phase.]{\label{fig:ConstPhaseDelta}The Efimov function $\Delta(\xi)$ as a function of $\xi$ computed using three-body bound states with constant phase. The difference $\Delta(\xi) - \Delta(-\pi/2)$ is plotted since a constant shift in the function is inconsequential.}
\end{figure}

Now that we have relations (\ref{eqn:LamStarDef}) and (\ref{eqn:DeltaDef}), we can numerically compute values for $\Delta(\xi)$.  One way to do this is to follow a state with constant $\Lambda_*$ as $B_2$ is changed.  For a fixed cutoff, constant $\Lambda_*$ implies a constant $\theta$.  The two Efimov states in Fig.~\ref{fig:EfimovStates} are for constant phase, so they may be used to calculate $\Delta(\xi)$.  The results are shown in Fig.~\ref{fig:ConstPhaseDelta}, where we have plotted $\Delta(\xi) - \Delta(-\pi/2)$ for both states.  The constant shift in the function does not matter.

\begin{figure}
\begin{center}
\includegraphics[height=4.0in]{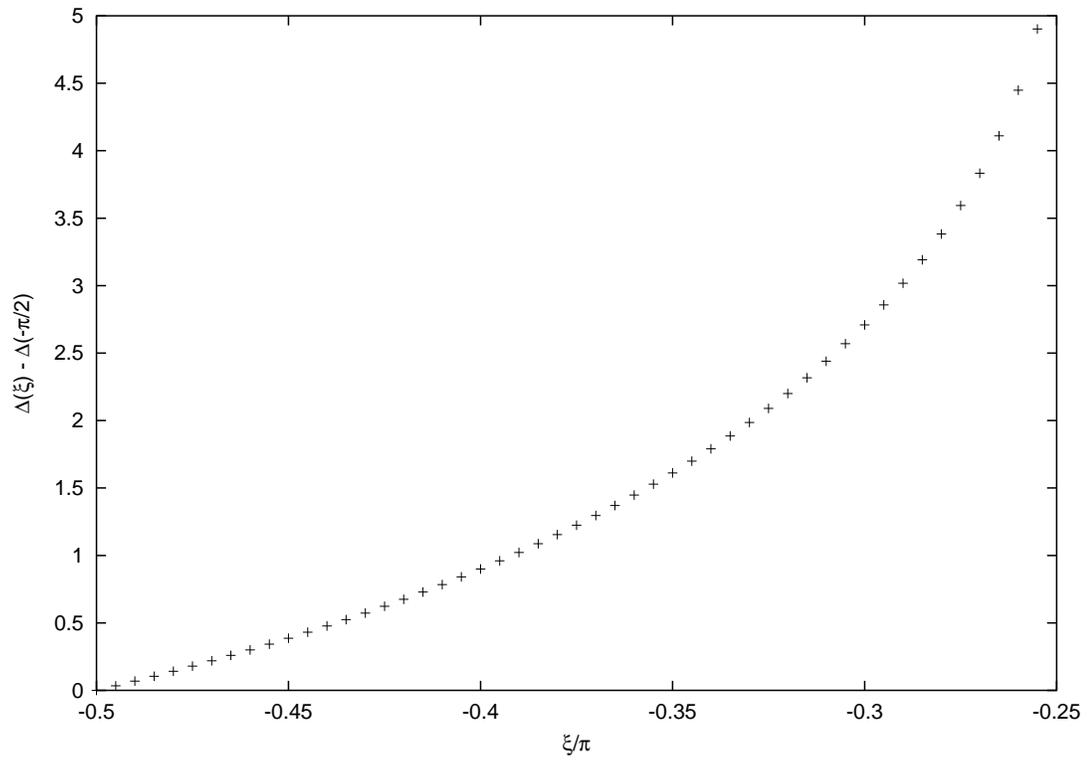}
\end{center}
\caption{\label{fig:ConstEnergyDelta}The Efimov function $\Delta(\xi)$ as a function of $\xi$ computed using a three-body bound state with constant binding energy $B_3 = 1$.}
\end{figure}

A second alternative is to keep $B_3$ constant and follow any changes in the phase as $B_2$ changes.  The function computed from this approach is shown in Fig.~\ref{fig:ConstEnergyDelta} for the case of $B_3 = 1.0$.  This method makes it easier to compute $\Delta(\xi)$ for any given $\xi$ since we do not need to search for the value of the three-body binding energy.

\begin{figure}
\begin{center}
\includegraphics[height=4.0in]{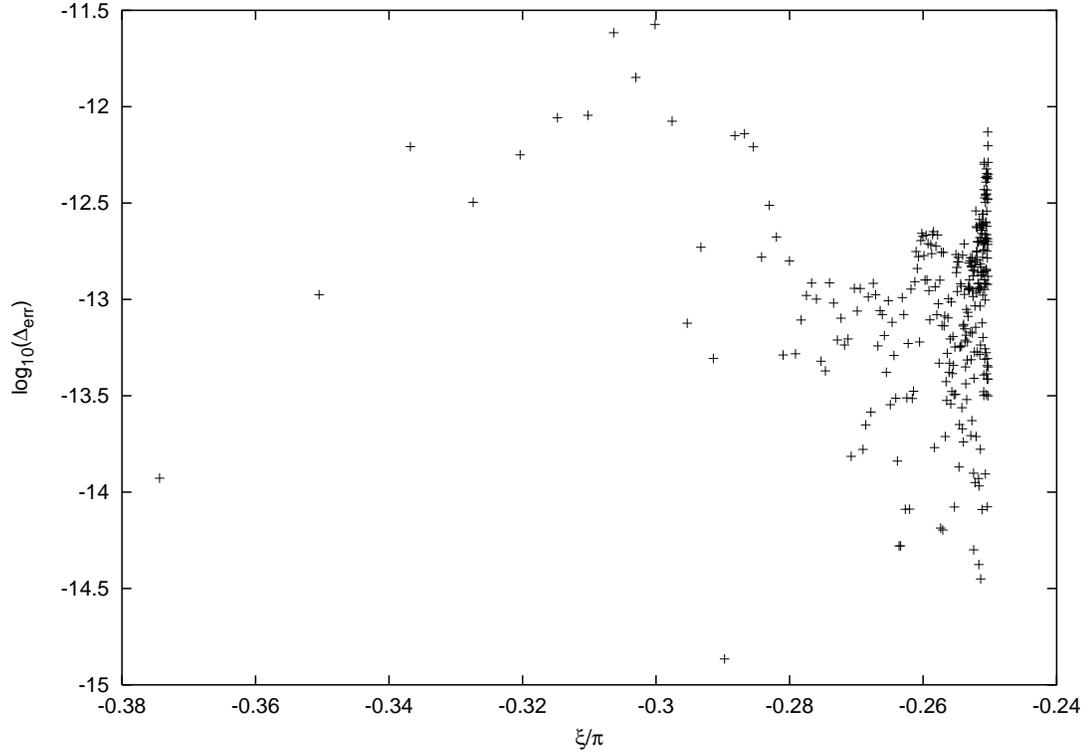}
\end{center}
\caption[Difference $\Delta_{err}$ between the Efimov functions calculated using constant phase and constant energy as a function of $\xi$.]{\label{fig:DeltaDiff-1}Difference $\Delta_{err}$ between the Efimov functions calculated using constant phase and constant energy as a function of $\xi$.  The constant phase Efimov state with $B_3 = 1.0$ at $B_2 = 0.0$ is compared to the constant energy state $B_3 = 1.0$.  The difference in the values suggests that our calculation is accurate to almost 12 digits.}
\end{figure}

Since the energy and phase values are accurate to about 12 digits, we expect similar accuracy in our calculation of $\Delta(\xi)$.  Figure \ref{fig:DeltaDiff-1} shows the difference between the Efimov function values generated from the Efimov state in Fig.~\ref{fig:EfimovStates} ($B_2 = 0.0, B_3 = 1.0$) and the constant energy $B_3 = 1.0$ state.  The errors support the statement that our calculation of $\Delta(\xi)$ is accurate to almost 12 digits, which is much higher than previous calculations \cite{Braaten:uefes}.  Table~\ref{tab:DeltaCompare} compares those previous values, labelled $\Delta(\xi)_{\mathrm{BHK}}$, to the ones obtained here.  The definition of these functions differ by an additive constant: $\Delta(\xi) = \Delta(\xi)_{\mathrm{BHK}} + C$.  We have fixed the constant $C$ by demanding that $\Delta(\xi) = \Delta(\xi)_{\mathrm{BHK}}$ when $\xi = -1.502$.  The discrepancies are less than 0.02 with the exception of the final point at $\xi = -0.787$ where the discrepancy increases to about 0.06.

%%%%%%%%%%%%%%%%%%%%%%%%
%%  NUMBER OF STATES  %%
%%%%%%%%%%%%%%%%%%%%%%%%
\section{Number of Efimov States}

For any given value of $B_2$, there will be a certain number of Efimov states that lie within the range of validity for this approximation.  Efimov's estimate for the number of states $N$ is related to the two-body scattering length $a$ and the range of the interaction $r_0$ by the simple formula
\begin{equation}
N = \frac{s_0}{\pi} \, \ln\left( a/r_0 \right) .
\end{equation}
\noindent This is easily derived from Eq.~(\ref{eqn:myefunc}).  Assume a non-zero value for $B_2$, which is related to the scattering length by $B_2 \propto a^{-2}$.  The deepest bound-state energy $B_3$ can only be of order $\Lambda^2$ before the conditions of our approximation are violated.  This state will correspond to $n = N$, and we will write the energy as $B_3^{(N)} \sim \Lambda^2$.  The shallowest bound-state energy, corresponding to $n = 0$,  must be greater than or equal to $B_2$ and should be of the same order.  We write this as $B_3^{(0)} \sim B_2$.  Therefore, we have the relations
\begin{eqnarray}
B_3^{(N)} + B_2 & = & \Lambda_*^2 \me^{2 \pi N/s_0} \exp(\Delta(\xi_N)/s_0) \label{eqn:Nstate1},
\\
B_3^{(0)} + B_2 & = & \Lambda_*^2 \, \exp(\Delta(\xi_0)/s_0) \label{eqn:0state1},
\end{eqnarray}
\noindent where $\xi_N = -\arctan\left(\sqrt{B_3^{(N)}/B_2}\right)$ and $\xi_0 = -\arctan\left(\sqrt{B_3^{(0)}/B_2}\right)$.  Since these states are part of the same spectrum, the value of $\Lambda_*$ in both equations must be identical.

Because we are assuming that $B_3^{(0)} \sim B_2$, we shall simplify Eq.~(\ref{eqn:0state1}) by making the substitution $B_3^{(0)} + B_2 \simeq 2 B_2$.  Upon taking the ratio of (\ref{eqn:Nstate1}) and (\ref{eqn:0state1}), we obtain
\begin{equation}
\frac{B_3^{(N)}}{2 B_2} \simeq \me^{2 \pi N/s_0} \exp\left[\left(\Delta(\xi_N) - \Delta(\xi_0)\right)/s_0\right] .
\end{equation}
\noindent The exponential will be $\mathcal{O}(1)$, so we may write
\begin{equation}
N \simeq \frac{s_0}{2 \pi} \ln\left(\frac{B_3^{(N)}}{2 B_2}\right) .
\end{equation}
\noindent The $2 B_2$ term in the logarithm can be approximated by $a^{-2}$.  The cutoff $\Lambda$ is related to the interaction range by $\Lambda \sim r_0^{-1}$, which allows us to estimate $B_3^{(N)} \sim r_0^{-2}$.  This makes our estimate of the number of states
\begin{equation}
N \simeq \frac{s_0}{2 \pi} \ln\left(\frac{a^2}{r_0^2}\right) = \frac{s_0}{\pi} \ln\left(a/r_0\right) 
\end{equation}
\noindent which matches the estimate given by Efimov.

The source of this estimate can also be seen graphically.  Recall Figs.~\ref{fig:FL0-Spectrum2} and \ref{fig:FL0-Spectrum1} where it is shown that successive bound states match onto the cosine behavior at successive peaks.  The x-axis in these figures is labelled by $x = s_0 \ln(p/\Lambda)$ so that the cosine behavior of $f_{l0}$ as it approaches $f_{d0}$ can easily be seen.  Each bound state joins the cosine curve at roughly $x = s_0 \ln(\eta_3/\Lambda)$.  If $\eta_3^{(0)}$ represents the shallowest state, then it joins the peak at $ x_0 \simeq s_0 \ln(\eta_3^{(0)}/\Lambda)$.  From Fig.~\ref{fig:FH0-2}, we see that the function $f_{h0}$ decays exponentially when $p \sim \Lambda$.  The last cosine peak is around $x_N \simeq s_0 \ln(\Lambda/\Lambda) = 0$.  The number of peaks between $x_0$ and $x_N$ then gives the number of Efimov states that can exist:
\begin{equation}
N = \frac{1}{\pi} (x_N - x_0) = - \frac{s_0}{\pi} \ln(\eta_3^{(0)}/\Lambda) .
\end{equation}
\noindent Using $\Lambda \sim r_0^{-1}$ and $\eta_3^{(0)} \sim a^{-1}$ leads to the the same estimate $N \simeq (s_0/\pi)\ln(a/r_0)$.

%%%%%%%%%%%%%%%
%%  FIGURES  %%
%%%%%%%%%%%%%%%
\begin{table}
\begin{center}
\begin{tabular}{|c|c|c|}
\hline
$\xi$ & $\Delta(\xi) + C$ & $\Delta(\xi)_{\mathrm{BHK}}$ \\
\hline
\hline
-0.787 & -2.59833945225 & -2.539 \\
-0.791 & -2.91699158253 & -2.897 \\
-0.797 & -3.21158697182 & -3.194 \\
-0.804 & -3.46384093607 & -3.448 \\
-0.820 & -3.88981806008 & -3.864 \\
-0.836 & -4.21442979183 & -4.196 \\ 
-0.852 & -4.48377281174 & -4.469 \\
-0.868 & -4.71685784858 & -4.701 \\
-0.899 & -5.09933733420 & -5.076 \\
-0.933 & -5.44804603324 & -5.434 \\
-0.965 & -5.72971959109 & -5.712 \\
-1.019 & -6.13316218831 & -6.123 \\
-1.065 & -6.42475682784 & -6.415 \\
-1.104 & -6.64354741727 & -6.634 \\
-1.166 & -6.94919952087 & -6.943 \\
-1.214 & -7.15708283017 & -7.151 \\
-1.296 & -7.46636472602 & -7.461 \\
-1.347 & -7.63469982709 & -7.632 \\
-1.408 & -7.81593232651 & -7.814 \\
-1.443 & -7.91119231675 & -7.910 \\
-1.482 & -8.01059689781 & -8.009 \\
-1.502 & -8.05900000000 & -8.059 \\
\hline
\end{tabular}
\end{center}
\caption[Comparison of Efimov function $\Delta(\xi)$ to previously computed values.]{\label{tab:DeltaCompare}Comparison of Efimov function $\Delta(\xi)$ to previously computed values.  The values $\Delta(\xi)_{\mathrm{BHK}}$ are those computed by Braaten, Hammer, and Kusunoki \cite{Braaten:uefes}.  The additive constant $C$ was chosen so that the two functions agree at $\xi = -1.502$.}
\end{table}

\chapter{Conclusion}
\label{ch:conclusion}

We have studied the three-body quantum mechanical bound state in the presence of short-range two- and three-body interactions.  These short-range interactions allow us to study low-energy systems and compute results that are insensitive to the details of the inter-particle potentials.  The equations describing the two- and three-body bound-state equations were derived using a regulated form of the delta function that removes divergences and lends itself to inclusion of higher order operators.

To study the cutoff dependence of these equations perturbatively, we developed a method for uniformly expanding any function of three or four variables that possesses two widely-separated energy scales.  This method was then applied to expand the three-body bound-state equation.  Two sets of integral equations were derived: one for the leading order behavior, and one for the first order corrections.  The solutions to these equations provide results that are accurate to $\mathcal{O}(1)$ and $\mathcal{O}(\sqrt{B_3}/\Lambda)$ respectively.

We were able to derive several analytic results from the leading order equations, including a relation that allowed us to compare the parameters in our bound-state equation to those for Efimov's universal function.  We proved the periodic behavior of the coupling $\delta_0$ and used this to argue that $G_3$ is periodic.  The cutoff independence of all bound-state energies to leading order was also shown.

Several techniques were shown that helped to improve the efficiency of our numerical computations.  New limits on the integral equations were found, part of which involved substituting the analytic form for the asymptotic behavior in the low- and high-momentum regions.  A method for achieving exponential convergence in the discretized integrals was given.

The leading order equations were then solved numerically to high accuracy.  We saw the limit-cycle behavior of $G_3$, and displayed the divergent bound-state energies that make this coupling necessary.  Solutions for the function were plotted to verify some of the assumptions we made in deriving the integral equations.  We showed that our equation satisfied the conditions in Efimov's work, and proceeded to compute his universal function for three-body bound states to about 12 digits of accuracy.

There are a few areas available for possible future work.  The integral equations for the leading order corrections have been derived, but they need to be solved numerically in order to study the leading cutoff dependence.  The expansion method could also be taken further, and equations for $\mathcal{O}(B_3/\Lambda^2)$ terms could be derived.  It may also be worthwhile seeing if the expansion method and numerical techniques used here can be of use in other non-related problems.

%
% If you have appendices in your dissertation, you will need the
% following, else keep it commented. The following appendices are in
% files called ``app1.tex'', and ``app2.tex'', and they
% look just like any chapter.
%

\appendix

\chapter{Analytic Form of Mid-Momentum Function}
\label{app:fd1form}

This appendix derives the analytic form of $f_{d1}$,\footnote{This derivation is an expanded version of one provided by Wilson \cite{Wilson:notes}.} which satisfies the integral equation
\begin{eqnarray}
\hspace{-0.4in} f_{d1}(p) & = & \frac{p}{4 \pi^2 D_{d0}(p)} \int_0^{\infty} \frac{dq}{q} \, \ln\left(\frac{p^2 + q^2 + pq}{p^2 + q^2 - pq}\right) f_{d1}(q) \nonumber
\\
&& - \frac{p}{4 \pi^2 D_{d0}(p)} \left(\frac{D_{d1}(\eta_2,p,\Lambda)}{D_{d0}(p)} \right) \int_0^{\infty} \frac{dq}{q} \, \ln\left(\frac{p^2 + q^2 + pq}{p^2 + q^2 - pq}\right) f_{d0}(q) \nonumber
\\
&& - \frac{p}{4 \pi^2 D_{d0}(p)} \left(\frac{D_{d1}(\eta_2,p,\Lambda)}{D_{d0}(p)} \right) \int_0^{\infty} \frac{dq}{q} \, \ln\left(\frac{p^2 + q^2 + pq}{p^2 + q^2 - pq}\right) f_{d1}(q) ,
\end{eqnarray}
\noindent with the $D_d$ functions reproduced for convenience:
\begin{equation}
D_{d0}(p) = \frac{\sqrt{3} \, p}{16 \pi} ,
\end{equation}
\begin{equation}
\frac{D_{d1}(\eta_2,p,\Lambda)}{D_{d0}(p)} = - \frac{\sqrt{2} \, \eta_2}{\sqrt{3} \, p} +  \frac{\sqrt{3} \left( h_2^2 + 8 h_2 - 16 \right) p}{8 \sqrt{2\pi} \, \Lambda} .
\end{equation}
\noindent The key to solving this equation is understanding the solutions to
\begin{equation}
\nu(n,s,p) = \int_0^{\infty} dq \: K(p,q) \nu(n,s,q) + p^{s} \, \left[ \ln(p) \right]^{n} \label{eqn:nudef},
\end{equation}
\noindent where
\begin{equation}
K(p,q) = \frac{4}{\sqrt{3} \, \pi \,q} \ln\left(\frac{p^2 + q^2 + pq}{p^2 + q^2 - pq}\right) .
\end{equation}
\noindent The solution for $n = 0$ is
\begin{eqnarray}
\nu(0,s,p) & = & \frac{p^s}{1 - \beta_0(s)} ,
\\
\beta_0(s) & = & \frac{8 \sin\left(\pi \, s/6\right)}{\sqrt{3} \, s \, \cos\left(\pi \, s/2\right) } ,
\end{eqnarray}
\noindent except when $s$ is an odd integer.  For the special case of $s = \pm i s_0$, the function $\beta_0(s)$ is one, and the solutions are then proportional to $p^s \, \ln(p)$ \cite{Wilson:notes}.

Solutions for non-zero values of $n$ can be obtained by differentiation.  For example, taking the derivative $d/ds$ of Eq.~(\ref{eqn:nudef}) for $n = 0$ yields
\begin{equation}
\frac{d \, \nu(0,s,p)}{ds} = \int_0^{\infty} dq \: \left[ K(p,q) \, \frac{d \, \nu(0,s,q)}{ds} \right] + p^s \, \left[ \ln(p) \right] ,
\end{equation}
\noindent which is the same equation satisfied by $\nu(1,s,p)$.  Thus,
\begin{equation}
\nu(1,s,p) = \frac{d \, \nu(0,s,p)}{ds} = \frac{p^s \, \ln(p)}{1 - \beta_0(s)} + \frac{p^s}{\left(1 - \beta_0(s) \right)^2}\frac{d \, \beta_0(s)}{ds} .
\end{equation}
\noindent The same procedure can be followed to obtain solutions for higher values of $n$.

To make the connection to Eq.~(\ref{eqn:nudef}) clearer, the equation for $f_{d1}$ is rewritten as
\begin{eqnarray}
f_{d1}(p) & = & \int_0^{\infty} dq \: K(p,q) \, f_{d1}(q) \nonumber
\\
&& + \left( \frac{\sqrt{2} \, \eta_2}{\sqrt{3} \, p} -  \frac{\sqrt{3} \left( h_2^2 + 8 h_2 - 16 \right) p}{8 \sqrt{2\pi} \, \Lambda} \right) \int_0^{\infty} dq \: K(p,q) \, f_{d0}(q) \nonumber
\\
&& + \left( \frac{\sqrt{2} \, \eta_2}{\sqrt{3} \, p} -  \frac{\sqrt{3} \left( h_2^2 + 8 h_2 - 16 \right) p}{8 \sqrt{2\pi} \, \Lambda} \right) \int_0^{\infty} dq \: K(p,q) \, f_{d1}(q) \label{eqn:newfd1int}.
\end{eqnarray}
\noindent We will also make use of the fact that $f_{d0}$ satisfies the equation
\begin{equation}
f_{d0}(p) = \int_0^{\infty} dq \: K(p,q) \, f_{d0}(q) \label{eqn:fd0int},
\end{equation}
\noindent which has the solution
\begin{equation}
f_{d0}(p) = A \, \cos\left( s_0 \, \ln\left( p/\Lambda \right) + \theta \right) .
\end{equation}
\noindent A solution for $f_{d1}$ shall be constructed in a series of steps, ensuring self-consistency at each step. 

Consider Eq.~(\ref{eqn:newfd1int}) with only the first term on the right-hand side included.  This equation is the same as the one for $f_{d0}$.  Thus, $f_{d1}$ contains a term proportional to $f_{d0}$.  However, it must be of order $\eta/\Lambda$, so we shall write this term as
\begin{equation}
A_{d10} \, \frac{\eta_2}{\Lambda} \, \cos\left( s_0 \, \ln\left(p/\Lambda \right) + \theta_{d10} \right) ,
\end{equation}
\noindent where $A_{d10}$ is an arbitrary coefficient of $\mathcal{O}(1)$, and $\theta_{d10}$ is an unknown phase.

Using Eq.~(\ref{eqn:fd0int}), the second integral in (\ref{eqn:newfd1int}) can be replaced by
\begin{equation}
\left( \frac{\sqrt{2} \, \eta_2}{\sqrt{3} \, p} -  \frac{\sqrt{3} \left( h_2^2 + 8 h_2 - 16 \right) p}{8 \sqrt{2\pi} \, \Lambda} \right) \, f_{d0}(p) \label{eqn:fd0pieces}.
\end{equation}
\noindent If we added only the first of the two pieces, the integral equation for $f_{d1}$ would become
\begin{equation}
f_{d1}(p) = \int_0^{\infty} dq \: K(p,q) \, f_{d1}(q) + \frac{\sqrt{2} \, \eta_2}{\sqrt{3} \, p} \, f_{d0}(p) .
\end{equation}
\noindent We extend this equation into the complex plane by writing it as
\begin{equation}
f_{d1}(p) = \int_0^{\infty} dq \: K(p,q) \, f_{d1}(q) + \frac{\sqrt{2} \, \eta_2}{\sqrt{3} \, p} \, A \, \me^{ \mi \left( s_0 \, \ln\left( p/\Lambda \right) + \theta \right)} .
\end{equation}
\noindent The inhomogeneous term is proportional to $p^{-1 + \mi s_0}$, so we may use Eq.~(\ref{eqn:nudef}) to obtain the solution
\begin{equation}
f_{d1}(p) = \frac{1}{1 - \beta_0(-1+is_0)} \left(\frac{\sqrt{2} \, \eta_2}{\sqrt{3} \, p}\right) \, A \, \me^{\mi \left( s_0 \, \ln\left( p/\Lambda \right) + \theta \right)} .
\end{equation}
\noindent The solution we desire is actually the real part of the previous solution.  We shall write it as
\begin{equation}
A_{l1} \, \frac{\eta_2}{p} \, \cos\left( s_0 \, \ln\left(p/\Lambda \right) + \theta_{l1} \right) .
\end{equation}
\noindent The constants $A_{l1}$ and $\theta_{l1}$ are related to $A$ and $\theta$ by the relation
\begin{equation}
\left( 1 - \beta_0(-1 + \mi s_0) \right) \, A_{l1} \, \me^{\mi \theta_{l1}} = \sqrt{\frac{2}{3}} \, A \, \me^{\mi \theta} .
\end{equation}

If we had added the second piece of Eq.~(\ref{eqn:fd0pieces}) instead of the first, our equation would have been
\begin{equation}
f_{d1}(p) = \int_0^{\infty} dq \: K(p,q) \, f_{d1}(q) -  \frac{\sqrt{3} \left( h_2^2 + 8 h_2 - 16 \right) p}{8 \sqrt{2 \pi} \, \Lambda}  \, f_{d0}(p) .
\end{equation}
\noindent The steps for solving the equation for this case are nearly identical to the previous one.  We extend the equation into the complex plane, noting that the inhomogeneous term is proportional to $p^{1 + \mi s_0}$.  This solution is
\begin{equation}
f_{d1}(p) = -  \frac{1}{1 - \beta_0(1+is_0)} \left( \frac{\sqrt{3} \left( h_2^2 + 8 h_2 - 16 \right) p}{8 \sqrt{2\pi} \, \Lambda} \right) \, A \, \me^{\mi \left( s_0 \, \ln\left( p/\Lambda \right) + \theta \right)} ,
\end{equation}
\noindent and we choose to write the real part as
\begin{equation}
A_{h1} \, \frac{p}{\Lambda} \, \cos\left( s_0 \, \ln\left(p/\Lambda \right) + \theta_{h1} \right) .
\end{equation}
\noindent The constants $A_{h1}$ and $\theta_{h1}$ are determined by
\begin{equation}
\left( 1 - \beta_0(1 + \mi s_0) \right) \, A_{h1} \, \me^{\mi \theta_{h1}} = -  \frac{\sqrt{3} \left( h_2^2 + 8 h_2 - 16 \right)}{8 \sqrt{2 \pi}} \, A \, \me^{\mi \theta} .
\end{equation}

At this point, we have three terms that contribute to $f_{d1}$.  Even though they were derived individually, they may simply be added together.  The reason is that each term is ``self-contained'' in the sense that integrating it results only in a term proportional to itself.  There are no cross-terms generated in the equation.  The solution, as it now stands, is
\begin{eqnarray}
f_{d1}(\eta_2,p,\Lambda) & = & A_{l1} \frac{\eta_2}{p} \cos\left(s_0 \ln\left(p/\Lambda\right) + \theta_{l1} \right) \nonumber
\\
&& +~A_{h1} \frac{p}{\Lambda} \cos\left(s_0 \ln\left(p/\Lambda\right) + \theta_{h1} \right) \nonumber
\\
&& +~A_{d10} \frac{\eta_2}{\Lambda} \cos\left(s_0 \ln\left(p/\Lambda\right) + \theta_{d10} \right) \label{eqn:prefd1form}.
\end{eqnarray}

The only thing remaining that affects the solution is the third integral from Eq.~(\ref{eqn:newfd1int}).  Using the version of $f_{d1}$ from (\ref{eqn:prefd1form}), its contribution is
\begin{eqnarray}
\lefteqn{ \left( \frac{\sqrt{2} \, \eta_2}{\sqrt{3} \, p} -  \frac{\sqrt{3} \left( h_2^2 + 8 h_2 - 16 \right) p}{8 \sqrt{2 \pi} \, \Lambda} \right) \int_0^{\infty} dq \: K(p,q) \, f_{d1}(q) } \nonumber
\\
& = & \left( \frac{\sqrt{2} \, \eta_2}{\sqrt{3} \, p} - \frac{\sqrt{3} \left( h_2^2 + 8 h_2 - 16 \right) p}{8 \sqrt{2 \pi} \, \Lambda} \right) \left( \left[ A_{d10} \frac{\eta_2}{\Lambda} \cos\left(s_0 \ln\left(p/\Lambda\right) + \theta_{d10} \right) \right] \right. \nonumber
\\
&& + \left[ A_{l1} \frac{\eta_2}{p} \cos\left(s_0 \ln\left(p/\Lambda\right) + \theta_{l1} \right) - \frac{\sqrt{2} \, \eta_2}{\sqrt{3} \, p} \, f_{d0}(p) \right] \nonumber
\\
&& + \left .\left[ A_{h1} \frac{p}{\Lambda} \cos\left(s_0 \ln\left(p/\Lambda\right) + \theta_{h1} \right) + \frac{\sqrt{3} \left( h_2^2 + 8 h_2 - 16 \right) p}{8 \sqrt{2 \pi} \, \Lambda} \, f_{d0}(p) \right] \right) \nonumber
\\
& = & - \frac{\sqrt{3} \left( h_2^2 + 8 h_2 - 16 \right) \eta_2}{8 \sqrt{2 \pi} \, \Lambda} \, A_{l1} \, \cos\left(s_0 \ln\left(p/\Lambda\right) + \theta_{l1} \right) \nonumber
\\
&& + \frac{\sqrt{2} \, \eta_2}{\sqrt{3} \, \Lambda} \, A_{h1} \, \cos\left(s_0 \ln\left(p/\Lambda\right) + \theta_{h1} \right) \nonumber
\\
&& + \frac{\left( h_2^2 + 8 h_2 - 16 \right) \eta_2}{4 \sqrt{\pi} \, \Lambda} \, A \, \cos\left(s_0 \ln\left(p/\Lambda\right) + \theta \right) \label{eqn:int3}.
\end{eqnarray}
\noindent Note that we have dropped all terms that are $\mathcal{O}(\eta^2/\Lambda^2)$ or higher.  Although this contribution can easily be absorbed into the $A_{d10}$ term, we must also consider the terms it may generate by acting as an inhomogeneous term in our integral equation.

To keep the equations simple, we will combine these three terms into one.  Since we will be extending them into the complex plane, let us define
\begin{eqnarray}
\frac{\eta_2}{\Lambda} \, A_{d1} \, \me^{\mi \left( s_0 \, \ln\left( p/\Lambda \right) + \theta_{d1} \right)} & \equiv & - \frac{\sqrt{3} \left( h_2^2 + 8 h_2 - 16 \right) \eta_2}{8 \sqrt{2\pi} \, \Lambda} \, A_{l1} \, \me^{\mi\left(s_0 \ln\left(p/\Lambda\right) + \theta_{l1} \right)} \nonumber
\\
&& + \frac{\sqrt{2} \, \eta_2}{\sqrt{3} \, \Lambda} \, A_{h1} \, \me^{\mi\left(s_0 \ln\left(p/\Lambda\right) + \theta_{h1} \right)} \nonumber
\\
&& + \frac{\left( h_2^2 + 8 h_2 - 16 \right) \eta_2}{4 \sqrt{\pi} \, \Lambda} \, A \, \me^{\mi\left(s_0 \ln\left(p/\Lambda\right) + \theta \right)} .
\end{eqnarray}
\noindent We must now solve the equation
\begin{equation}
f_{d1}(p) = \int_0^{\infty} dq \: K(p,q) \, f_{d1}(q) + \frac{\eta_2}{\Lambda} \, A_{d1} \, \me^{\mi \left( s_0 \, \ln\left( p/\Lambda \right) + \theta_{d1} \right)} .
\end{equation}
\noindent The solution takes the form
\begin{equation}
f_{d1}(p) = \left[ - \frac{d}{ds} \beta_0(s) \right]_{s = \mi s_0} \ln\left(\frac{p}{\Lambda}\right) \frac{\eta_2}{\Lambda} \, A_{d1} \, \me^{\mi \left( s_0 \, \ln\left( p/\Lambda \right) + \theta_{d1} \right)} .
\end{equation}
\noindent The real part will be written as
\begin{equation}
A_{d11} \, \frac{\eta_2}{\Lambda} \, \ln\left(p/\Lambda\right) \, \cos \left( s_0 \, \ln\left( p/\Lambda \right) + \theta_{d1} \right) \label{eqn:Ad11},
\end{equation}
\noindent subject to the relation
\begin{equation}
\left[ - \frac{d}{ds} \beta_0(s) \right]_{s = \mi s_0} A_{d11} \me^{\mi \theta_{d11}} = A_{d1} \me^{\mi \theta_{d1}} .
\end{equation}

Equation (\ref{eqn:Ad11}) adds a new term to the solution.  The only remaining consideration is whether or not it affects the results of Eq.~(\ref{eqn:int3}) since the value of $f_{d1}$ used in that equation did not contain this new term.  Fortunately, the new term would only contribute at $\mathcal{O}(\eta^2/\Lambda^2)$ so the result remains unaffected.

We now have the self-consistent, analytic solution shown previously in Chapter \ref{ch:expand3body}:
\begin{eqnarray}
f_{d1}(\eta_2,p,\Lambda) & = & A_{l1} \frac{\eta_2}{p} \cos\left(s_0 \ln\left(p/\Lambda\right) + \theta_{l1} \right) \nonumber
\\
&& +~A_{h1} \frac{p}{\Lambda} \cos\left(s_0 \ln\left(p/\Lambda\right) + \theta_{h1} \right) \nonumber
\\
&& +~A_{d10} \frac{\eta_2}{\Lambda} \cos\left(s_0 \ln\left(p/\Lambda\right) + \theta_{d10} \right) \nonumber
\\
&& +~A_{d11} \frac{\eta_2}{\Lambda} \ln\left(p/\Lambda\right) \cos\left(s_0 \ln\left(p/\Lambda\right) + \theta_{d11} \right) .
\end{eqnarray}

\chapter{Proof of Exponential Convergence}
\label{app:kenproof}

The proof given here is based on notes provided by Kenneth Wilson.  It is meant to give the reader a justification as to why the uniform spacing of integration points on a logarithmic scale gives exponential convergence, but it will not be entirely rigorous.

We start by making the following assumptions:
\begin{itemize}
\item $f(z)$ is analytic in the region $-\infty < \mathrm{Re}(z) < \infty$ and $-r < \mathrm{Im}(z) < r$ for some value of $r$.
\item $f(x)$ is real for any real value $x$.
\item The integral $\int_{-\infty}^{\infty} f(x) \, dx$ is finite.
\end{itemize}
\noindent The first two assumptions imply (via the reflection principle) that $f(z^*) = f(z)^*$ in this region \cite{Brown:complex}.  With this in mind, consider
\begin{equation}
\int_{-\infty}^{\infty} dx \left[ \frac{f(x+\mi r)}{1 - \exp(-2\pi\mi(x+\mi r)/a)} + \frac{f(x-\mi r)}{1 - \exp(2\pi\mi(x-\mi r)/a)} \right].
\end{equation}
\noindent This must be real since the second term is the complex conjugate of the first.  Therefore,
\begin{eqnarray}
&& \hspace{-0.3in} \int_{-\infty}^{\infty} dx \left[ \frac{f(x+\mi r)}{1 - \exp(-2\pi\mi(x+\mi r)/a)} + \frac{f(x-\mi r)}{1 - \exp(2\pi\mi(x-\mi r)/a)} \right] \nonumber
\\
& = & \left| \int_{-\infty}^{\infty} dx \left[ \frac{f(x+\mi r)}{1 - \exp(-2\pi\mi(x+\mi r)/a)} + \frac{f(x-\mi r)}{1 - \exp(2\pi\mi(x-\mi r)/a)} \right] \right| \nonumber
\\
& \le & \int_{-\infty}^{\infty} dx \left| \frac{f(x+\mi r)}{1 - \exp(-2\pi\mi(x+\mi r)/a)} + \frac{f(x-\mi r)}{1 - \exp(2\pi\mi(x-\mi r)/a)} \right| \nonumber
\\
& \le & \me^{-2 \pi r/a} \int_{-\infty}^{\infty} dx \left| \frac{f(x+\mi r)}{\exp(-2 \pi r/a) - \exp(-2\pi \mi x)} \right| + \left| \frac{f(x-\mi r)}{\exp(-2 \pi r/a) - \exp(-2\pi \mi x)} \right| \nonumber
\\
& \le & \frac{\me^{-2 \pi r/a}}{1 - \me^{-2 \pi r/a}} \int_{-\infty}^{\infty} dx \left| f(x+\mi r) \right| + \left| f(x-\mi r) \right| \label{eqn:doubleterm}.
\end{eqnarray}
\noindent However, if $f(z)$ is such that it tends to zero as $|z| \rightarrow \infty$, then from the analyticity of $f(z)$ we know that
\begin{equation}
\int_{-\infty}^{\infty} dx \, f(x) = \int_{-\infty}^{\infty} dx \, f(x + \mi r) = \int_{-\infty}^{\infty} dx \, f(x - \mi r) .
\end{equation}
\noindent Now
\begin{eqnarray}
\int_{-\infty}^{\infty} dx \left| f(x+\mi r) \right| + \left| f(x-\mi r) \right| & \le & \int_{-\infty}^{\infty} dx \, \mathrm{Re}[f(x + \mi r)] + \int_{-\infty}^{\infty} dx \, \mathrm{Im}[f(x + \mi r)] \nonumber
\\
&& +\int_{-\infty}^{\infty} dx \, \mathrm{Re}[f(x - \mi r)] + \int_{-\infty}^{\infty} dx \, \mathrm{Im}[f(x - \mi r)] \nonumber
\\
& = & \int_{-\infty}^{\infty} dx \, \mathrm{Re}[f(x + \mi r)] + \int_{-\infty}^{\infty} dx \, \mathrm{Re}[f(x - \mi r)] \nonumber
\\
& = & 2 \int_{-\infty}^{\infty} dx \, f(x) .
\end{eqnarray}
\noindent Combining this with Eq.~(\ref{eqn:doubleterm}) yields
\begin{equation}
\int_{-\infty}^{\infty} dx \left[ \frac{f(x+\mi r)}{1 - \me^{-2\pi\mi(x+\mi r)/a}} + \frac{f(x-\mi r)}{1 - \me^{2\pi\mi(x-\mi r)/a}} \right] \le \frac{2 \me^{-2 \pi r/a}}{1 - \me^{-2 \pi r/a}} \int_{-\infty}^{\infty} dx \, f(x) \label{eqn:errorterm}.
\end{equation}
\noindent If $\exp(-2 \pi r/a)$ is small, then the above term is small provided that the integral is not unusually large.  We use this to write
\begin{equation}
\int_{-\infty}^{\infty} dx \, f(x) \simeq \int_{-\infty}^{\infty} dx \, f(x) - \int_{-\infty}^{\infty} dx \left[ \frac{f(x+\mi r)}{1 - \me^{-2\pi\mi(x+\mi r)/a}} + \frac{f(x-\mi r)}{1 - \me^{2\pi\mi(x-\mi r)/a}} \right] \label{eqn:approx1}.
\end{equation}
\noindent The second and third terms have poles on the real axis at $na$ where $n = 0, \pm 1, \pm 2,$ etc.  This means we can move their contour integrations from $x + \mi r$ and $x - \mi r$ to $x + \mi \epsilon$ and $x - \mi \epsilon$ respectively (where $\epsilon$ is some small quantity) without crossing any poles.  Therefore, Eq.~(\ref{eqn:approx1}) becomes
\begin{equation}
\int_{-\infty}^{\infty} dx \, f(x) \simeq \int_{-\infty}^{\infty} dx \, f(x) - \int_{-\infty}^{\infty} dx \left[ \frac{f(x+\mi \epsilon)}{1 - \me^{-2\pi\mi(x+\mi \epsilon)/a}} + \frac{f(x-\mi \epsilon)}{1 - \me^{2\pi\mi(x-\mi \epsilon)/a}} \right] \label{eqn:approx2}.
\end{equation}
\noindent Now consider a very small loop around the entire real axis.  The Residue Theorem implies that
\begin{equation}
\int_{-\infty}^{\infty} dx \, \frac{f(x - \mi \epsilon)}{1 - \me^{2\pi\mi(x - \mi \epsilon)/a}} - \int_{-\infty}^{\infty} dx \, \frac{f(x + \mi \epsilon)}{1 - \me^{2\pi\mi(x + \mi \epsilon)/a}} = 2 \pi \mi \sum_{n = -\infty}^{\infty} - \frac{a \, f(na)}{2 \pi \mi} \label{eqn:residues},
\end{equation}
\noindent where $-a\,f(an)/2\pi\mi$ is the residue from the pole at $x = na$.  Combining Eq.~(\ref{eqn:approx2}) and Eq.~(\ref{eqn:residues}) gives us
\begin{equation}
\int_{-\infty}^{\infty} dx \, f(x) \simeq \int_{-\infty}^{\infty} dx \, f(x + \mi\epsilon) \left[ 1 -  \frac{1}{1 - \me^{-2\pi\mi(x+\mi \epsilon)/a}} - \frac{1}{1 - \me^{2\pi\mi(x+\mi \epsilon)/a}} \right] + \sum_{n = -\infty}^{\infty} a f(an) .
\end{equation}
\noindent Here we have replaced $\int dx \, f(x)$ with $\int dx \, f(x + \mi\epsilon)$ which makes a small error of $\mathcal{O}(\epsilon)$.  The term in brackets is equal to zero, leaving us with
\begin{equation}
\int_{-\infty}^{\infty} dx \, f(x) \simeq \sum_{n = -\infty}^{\infty} a f(an) .
\end{equation}
\noindent The error in this approximation is equal to the magnitude of Eq.~(\ref{eqn:errorterm}), which goes like $\exp(-2 \pi r/a)$.  The parameter $a$ represents the spacing used in discretizing the integral.  As it approaches zero, the error decays exponentially.
\chapter{Error Analysis}
\label{app:erroranalysis}

This appendix provides a simple error analysis for the integration technique we use.  It is meant to justify the claim of exponential convergence.  If the answers truly converge exponentially, then we expect the approximation using $n$ integration points to behave like
\begin{equation}
A_n = A + \alpha \me^{-\beta n} ,
\end{equation}
\noindent where $A$ represents the exact value.  By taking some large value  $n=N$ and using $A_N$ as the value for $A$, we should find that
\begin{equation}
\epsilon_n \equiv \log_{10}\left(\frac{|A_n - A_N|}{A_N}\right) \simeq -\frac{\beta \log_{10}(\me)}{A_N} n + \frac{\log_{10}(\alpha)}{A_N} ,
\end{equation}
\noindent where $\epsilon_n$ is the relative error in $A_n$.  Plotting $\epsilon_n$ as a function of $n$ should show a straight line.  At some value of $n$, other sources of error, such as machine error, will dominate and $\epsilon_n$ will become roughly constant.  The value of $\epsilon_n$ in this flat region tells us the number of digits of accuracy in our result.

\begin{figure}
\begin{center}
\includegraphics[height=4.0in]{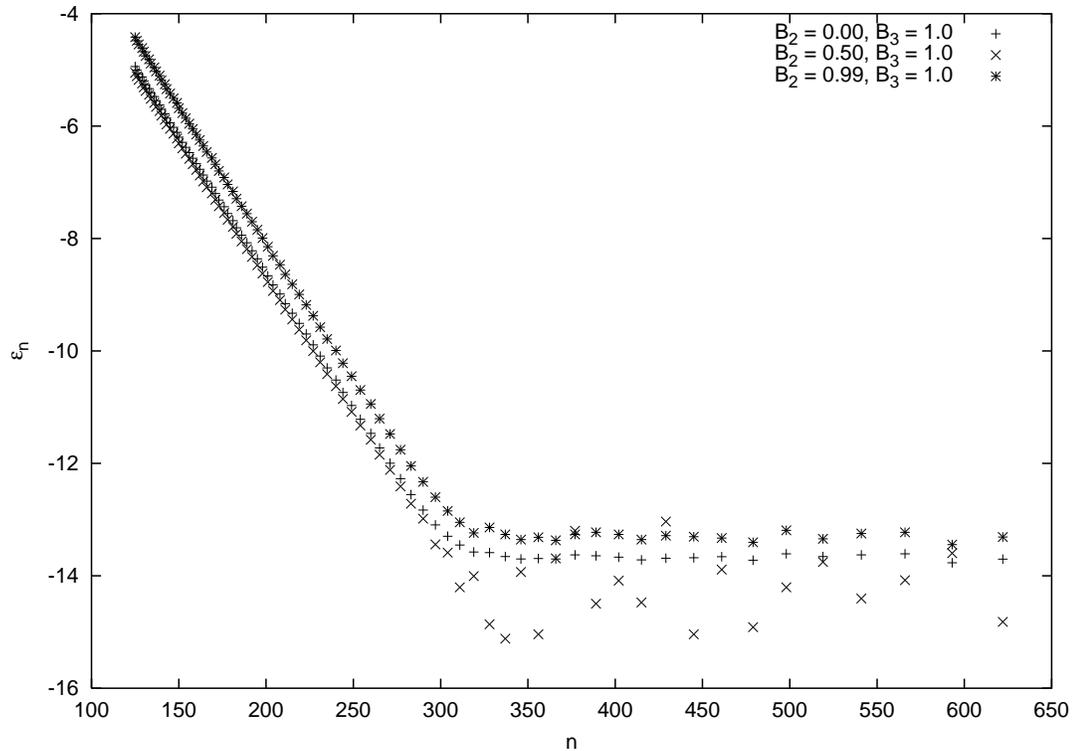}
\end{center}
\caption[Relative error $\epsilon_n$ in the phase calculation for different values of $B_2$ and $B_3$ as a function of the number of integration points $n$.]{\label{fig:PhaseError}Relative error $\epsilon_n$ in the phase calculation for different values of $B_2$ and $B_3$ as a function of the number of integration points $n$.  Three different sets of energies were used.  Notice the initial linear behavior indicating exponential convergence, and the plateau indicating the limit of accuracy (approximately 13 digits).}
\end{figure}

The integral equation for $f_{l0}$ was used to calculate the phase using different numbers of integration points.  This was done for three sets of $B_2$ and $B_3$ values, and the results for the relative errors are shown in Fig.~\ref{fig:PhaseError}.  The initial linear behavior indicates exponential convergence, which is then dominated by other errors resulting in a plateau.  The plot indicates an accuracy of about 13 digits.

%%%%%%%%%%%%%%%
%%  FIGURES  %%
%%%%%%%%%%%%%%%

\backmatter

%
% The all important bibliography file at the end of your document!! Use
% the bibstyle you (your department) like in the \bibliographystyle{}
% statement and list the name of your bibliography database file in
% the \bibliography{} statement.  In this example, ``bibfile.bib'' is
% the name of the database.
%
\bibliographystyle{physrev}
\bibliography{thesis}

\end{document}